\documentclass{aa}

\usepackage{txfonts}
\usepackage{textcomp}
\usepackage[T1]{fontenc}

\usepackage{graphicx}	
\usepackage{amsmath}	
\usepackage{amssymb}	
\usepackage[alsoload=hep]{siunitx}

\usepackage[caption=false]{subfig}
\usepackage{booktabs}

\usepackage[normalem]{ulem}

\usepackage{xcolor}

\usepackage{footmisc}

\usepackage{natbib}
\usepackage{hyperref}
\usepackage{cleveref}

\graphicspath{%
    {plots/}
    {plots/profiles/}
    {plots/Halpha_maps/}
}

\definecolor{alizarin}{rgb}{0.82, 0.1, 0.26}
\definecolor{blue (munsell)}{rgb}{0.0, 0.5, 0.69}
\definecolor{asparagus}{rgb}{0.53, 0.66, 0.42}

\newcommand{\alphaFe}[0]{[$\alpha$/Fe]}
\newcommand{\steckmap}[0]{\texttt{STECMAP}}
\newcommand{\ppxf}[0]{\texttt{pPXF}}
\newcommand{\Ha}[0]{H$\alpha$}
\newcommand{\Hb}[0]{H$\beta$}

\DeclareSIUnit\Gyr{Gyr}
\DeclareSIUnit\Myr{Myr}
\DeclareSIUnit\Mpc{Mpc}
\DeclareSIUnit\kpc{kpc}
\DeclareSIUnit\pc{pc}
\DeclareSIUnit\rkin{\text{\ensuremath{\textup{R}_{\textup{kin}}}}}
\DeclareSIUnit\msol{\text{\ensuremath{\textup{M}_{\odot}}}}
\DeclareSIUnit\riso{\text{\ensuremath{\textup{R}_{25.5}}}}
\DeclareSIUnit\dex{dex}
\DeclareSIUnit\mag{mag}
\DeclareSIUnit\AB{AB}
\DeclareSIUnit\erg{erg}
\DeclareSIUnit\arcsec{arcsec}
\DeclareSIUnit\arcmin{arcmin}

\begin{document}

    \title{The inside-out formation of nuclear discs and the absence of old central spheroids in barred galaxies of the TIMER survey}
    \titlerunning{Stellar populations of nuclear discs}

    \author{%
        Adrian Bittner
        \inst{1,2}\and
        Patricia Sánchez-Blázquez
        \inst{3,4}\and
        Dimitri A. Gadotti
        \inst{1}\and
        Justus Neumann
        \inst{5}\and
        Francesca Fragkoudi
        \inst{6}\and
        Paula Coelho
        \inst{7}\and
        Adriana de Lorenzo-Cáceres
        \inst{8,9}\and
        Jesús Falcón-Barroso
        \inst{8,9}\and
        Taehyun Kim
        \inst{10}\and
        Ryan Leaman
        \inst{11}\and
        Ignacio Martín-Navarro
        \inst{8,9}\and
        Jairo Méndez-Abreu
        \inst{8,9}\and
        Isabel Pérez
        \inst{12,13}\and
        Miguel Querejeta
        \inst{14}\and
        Marja K. Seidel
        \inst{15}\and
        Glenn van de Ven 
        \inst{16}
    }

    \authorrunning{A. Bittner et al.}

    \institute{%
        European Southern Observatory, Karl-Schwarzschild-Str. 2, D-85748 Garching bei München, Germany\\ \email{adrian.bittner@eso.org}\and
        Ludwig-Maximilians-Universität, Professor-Huber-Platz 2, 80539 München, Germany\and
        Departamento de Física de la Tierra y Astrofísica, Universidad Complutense de Madrid, E-28040 Madrid, Spain\and
        Instituto de Física de Partículas y del Cosmos, Universidad Complutense de Madrid, E-28040 Madrid, Spain\and
        Institute of Cosmology and Gravitation, University of Portsmouth, Burnaby Road, Portsmouth PO1 3FX, United Kingdom\and
        Max-Planck-Institut für Astrophysik, Karl-Schwarzschild-Str. 1, D-85748 Garching bei München, Germany\and
        Instituto de Astronomia, Geofísica e Ciências Atmosféricas, Universidade de São Paulo, \\ R. do Matão 1226, 05508-090 São Paulo, Brazil\and
        Instituto de Astrofísica de Canarias, Calle Vía Láctea s/n, E-38205 La Laguna, Tenerife, Spain\and
        Departamento de Astrofísica, Universidad de La Laguna, E-38200 La Laguna, Tenerife, Spain\and
        Department of Astronomy and Atmospheric Sciences, Kyungpook National University, Daegu 702-701, Korea\and
        Max-Planck Institut für Astronomie, Königstuhl 17, D-69117 Heidelberg, Germany\and
        Departamento de Física Teórica y del Cosmos, Universidad de Granada, Facultad de Ciencias, E-18071, Granada, Spain\and
        Instituto Universitario Carlos I de Física Teórica y Computacional, Universidad de Granada, E-18071 Granada, Spain\and
        Observatorio Astronómico Nacional, C/Alfonso XII 3, Madrid E-28014, Spain\and
        Caltech-IPAC, MC 314-6, 1200 E California Blvd, Pasadena, CA, 91125, USA\and
        Department of Astrophysics, University of Vienna, Türkenschanzstraße 17, 1180 Wien, Austria
    }

    \date{Received; accepted}

    \abstract{%
    The centres of disc galaxies host a variety of structures built via both internal and external processes.  In this
    study, we constrain the formation and evolution of these central structures, in particular nuclear rings and nuclear
    discs, by deriving maps of mean stellar ages, metallicities and {\alphaFe} abundances. We use observations obtained
    with the MUSE integral-field spectrograph for the TIMER sample of 21 massive barred galaxies.  Our results indicate
    that nuclear discs and nuclear rings are part of the same physical component, with nuclear rings constituting the
    outer edge of nuclear discs. All nuclear discs in the sample are clearly distinguished based on their stellar
    population properties. As expected in the picture of bar-driven secular evolution, nuclear discs are younger, more
    metal-rich, and show lower {\alphaFe} enhancements, as compared to their immediate surroundings. Moreover, nuclear
    discs exhibit well-defined radial gradients, with ages and metallicities decreasing, and {\alphaFe} abundances
    increasing with radius out to the nuclear ring. Often, these gradients show no breaks from the edge of the nuclear
    disc until the centre, suggesting that these structures extend to the very centres of the galaxies. We argue that
    continuous (stellar) nuclear discs may form from a series of bar-built (initially gas-rich) nuclear rings that grow
    in radius, as the bar evolves. In this picture, nuclear rings are simply the (often) star-forming outer edge of
    nuclear discs.  Finally, by combining our results with those from a accompanying kinematic study, we do not find
    evidence for the presence of large, dispersion-dominated components in the centres of these galaxies. This could be
    a result of quiet merger histories, despite the large galaxy masses, or perhaps high angular momentum and strong
    feedback processes preventing the formation of these kinematically hot components.  
    }


    \maketitle

\section{Introduction}%
\label{sec:introduction}
Bars are prominent stellar structures frequently found in disc galaxies. Approximately 2/3 of all local disc galaxies
exhibit a bar and this fraction is monotonically decreasing with increasing redshift, down to bar fractions of
\SIrange{10}{15}{\percent} at $z>1$ \citep[see e.g.][]{eskridge2000, menendezDelmestre2007, sheth2008, aguerri2009,
mendezAbreu2010, kraljic2012, sheth2012, melvin2014}.  Nonetheless, in massive galaxies strong bars have been identified
up to higher redshifts \citep[$z\sim2$; ][]{simmons2014} and their existence at these redshifts has been inferred from
studies of their stellar age distribution \citep[see e.g.][]{gadotti2015, perez2017}.  In addition, it appears difficult
to destroy bars, at least once they grew sufficiently strong \citep{athanassoula2005b}.  Altogether, these studies
suggest that bars influence the evolution of their host galaxies over timescales as long as about \SI{10}{\Gyr}.

Bars evolve and influence galaxies in a variety of ways, which becomes clear when comparing barred and unbarred
galaxies. For instance, the inner regions of barred galaxies show systematically higher metallicities and star formation
rates \citep{ellison2011} as well as increased nuclear activity and accretion onto central black holes
\citep{alonso2018}. Moreover, bars typically exhibit flat age and metallicity gradients along their major axis, a clear
indication of their influence on the inner discs of galaxies \citep[see e.g.][]{sanchezBlazquez2011,
fraser-McKelvie2019, neumann2020}. This influence is also evident in the fact that barred galaxies often show a light
deficit in the disc surrounding the bar, an effect that is absent in unbarred galaxies and thought to be caused by the
capture of disc stars \citep[see e.g.][]{james2009, james2016, kim2016, donohoe-Keyes2019}. 

One particularly interesting effect of bars is the creation of substructures in the nuclear region of disc galaxies such
as nuclear rings and nuclear discs by redistributing angular momentum \citep{combes1985}.  More specifically, the
non-axisymmetric potential of the bar exerts strong tangential forces in the main disc which cause interstellar gas to
shock and lose angular momentum. As a result, the gas streams inward along the leading edges of the bar. These
large-scale streaming motions are typically highlighted by prominent dust lanes and are clearly evident in both
numerical and observational studies \citep[see e.g.][]{athanassoula1992, knapen2007, cole2014, fragkoudi2016}.  This
inward gas flow is halted in the nuclear region of the galaxy, where the gas, due to its collisional nature, often
settles in a nuclear ring where star formation proceeds.  It has also been suggested that nuclear discs could be formed
via the same mechanism, but extend to smaller radii \citep{piner1995, sakamoto1999, sheth2005, sormani2015}.  

While there is convincing evidence that these nuclear rings and nuclear discs are built from gas that was funnelled to
the centre by the bar, it remains unclear what physical mechanism determines the size of these structures. It has been
suggested that the radius of nuclear rings is related to the Inner Lindblad Resonance (ILR) of the bar.
Observationally, nuclear rings are often found close to the ILR and therefore it is argued that nuclear rings are a
result of bar resonances\footnote{However, the ILR used in these studies only strictly holds in the mildly
non-axisymmetric regime, and therefore is ill-defined for strong bars.} \citep[see e.g.][]{combes1985, knapen2005,
comeron2010}. More precisely, \citet{athanassoula1992a, athanassoula1992} argues that the size of the nuclear ring is
limited by the radial extent of the $x_2$ orbit family.  However, \citet{kim2012} suggests that the size of the nuclear
ring is not determined by bar resonances, but instead given by the residual angular momentum of the inflowing gas.
Another scenario was given by \citet{sormani2018} who presents a mechanism for the origin of nuclear rings and finds
that the size of the ring is set by the effect of viscous shear forces. Using numerical simulations, \citet{seo2019}
show that nuclear rings grow in size as the bar grows longer and funnels in gas from larger radii in the galactic disc.
In line with this result, \citet{knapen2005}, \citet{comeron2010}, and Gadotti et al. (subm., hereafter G20) find that
the radii of nuclear rings and nuclear discs correlate with the bar length

Due to the collisional nature of the gas, the subsequent star formation generates stars in (near) circular orbits.  More
precisely, stellar nuclear discs are expected to be characterised by high rotational velocities and low velocity
dispersions \citep[see e.g.][]{cole2014}.  In an accompanying study, G20 uses the same integral-field spectroscopic
observations of the TIMER survey (Time Inference with MUSE in Extragalactic Rings) employed in this study to confirm
these expectations. In addition, they show that nuclear discs have exponential surface brightness profiles and dominate
the stellar light in the centre of the galaxy. Similarly, numerical simulations expect nuclear discs to be younger and
more metal-rich than the bar. In particular in the framework of secular evolution, nuclear discs form only after the
formation of the bar and are thus expected to have stellar populations younger than those found in the bar. As star
formation in these central regions continues one also expects their metal content to increase \citep{cole2014}. 

Numerical simulations show that the formation of nuclear discs can also be initiated by galaxy mergers \citep[see
e.g.][]{mayer2008, chapon2013}. However, the nuclear discs in these simulations are at least one order of magnitude
smaller than those commonly produced in the bar-driven formation scenario. Interestingly, \citet{comeron2010} finds that
\SI{19}{\percent} of all nuclear rings occur in unbarred galaxies. However, they show that in most of these cases there
is evidence of some non-axisymmetry in the potential of the galaxy which might cause the formation of nuclear rings
through mechanisms similar to those in the bar-driven scenario, albeit weaker. 

Secular evolution continues to take effect within these stellar nuclear discs themselves. In fact, some nuclear discs
develop bars themselves, resulting in the remarkable situation of having a small disc with a small bar embedded within a
large disc with a large bar.  These inner bars do not only resemble the shape of regular bars, but also seem to form and
evolve in the same way main bars do \citep{deLorenzoCaceres2019, deLorenzoCaceres2020}. In fact, inner bars buckle
vertically just as main bars \citep{mendezAbreu2019} and even exhibit the same $v-h_3$ correlation typically associated
with bars \citep{bittner2019}. 

A variety of nomenclatures for these central substructures of galaxies has been established. Particularly common is the
term ``bulge'' and its variations, as for instance ``pseudo-bulge'' and ``disc-like bulge''.  In this paper, we avoid
the term ``bulge'' and instead use more physical descriptions of the stellar structures. In particular, we refer to a
\textsl{kinematically hot spheroid} instead of using the term ``classical bulge''.  Bar-built central discs with typical
sizes of hundreds of \si{\pc} that are rotationally supported but kinematically distinct from the main galactic disc are
denoted \textsl{nuclear discs} (see G20). This choice is made to avoid confusion with inner and outer discs in the
context of breaks in the light profile of main discs of galaxies. Rings associated with the outermost edge of nuclear
discs are named \textsl{nuclear rings}, in order to clearly distinguish them from inner and outer rings typically found
close to and outside of the bar radius, respectively. Similarly, large scale bars found in main stellar discs of
galaxies are simply denoted \textsl{bars} while smaller bars that form and evolve within nuclear discs themselves are
called \textsl{inner bars}\footnote{In the literature inner bars are often also referred to as \textsl{nuclear bars}.}. 

While the formation of nuclear rings and nuclear discs is thought to be bar-driven, the connection between nuclear rings
and nuclear discs is still elusive. In particular, little is known about how the settling of gas near the ILR can
originate a stellar disc that seems to extend from the nuclear ring inwards. In the present study, we build upon an
accompanying study which investigates the stellar kinematics of nuclear discs (G20).  Here we characterise nuclear discs
based on their spatially resolved, mean stellar population properties. For the first time we observe nuclear discs in
sufficiently high spatial resolution (\SI{\sim100}{\pc} or less) to investigate detailed spatial changes in their
population properties, even including {\alphaFe} abundances.  In addition, we aim to establish if nuclear discs extend
all the way to the galactic centre and how their properties compare to those of nuclear rings. Thanks to the superb
quality of the data, we further explore the presence of composite structures consisting of nuclear discs and
kinematically hot spheroids in the centres of these galaxies \citep[see e.g.][]{mendezAbreu2014, erwin2015}. 

This paper is organised as follows. In Sect.~\ref{sec:observations} we introduce the TIMER survey and summarise its
observation and data reduction strategy. The measurement of mean stellar population properties is described in
Sect.~\ref{sec:analysis} and its reliability and uncertainties discussed in Sect.~\ref{sec:comparisons}. Our main
observational results are presented in Sect.~\ref{sec:results} and we discuss their physical implications in
Sect.~\ref{sec:discussion}. We close with a summary of our findings in Sect.~\ref{sec:summary}.  


\section{The TIMER survey}%
\label{sec:observations}
The TIMER project is a survey aiming to reconstruct the star formation histories of nuclear structures in order to
constrain the formation time of bars and establish when the main discs of galaxies became dynamically mature.  Building
upon the results of a pilot study of NGC\,4371 \citep{gadotti2015}, the current TIMER sample consists of 24 barred
galaxies with a large variety of bar-built central structures, such as nuclear rings, nuclear discs, and inner bars
\citep{gadotti2019}. To date, 21 galaxies have been observed with the Multi-Unit Spectroscopic Explorer
\citep[MUSE;][]{bacon2010} on the Very Large Telescope. 

The TIMER sample has been selected from the Spitzer Survey of Stellar Structure in Galaxies
\citep[S$^4$G;][]{sheth2010}, thus naturally constraining the sample to nearby ($d < \SI{40}{\Mpc}$), bright ($m_B <
\SI{15.5}{\mag}$), and large ($D_{25} > \SI{1}{\arcmin}$) objects. In addition, we required all galaxies to have
inclinations below \SI{60}{\deg} and central substructures as classified by \citet{buta2015}.  The resulting sample
covers a range in stellar mass from \SIrange{2.0e10}{17.4e10}{\msol}. 

All observations were taken in ESO's period 97 between April and September 2016. Using the wide-field-mode of the MUSE
spectrograph, we obtained observations covering a wavelength range from \SIrange{4750}{9350}{\angstrom} with a spectral
sampling of \SI{1.25}{\angstrom} and a field of view of \SI{1}{\arcmin\squared} at a spatial sampling of
\SI{0.2}{\arcsec}. The typical seeing of the observations was \SIrange{0.8}{0.9}{\arcsec}. Each galaxy was observed with
approximately one hour of integration on source and dedicated sky exposures. 

The data reduction is based on version 1.6 of the MUSE data reduction pipeline \citep{weilbacher2012}. In summary, bias,
flat-fielding, and illumination corrections are applied, the data is calibrated in flux and wavelength, and telluric
features are removed. Thanks to the dedicated sky exposures, the sky background is removed exploiting a principal
component analysis.  Finally, the observations are accurately registered astrometrically. A detailed accounting of the
physical properties of the TIMER sample, observations, and data reduction is presented in \citet{gadotti2019}. 


\section{Data analysis}%
\label{sec:analysis}
The analysis of the data, as reviewed in detail below, is conducted within the modular analysis framework of the
\texttt{GIST} pipeline\footnote{\url{http://ascl.net/1907.025}} \citep[Galaxy IFU Spectroscopy Tool;][]{bittner2019}.
More specifically, this software provides an all-in-one framework for the analysis of spectroscopic data, including all
tasks from the preparation of the input data, over its scientific analysis, to the generation of publication quality
plots. 

\begin{figure}
    \centering{%
      \includegraphics[width=0.5\textwidth]{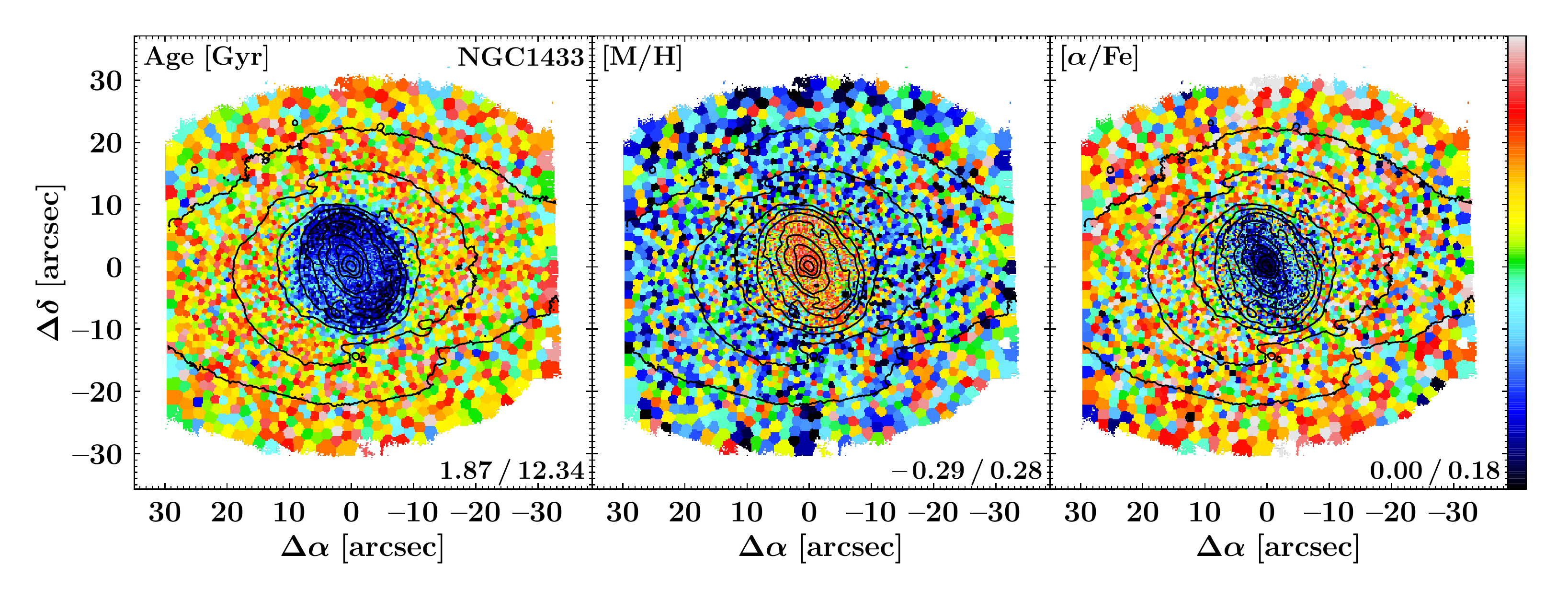} \\
      \includegraphics[width=0.5\textwidth]{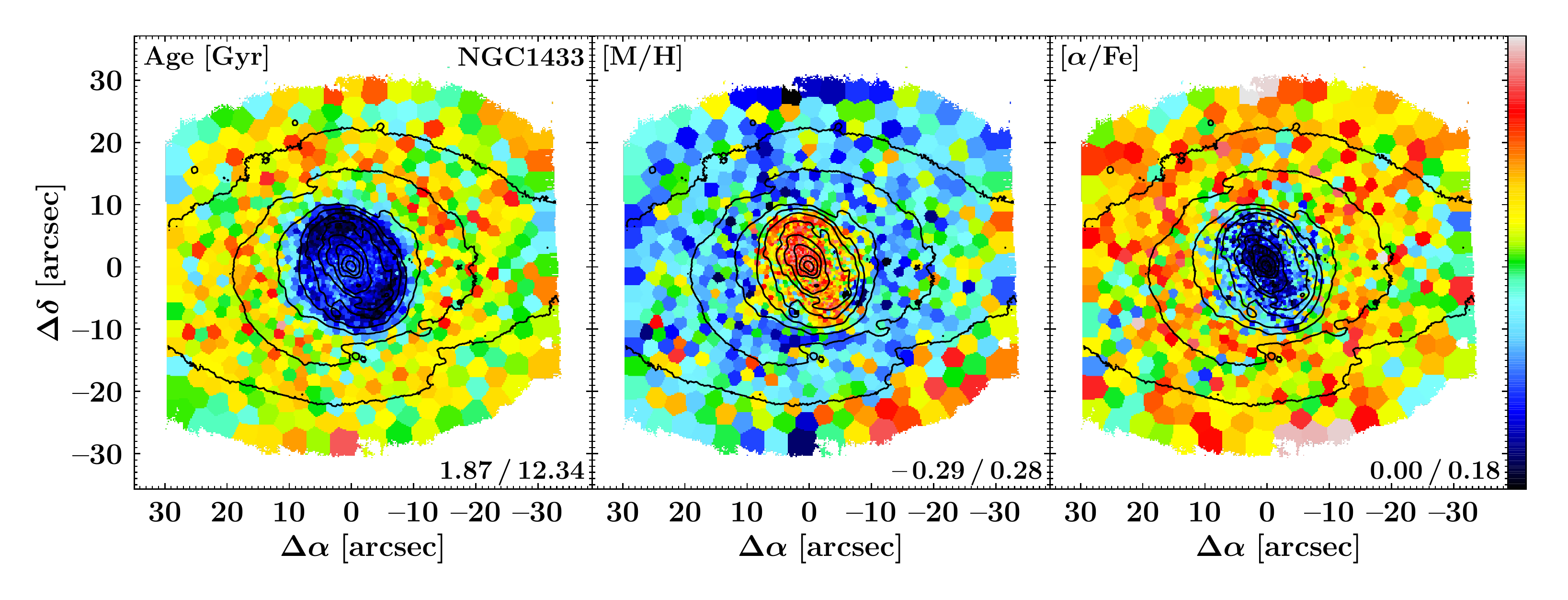} \\
      \includegraphics[width=0.5\textwidth]{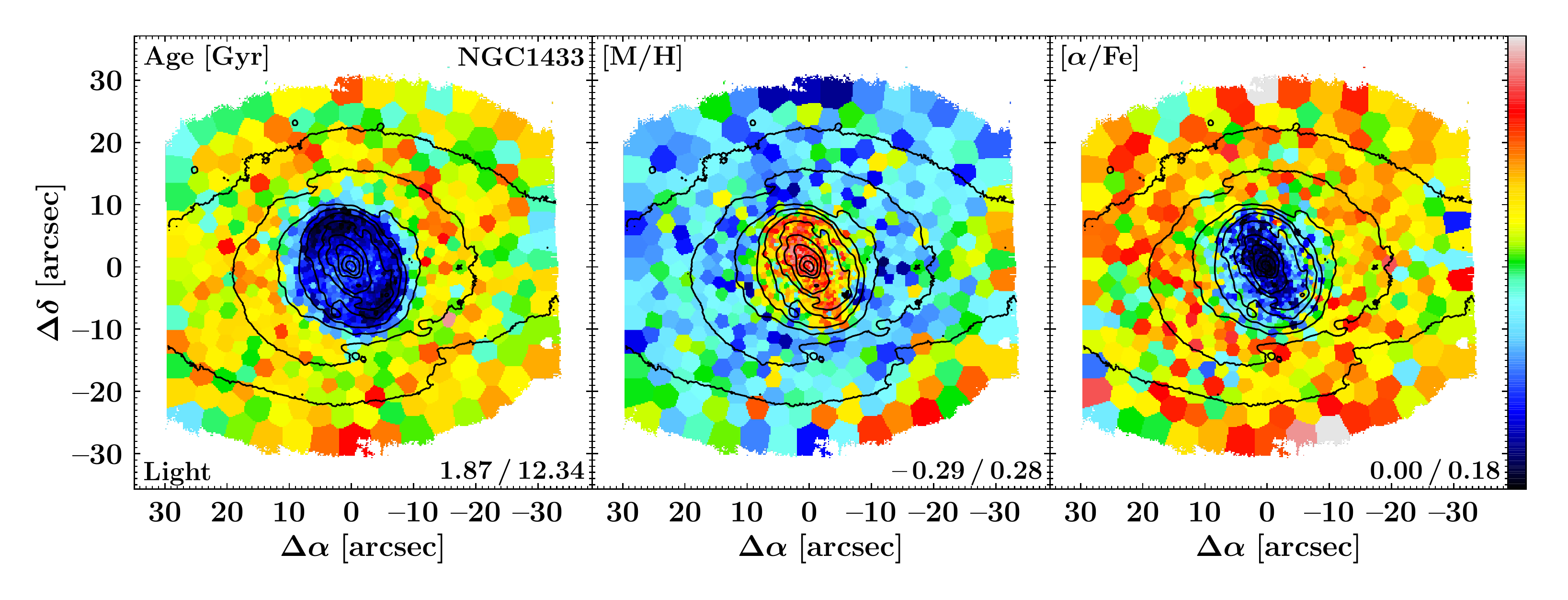}
    }
    \caption{%
        Comparison of light-weighted, mean stellar population properties of NGC\,1433, derived at a signal-to-noise
        ratio of 40 (upper panels), 80 (centre panels), and 100 (lower panels). Each set of panels displays age,
        metallicity, and {\alphaFe} enhancements.  The limits of the colour bar are stated in the lower-right corner of
        each panel.  Based on the reconstructed intensities from the MUSE cube, we display isophotes in steps of
        \SI{0.5}{\mag}. North is up; east is to the left.
    }%
    \label{fig:snrComparison}
\end{figure}

We spatially bin the data to a signal-to-noise ratio of approximately 100 per bin, exploiting the adaptive Voronoi
tessellation routine of \citet{cappellari2003}. The signal-to-noise ratio per spaxel is measured within the wavelength
range of \SIrange{4800}{5800}{\angstrom}, identical to the fitted wavelength range.  The notably high signal-to-noise
ratio is chosen to assure the robustness of the analysis, in particular as the derivation of stellar population
properties is a signal-to-noise sensitive measurement.  In a series of tests we found that increasing the
signal-to-noise ratio from 40 to 80 results in more homogeneous stellar populations across contiguous spatial bins, as
illustrated in Fig.~\ref{fig:snrComparison}.  In other words, using a higher signal-to-noise ratio reduces the level of
stochasticity in the stellar population properties in adjacent bins. This is at least partly due to an improved accuracy
in the subsequent emission-line subtraction (see below).  Further increasing the signal-to-noise ratio from 80 to 100
has only little effect on the obtained population properties. Nonetheless, we prefer to follow the more conservative
approach of using a signal-to-noise level of 100. Owing to the high quality of our data, this does not significantly
reduce the obtained spatial resolution in the nuclear discs. 
We further note that spaxels which surpass this signal-to-noise threshold, as commonly found in the centre of our
fields, remain unbinned. Spaxels below the isophote level which has an average signal-to-noise level of 3 are excluded
from the analysis, in order to avoid systematic effects in the low surface brightness regime. 

Prior to any analysis, we de-redshift the spectra to rest-frame based on an initial guess of the systemic redshift.  We
further adopt as line-spread function the udf-10 parametrisation by \citet{bacon2017} and broaden all template spectra
to this resolution before conducting any fits. 
In a series of tests we explore the effect of different wavelength ranges on the derivation of stellar population
properties. The results indicate that the red part of the spectra (beyond \SI{5800}{\angstrom}) is, in contrast to the
derivation of stellar kinematics, not suited for the measurement of the stellar population content. In particular, the
lower sensitivity to stellar population properties in this part of the spectrum, small residuals from the sky
subtraction, and absorption features from the interstellar medium add difficulties to the analysis. Therefore, we
restrict the analysis to the wavelength range from \SIrange{4800}{5800}{\angstrom}. 

The actual analysis is conducted in three separate steps. Firstly, we derive the stellar kinematics by performing an
unregularised run of the {\ppxf} routine \citep{cappellari2004,cappellari2017}. In order to account for small
differences in the shape of the continuum between spectra and templates, we include a low order multiplicative Legendre
polynomial in the fit. At this stage, all emission lines are masked. 
In the next step, we model emission lines by fitting single Gaussian templates with \texttt{pyGandALF}
\citep{bittner2019}, a new Python implementation of the original \texttt{GandALF} routine \citep{sarzi2006,jfb2006}.
The fitting routine linearly combines Gaussian emission-line templates with a set of spectral templates to obtain the
emission-line properties. In this process, the stellar continuum and emission-lines are fit simultaneously, while
keeping the stellar kinematics of the continuum fixed to those obtained before.  Instead of a Legendre polynomial,
\texttt{pyGandALF} exploits a two-component reddening correction. This correction accounts for ``screen-like'' dust
extinction of the entire spectrum as well as for reddening that affects only the emission line regions.  We note that
the kinematics of the {\Hb} and [OIII] lines are kept as free parameters in the fit, while the kinematics of [NI] are
tied to that of [OIII].  If the amplitude-to-residual noise ratio of a measured emission line is at least four, we
consider the line detection significant and subtract the emission line from the observed spectrum. In this way, we
obtain emission-subtracted spectra. 
In addition to this quantitative check, the quality of the emission-line modelling is inspected visually. While the
quality of the fits is good for the majority of the spectra, deviations are found in some regions that show starbursts
or a significant AGN contribution. This is a result of the different dynamics and the superposition of distinct
components (star-forming regions, AGN, etc.) in these regions which therefore cannot be modelled by a single Gaussian
template. However, performing a detailed multi-component emission-line analysis for the entire sample is beyond the
scope of this study, in particular as these small deviations in starbursting and AGN-affected regions do not affect our
general conclusions on nuclear discs. 

Finally, based on these emission-subtracted spectra, we perform a regularised run of the {\ppxf} routine to estimate the
mean stellar population properties. In order to avoid possible degeneracies between velocity dispersion and metallicity
\citep[see e.g.][]{sanchezBlazquez2011}, we fix the stellar kinematic to those obtained before. In addition, we apply a
8th order multiplicative Legendre polynomial in the fit that also accounts for extinction and other continuum effects. 
%
{\ppxf} estimates non-parametric star formation histories by assigning weights to the spectral templates such that the
observed spectrum is best reproduced. However, this measurement represents an inverse, ill-conditioned problem.
Therefore, in order to obtain a physically meaningful solution, {\ppxf} applies a regularisation during the fit
\citep{press1992,cappellari2017}. Thus, of all equally consistent solutions, the regularised run of {\ppxf} returns the
smoothest solution that is still statistically consistent with the data in consideration. 
While the strength of the regularisation can have a substantial impact on the shape of individual star formation
histories, the derived mean population properties we consider in this study show only little dependence on the chosen
regularisation. Nonetheless, we follow the procedure of determining the maximum allowed regularisation parameter, as
described, for instance, in \citet{mcdermid2015}.  Firstly, the noise is rescaled in such a way that the resulting
$\chi^2$ of the unregularised run is unity.  Subsequently, the regularisation strength is increased iteratively until
the $\chi^2$ of the regularised run exceeds that of the unregularised run by approximately $\sqrt{2N_{\mathrm{pix}}}$,
with $N_{\mathrm{pix}}$ being the number of spectral pixels included in the fit. 
This procedure is applied to the bin with the highest signal-to-noise ratio in each cube.  In case of contamination of
this spectrum by strong extinction, features from active galactic nuclei, or intense star formation, a spatial bin with
similarly high signal-to-noise ratio in its close vicinity is used instead.  The obtained regularisation strength is
then applied to the entire galaxy.  We note that for NGC\,1365 and NGC\,5728 the above procedure allowed conspicuously
high regularisation strengths, possibly due to large-scale outflows from the active galactic nuclei. For these cases we
therefore chose a lower regularisation strength of \num{10}, similar to that obtained for the rest of the sample.

In Sect.~\ref{sec:comparisons}, we investigate if the derived population properties depend on whether the {\alphaFe}
enhancement is modelled in the {\ppxf} fit or not. Therefore we repeat the analysis with two variants of the MILES
single stellar population (SSP) models. Firstly, we use the MILES ``base models'' which follow the abundance pattern of
stars in the solar neighbourhood \citep{vazdekis2010}. At low metallicities these models show elevated values of
{\alphaFe} enhancements, although the used isochrones are scaled-solar, while at high metallicities the {\alphaFe}
values resemble the solar abundance. Secondly, we employ a combination of scaled-solar and {\alphaFe} enhanced MILES
models \citep{vazdekis2015}. These provide two values of {\alphaFe}, namely \num{0.00} (solar abundance) and \num{0.40}
(supersolar abundance). While the use of a SSP model library that covers only two values of {\alphaFe} is not optimal,
{\ppxf} is capable of interpolating between these two values and returns intermediate {\alphaFe} enhancements \citep[see
also][]{pinna2019}.  We note that full-spectral fitting codes are typically measuring an average {\alphaFe} ratio. In
other words, in the models all $\alpha$ elements are increased/decreased while in realistic galaxies different $\alpha$
elements are decoupled from each other. In addition, different wavelength ranges are more sensitive to different
$\alpha$ elements.  Therefore, the {\alphaFe} abundances returned by {\ppxf} are a convoluted average of the underlying
abundances of individual $\alpha$ elements. For the sake of clarity, we choose to refer to {\alphaFe} abundances
throughout this study. 

Both sets of SSP models assume a Kroupa Revised IMF with a slope of 1.30 \citep{kroupa2001}, use BaSTI isochrones
\citep{pietrinferni2004, pietrinferni2006, pietrinferni2009, pietrinferni2013}, and have a spectral resolution of
\SI{2.51}{\angstrom} \citep{jfb2011}. The parameter space is sampled in 53 values of age between \SI{0.03}{\Gyr} and
\SI{14.0}{\Gyr} and 12 values of stellar metallicity ([M/H]) between \SI{-2.27}{\dex} and \SI{0.40}{\dex}. 

Over the past years, several fitting routines for the measurement of stellar population properties have been
implemented. In order to assess the systematic effects related to different software implementations, we again repeat
the analysis with {\steckmap} \citep{ocvirk2006a,ocvirk2006b}.  To this end, we employ the same emission-subtracted
spectra and fix the stellar kinematics to those initially obtained with {\ppxf}.  Since {\steckmap} is not designed to
model {\alphaFe} abundances, we only use the base models here. 

The above analysis is performed in such a way that light-weighted results are obtained. This is achieved by normalising
each spectral template by its own mean flux within the used wavelength range. In order to convert those light-weighted
stellar population properties to mass-weighted ones, we employ the mass-to-light ratio predictions of the SSP models.
In particular, this mass-to-light ratio includes not only the remaining mass in the stellar component but also the mass
of all resulting stellar remnants. 

The light- and mass-weighted population properties are averaged via
\begin{eqnarray}
    \langle \mathrm{t}           \rangle =& \dfrac{\sum_i \: w_i \: \mathrm{t}_{\mathrm{SSP, i}}}{\sum_i w_i}  \\
    \langle \mathrm{[M/H]}       \rangle =& \dfrac{\sum_i \: w_i \: \mathrm{[M/H]}_{\mathrm{SSP, i}}}{\sum_i w_i}   \\
    \langle \mathrm{[\alpha/Fe]} \rangle =& \dfrac{\sum_i \: w_i \: \mathrm{[ \alpha /Fe]}_{\mathrm{SSP, i}}}{\sum_i w_i} 
\end{eqnarray}
with the weight $w_i$ assigned to the $i$th template with age $\mathrm{t}_{\mathrm{SSP, i}}$, metallicity
$\mathrm{[M/H]}_{\mathrm{SSP, i}}$, and an {\alphaFe} enhancement of $\mathrm{[ \alpha /Fe]}_{\mathrm{SSP, i}}$. Hence,
stellar ages are averaged in linear scale while metallicities and {\alphaFe} abundances are averaged logarithmically. 


\section{Stability and errors of the measurements}%
\label{sec:comparisons}
In this section we further explore the reliability of the derived mean stellar population properties. First, we discuss
the effect that modelling the {\alphaFe} abundances has on ages and metallicities, before comparing results obtained
with the {\ppxf} and {\steckmap} routines.  We do not intend to provide a thorough software comparison here, but simply
mean to check our results with a second, independent analysis.  This also allows us to better understand error estimates
on the derived stellar populations properties.

\subsection{Population properties with and without {\alphaFe} modelling}%
\label{subsec:templateComparison}
\begin{figure}
    \includegraphics[width=\hsize]{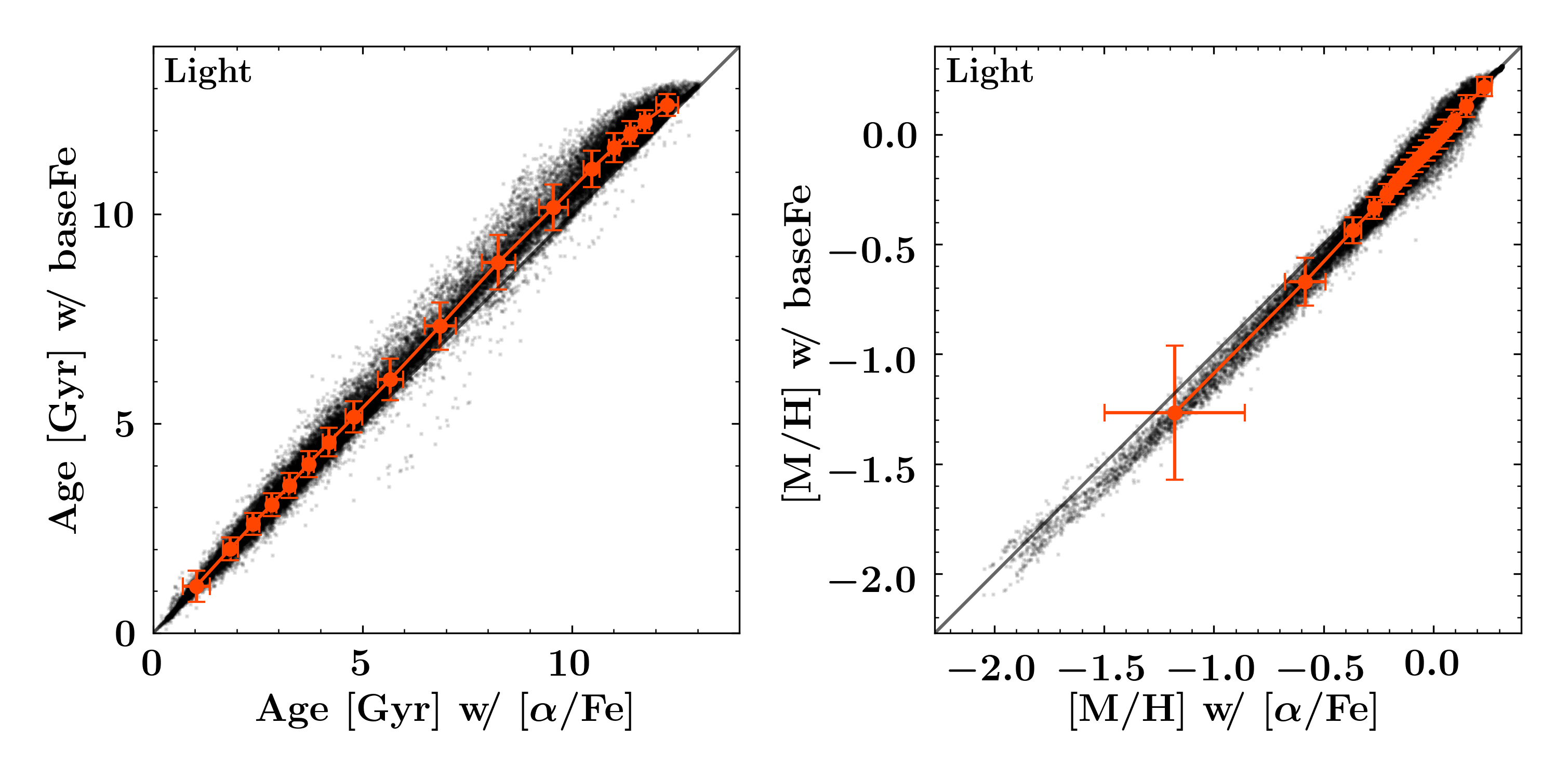}\\
    \includegraphics[width=\hsize]{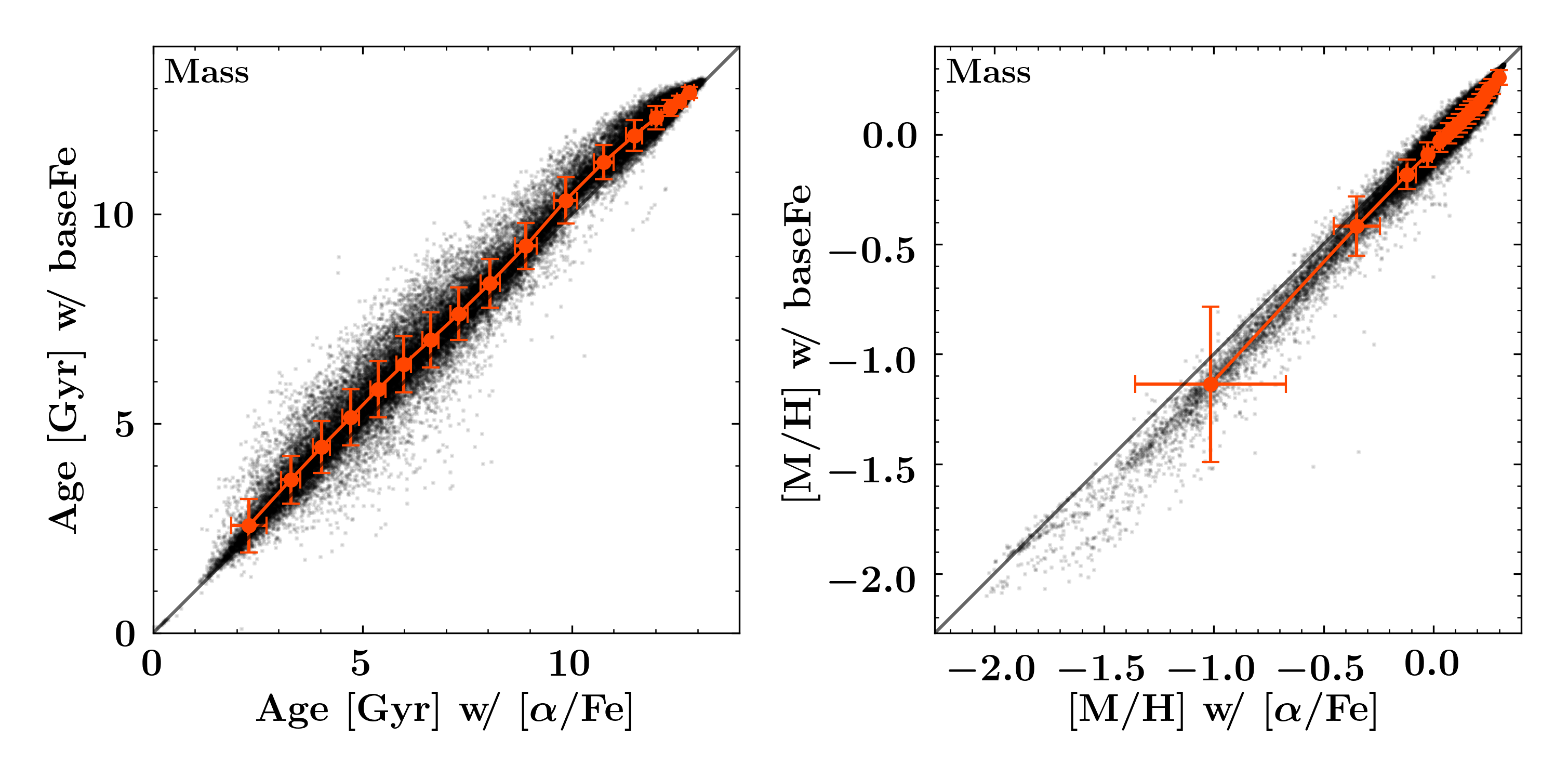}
    \caption{%
      Light-weighted (upper panels) and mass-weighted (lower panels) population properties derived with {\ppxf} without
      {\alphaFe} modelling as a function of those derived with {\alphaFe} modelling.  This figure includes all spatial
      bins from all galaxies in the present TIMER sample.  Highlighted in orange are the means and standard deviations
      in 17 bins, each combining the results from approximately 2600 observed spectra.  We note that the large standard
      deviation in the bin with the lowest metallicity is a result of the large range of metallicities included in this
      bin and does not necessarily result from an increased scatter.
    }%
    \label{fig:alphaComparison}
\end{figure}
To date, most stellar population spectral libraries do not provide spectra with varying {\alphaFe} abundances.
Similarly, not all full spectral fitting codes are designed to model {\alphaFe} enhancements in the fitting process.
Here we test whether stellar ages and metallicities obtained with or without the modelling of {\alphaFe} abundances are
consistent. In Fig.~\ref{fig:alphaComparison} we plot population properties derived with \texttt{pPXF} without the
modelling of {\alphaFe} and using base models as a function of those derived with {\alphaFe} modelling and the enhanced
templates.  

In the case of the light-weighted results, we find a very good agreement between the two runs, with the median standard
deviation of all 17 bins being \SI{0.34}{\Gyr} and \SI{0.05}{\dex} in age and [M/H], respectively. For the mass-weighted
results, the correspondence is good as well, with slightly older ages and lower metallicities being obtained without the
{\alphaFe} modelling. The corresponding median standard deviation of all bins is \SI{0.57}{\Gyr} in age and
\SI{0.04}{\dex} in metallicity.  We thus conclude that ages and metallicities obtained with {\ppxf} depend only little
on whether {\alphaFe} abundances are included in fitting processes or not. 

Nonetheless, we find a small systematic offset in the measured metallicities. At the lowest metallicities, the fit that
uses the MILES base models returns systematically lower metallicities. This effect seems to be related to the chemical
composition of the base models that might differ from the abundance pattern of the observed spectra. This is confirmed
by repeating the analysis of NGC\,1097, now using models with {\alphaFe} enhancements of \SI{0.00}{\dex} and
\SI{0.40}{\dex} separately. While the analysis using models with {\alphaFe} = 0.00 returns slightly lower metallicities
compared to the run with varying {\alphaFe} abundances, the fits with {\alphaFe} = 0.40 return systematically higher
metallicities. In fact, this behaviour is expected, as the typical {\alphaFe} abundances measured in the low metallicity
bins of NGC\,1097 is approximately \SI{0.2}{\dex}.  Nonetheless, the measured systematic differences of approximately
\SI{0.1}{\dex} are slightly smaller than the typical errors estimated for this measurement (\SI{0.14}{\dex}, see
Sect.~\ref{subsec:errorEstimates}) and only a relatively small number of bins in the TIMER sample have such low
metallicities.

\subsection{Population properties from {\ppxf} and {\steckmap}}%
\label{subsec:codeComparison}
\begin{figure}
    \includegraphics[width=\hsize]{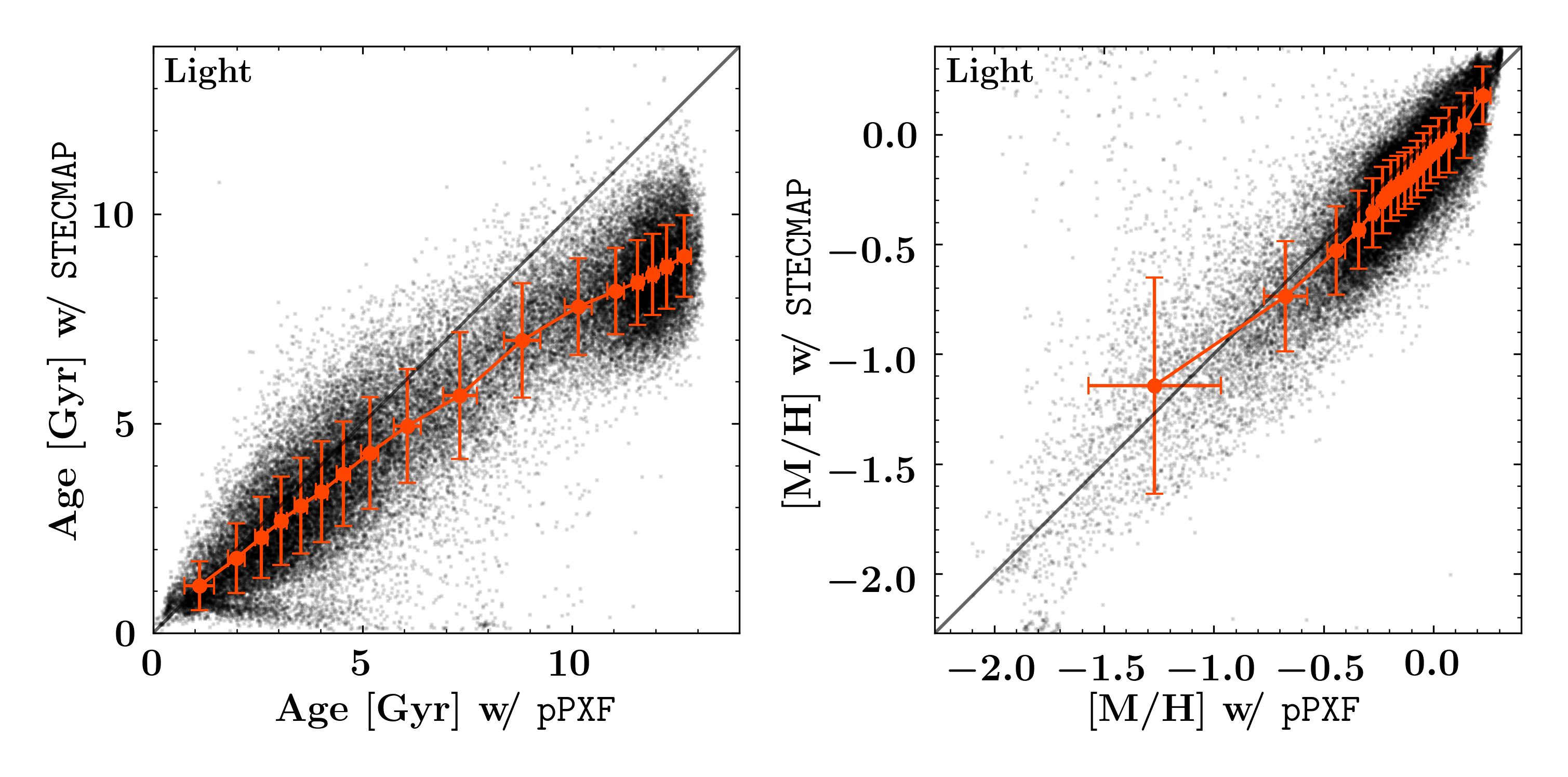}\\
    \includegraphics[width=\hsize]{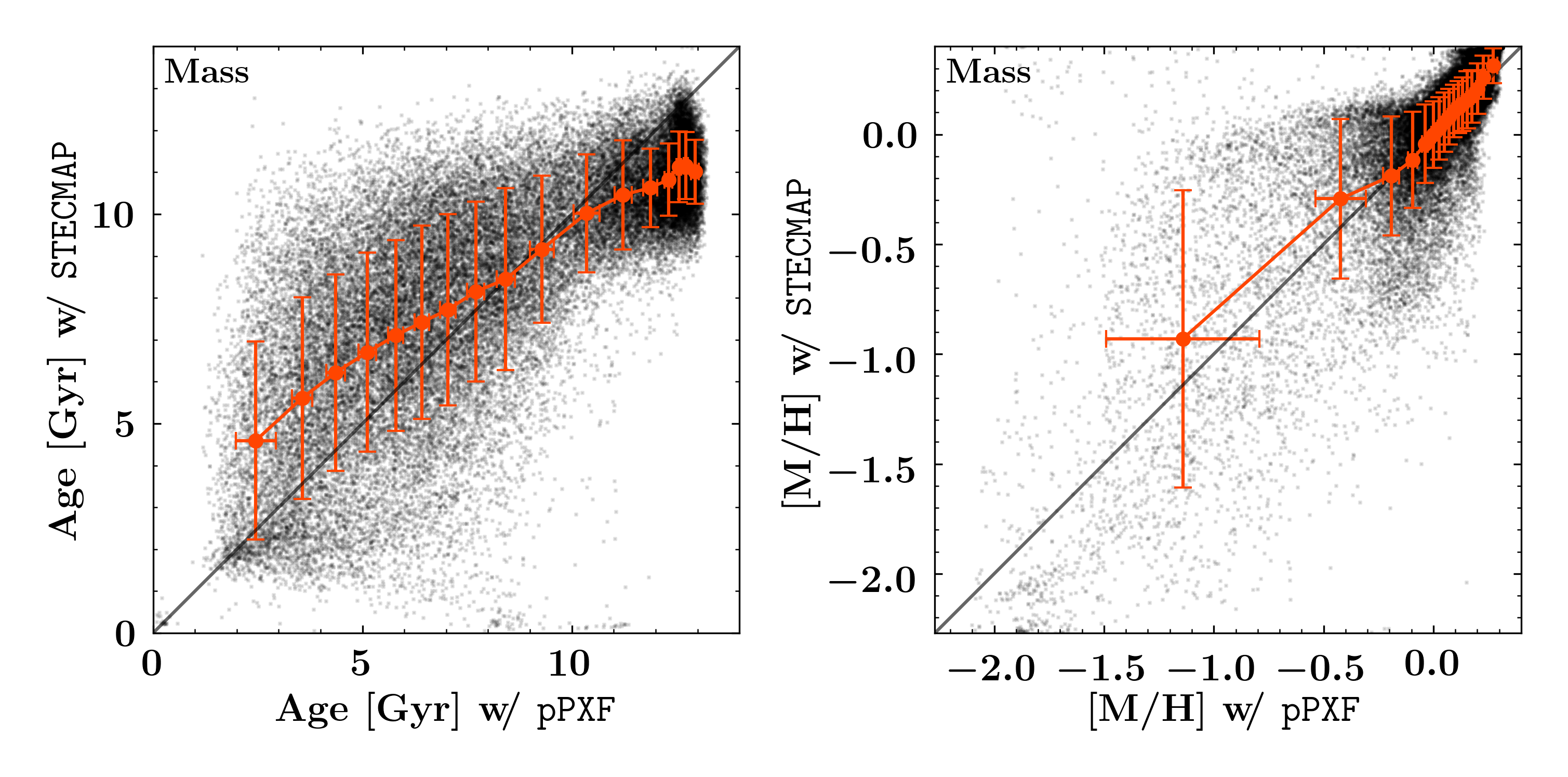}
    \caption{%
      Same as Fig.~\ref{fig:alphaComparison}, but showing mean stellar population properties derived with {\steckmap} as
      a function of those derived with {\ppxf}. 
    }%
    \label{fig:codeComparison}
\end{figure}

Over time, a large set of well-known fitting routines for the derivation of non-parametric star formation histories was
developed, for instance \texttt{MOPED} \citep{heavens2000}, \texttt{pPXF} \citep{cappellari2004,cappellari2017},
\texttt{STARLIGHT} \citep{cidFernandes2005}, \texttt{STECKMAP} \citep{ocvirk2006a,ocvirk2006b}, \texttt{VESPA}
\citep{tojeiro2007}, \texttt{ULySS} \citep{koleva2009}, \texttt{FADO} \citep{gomes2017}, or \texttt{FIREFLY}
\citep{wilkinson2017}.  As different routines are based on distinct fitting methodologies with diverse code
implementations, the results obtained with different software might vary. Therefore, it is important to repeat the data
analysis with various routines, to better understand the involved systematics. In the following we compare the stellar
population properties obtained with the {\ppxf} and {\steckmap} routines. 

In Fig.~\ref{fig:codeComparison} we plot population properties derived with {\steckmap} as a function of those measured
with {\ppxf}.  We emphasise that we perform both runs using exactly the same emission-subtracted spectra, MILES base
models, and stellar kinematics, in order to isolate the differences due to the fitting algorithms. 
%
The light-weighted age results obtained with {\ppxf} and {\steckmap} are in reasonable agreement, as indicated by the
median standard deviation of all 17 bins being \SI{1.06}{\Gyr}.  However, {\steckmap} returns systematically lower
stellar ages. More precisely, the deviation between {\ppxf} and {\steckmap} is increasing with age: little deviation is
found at low ages ($\Delta t \approx \SI{0}{\Gyr}$ at \SI{1}{\Gyr}) while the discrepancy for old populations is the
largest ($\Delta t \approx \SI{4}{\Gyr}$ at \SI{13}{\Gyr}).  The measured metallicities conform well, with {\steckmap}
providing slightly lower [M/H] values at all but the lowest metallicities. The median standard deviation of all bins is
\SI{0.14}{\dex}. 

We find only a few outliers at low ages, in particular at {\steckmap} ages below \SI{1}{\Gyr} and {\ppxf} ages of
\SIrange{1}{5}{\Gyr}, which is caused by two strongly starbursting regions in NGC\,5236. A visual inspection of the
corresponding spectra and fits has shown complex H$\beta$ profiles for which the emission line modelling with
\texttt{pyGandALF} is imperfect. In fact, this discrepancy is absent when repeating the analysis of this galaxy while
masking all wavelength regions affected by emission lines. We emphasise that the derived stellar population quantities
in all other parts of this galaxy are identical, whether or not emission-lines are masked or subtracted. A more detailed
modelling of such complex emission-line profiles for the entire sample is beyond the scope of this study. However, as
there are no other clear discrepancies found, the emission-line modelling appears to be sufficiently good for all other
galaxies. 

For mass-weighted stellar population properties the scatter is substantially higher, as compared to the light-weighted
results. However, mean ages and metallicities within each bin are still in reasonable agreement. Similarly to the
comparison of light-weighted ages, a systematic deviation of the measured ages is found. At young ages {\steckmap}
detects older stellar populations while for old ages it derives younger stellar ages, as compared to {\ppxf}.  The
median standard deviation of all bins amounts to \SI{2.15}{\Gyr}.  Metallicities again conform well, with a similar
median standard deviation of \SI{0.13}{\dex}. 
We speculate that this increased deviation between {\ppxf} and {\steckmap} might be a result of the conversion from
light- to mass-weighted population properties. This conversion is calculated using mass-to-light ratio predictions of
the employed SSP models. These predictions have intrinsic uncertainties which result from the used stellar evolution
models, but are also a function of both the derived stellar age and metallicity. Thus, the mass-weighted ages and
metallicities depend on the uncertainty in both light-weighted ages and metallicities, as well as the mass-to-light
ratio predictions themselves. Therefore, it is not surprising that the obtained uncertainties are significantly higher
for mass-weighted results.

\subsection{Error estimates on stellar population properties}%
\label{subsec:errorEstimates}
Estimating measurement uncertainties on mean stellar population properties or non-parametric star formation histories
derived with full spectral fitting is not trivial.  Firstly, it is not possible to assign formal errors, as the
population properties are simply averaged from the used subset of spectral templates and not measured in the actual
fitting process.  Secondly, errors derived by repeating the measurement on multiple Monte-Carlo realisations of the
data, as commonly used for the estimation of errors on the stellar kinematics, might be significantly underestimated due
to the effect of regularisation. 

Here we propose to use the median standard deviations from the comparisons between {\ppxf} and {\steckmap}, as stated in
Sect.~\ref{subsec:codeComparison}, as error estimates. While such an uncertainty does not represent a real measurement
error, the scatter between two different fitting routines does provide a realistic estimate.  Of course, this approach
cannot account for errors introduced by other systematics, such as intrinsic problems with the spectral models, template
mismatch, or IMF variations.  Nonetheless, these error estimates compare well to those discussed by other authors. 

For instance, \citet{gadotti2019} repeat the analysis on 100 Monte-Carlo realisations, in order to estimate errors.
Using {\steckmap} at a signal-to-noise level of 40, they find typical errors in age of \SIrange{0.5}{1}{\Gyr} and
\numrange{0.005}{0.010} in metallicities (Z) which translate to errors of approximately \SIrange{0.11}{0.23}{\dex} in
[M/H] when evaluated at solar metallicity.  In addition, their error estimates depend on both age and metallicity, and
whether light- or mass-weighted results are considered. They further conclude that for {\steckmap} the chosen initial
conditions and smoothness parameter have no effects. 
%
In contrast, \citet{pinna2019} use 50 Monte-Carlo realisations and also investigate how much the applied corrective
polynomial and regularisation parameter of {\ppxf} affect the resulting error estimates. Fitting stellar kinematics and
population properties simultaneously at a signal-to-noise level of 40 they find typical errors of \SI{3}{\Gyr},
\SI{0.1}{\dex}, and \SI{0.06}{\dex} for mass-weighted age, [M/H], and {\alphaFe}. 
%
\citet{ruiz-lara2015} performs an extensive comparison of star formation histories derived from colour-magnitude
diagrams of resolved stars and the analysis of integrated light spectroscopy with {\steckmap}, \texttt{ULySS}, and
\texttt{STARLIGHT} and reach similar conclusions. 


\section{Results}%
\label{sec:results}
In this section we present our general findings concerning the mean stellar population properties in the central regions
of all TIMER galaxies. In addition, we complement these results by presenting maps of {\Ha} emission-line fluxes, which
help inform our analysis.  We do not focus on the properties of individual objects here but instead intend to provide an
overview of the common characteristics of nuclear discs. We refer the reader to Appendix~\ref{app:mapdescriptions} for
more detailed descriptions of each observed galaxy. 

Observationally it seems that the gaseous nuclear ring simply highlights the outer edge of the stellar nuclear disc. In
this section we will provide further evidence supporting this picture. We remind the reader that the term nuclear disc
refers to the kinematically cold and regularly rotating stellar discs in the central regions of the galaxies. According
to the kinematic analysis of the TIMER sample in an accompanying paper (see G20), these nuclear discs extend to the
centres of the galaxies and have a well-defined outer edge. In contrast, we use the term nuclear ring only to describe
the outermost part of these nuclear discs, as these regions are often highlighted by gaseous nuclear rings. Therefore,
the denomination nuclear disc includes both the stellar nuclear disc and the gaseous nuclear ring, while the term
nuclear ring refers only to the outer edge of the nuclear disc.

\begin{table*}
    \centering
    \begin{tabular}{lcccccccc}
        \toprule
        \toprule
        Galaxy      & $i$       & PA              & $M_{\mathrm{stellar}}$ & Spatial scale     & $R_{kin}$ & {\Ha}         & Central   & Age       \\
                    & \si{\deg} & \si{\deg}       & \SI{e10}{\msol}        & \si{\pc/\arcsec}  & \si{\pc}  & morphology    & emission  & gradient  \\
        (1)         & (2)       & (3)             & (4)                    & (5)               & (6)       & (7)           & (8)       & (9)       \\
        \midrule                                                                                                                
        \addlinespace[0.5em]                                                                                                
        \multicolumn{4}{l}{Non-star-forming nuclear rings} \\                                                               
        \cmidrule(lr{5em}){1-4}                                                                                             
        IC\,1438    & 24        & -25.4           & 3.1                    & 164               &  604      & NR            & LINER     & SYM       \\
        NGC\,1291   & 11        & -8.9            & 5.8                    & 42                &  ---      & IRR           & LINER     & ---       \\
        NGC\,1300   & 26        & -45.9           & 3.8                    & 87                &  332      & NR            & LINER     & FLAT      \\
        NGC\,1433   & 34        & 18.2            & 2.0                    & 49                &  381      & IRR           & LINER     & FLAT      \\
        NGC\,4371   & 59        & 88.1            & 3.2                    & 82                &  952      & NO            & NO        & ---       \\
        NGC\,4643   & 44        & 55.5            & 10.7                   & 125               &  495      & C             & LINER     & SYM       \\
        NGC\,5248   & 41        & -75.6           & 4.7                    & 82                &  489      & U             & SF/LINER  & SYM       \\
        NGC\,5850   & 39        & -26.5           & 6.0                    & 112               &  796      & C             & NO        & SYM       \\
        NGC\,7140   & 51        & 4.1             & 5.1                    & 180               &  634      & NR            & SF        & FLAT      \\
        NGC\,7755   & 52        & 23.9            & 4.0                    & 153               &  466      & NR            & LINER     & SYM       \\
        \addlinespace[0.5em]                                                                                                     
        \multicolumn{4}{l}{Star-forming nuclear rings} \\                                                                        
        \cmidrule(lr{5em}){1-4}                                                                                                  
        NGC\,613    & 39        & -50.1           & 12.2                   & 120               &  590      & NR            & LINER     & ---       \\
        NGC\,1097   & 51        & -52.1           & 17.4                   & 100               & 1072      & NR            & LINER     & ---       \\
        NGC\,3351   & 42        & 11.2            & 3.1                    & 49                &  236      & NR            & SF        & ---       \\
        NGC\,4303   & 34        & -36.7           & 7.2                    & 80                &  214      & NR            & LINER     & ---       \\
        NGC\,4981   & 54        & -28.2           & 2.8                    & 120               &  139      & NR            & LINER     & ---       \\
        NGC\,4984   & 53        & 29.6            & 4.9                    & 103               &  491      & NR/C          & AGN       & ---       \\
        NGC\,5236   & 21        & 47.0            & 10.9                   & 34                &  368      & IRR           & SF        & ---       \\
        NGC\,7552   & 14        & 54.9            & 3.3                    & 83                &  332      & NR            & SF        & ---       \\
        \addlinespace[0.5em]                                                                                                     
        \multicolumn{4}{l}{Peculiar nuclear regions} \\                                                                          
        \cmidrule(lr{5em}){1-4}                                                                                                  
        NGC\,1365   & 52        & 42.0            & 9.5                    & 87                &  ---      & NR/IRR        & AGN       & ---       \\
        NGC\,5728   & 44        & 1.1             & 7.1                    & 149               &  628      & NR/C          & AGN       & ---       \\
        NGC\,6902   & 37        & -49.6           & 6.4                    & 187               &  ---      & PECULIAR      & NO        & ---       \\
        \bottomrule
    \end{tabular}
    \caption{%
        Overview of the different subsamples and some general properties of nuclear rings, nuclear discs, and their host
        galaxies. Column (1) states the galaxy name, while columns (2) and (3) provide the inclination and position
        angle of the galaxy disc \citep{munozMateos2015}. Columns (4) to (6) state the total stellar mass of the galaxy
        derived within S$^4$G, the spatial scale of the observations, and the kinematic radius of the nuclear discs (see
        G20), respectively. Column (7) states whether the morphology of the {\Ha} emission is dominated by a nuclear
        ring (NR), central emission (C), a uniform nuclear disc (U), an irregular emission pattern (IRR), or no ionised
        gas emission throughout the field of view (NO).  In column (8) we provide the ionisation source in the innermost
        region of the galaxy (at $r<<R_{\mathrm{kin}}$), as determined with BPT-diagrams (note notwithstanding that the
        main ionisation source in the nuclear rings the majority of the radial extent of the nuclear disc is star
        formation).  Column (9) describes the shape of the age profile inside/outside of the nuclear disc as symmetric
        (``SYM'') or flatter within the nuclear disc (``FLAT''). This classification is given only for the
        non-star-forming galaxies, as the age profiles are often hard to distinguish in the other subsamples. In
        NGC\,1291 and NGC\,4371 the age profiles are dominated by the inner bar and projection effects, respectively,
        and thus we omit these galaxies in this classification. 
    }%
    \label{tab:overview}
\end{table*}

\subsection{Maps of dust-corrected {\Ha} emission-line fluxes}%
\label{subsec:results_HalphaMaps}
We use {\Ha} emission-line fluxes as a tracer of HII regions, in order to reliably distinguish star-forming from
non-star-forming nuclear rings/discs in the TIMER sample. This distinction allows to investigate their stellar
population properties separately and detect areas in which the derived stellar population content might have been
affected by strong star formation. 

For the purpose of deriving emission-line fluxes, it is not necessary to spatially bin the data to high signal-to-noise
ratios, as the signal-to-noise ratios of individual emission lines are typically higher compared to that of the stellar
continuum. Therefore, we prefer to consider {\Ha} maps on a spaxel-by-spaxel basis. To this end, we use results derived
in previous TIMER papers \citep{gadotti2019, neumann2020}, exploiting the software \texttt{PyParadise}, an extended
Python implementation of \texttt{Paradise} \citep{walcher2015}.  The obtained emission-line fluxes are corrected for
dust extinction by measuring the Balmer decrement and applying the models of \citet{calzetti2000} to account for the
wavelength dependency. 

In Fig.~\ref{fig:HalphaMaps} we present the maps of the dust-corrected {\Ha} fluxes for all TIMER galaxies.  The maps
reveal a large variety of ionised gas structures in the galaxy centres.  While some galaxies exhibit little to no {\Ha}
emission, other objects show large amounts of {\Ha} emission. In most galaxies, in particular in the cases with
significant {\Ha} emission, the star formation is concentrated in a well-defined nuclear ring.  In order to compare the
stellar population properties of galaxies with and without such star-forming nuclear rings, we split the sample in two
groups of galaxies. All nuclear rings with {\Ha} emission-line fluxes above \SI[product-units=single]{5 x
d-10}{\erg\per\second\per\cm\squared\per\arcsec\squared} are classified as star-forming, while galaxies with lower {\Ha}
emission-line fluxes are classified to host non-star-forming nuclear rings. We note that this threshold is an empirical
finding chosen to reproduce the morphological differences evident in the galaxies. The classification in star-forming
and non-star-forming nuclear rings does not depend on the precise value of this threshold. An overview about the
different subsamples and the basic properties of the respective galaxies, in particular their {\Ha} morphology, is
provided in Table~\ref{tab:overview}. 

A total of 8 galaxies is found to have star-forming nuclear rings (NGC\,613, NGC\,1097, NGC\,3351, NGC\,4303, NGC\,4981,
NGC\,4984, NGC\,5236, NGC\,7552). The high {\Ha} emission-line fluxes in the nuclear rings of these galaxies suggest
strong ongoing star formation or even a starburst episode. Only in the case of NGC\,5236 the ionised gas emission is not
concentrated to the nuclear ring, but irregularly distributed in the centre of the galaxy. We note that the derived
stellar population properties might be affected by the strong star formation in these regions, for instance by residuals
in the emission-line subtraction of the {\Hb} line.  Moreover, the measurement of stellar population properties might be
affected by the nebular continuum and contributions from active galactic nuclei (AGN). However, the nebular continuum is
only important in very young, star-forming regions, while possible AGN contributions are restricted to the centremost
spaxels.  In addition, unphysical results might be obtained if the stellar light is dominated by a population that is
not represented in the employed set of SSP models. In our set-up this could be the case for stellar populations with
ages below \SI{30}{\Myr}, as this is the youngest population included in the MILES models. In any case, distinguishing
the ages of stellar populations younger than \SI{1}{\Gyr} is not relevant for our conclusions presented in
Sect.~\ref{sec:discussion}. 

The subsample of non-star-forming nuclear rings consists of 10 galaxies (IC\,1438, NGC\,1291, NGC\,1300, NGC\,1433,
NGC\,4371, NGC\,4643, NGC\,5248, NGC\,5850, NGC\,7140, NGC\,7755). The {\Ha} morphology in the centres of these galaxies
shows a larger variety. While NGC\,4371 does not show any ionised gas emission, other galaxies show irregular emission
patterns (NGC\,1291, NGC\,1433), centrally concentrated emission (NGC\,4643, NGC\,5850), or a more uniform gas disc
(NGC\,5248). Nonetheless, IC\,1438, NGC\,1300, NGC\,7140, and NGC\,7755 still show some concentration of gas in a
nuclear ring. We note that in these galaxies the derived stellar populations are presumably not affected by low
star-formation activity.

Three galaxies are excluded from above subsamples. NGC\,1365 and NGC\,5728 are significantly affected by outflows from
their AGN \citep[see e.g.][]{venturi2018, durre2018} on large spatial scales. However, we note that other galaxies in
the sample, for instance NGC\,613 \citep{jfb2014}, host AGN outflows as well, but these appear weaker and do not
significantly affect the derived stellar population properties (see also Appendix~\ref{app:mapdescriptions}).  The third
excluded galaxy is NGC\,6902 which is only weakly barred and does not show unequivocal signatures of a nuclear disc.
Unless specifically noted, these galaxies are excluded from the discussions below. 

While star formation is an efficient ionisation source, the observed {\Ha} emission could also be triggered by AGN\@.
In order to distinguish between ionisation from star formation and AGN, we apply the standard BPT methodology
\citep{baldwin1981}.  We find that the {\Ha} emission in the very centres of many galaxies, at spatial scales much
smaller than that of the nuclear disc, is often generated by AGN or LINER (see Table~\ref{tab:overview} for an
overview).  However, we also confirm that the {\Ha} emission observed in the nuclear rings and the large majority of the
nuclear discs indeed originates from star formation.  We refer the reader to \citet{gadotti2019} for a detailed analysis
and classification of the emission-line ratios and ionisation sources of all TIMER galaxies.

\subsection{Maps of mean stellar population properties}%
\label{subsec:results_sppMaps}
\begin{figure*}
    \begin{minipage}[c]{\textwidth}
        \centering
        \begin{minipage}[c]{0.55\textwidth}
            \centering
            \includegraphics[width=0.95\textwidth]{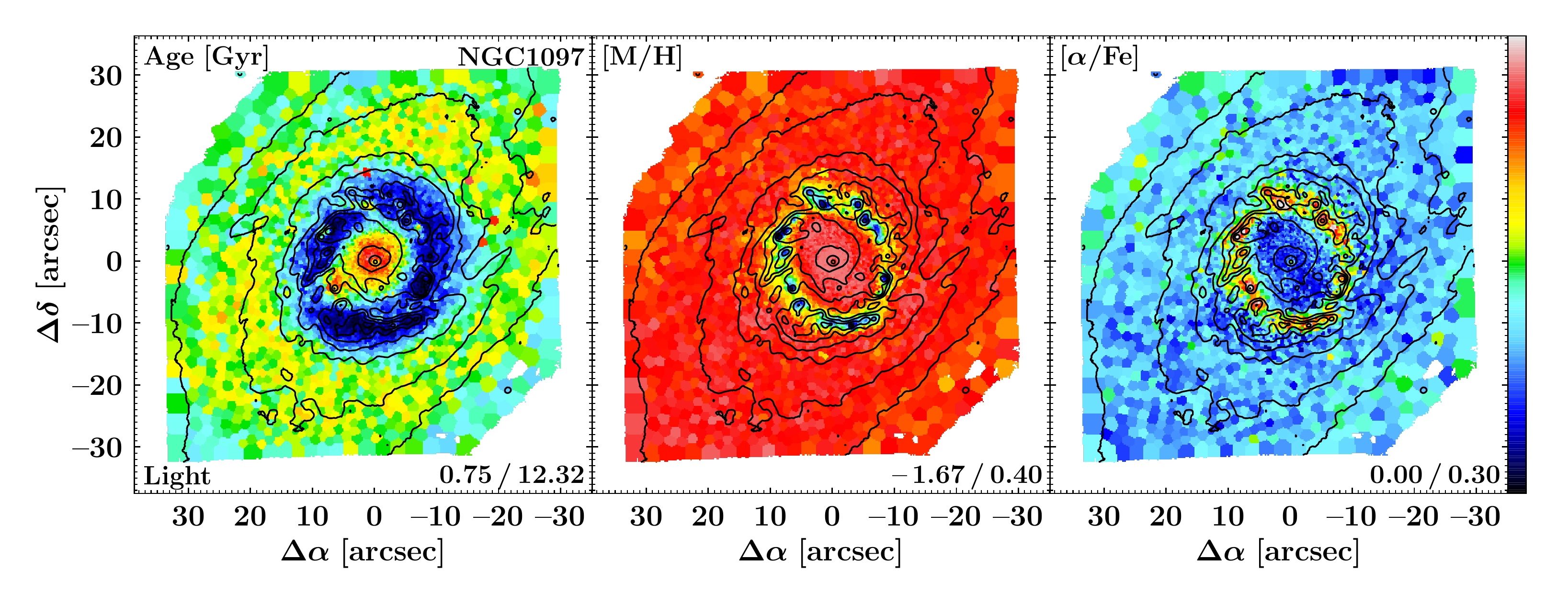}\\
            \includegraphics[width=0.95\textwidth]{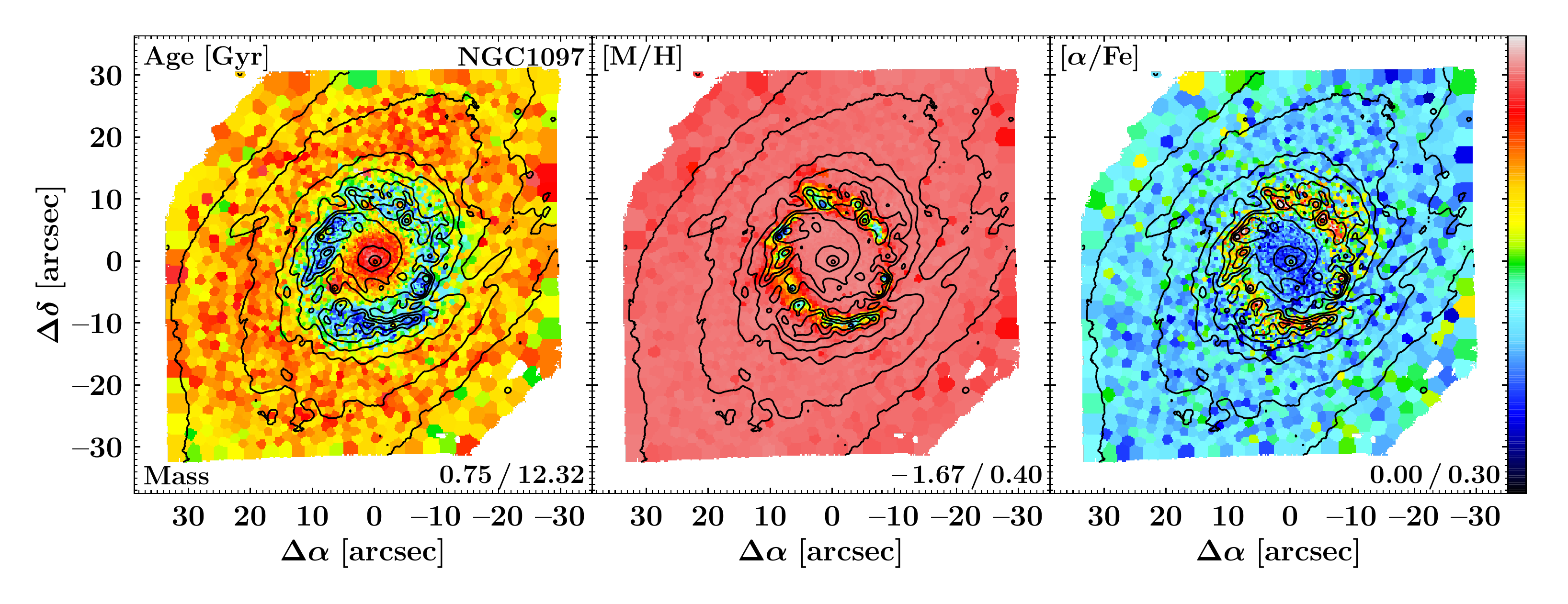}
        \end{minipage}%
        \begin{minipage}[c]{0.35\textwidth}
            \centering
            \includegraphics[width=0.95\textwidth]{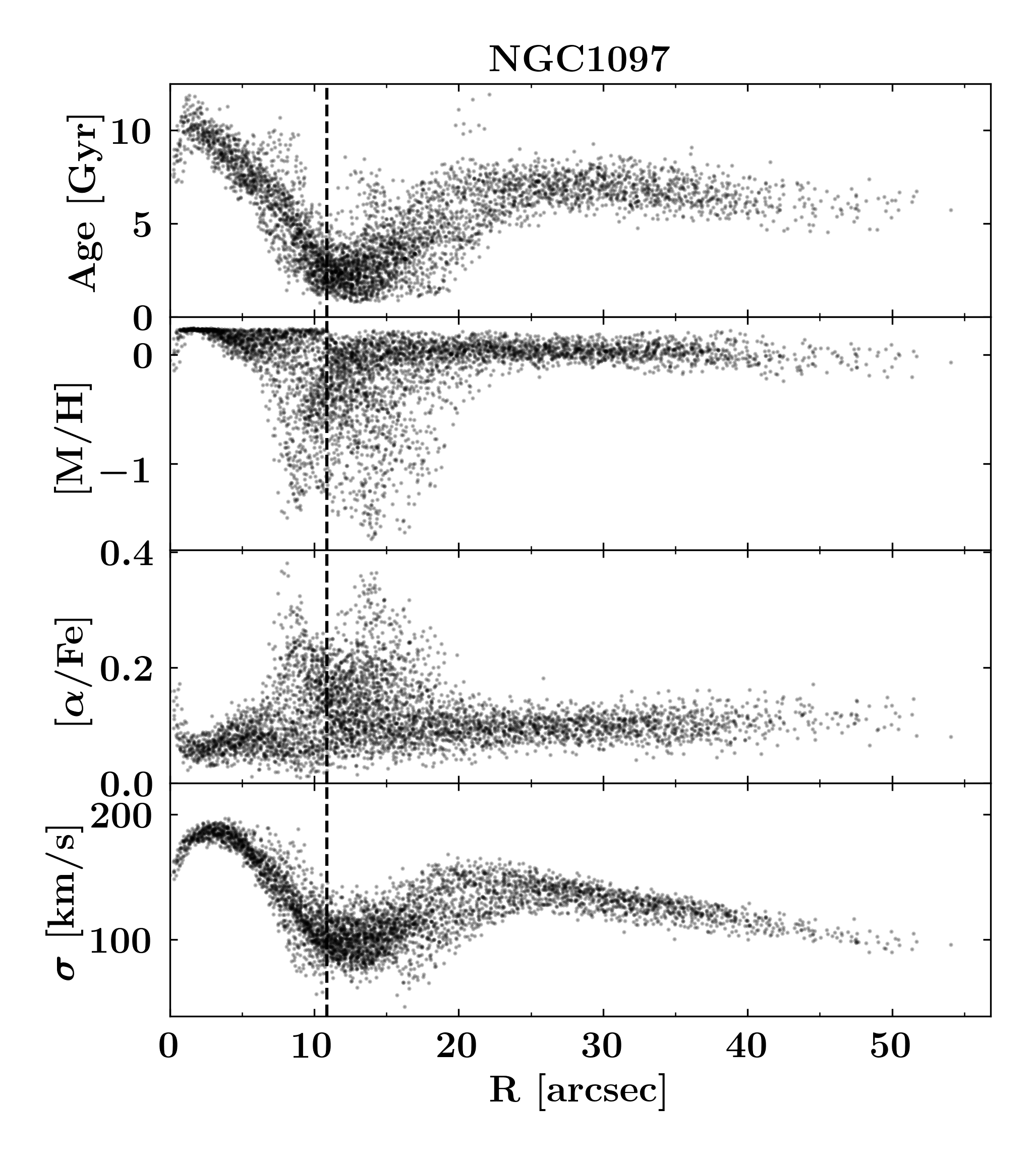}
        \end{minipage}
    \end{minipage}
    \rule{\textwidth}{0.6pt}
    \begin{minipage}[c]{\textwidth}
        \centering
        \begin{minipage}[c]{0.55\textwidth}
            \centering
            \includegraphics[width=\textwidth]{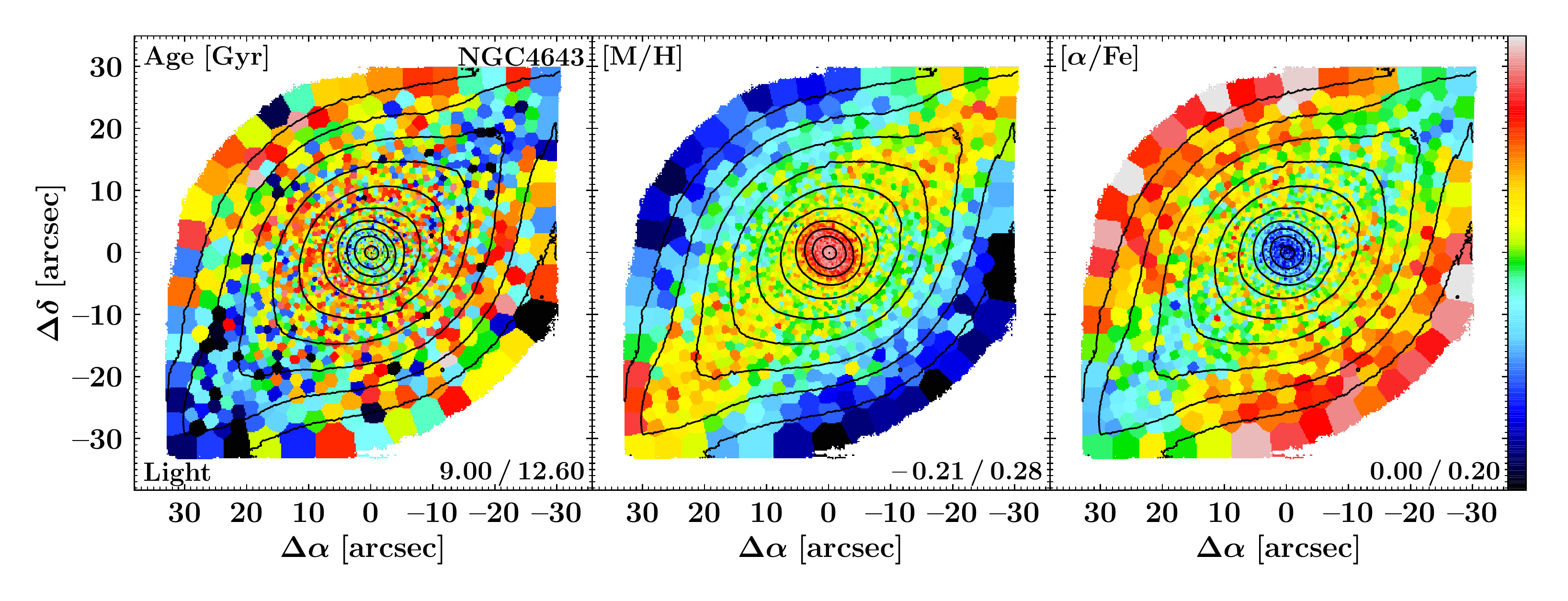}\\
            \includegraphics[width=\textwidth]{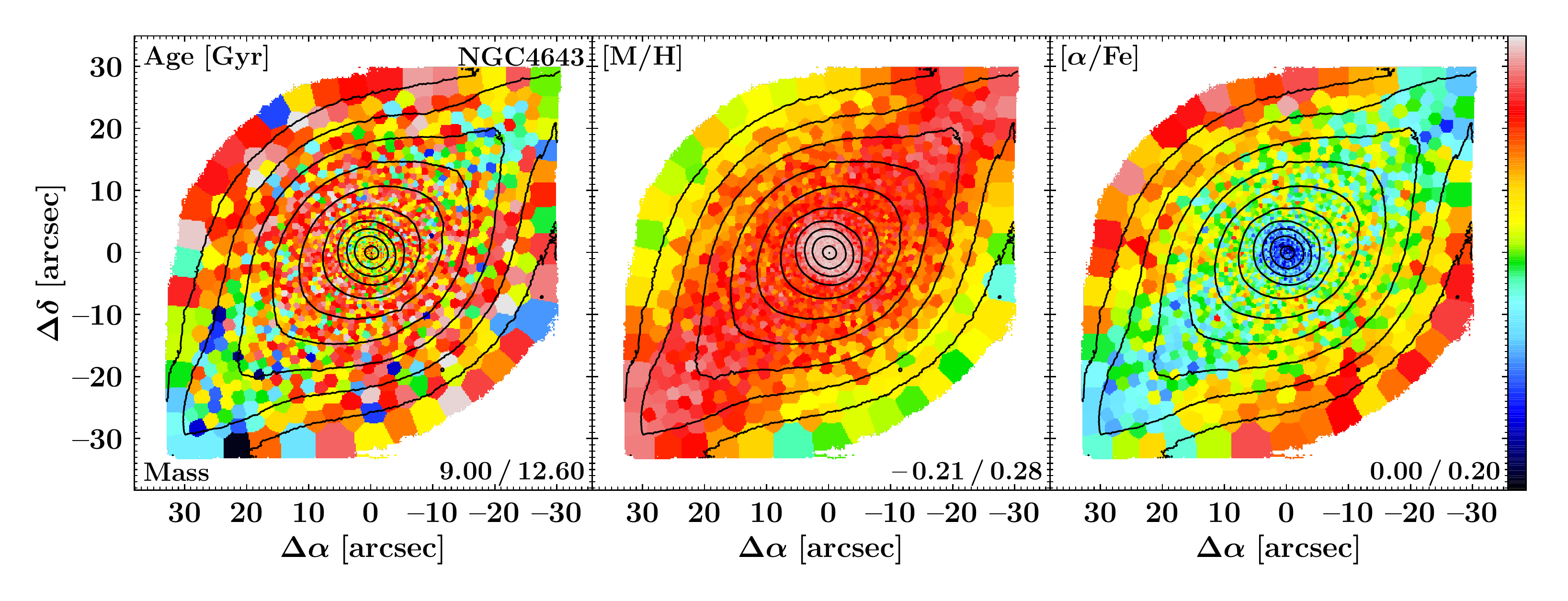}
        \end{minipage}%
        \begin{minipage}[c]{0.35\textwidth}
            \centering
            \includegraphics[width=\textwidth]{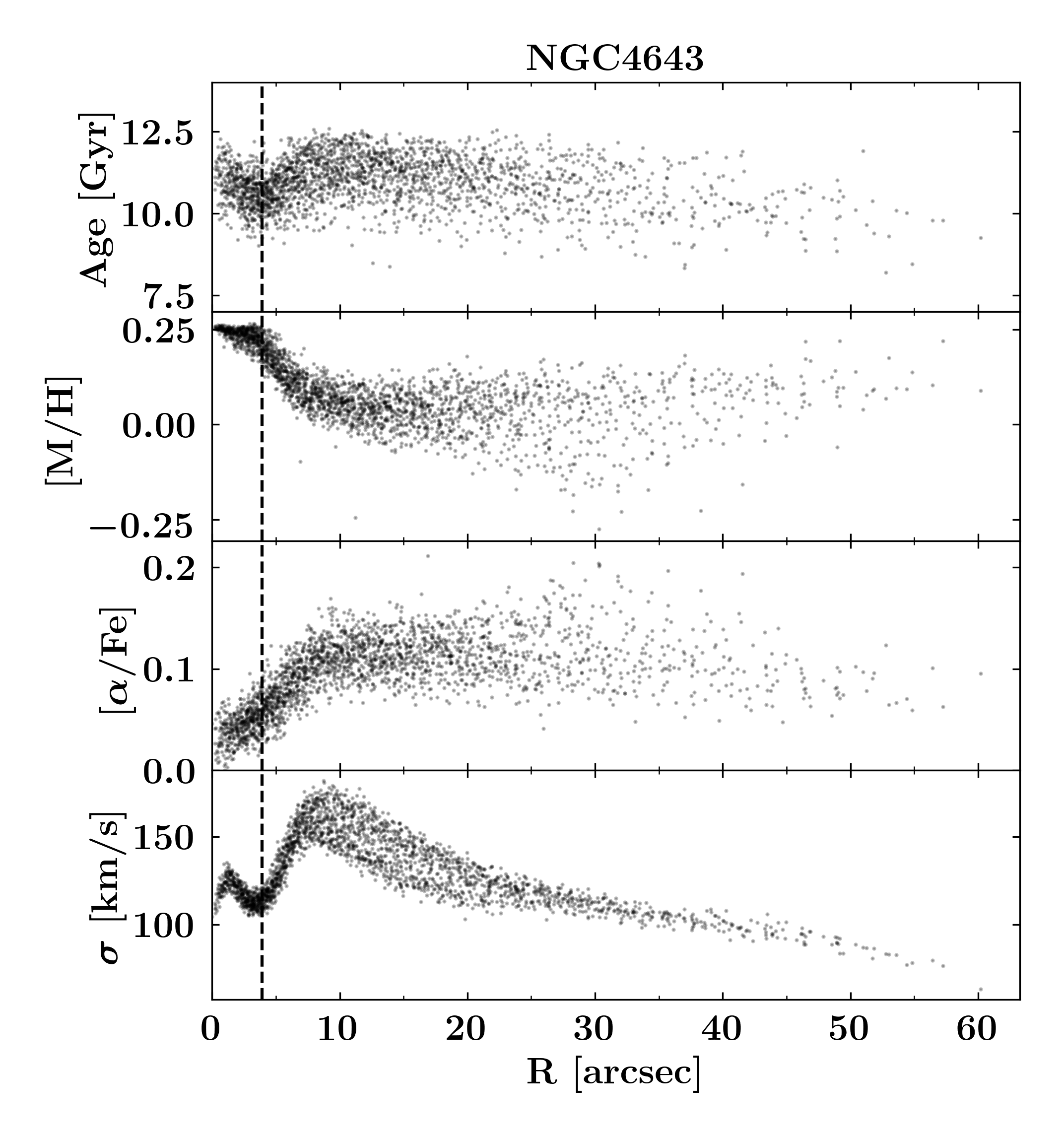}
        \end{minipage}
    \end{minipage}
    \caption{%
        \emph{Left:} Maps of the mean stellar population properties of NGC\,1097 from the star-forming subsample (upper
        half) and NGC\,4643 from the non-star-forming subsample (lower half). We present for each galaxy the
        light-weighted (upper panels) and mass-weighted (lower panels) population content.  The maps have been generated
        with {\ppxf} and include the modelling of {\alphaFe} abundances.  The figures display age, [M/H], and {\alphaFe}
        enhancements in the left-hand, centre, and right-hand panels, respectively.  The limits of the colour bar are
        stated in the lower-right corner of each panel.  Based on reconstructed intensities from the MUSE cube, we
        display isophotes in steps of \SI{0.5}{\mag}. North is up; east is to the left. 
        \emph{Right:} Radial profiles of light-weighted stellar ages (first panels), metallicities (second panels),
        {\alphaFe} enhancements (third panels), and velocity dispersions (fourth panels) as a function of the
        galactocentric radius of all spatial bins in the field of view.  The profiles have been deprojected using
        inclinations and position angles derived in S$^4$G \citep{munozMateos2015}, as presented in
        Table~\ref{tab:overview}. The vertical dashed lines represent the kinematic radii of the nuclear discs, which
        was defined in G20 as the radius at which $V/\sigma$ reaches its maximum in the region dominated by the nuclear
        disc. 
    }%
    \label{fig:sppMapsMaintext}
\end{figure*}

The kinematic analysis of the sample in an accompanying study (see G20) has already shown that nuclear discs are clearly
present in all TIMER galaxies except perhaps NGC\,6902, which is a rather peculiar object. These nuclear discs are also
found in the maps of the mean stellar population properties (see left columns of Figs.~\ref{fig:sppMapsMaintext} and
Appendix~\ref{sec:maps}). While these structures are unambiguously present, the general appearance of nuclear discs and
nuclear rings varies significantly between the star-forming and non-star-forming subsample. 

Nuclear discs in the non-star-forming subsample appear as well-defined discs with mean stellar populations that are
younger, more metal-rich, and less {\alphaFe} enhanced as compared to their immediate surroundings. Since these galaxies
do not host a starbursting nuclear ring, the maps of the mean stellar population properties do also not show signs of
such nuclear rings, i.e.\ significantly different populations at the outer edges of the nuclear discs.

A particularly notable nuclear disc in the non-star-forming subsample is hosted by NGC\,1291. Previous studies of the
TIMER team have already claimed that this galaxy hosts a nuclear disc with an effective radius of \SI{15.6}{\arcsec}, as
determined through photometric decompositions \citep{deLorenzoCaceres2019, mendezAbreu2019}, and the stellar populations
in its centre are dominated by an inner bar. This is most clear in the maps of metallicities and {\alphaFe} enhancements
where the inner bar is clearly visible. In addition, the ends of this inner bar exhibit younger stellar populations, as
compared to the rest of the field of view. \citet{deLorenzoCaceres2019} already discussed the different stellar
populations in the inner bar, nuclear disc, and all other components in this complex galaxy and we will complement this
analysis in a dedicated, forthcoming study. 
Only in NGC\,4371 the nuclear disc is not immediately apparent, but this is a result of projection effects.  The high
inclination of \SI{59}{\deg} together with the fact that the bar is seen almost end-on makes the detection of the
nuclear disc challenging. Nonetheless, a inspection of the stellar population map reveals the same trends found for the
other non-star-forming galaxies. The nuclear disc appears almost edge-on, with young, metal-rich, and {\alphaFe}
depleted stellar populations observed approximately \SI{10}{\arcsec} east and west of the galaxy centre. For a detailed
analysis of this galaxy we refer the reader to \citet{gadotti2015}. 

Nuclear discs in the star-forming subsample follow, in general, the same trends. Over the majority of the radial extent
of the nuclear discs these appear as well younger, more metal-rich, and {\alphaFe} depleted. However, the star-forming
nuclear rings, i.e.\ the outer edges of the nuclear discs, show significantly different stellar population properties.
These nuclear rings are often characterised by very low ages (often below \SI{1}{\Gyr}), exceptionally low metallicities
(even below [M/H] = -1.5), and significantly enhanced {\alphaFe} abundances (up to {\alphaFe} = 0.30).  In fact, these
distinctive nuclear rings spatially coincide with the regions of elevated {\Ha} emission-line fluxes and, based on the
standard BPT methodology, are attributed to effects from ongoing star formation or a starburst episode (see
Sect.~\ref{subsec:results_HalphaMaps}).  As discussed above, in these cases, the derived population properties might be
unreliable, as the light in the star-forming regions could be dominated by very young stellar populations (below
\SI{30}{\Myr}) which are not included in the employed set of SSP models, or affected by uncertainties in the {\Hb}
emission-line subtraction. 

In the surroundings of the nuclear discs other structural components  of the galaxies are evident. In some cases
(almost) the entire bar is covered by the MUSE field of view. In these galaxies (see e.g. IC\,1438, NGC\,4643,
NGC\,7755) the bar is visible through its elevated metallicities and low {\alphaFe} enhancements \citep[see
also][]{neumann2020}. Nonetheless, the metallicities and {\alphaFe} abundances of the bars are not as elevated/depleted
as those detected in the nuclear discs.  In almost all galaxies the nuclear discs are also surrounded by a region of
older stellar populations. A comparison with the kinematic analysis of G20 shows that these areas spatially coincide
with regions showing a correlation between radial velocity and the higher order moment $h_3$ of the line-of-sight
velocity distribution or drops in the higher order moment $h_4$ along the bar major axis. These are kinematic signatures
of the strongly elongated stellar orbits in bars and their vertically thickened, box/peanut structure. Therefore, the
regions of old stellar populations surrounding the young nuclear discs are probably related to the main bars in these
galaxies. 

The light- and mass-weighted stellar population maps compare generally well and show qualitatively the same results.
Mass-weighted maps indicate, as expected, systematically higher ages and metallicities, but similar (or in some cases
lower) {\alphaFe} abundances. Nuclear discs often appear less pronounced in mass-weighted age maps; however, this is
expected, as mass-weighted results highlight the old stellar component while nuclear discs are generally found to be
comparatively young.

Interestingly, the orientation of the nuclear ring in NGC\,1097 seems to vary between the map of stellar age and those
of metallicity and {\alphaFe} enhancement. While the age map the nuclear ring is elongated in the north-west direction
(along the bar major axis), the metallicity and {\alphaFe} maps indicate an elongation towards the north-east
(perpendicular to the bar major axis). However, this apparent discrepancy is an effect of the chosen colourbar limits in
the maps and the fact that ages are displayed linearly while metallicities are plotted on a logarithmic scale.  In fact,
the regions with the youngest stellar populations correspond precisely to the regions with the lowest metallicities and
{\alphaFe} abundances. Hence, there is no physical difference in the orientation of the nuclear between the age and
metallicity map.

\subsection{Radial profiles of mean stellar population properties}%
\label{subsec:results_sppProfiles}
Spatially resolved maps are an indispensable tool to investigate how stellar population properties vary between
different structural components of a galaxy.  Nevertheless, it can be instructive to reduce those maps to one dimension,
in order to emphasise, for instance, the dependency of the population properties on the galactocentric radius.  In the
right-hand side of Fig.~\ref{fig:sppMapsMaintext} and Appendix~\ref{sec:maps} we plot light-weighted stellar ages,
metallicities, {\alphaFe} enhancements, and velocity dispersions as a function of the deprojected galactocentric radius
of the respective spatial bin. 

All galaxies in the non-star-forming subsample exhibit well-defined radial profiles in the regions of the nuclear discs.
More specifically, we find that ages and metallicities are radially decreasing while {\alphaFe} abundances are
increasing with radius.  Interestingly, the lowest {\alphaFe} enhancements are always found in the very centre of the
galaxies and almost always reach values of \num{0}. Outside the nuclear disc the stellar ages increase again, and the
kinematic radii coincide with this turn-around point in age.  Similarly, metallicities and {\alphaFe} abundances
frequently show changes in their radial profiles close to the kinematic radius, often exhibiting flat profiles outside
the nuclear discs. The fact that the slopes of these profiles appear to be constant throughout the nuclear discs, and in
most cases to the very centre, supports the idea that nuclear discs are radially continuous components often extending
all the way to the centres of the galaxies.  Only NGC\,1291 and NGC\,4371 show more complicated radial profiles, but
this is due to the prominent inner bar in NGC\,1291 and the high inclination and projection effects in NGC\,4371 (see
above). 

In contrast, galaxies in the star-forming subsample show a more complicated behaviour. The galaxies show distinctive
extrema of low ages, low metallicities, and high {\alphaFe} enhancements, in particular at the outer edge of the nuclear
discs, i.e.\ their kinematic radii. These are again the nuclear rings with distinct stellar populations and high {\Ha}
emission-line fluxes discussed above. These features appear to be a result of ongoing star formation or a present
starburst. Thus, the well-defined radial trends found in the non-star-forming subsample might actually be present in all
galaxies but could be temporarily outshone by current star formation.  In fact, in various galaxies, for instance
NGC\,1097, the radial profiles inside of the starbursting nuclear ring appear similar to the well-defined gradients
observed in galaxies unaffected by star formation.  Only NGC\,5236 exhibits strongly irregular radial profiles in which
extrema of young ages, low metallicities, and increased {\alphaFe} abundances are not confined to the nuclear ring.
Instead, these are widely distributed within the nuclear disc, in line with our findings from the maps of the stellar
population content and {\Ha} emission-line fluxes. 

A few galaxies (e.g NGC\,1097, NGC\,4303, NGC\,4984) show abrupt changes of their population content in the very centres
of these galaxies. More precisely, these changes appear on radial scales much smaller than that of the nuclear disc,
i.e.\ are confined to the innermost spaxels. We speculate that these changes are connected to AGN activity and we will
explore this aspect in a dedicated TIMER study.

\subsection{Mean stellar population content in nuclear discs and nuclear rings}%
\begin{figure}
    \includegraphics[width=0.5\textwidth]{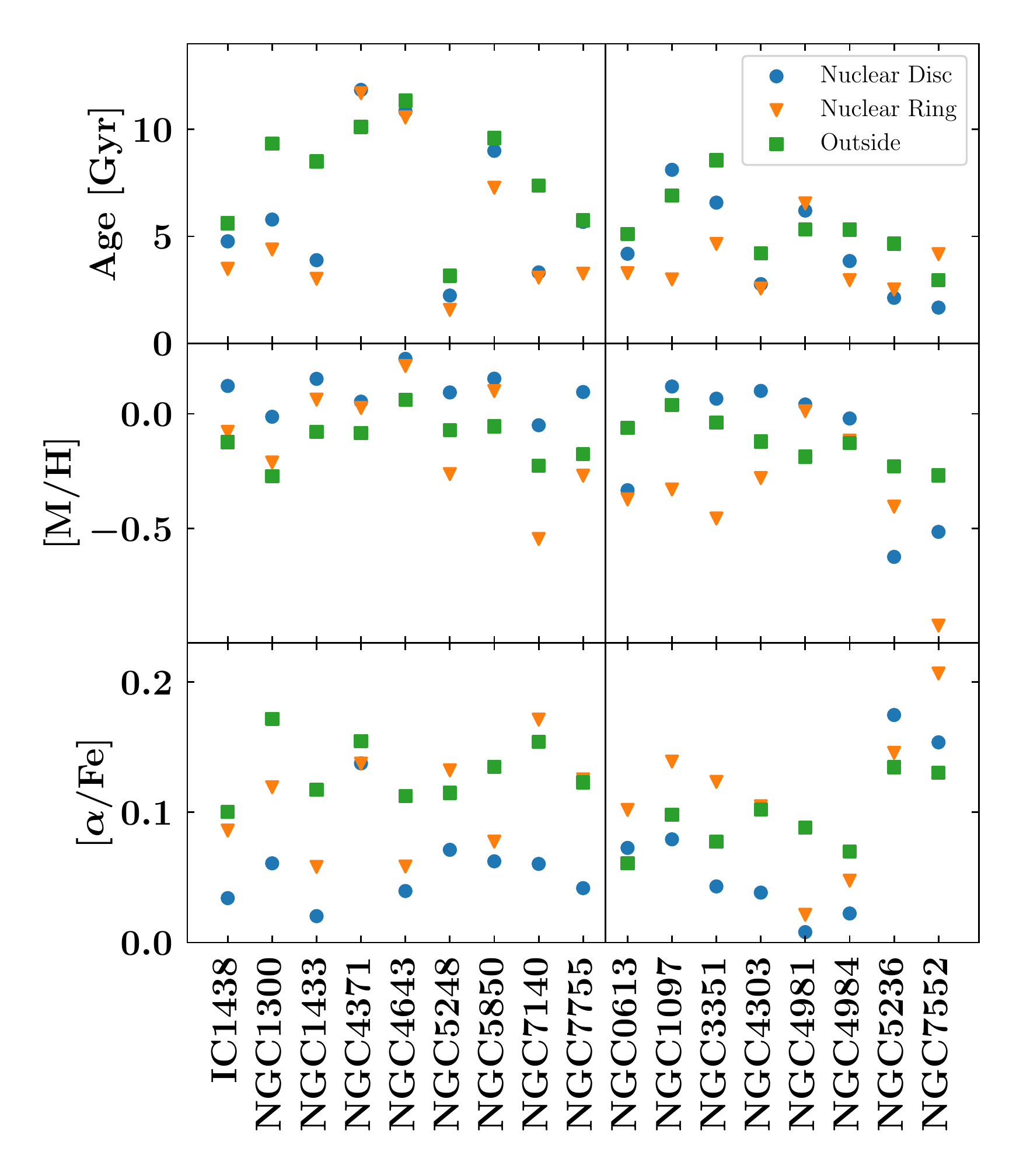}
    \caption{%
        Overview of light-weighted mean ages (upper panel), metallicities (central panel), and {\alphaFe} abundances
        (lower panel) in the nuclear discs (blue circles), nuclear rings (orange triangles), and outside of the central
        regions (green square). The light-weighted averages are calculated in the radial range from
        \SIrange{0.1}{0.7}{\rkin} for the nuclear discs, \SIrange{0.8}{1.2}{\rkin} for the nuclear rings, and
        \SIrange{2.0}{3.0}{\rkin} for the surrounding regions. The non-star-forming subsample is displayed in the left,
        separated by a vertical line from the galaxies in the star-forming subsample. As no kinematic radius could be
        determined for NGC\,1291, this galaxy is not included in this plot. 
    }%
    \label{fig:sppVsName}
\end{figure}

\label{subsec:meanPops_nuclearDiscs}
The maps and radial profiles presented above indicate that the stellar population content of nuclear discs and nuclear
rings are significantly different from the populations detected in their direct surroundings (e.g.\ the inner part of
the main bar). To better quantify these differences, we calculate light-weighted mean ages, metallicities, and
{\alphaFe} abundances in the radial region of the nuclear disc (\SIrange{0.1}{0.7}{\rkin}), nuclear ring
(\SIrange{0.8}{1.2}{\rkin}), and outside of these structures (\SIrange{2.0}{3.0}{\rkin}). Figure~\ref{fig:sppVsName}
illustrates the results. 

The figure highlights the large range of mean ages covered by nuclear discs, especially when considering the
non-star-forming subsample. While some galaxies (e.g. NGC\,5248) have young nuclear discs with ages around \SI{2}{\Gyr},
other galaxies (e.g. NGC\,4371, NGC\,4643) have nuclear discs which are dominated by old stellar populations with ages
above \SI{10}{\Gyr}. 

Regardless of the observed absolute mean ages, all nuclear discs in the non-star-forming subsample are younger, more
metal-rich, and less {\alphaFe} enhanced as compared to their immediate surroundings. Nuclear rings, as the outer edges
of the nuclear discs, often show slightly younger ages compared to the nuclear discs, as expected from the radial
profiles, and intermediate values of metallicities and {\alphaFe} abundances. Only in three cases (NGC\,5248, NGC\,7140,
and NGC\,7755) nuclear rings exhibit lower metallicities and elevated {\alphaFe} abundances, probably a result of weak
star-formation activity.  The only galaxy with a deviating behaviour is NGC\,4371, but this is a result of projection
effects, as discussed above.  The same trends are found for the star-forming subsample. In general, nuclear discs are
found to be younger, more metal-rich, and less {\alphaFe} enhanced. Nuclear rings often exhibit low metallicities and
elevated {\alphaFe} abundances. This dichotomy is a result of the ongoing star formation that is concentrated in the
nuclear rings but relatively low throughout the nuclear discs (see also Sect.~\ref{subsec:results_HalphaMaps}).  In
contrast to the non-star-forming subsample, more exceptions are evident, for instance the metal-poor and {\alphaFe}
enriched nuclear discs of NGC\,5236 and NGC\,7552, and the old nuclear discs in NGC\,1097 and NGC\,4981. In fact, the
mean values substantially depend on where and how violent star formation proceeds in the nuclear rings of these
galaxies.  Therefore we urge the reader to carefully inspect both maps and profiles of these galaxies in addition to the
mean values presented in Fig.~\ref{fig:sppVsName}. 


\section{Discussion}%
\label{sec:discussion}
In this section we put our observational results in the context of secular evolution. In addition, we discuss the
connection between nuclear rings and nuclear discs, the coevolution of bars and nuclear discs, as well as the possible
rejuvenation of an old nuclear disc in NGC\,1097. We further investigate the absence of central spheroids in the TIMER
sample in the context of galaxy formation.

\subsection{The connection between nuclear rings and nuclear discs}%
\begin{figure*}
    \centering
    \includegraphics[width=0.9\textwidth]{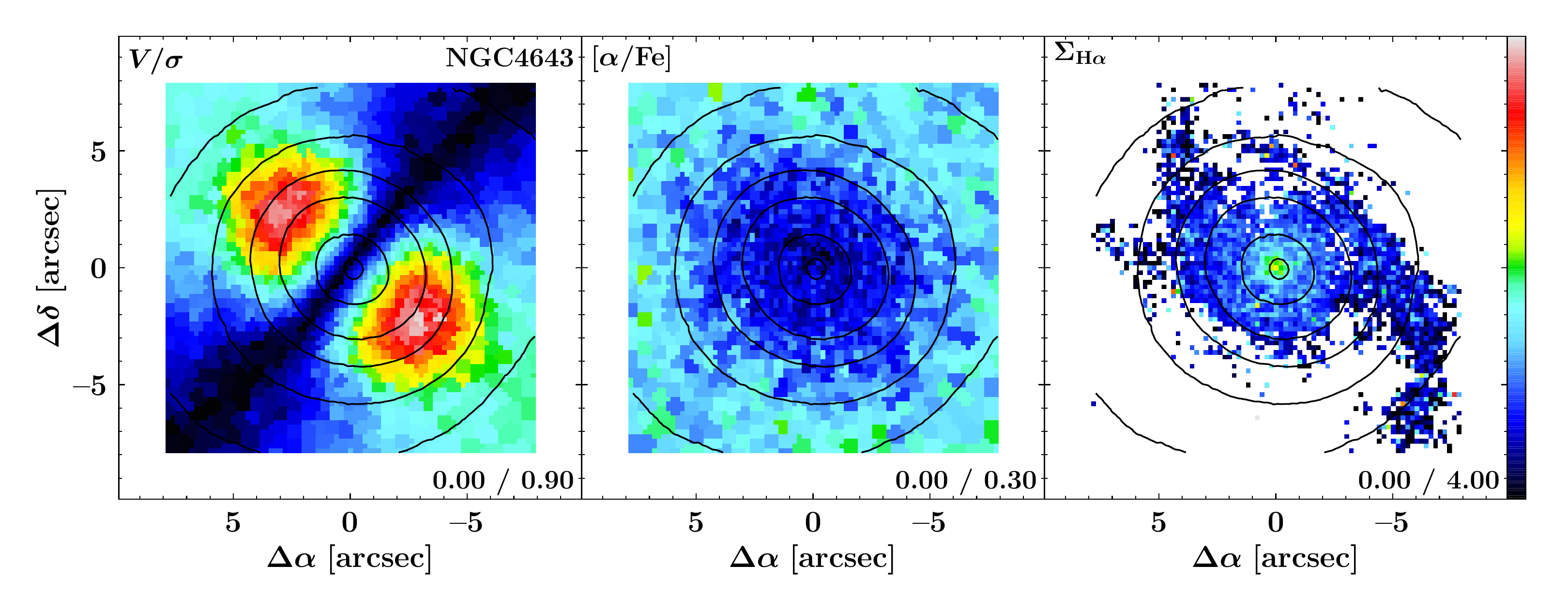}\\
    \includegraphics[width=0.9\textwidth]{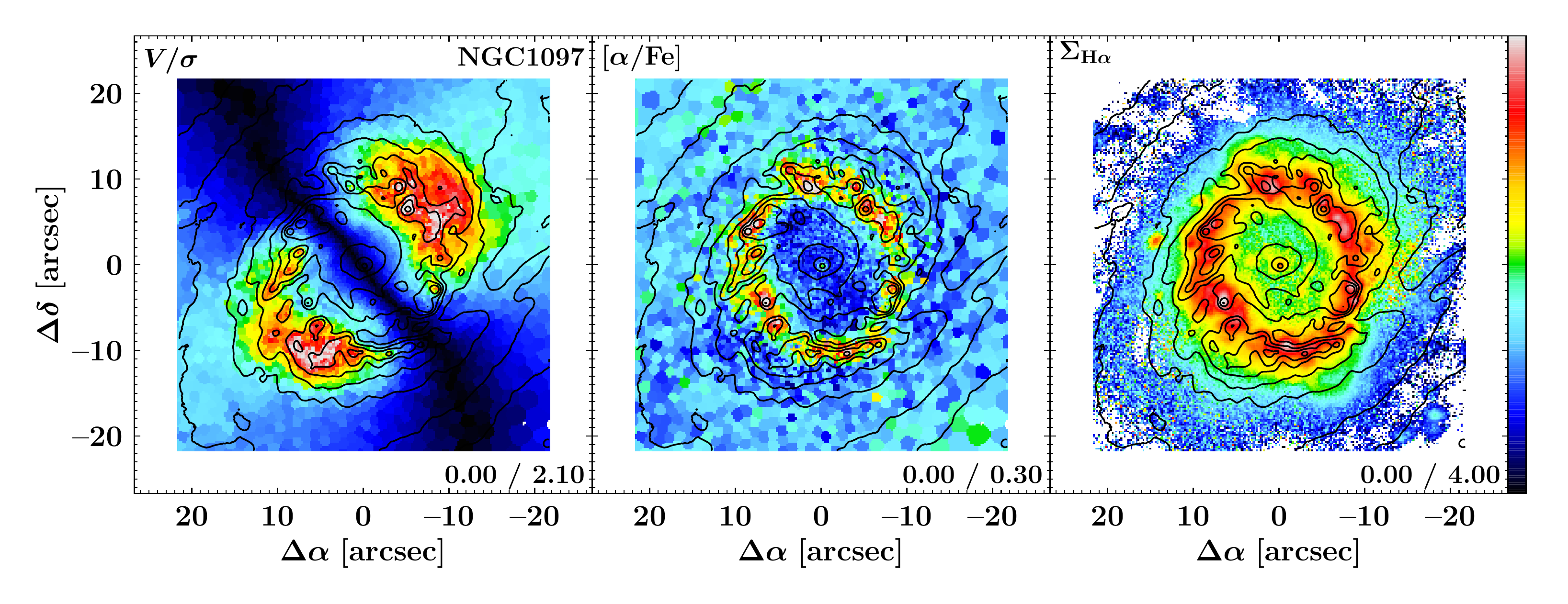}
    \caption{%
        Maps of $V/\sigma$ (left-hand panels), {\alphaFe} abundances (centre panels), and {\Ha} emission-line fluxes
        (right-hand panels) of the galaxies NGC\,4643 from the non-star-forming subsample (upper panels) and NGC\,1097
        from the star-forming subsample (lower panels). The maps do not display the full MUSE field of view, but focus
        on the region of the nuclear discs. The limits of the colour bar are stated in the lower- right corner of each
        panel.  Based on reconstructed intensities from the MUSE cube, we display isophotes in steps of \SI{0.5}{\mag}.
        North is up; east is to the left.
    }%
    \label{fig:overview_NGC1097_NGC4643}
\end{figure*}

\label{subsec:nuclearRingsVersusNuclearDiscs}
Nuclear discs and nuclear rings are intimately connected through their bar-driven formation histories. Nonetheless,
these structures are more or less prominent in different galaxies. Based on our high-resolution observations of stellar
population properties, and in the context of {\Ha} emission-line fluxes, as well as previously derived kinematic maps,
we explore similarities and differences between these two components in greater detail.
In Fig.~\ref{fig:overview_NGC1097_NGC4643} we illustrate these similarities and differences by presenting maps of
$V/\sigma$, {\alphaFe} abundances, and {\Ha} emission-line fluxes focussed on the spatial region of the nuclear disc.
We show the maps of the galaxy NGC\,4643 from the non-star-forming subsample and NGC\,1097 which is a poster child
example of a galaxy with a prominent, starbursting nuclear ring. 

In G20, we have shown that the kinematic signatures of nuclear discs in the TIMER sample are strikingly clear. Nuclear
discs are well aligned with the main disc, show stellar kinematics with a strong rotational support and low velocity
dispersion, and thus appear as kinematically distinct components in $V/\sigma$ maps. Based on such $V/\sigma$ maps, no
specific signatures of nuclear rings as a separate stellar component are evident, and thus the two structures appear
indistinguishable regardless of whether the galaxy is part of the star-forming or a non-star-forming subsample.  This is
further highlighted in the left-hand panels of Fig.~\ref{fig:overview_NGC1097_NGC4643} where we show maps of $V/\sigma$
for two TIMER galaxies, one with and the other without a star-forming nuclear ring. 

As shown above, nuclear discs are also spatially well-defined features in maps of mean stellar population properties,
show clear radial gradients in their stellar populations, and extend to the very centre of their host galaxy. However,
the stellar population properties at the outer edge of the nuclear discs depend significantly on whether a star-forming
nuclear ring is present or not (see central and right-hand panels of Fig.~\ref{fig:overview_NGC1097_NGC4643}). In the
cases of non-star-forming nuclear regions, one cannot distinguish nuclear rings and nuclear discs.  The outer edges of
these nuclear discs are their youngest, most metal-poor, and highest {\alphaFe} region, which only sometimes show faint
{\Ha} emission. Nevertheless, these features are smoothly connected to the radial gradients of the nuclear disc. 
In contrast, in the star-forming subsample nuclear rings are much more significantly young, metal-poor, and show very
high {\alphaFe} abundances, features that are also clearly visible as discontinuities in the radial profiles.  While
some of these findings might be spurious due to the fact that young stellar populations (\SI{< 30}{\Myr}) are not
included in the employed set of SSP models, it still illustrates how starbursting nuclear rings are distinguished in
their measured mean population properties.

These observations suggest that nuclear rings and nuclear discs should not be referred to as two separate physical
components, nor should their denomination be used interchangeably. Instead, nuclear rings are simply the outermost part
of nuclear discs. In galaxies with non-star-forming nuclear regions, the nuclear ring represents the youngest and
outermost part of the nuclear disc. In the galaxies of the star-forming subsample however, the nuclear ring merely
highlights the region in which the majority of the gas is located and efficiently forms stars. As above, this region is
the outermost part of the nuclear disc.

\subsection{Nuclear discs in the global context of secular evolution}%
\label{subsec:globalFormation}
A number of studies has established a theoretical framework for the formation of nuclear structures. If nuclear rings
and discs are indeed built by secular, bar-driven processes, this provides theoretical expectations on the stellar
kinematics and population properties of these structures. While in G20 we have recently shown that the kinematics of
nuclear discs, in particular their high rotational velocities and low velocity dispersions, are consistent with the
picture of bar-driven secular evolution, here we further consider their stellar population properties. 

To some extent, secular evolution in disc galaxies first requires a sequential process: a bar can only arise after the
main stellar disc becomes dynamically unstable to bar formation, while the formation of a nuclear disc fundamentally
depends on the presence of such a bar. More precisely, stars in the nuclear disc/ring form in-situ from gas that has
been funnelled to the centre by the bar (see e.g. \citealt{athanassoula1992a, athanassoula1992, piner1995}; but also
\citealt{kormendy2004} for a review). Following this sequential process, one naturally expects that nuclear discs are on
average younger compared to the other galaxy components. The finding that this holds for the mean stellar population
properties, as observed in Sect.~\ref{subsec:results_sppMaps}, is consistent with this picture.  More precisely, the bar
is efficient in transporting gas from the main galaxy disc to its centre. However, at least at low redshifts, there is
typically no star formation observed along the bar, as strong shear forces suppress the collapse of gas clouds
\citep[see e.g.][]{reynaud1998, emsellem2015, neumann2019}. Instead, the gas aggregates in a small volume in the centres
of the galaxies, which typically constitutes a nuclear ring (i.e.\ the outer edge of the nuclear disc). Star formation
progresses in this nuclear ring and, depending on the amount of available gas, can be very intense.  While star
formation might also proceed in other parts of the galaxy, in particular the main disc, at least in some cases the
star-formation density is higher in the nuclear rings.  This would result in a nuclear ring whose mean ages appear
younger compared to the main disc. Even if star-formation proceeds at similar rates in the nuclear ring and the main
disc, one expects at least similar ages in these structures.  In fact, similar ages in the nuclear ring and main disc
are, for instance, observed in IC\,1438. Both at the radius of the nuclear ring (\SI{3.7}{\arcsec}) and at the largest
radii in the field of view (e.g. \SIrange{20}{25}{\arcsec}) stellar ages of approximately \SI{3}{\Gyr} are evident.
However, we note that the present TIMER observations sample the main discs of the galaxies only partially. In fact, the
outer parts of the main disc might still appear younger than the nuclear discs.

The mean ages of the nuclear discs and rings vary significantly across different galaxies. In fact, the nuclear rings of
some nuclear discs are as young as \SI{\sim2}{\Gyr}, while other nuclear discs are older than \SI{10}{\Gyr} in their
entire radial extent. This is not surprising as the initial formation time of the nuclear disc depends on various
factors, such as the availability of gas, and is further limited by that of the bar itself and, thus, the cosmic epoch
at which the main galaxy disc first settled. In fact, according to this scenario, the oldest nuclear discs indicate an
early formation of the bar. 

The increased metallicities detected throughout all non-star-forming nuclear discs are also in agreement with our
present understanding of bar-driven secular evolution. As the nuclear disc is built from gas brought there from other
parts of the galaxy, in particular the main disc, this gas should typically be as metal-enriched as the gas in these
other galaxy components.  With the formation of the nuclear disc and subsequent generations of stars, the metallicity in
this central component should continuously increase. This is further supported by the fact that nuclear discs are
located in the centres of their host galaxies where the deep potential well confines the produced metals.  This result
also corroborates previous numerical studies. For instance, \citet{cole2014} finds increased values of [Fe/H] in nuclear
discs compared to the surrounding bar, as a result of continuing star formation in the nuclear discs. 
 
We have also, for the first time, derived {\alphaFe} element abundances in spatially resolved nuclear discs. We find
that all nuclear discs in the non-star-forming subsample have low {\alphaFe} enhancements, indicating that their
build-up takes a long time and is indeed a slow and continuous process, just as expected in the context of bar-driven
secular evolution. In fact, if nuclear discs were built by more violent processes such as mergers \citep{davies2007,
chapon2013}, one would expect a singular, rapid period of star formation that results in more elevated values of
{\alphaFe}, similar to those found in elliptical galaxies. In addition, the entire nuclear disc would be formed at the
same time, either through star formation or the violent re-distribution of existing stars. In both cases, one would
expect to find flat age profiles, in contrast to the observations presented here \citep[see][for a detailed account of
radial age profiles in galaxy bulges]{breda2020a}.  

The extremely low metallicities and increased {\alphaFe} abundances detected in the nuclear rings of the star-forming
subsample could result from uncertainties in the measurement (see Sect.~\ref{sec:analysis}).  However, such
low-metallicity populations in the nuclear rings could also be consistent with the bar-driven formation scenario, if
external, low metallicity gas is accreted onto the galaxy. Such low metallicity gas could originate from the
circum-galactic-medium of the galaxy, or encounters with dwarf galaxies.  If the subsequent bar-driven gas inflow to the
nuclear ring proceeds rapidly enough to avoid the enrichment of the gas in the main galaxy disc, a nuclear ring showing
very low metallicities can be formed.  The elevated {\alphaFe} enhancements in the nuclear rings are probably just a
transient signatures of the strong, ongoing star formation from low metallicity gas and not a persistent property of the
underlying stellar nuclear disc, in particular since all nuclear discs in the non-star-forming subsample show low
{\alphaFe} abundances.

\subsection{Do nuclear discs and bars grow simultaneously?}%
\label{subsec:nuclearDiscsGrowth}
The bar-driven transport of gas to the centre of a galaxy is a rather well-understood process. In the galaxy centre the
gas is usually deposited in a nuclear ring with a well-defined inner and outer edge, although it remains unclear what
physical process determines the width of these nuclear rings. In fact, the presence of gaseous nuclear discs without
more prominent gaseous nuclear rings is uncommon. However, the results above suggest that often stellar nuclear discs
extend from the radial region of the nuclear ring all the way to the very centre of the galaxy. These extended nuclear
discs are not only detected in the maps of stellar population properties presented here, but also found based on stellar
kinematics and photometric studies (see G20, and references therein).  To date, it remains unclear if and how initially
gaseous nuclear rings with well-defined inner and outer edges are transformed into the observed stellar nuclear discs
that extend to the very centres of the galaxies. 

It is also not clear what physical process determines the radius of the gaseous nuclear rings. Studies suggest that the
sizes of nuclear rings are related to bar resonances \citep[see e.g.][]{piner1995}, determined by viscous shear forces
\citep{sormani2018, sormani2020}, or simply set by the residual angular momentum of the inflowing gas
\citep{kim2012,seo2019}.  In Sect.~\ref{subsec:results_sppProfiles} we have shown that the kinematic radii are a good
tracer of the radius of nuclear discs and, thus, their nuclear rings. In an accompanying study, G20 show that these
kinematic radii correlate well with the bar length and other properties, corroborating the scenario in which nuclear
discs are built by bars.

Interestingly, many studies argue that bars grow longer and stronger as they evolve (see e.g. \citealt{athanassoula2003,
martinezValpuesta2006, gadotti2011}, but also \citealt{deLorenzoCaceres2020} for inner bars). If indeed bars grow longer
with time and the radii of nuclear rings depend in some way on the bar length, one expects that bars and nuclear rings
evolve simultaneously.  Following the mechanism suggested by \citet{seo2019}, as the bar grows longer it triggers the
inflow of gas from larger radii in the main galaxy disc. This gas has a larger residual angular momentum and, thus,
settles at larger radii in the galaxy centre resulting in a larger nuclear ring. 
%
However, the residual angular momentum of the inflowing gas might not only depend on the length of bar, but also on
various other bar properties, for instance the axial ratio of the bar. Therefore, one might question a direct causal
connection between the bar length and the radius of the gaseous nuclear ring. Other studies argue that the size of the
nuclear ring is not linked to the residual angular momentum of the gas, but instead limited by the radial extent of the
bar $x_2$ orbits, a parameter that fundamentally depends on the axisymmetric central mass concentration and the bar
pattern speed \citep[see e.g.][]{athanassoula1992a, athanassoula1992, fragkoudi2017a}. In fact, the bar pattern speed
typically decreases, as bars are transferring angular momentum to the disc and halo, while simultaneously increasing the
central mass concentration through gas inflow, thus allowing more extended $x_2$ orbits and the nuclear ring to grow
with time. In this framework, a correlation between bar length and nuclear ring radius might naturally arise as well,
without requiring a direct causal connection between nuclear ring radius and bar length. While it remains unclear which
physical mechanism determines the size of the nuclear rings, both mechanisms hint towards a scenario in which nuclear
rings continuously increase their radius as bars evolve. 

In this framework, star formation only needs to proceed in the gaseous nuclear ring in order to produce a continuous
stellar nuclear disc. As the radius of the location of the nuclear ring increases, a star-forming nuclear ring is
located at a given point in time at a range of radii, thus producing stars on near-circular orbits at all such radii,
or, in other words, a disc. In this way a continuous stellar nuclear disc can be formed without the necessity of forming
a gaseous nuclear disc. 

This possible scenario translates into predictions that are in good agreement with the nuclear discs found in the TIMER
sample.  While {\Ha} emission is predominantly detected in nuclear rings at the outer edge of the nuclear discs, the
nuclear discs themselves are continuous and extend to the centres of the galaxies.  It is also in agreement with the
fact that nuclear discs are rapidly rotating and exhibit low velocity dispersions. Similarly, it predicts that nuclear
discs are relatively old in the centre and become increasingly younger towards their outer edge. This behaviour is
evident in the radial gradients presented in Sect.~\ref{subsec:results_sppProfiles}.  We note that the negative
gradients in [M/H] do not necessarily contradict this picture. In fact, old stars are not always less metal-enriched
than young stars, in particular since metallicities can increase very rapidly in star-forming systems. In addition, the
more the gas is bound in the potential well of the galaxy, the more difficult it is for feedback processes to expel this
gas.  Therefore, it is expected that in most evolved systems, regardless of their formation history, the metallicity
increases towards the centre. 

This formation mechanism still predicts the presence of an inner edge of the stellar nuclear disc, in particular at the
radius at which the first gaseous nuclear ring formed.  Such an inner edge is not obvious from the observations
presented here. While a few galaxies show systematically different population properties in their very centre, these
features might well be caused by contamination of the stellar continuum from an AGN\@. In addition, such an inner edge
might be well beyond the resolution limit of our MUSE observations which is typically around \SI{100}{\pc} or less. 

\citet{seo2019} suggest the same mechanism for the formation of nuclear discs, based on a numerical study. They perform
simulations of individual Milky Way-like galaxies ($M_{\mathrm{disc}} = \SI{5e10}{\msol}$) with varying gas fractions
(between \SIlist{0;10}{\percent}) and velocity anisotropy parameters. Depending on the simulation set-up, they find that
the initial nuclear ring is very small (down to \SI{40}{\pc}) and subsequently grows as the bar grows longer and funnels
in gas from larger radii. Gas located in preceding nuclear rings is quickly consumed by star formation. In this way an
increase in radius of the nuclear ring of up to a factor of \num{10} is found, which matches the typical sizes of the
nuclear discs we find in TIMER\@. 

The proposed mechanism of nuclear disc formation is based on the assumption that gas and star formation are
predominantly located in the nuclear ring. While this is in agreement with the observational appearance of nuclear discs
and nuclear rings, the presence of gas within the nuclear disc itself is not surprising and does not contradict the
above picture. Star formation in the nuclear rings is often very strong and thus it is expected that stellar feedback
significantly affects the gas. More precisely, a fraction of the gas should be expelled from within the nuclear ring to
both larger and smaller radii, a process studied in detail for NGC\,3351 by \citet{leaman2019}. Therefore, the presence
of gas and continuing star formation within the nuclear disc is expected, although in small amounts as compared to the
nuclear ring. This further supports the continuing enrichment of the nuclear disc with metals, as evident in the
metallicity maps. In addition, this also explains why there is no sharp, step-like transition between the stellar
populations of the nuclear discs and the regions outside. Instead a gradual change in stellar populations is observed,
e.g.\ the typical V-shape in age, that might result from the contamination of these regions by gas expelled from the
star-forming nuclear ring. 

Our observational results in combination with the current theoretical framework of bar evolution hint towards an
inside-out formation of nuclear discs through a series of star-forming nuclear rings. Nonetheless, various crucial
questions related to their bar-driven formation remain unanswered, in particular which physical processes determine the
size and the width of nuclear rings. Further studies are needed to answer these questions, especially numerical models
of nuclear disc formation in a cosmological context. These studies will shed light on alternative scenarios (e.g.,
formation from a gaseous disc) that will have to reproduce the radial gradients of stellar age, metallicities, and
{\alphaFe} abundances reported in this study.

\subsection{The absence of kinematically hot spheroids}%
\label{subsec:absenceOfSpheroids}
In Sect.~\ref{subsec:results_sppProfiles} we find well-defined radial gradients in the stellar population properties of
nuclear discs. Above we argue that these gradients, in particular the negative gradient in stellar age, might suggest
that nuclear discs form out of consecutive generations of gaseous nuclear rings. However, one might speculate if such
gradients could also be caused by a superposition of physically different stellar components with the nuclear discs,
such as for instance a kinematically hot spheroid, the main galactic disc, or the bar and its box/peanut. 

In fact, a superposition of the nuclear disc and the main disc as well as bar is likely, in particular since these
components are dynamically expected to extend to the spatial region covered by the nuclear disc. Indeed, such a
superposition becomes evident in the observed kinematic properties (see G20), indicated by elevated values of the
higher-order moment $h_4$ of the Gauss-Hermite parametrisation of the line-of-sight velocity distribution.  This
kinematic analysis nevertheless indicates that the nuclear discs clearly dominate the stellar light, as rapidly rotating
discs with low velocity dispersions are evident. This shows that there is no major dispersion-dominated component in the
central region, excepts perhaps in NGC\,6902. A small, kinematically hot spheroid could, nevertheless, be located at the
very centre of the galaxies \citep[see][]{erwin2015}. In fact, due to the typically large photometric concentrations of
kinematically hot spheroids, one might expect that these components, if present, dominate the stellar light at least in
the innermost part of the nuclear discs. 

\begin{figure}
    \includegraphics[width=0.5\textwidth]{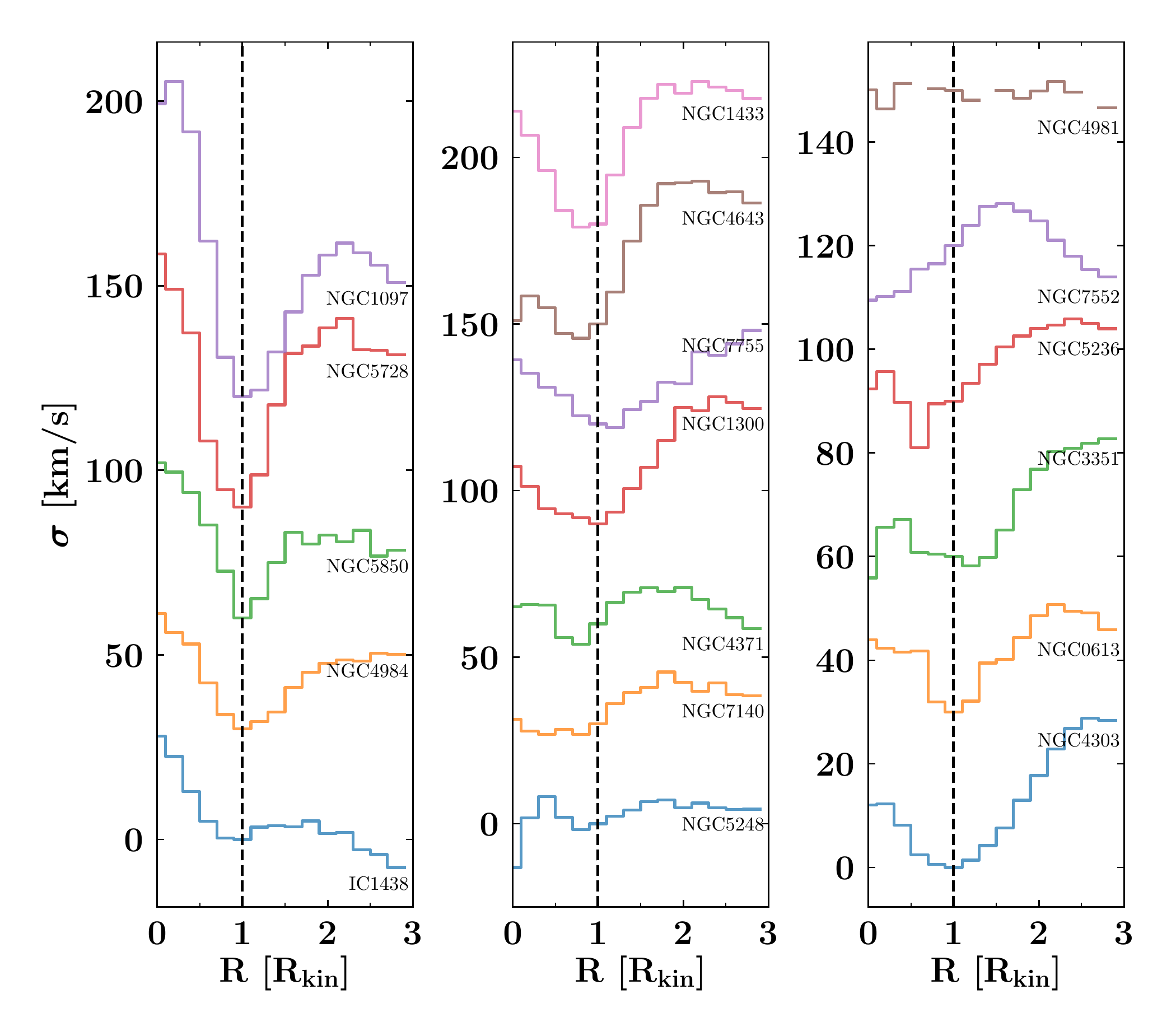}
    \caption{%
        Overview of the radial profiles of stellar velocity dispersions in the region of the nuclear disc. The left-hand
        panel shows all galaxies in which the velocity dispersions within the nuclear disc exceed the values observed in
        their direct surroundings. The central and right-hand panels display the remaining galaxies from the
        non-star-forming and star-forming subsample, respectively. The profiles are vertically offset by
        \SI{30}{\kilo\metre\per\second}, according to the values measured at the kinematic radius. Using the
        galactocentric radius, the data is binned in radial steps of \SI{0.2}{\rkin} and the median velocity dispersions
        displayed. The vertical dashed line highlights the kinematic radius of the nuclear discs. We note that
        NGC\,1291, NGC\,1365, and NGC\,6902 are not included in this figure, as no kinematic radii are available for
        these galaxies. Unbinned radial profiles without any vertical offset are provided in Appendix~\ref{sec:maps}. 
        }%
    \label{fig:overview_SigmaProfiles}
\end{figure}

To further assess the presence of small, kinematically hot spheroids, we carefully inspect radial profiles of the
stellar velocity dispersion. The majority of all galaxies show relatively low velocity dispersions throughout the
nuclear discs. Nonetheless, seven galaxies exhibit velocity dispersions in their centre which surpass the velocity
dispersions measured just outside of their nuclear discs.  These galaxies are IC\,1438, NGC\,1097, NGC\,1291, NGC\,4984,
NGC\,5728, NGC\,5850, and NGC\,6902, and their velocity dispersion profiles are shown in the left-hand panel of
Fig.~\ref{fig:overview_SigmaProfiles}. The central and right-hand panels display the remaining galaxies of the
non-star-forming and star-forming subsamples, respectively.
 
Although the large velocity dispersions in the centres of these galaxies can be connected to kinematically hot
spheroids, at least some of the signatures found here can be attributed to other galaxy components. More precisely,
NGC\,5728 is significantly influenced by AGN activity and a large-scale outflow. Thus, the increased central velocity
dispersion in this galaxy might not be a property of the underlying stellar component but related to the AGN\@. While
the other galaxies do not show such strong AGN activity, deviations from a single velocity dispersion profiles in the
innermost spaxels could indeed be related to AGN activity. 
In contrast, NGC\,6902 is only weakly barred and does not show clear signatures of a rapidly rotating nuclear disc (see
G20). The stellar velocity dispersion continuously increases towards its centre, consistent with the presence of a
kinematically hot spheroid that dominates the stellar light in the centre of this galaxy. On the other hand
\citet{deLorenzoCaceres2019} claim that NGC\,1291 and NGC\,5850 host a small, kinematically hot spheroid within their
(more prominent) nuclear discs \citep[see also][]{mendezAbreu2019, deLorenzoCaceres2019_2}, and thus constitute galaxies
with composite bulges.  Only the elevated central velocity dispersions in IC\,1438, NGC\,1097, and NGC\,4984 might
indeed be related to previously undetected kinematically hot spheroids.

\begin{figure}
    \includegraphics[width=0.5\textwidth]{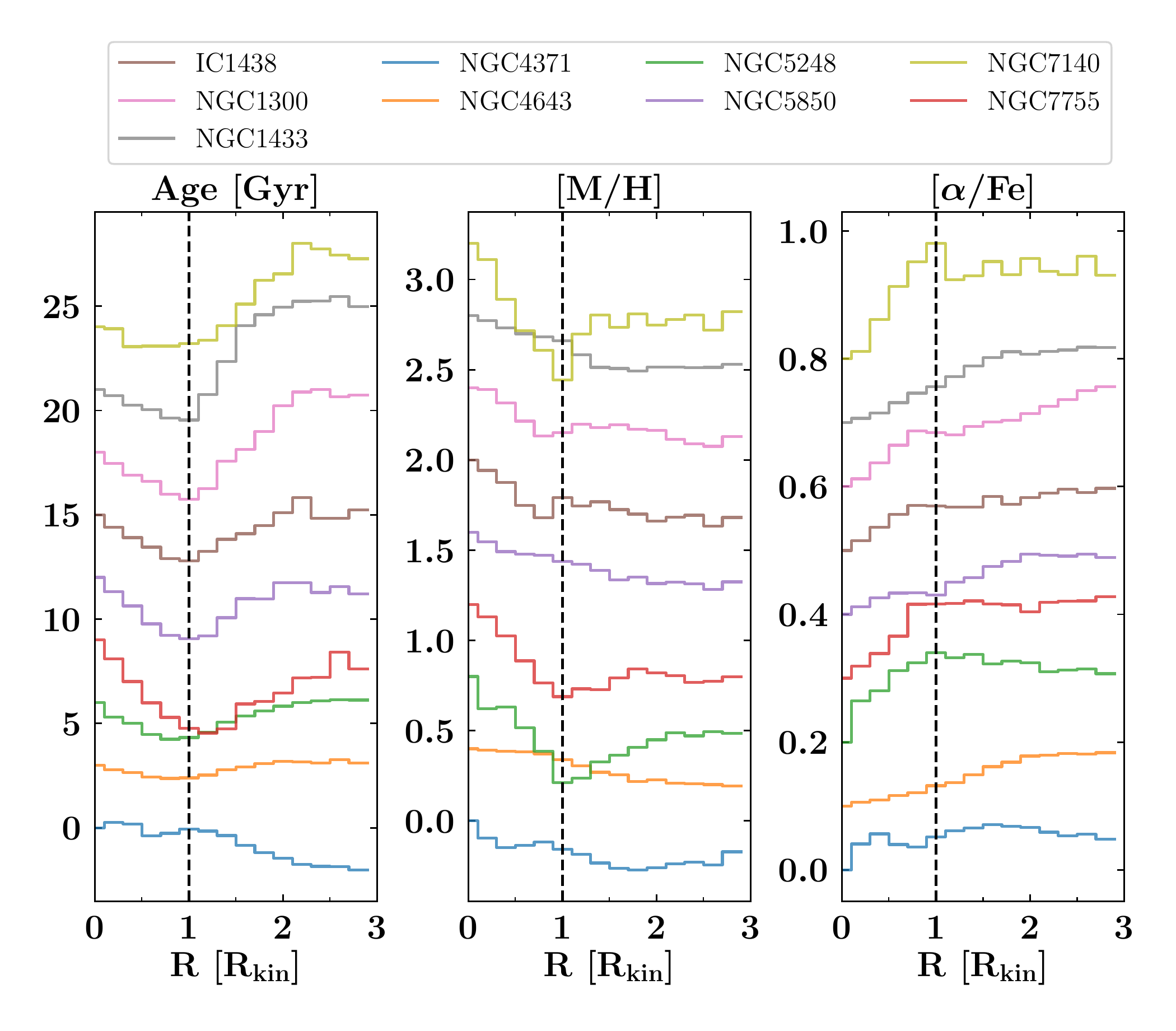}
    \caption{%
        Overview of the radial profiles of light-weighted ages (left-hand panel), metallicities (centre panel), and
        {\alphaFe} enhancements (right-hand panel) of the non-star-forming subsample. Using the galactocentric radius,
        the data is binned in radial steps of \SI{0.2}{\rkin} and the median stellar population properties displayed.
        The vertical dashed line highlights the kinematic radius of the nuclear discs.  The profiles are vertically
        offset by \SI{3}{\Gyr} for ages, and \SI{0.4}{\dex} and \SI{0.1}{\dex} for metallicities and {\alphaFe}
        abundances, according to the values measured in the innermost bin. We note that NGC\,1291 is not included in
        this figure, as no kinematic radius is available for this galaxy. Unbinned radial profiles without any vertical
        offset are provided in Appendix~\ref{sec:maps}.
    }%
    \label{fig:overview_sppProfiles}
\end{figure}

Signatures of kinematically hot spheroids are not only expected in the kinematics but also in the stellar population
properties. Therefore, we present an overview of the radial profiles of ages, metallicities, and {\alphaFe} enhancements
for the non-star-forming subsample in Fig.~\ref{fig:overview_sppProfiles}. 
These gradients, in particular that of stellar age, do not show breaks within the radial range of the nuclear disc. In
other words, in most galaxies with non-star-forming nuclear discs the slopes of the population gradients are constant
from the very centre to the nuclear ring. However, if there was a distinct physical component dominating in a given
radial range, one would expect the mean population properties to change there, in particular since nuclear discs are
relatively young components while kinematically hot spheroids are expected to be old. Such breaks should be further
enhanced by the different central concentrations of nuclear discs and hot spheroids. 
The only two galaxies in the non-star-forming subsample that show breaks in their age profiles are NGC\,4371 and
NGC\,1291. The deviations in NGC\,4371 may be a result of the peculiar geometrical projection of this galaxy, in which
the nuclear disc is seen almost edge-on (see \citealt{gadotti2015} for a detailed discussion, but see also
\citealt{erwin2015}). In NGC\,1291 the stellar populations in the nuclear disc are dominated by an inner bar and we
explore these stellar population profiles in a dedicated study.  Nonetheless, previous studies of the TIMER team suggest
that the centre of NGC\,1291 is dominated by a kinematically hot spheroid \citep{deLorenzoCaceres2019, mendezAbreu2019}.
This idea is consistent with the stellar population and velocity dispersion profiles presented here: the oldest ages,
highest metallicities, and highest velocity dispersion are detected in the centre of the galaxy. In addition, the radial
gradients of the hot spheroid clearly deviate from the flat gradients associated with the inner bar and nuclear disc. 

Identifying possible breaks in the stellar population profiles of the star-forming subsample is challenging, as any
gradients are often outshone or confused with the starbursting regions. Breaks in the stellar population profiles that
do not seem to be associated with star-forming regions are only detected in a few galaxies, namely NGC\,1097, NGC\,4303,
NGC\,4984, and NGC\,5728.  However, the breaks in NGC\,1097 and NGC\,4984 are confined to the innermost spaxels, hinting
towards a possible contamination with light from the AGN in these galaxies \citep{gadotti2019}. The same seems to be the
case for NGC\,5728 which is affected from a large-scale AGN outflow. These aspects will be explored further in a
forthcoming study of the TIMER team. Only NGC\,4303 shows a significant break in its age profile at around
\SI{0.5}{\rkin}, but not so in the profiles of [M/H], {\alphaFe} abundances, and velocity dispersions. In fact, the low
central velocity dispersion of approximately \SI{75}{\kilo\metre\per\second} does not hint towards the presence of a
kinematically hot spheroid. 

Interestingly, the inspection of the stellar population profiles of NGC\,6902 is consistent with the presence of a
massive, kinematically hot spheroid in this galaxy. In contrast to the non-star-forming nuclear discs, the stellar
populations in the centre of this galaxy are comparably old and show high velocity dispersions. In addition, no
well-defined, single gradients are observed. Moreover, this galaxy lacks clear kinematical signatures of a rapidly
rotating nuclear disc but shows a v-$h_3$ anticorrelation at all radii from the main disc of the galaxy to its central
region (see G20). In other words, NGC\,6902 exhibits properties that are consistent with a the presence of a
kinematically hot spheroid as well as a regularly rotating disc. 

In fact, kinematically hot spheroids are expected to be very old, as these are typically thought to form in the early
phases of galaxy formation\footnote{If a substantial merger had happened within the last few \si{\Gyr}, its effects on
the galaxy would be directly observable, for instance as tidal tails.}. But as described above, often we see that the
oldest stars in the TIMER galaxies reside in the bar and not in the centre. This fact, connected to the typical absence
of breaks in the stellar population gradients from the nuclear ring all the way to the centre and the relatively low
velocity dispersions in the galaxy centres, suggests that in almost all TIMER galaxies no kinematically hot spheroids
are present. More precisely, the stellar light in the centres of these galaxies is dominated by the nuclear disc and not
by kinematically hot spheroids.  Only NGC\,6902 shows some signatures that are consistent with the presence of a
dynamically hot spheroid and in NGC\,1291 and NGC\,5850 such a spheroid dominates only at radii much smaller than that
of the nuclear disc.  In a few cases the evidence is unclear, but the unambiguous presence of nuclear discs, as inferred
from the kinematic and stellar population analysis, shows that these components dominate the stellar light. 

Nonetheless, some minor contribution from kinematically hot spheroids might be present, but remain undetected in the
analysis of the mean stellar population properties performed here. A careful analysis of star formation histories could
provide further insights and show to what extent underlying, old stellar populations are present in the nuclear discs of
the TIMER galaxies. While this analysis is beyond the scope of this paper, we will report on this aspect in a
forthcoming study. 

At least in the context of a $\Lambda$CDM cosmology, kinematically hot spheroids are expected in the central regions of
massive disc galaxies. However, the considerations above show that in the large majority of the present TIMER sample no
kinematically hot spheroids, not even small ones, contribute substantially to the stellar light in the nuclear regions.
How much this is a challenge to $\Lambda$CDM cosmology is not clear.  For example, the large majority of the bulges
found in the Auriga cosmological simulations are classified as pseudo or composite bulges, while none are identified to
be kinematically hot spheroids \citep{gargiulo2019}. Moreover, in approximately \SI{20}{\percent} of the Auriga galaxies
the fraction of stars in the galaxy centres that formed ex-situ (and thus would end up in a spheroid) is below
\SI{1}{\percent} \citep[][]{fragkoudi2019}.  In addition, strong feedback processes may preferentially remove low
angular momentum gas in mergers, allowing for the formation of a disc but preventing the formation of central,
kinematically hot components \citep{brook2011, brook2012}. On the other hand, \citet{kormendy2010} also highlight on
statistical grounds that the large presence of massive bulgeless galaxies is a challenge in the current galaxy formation
paradigm \citep[see also][and references therein]{kormendy2016}. Clearly, this issue has to be addressed with further
analyses of the results from cosmological simulations and a TIMER-like study including a wider variety of galaxies.

\subsection{Rejuvenation of an old nuclear disc in NGC\,1097?}%
\label{subsec:rejuvenationOfNGC1097}
NGC\,1097 is a poster child example of a galaxy with a bright, starbursting nuclear ring with a clearly defined inner
and outer edge.  However, in the framework of the downsizing scenario it is a rather intriguing case. The downsizing
scenario \citep[see e.g.][]{cowie1996,thomas2010,sheth2012} predicts that more massive galaxies form earlier, i.e.\
their discs become dynamically mature at an earlier cosmic epoch resulting in an earlier bar formation \citep[see
e.g.][]{sheth2008}. Hence, considering the large stellar mass of NGC\,1097 \citep[\SI{17.4e10}{\msol}; see
e.g.][]{gadotti2019}, one would expect a comparably old bar and for the same reason also an old nuclear disc. However,
although this galaxy is the most massive one in the present sample, it exhibits a prominent and extremely young
starbursting ring, in contrast to the less massive but extremely old galaxies in TIMER \citep[see e.g. NGC\,4371, as
discussed by][]{gadotti2015}. In the following, we discuss how an interaction of NGC\,1097 with a small companion galaxy
could explain this peculiarity, consistent with the expectations from the downsizing scenario. 

As discussed above and illustrated in Fig.~\ref{fig:overview_NGC1097_NGC4643}, the starbursting ring of NGC\,1097 is
characterised by high {\Ha} fluxes, young ages, extremely low metallicities and elevated {\alphaFe} abundances. While
the {\Ha} measurements are robust, the derived population properties might be erroneous (see
Sect.~\ref{subsec:results_HalphaMaps}). In particular, the extremely low metallicity and elevated values of {\alphaFe}
could simply be a result from ongoing star formation.  Nevertheless, the galaxy seems to show signatures of an old
nuclear disc within the nuclear ring, in particular at radii below \SI{\sim7}{\arcsec}. This feature exhibits well
defined gradients in its population properties, just as found for many other nuclear discs in the sample. 

A possible interpretation for this peculiar composition of stellar population and galaxy properties could be the
rejuvenation of an old nuclear disc, in particular since NGC\,1097 is interacting with its companion NGC\,1097A
\citep[see e.g.][]{ondrechen1989,prieto2019}. 
In this picture, the bar forms very early, as expected within the downsizing scenario, and naturally triggers the
formation of a nuclear disc.  The subsequent evolution of the galaxy presumably continued with star formation in the
nuclear ring, as discussed in Sect.~\ref{subsec:nuclearDiscsGrowth}, or proceeded more quiescent, similar to the current
state of, for instance, NGC\,4643.
At a later point in time, NGC\,1097 started to interact with its small companion. In fact, the tidal forces exerted by
this companion might efficiently promote the inflow of gas. This inflowing gas could originate from the outskirts of the
main disc of NGC\,1097, the companions itself, or the surrounding circumgalactic medium. Our observations are not suited
to determine the precise origin of the gas, but if it indeed originated from outside of the main disc of NGC\,1097, it
is expected to be metal-poor. Although the measured low metallicities in the nuclear ring are somewhat unreliable, this
measurement is well consistent with this picture. If this is true, it would further indicate that the gas inflow
proceeds very rapidly hence preventing any significant metal enrichment of the gas in the main galaxy disc, before its
accumulation in the nuclear ring.

In summary, the set-up of population properties in the centre of NGC\,1097 might simply originate from the rejuvenation
of an old nuclear disc by an interaction-driven gas inflow. However, the discussed picture assumes that the inner region
of NGC\,1097 indeed contains an old nuclear disc. While the gradients of the population properties in this region look
very similar to those of other TIMER galaxies with nuclear discs unaffected by significant star formation, the kinematic
analysis indicates elevated velocity dispersions in the centre. This is rather untypical for nuclear discs, but there
might be mechanisms that efficiently heat them. It remains unclear if the passage of the companion could be responsible
for the heating of the nuclear disc. In fact, the high velocity dispersion might as well be explained by the presence of
a small, kinematically hot spheroid (see Sect.~\ref{subsec:absenceOfSpheroids}). 


\section{Summary and conclusions}%
\label{sec:summary}
We have used MUSE observations of the central regions of 21 massive, barred galaxies obtained within the TIMER survey.
The galaxies exhibit a large variety of bar-built central structures, such as nuclear rings, nuclear discs, and inner
bars. In this study we have derived high-resolution, spatially resolved maps of their mean stellar population properties
in order to determine present properties of nuclear rings and nuclear discs, and further investigate processes related
to their formation and evolution. To this end, we performed the analysis with both the {\ppxf} and {\steckmap} routines
and found that the obtained results are consistent within standard deviations of \SI{1.06}{\Gyr} in age and
\SI{0.14}{\dex} in metallicity, yet with {\steckmap} returning systematically lower stellar ages. In addition, we show
that derived ages and metallicities do not depend on whether {\alphaFe} enhancements are modelled in the fitting process
or not. Our main results are as follows:

\textbf{(i)}
  Nuclear discs are clearly distinguished from other galaxy components by their mean population properties. All
  non-star-forming nuclear discs appear younger, more metal-rich, and less {\alphaFe} enhanced compared to their
  immediate surroundings, as expected in the framework of bar-driven secular evolution. In particular, these findings
  corroborate that the formation of nuclear discs is a slow and continuous process thus clearly contradicting the idea
  of merger-built nuclear discs. 

\textbf{(ii)} 
  The mean ages of nuclear discs vary significantly across different galaxies. While some nuclear discs/rings are very
  young and star forming, other objects exhibit extremely old nuclear discs that appear to have evolved quiescently over
  the past \SI{\sim10}{\Gyr}. Based on the example of NGC\,1097, we discuss how such old nuclear discs can be
  rejuvenated by recently accreted gas. 

\textbf{(iii)} 
  In all non-star-forming nuclear discs, we detect exceptionally well-defined radial gradients of the mean population
  properties. More specifically, stellar ages and metallicities are decreasing with radius with a single slope within
  the nuclear disc while {\alphaFe} enhancements are increasing, again with a single slope. Interestingly, the lowest
  {\alphaFe} abundances are always found in the very centre. The absence of breaks in these gradients within the nuclear
  discs suggest that nuclear discs are extending all the way from the nuclear ring to the very centre of the galaxies.
  Breaks in these population profiles, associated with the outer edge of nuclear discs, are consistent with their
  kinematic radii determined by G20. 

\textbf{(iv)} 
  To date, it remains unclear how radially extended nuclear discs can be formed by star formation that typically
  proceeds in radially well-defined and relatively narrow nuclear rings. Based on the detected radial stellar population
  gradients and recent observational and numerical results, we argue that nuclear rings and bars evolve simultaneously,
  resulting in nuclear rings that grow in radius. In this way a continuous nuclear disc could simply be built by a
  series of nuclear rings that have increasing radii with time. 

\textbf{(v)} 
  Combining the results from this study with those in G20 we find no clear evidence for large, kinematically hot
  spheroids in most of the sample. This is indicated by the smooth stellar population gradients within the nuclear discs
  and the fact that the very centre is often dominated by a stellar population that is younger, or has the same age, as
  that in the bar and low stellar velocity dispersion. Most of the galaxies in fact show no signatures of even a small
  spheroid.


\begin{acknowledgements}
    We thank the referee for a prompt and constructive report.
    Based on observations collected at the European Southern Observatory under programmes 060.A-9313(A), 094.B-0321(A),
    095.B-0532(A), and 097.B-0640(A). 
    J. F-B, AdLC, and PSB acknowledge support through the RAVET project by the grants PID2019-107427GB-C31,
    AYA2016-77237-C3-1-P, and AYA2016-77237-C3-2-P from the Spanish Ministry of Science, Innovation and Universities
    (MCIU).  J. F-B and AdLC acknowledge support through the IAC project TRACES which is partially supported through the
    state budget and the regional budget of the Consejería de Economía, Industria, Comercio y Conocimiento of the Canary
    Islands Autonomous Community.  The Science, Technology and Facilities Council is acknowledged by JN for support
    through the Consolidated Grant Cosmology and Astrophysics at Portsmouth, ST/S000550/1.  PC acknowledges financial
    support from Fundação de Amparo à Pesquisa do Estado de São Paulo (FAPESP) process number 2018/05392-8 and Conselho
    Nacional de Desenvolvimento Científico e Tecnológico (CNPq) process number 310041/2018-0.  JMA acknowledges support
    from the Spanish Ministry of Economy and Competitiveness (MINECO) by grant AYA2017-83204-P. GvdV acknowledges
    funding from the European Research Council (ERC) under the European Union's Horizon 2020 research and innovation
    programme under grant agreement No 724857 (Consolidator Grant ArcheoDyn).  TK was supported by the Basic Science
    Research Program through the National Research Foundation of Korea (NRF) funded by the Ministry of Education (No.
    2019R1A6A3A01092024).
    This research has made use of the SIMBAD database, operated at CDS, Strasbourg, France \citep{wenger2000}; NASA's
    Astrophysics Data System (ADS); Astropy (\url{http://www.astropy.org}), a community-developed core Python package
    for Astronomy \citep{astropy2013,astropy2018}; NumPy \citet{numpy2006}; SciPy \citep{scipy2020}; and Matplotlib
    \citep{matplotlib2007}.
\end{acknowledgements}


\bibliographystyle{aa}
\bibliography{literature}


\begin{appendix}
\section{Descriptions of individual galaxies}%
\label{app:mapdescriptions}
\indent\indent
\textbf{IC\,1438:}
  This is a typical example of a barred galaxy with a nuclear disc. The bar is fully included in the field of view and
  characterised by high [M/H] and low {\alphaFe} abundances. The nuclear disc shows even higher metallicities and lower
  {\alphaFe} enhancements, as compared to the bar. A young nuclear disc is clearly visible and surrounded by a region of
  older stellar populations.

\textbf{NGC\,613:}
  The nuclear ring of this galaxy appears rather asymmetric, in particular in north-eastern part of the ring where a
  region of old stellar populations is detected. However, this detection of old stellar populations is probably not
  real: a visual inspection of the spectra show a broad emission-line component that is not included in the
  emission-line modelling performed here. According to high {\Ha} emission-line fluxes, the ring is starbursting and we
  detect low metallicities and high {\alphaFe} abundances in the ring. 

\textbf{NGC\,1097:} 
  A poster child example of a galaxy with a starbursting nuclear ring. Interestingly, the nuclear ring shows a
  significant width, especially in age, an exceptionally low metallicity, and elevated {\alphaFe} abundances. Since this
  galaxy is currently undergoing an interaction \citep[see e.g.][]{ondrechen1989, prieto2019}, one might suspect this
  interaction as origin of the low metallicity gas.  Within the starbursting nuclear ring, a nuclear disc with its
  typical gradients is evident. See Sect.~\ref{subsec:rejuvenationOfNGC1097} for a detailed discussion of this galaxy.

\textbf{NGC\,1291:}
  This quiescent galaxy highlights a very prominent inner bar that almost fills the field of view. While the main bar is
  almost not visible, the inner bar appears clearly distinguished by its increased metallicity and low {\alphaFe}
  enhancement.  Interestingly, the ends of the inner bar show slightly younger stellar populations as it is also seen in
  main bars \citep{neumann2020}. The inner bar of this galaxy is discussed in greater detail in \citet{mendezAbreu2019},
  \citet{deLorenzoCaceres2019}, and a forthcoming paper.

\textbf{NGC\,1300:}
  This galaxy exhibits a typical nuclear disc with young ages, elevated metallicities, and low {\alphaFe} abundances. We
  highlight the excellent agreement between the kinematic radius of the nuclear disc and the minimum of the age profile. 

\textbf{NGC\,1365:}
  The analysis of this galaxy is hampered by various effects: strong dust extinction is found along a spiral-like
  pattern, several regions of violent ongoing star formation are evident, and a significant, large-scale contribution
  from an AGN is found, especially towards the south-east \citep[see e.g.][]{venturi2018}.  In contrast to the other
  galaxies in the sample, the nuclear ring/disc is hard to distinguish. 

\textbf{NGC\,1433:}
  A noteworthy example of a young and metal-rich nuclear disc with low {\alphaFe} abundances.  In contrast to previously
  shown stellar population maps of this galaxy \citep[see][]{bittner2019}, the maps shown here highlight a strongly
  elongated feature of increased [M/H] and low {\alphaFe}, thanks to the higher signal-to-noise ratio employed in this
  paper. In fact, it has been proposed that this galaxy has an inner bar \citep{erwin2004, buta2015} and we will discuss
  this issue further in a forthcoming paper.  We highlight the excellent agreement between the kinematic radius and
  changes in the stellar population profiles in this galaxy. 

\textbf{NGC\,3351:}
  Similar to NGC\,1097, this galaxy hosts a starbursting nuclear ring with very low metallicities and increased
  {\alphaFe} enhancements. Intense stellar feedback originates from this starbursting ring \citep[see][]{leaman2019}.
  Interestingly, this galaxy is member of a group \citep[see e.g.][]{garcia1993} that might facilitate the accretion of
  low-metallicity gas. Encompassed within the nuclear ring, a nuclear disc consisting of comparably young and metal rich
  stellar populations with low {\alphaFe} enhancements is evident.  

\textbf{NGC\,4303:}
  In this galaxy, almost the entire bar is included in the field of view and characterised by high metallicities and low
  {\alphaFe} enhancements. In addition, a typical young nuclear disc with a star-forming nuclear ring, encompassed by a
  region of old stars, is found. The population gradients generally follow the typical well-defined profiles, except in
  the very centre of the galaxy where a clear break in these profiles is found which could result from effects of the
  AGN in the spectral analysis, or indicate the presence of an additional stellar component. 

\textbf{NGC\,4371:}
  This galaxy has a nuclear disc with the typical stellar population properties: young ages, low {\alphaFe} enhancement
  and high metallicity. However, it is peculiar in the sense that interior to that region one sees older ages and
  variations in metallicity and {\alphaFe} abundances. These may be produced by projection effects, since the
  inclination is \SI{\sim59}{\deg} and the line of nodes is almost perpendicular to the bar. Alternatively, it may be
  produced by additional stellar components \citep[see][]{erwin2015}.  For a detailed discussion of this galaxy, we
  refer the reader to \citet{gadotti2015}. 

\textbf{NGC\,4643:}
  Another typical example of a nuclear disc: a young, metal-rich nuclear disc with low {\alphaFe} enhancement is
  embedded in a region of older stellar populations. The bar of the galaxy is oriented from the south-east to north-west
  and prominently highlighted by its relatively young ages, high metallicities and low {\alphaFe} abundances.
  Interestingly, these trends along the bar seem to become stronger with increasing radius.  While {\Ha} emission is
  barely detected, it is arranged in a small, two-armed spiral structure. 

\textbf{NGC\,4981:}
  After applying the spatial binning, this is the galaxy with the lowest spatial resolution of the present TIMER sample.
  In this galaxy it is difficult to distinguish nuclear structures but nonetheless the highest metallicities and lowest
  {\alphaFe} enhancements are observed in the centre of the galaxy. 

\textbf{NGC\,4984:}
  This galaxy hosts a typical nuclear disc, showing radial population gradients that appear similar but less
  well-defined than in other galaxies, most likely due to an inclination effect. The populations found in the very
  central bins deviate strongly from these profiles. In particular, old ages, low metallicities and high {\alphaFe}
  abundances are detected. We speculate that these deviations might be from effects on the spectral analysis caused by
  the AGN in this galaxy.

\textbf{NGC\,5236:}
  This galaxy shows a highly irregular nuclear structure. Star formation, as traced by the {\Ha} emission, is
  irregularly distributed in the central region, not forming a nuclear ring or nuclear disc. It is also heavily obscured
  by dust. As a result, also the stellar population maps and gradients do not show regular features. 

\textbf{NGC\,5248:}
  This galaxy has a typical nuclear disc. While the radial population gradients are well-defined, a slight deviation is
  evident in the central bins which, again, can be an AGN effect. In the region of the nuclear ring a few small,
  star-forming spots are found, showing significantly lower ages and metallicities and increased {\alphaFe} abundances.
  Interestingly, the {\Ha} emission is not restricted to the nuclear ring but appears rather smoothly distributed over
  the entire nuclear disc. 

\textbf{NGC\,5728:}
  While this galaxy seems to host a common nuclear disc, the observed stellar population properties are severely
  contaminated by an AGN jet. This jet is strikingly visible in the maps as an elongated feature of old ages and low
  metallicities crossing the nuclear disc from north-west to south-east and extending into an even larger cone outside
  of the nuclear region \citep[see e.g.][]{durre2018}. 

\textbf{NGC\,5850:}
  A prominent example of a galaxy hosting an inner bar. The inner bar is clearly visible by its elevated [M/H] and low
  {\alphaFe} enhancements, and shows regions with young stellar populations at its ends.  We refer the reader to
  \citet{deLorenzoCaceres2019} for a detailed discussion of the double-barred structure of this galaxy.  The nuclear
  disc shows the typical aforementioned properties. 

\textbf{NGC\,6902:}
  The galaxy NGC\,6902 is a peculiar object. Combining the results from this study and G20, it is unclear whether the
  galaxy hosts a nuclear disc. A weak bar with slightly elevated metallicities and low {\alphaFe} abundances is found,
  encompassed by a star-forming inner ring at the bar radius. Within this inner ring, spatially coinciding with the bar,
  a kinematically hot spheroid is detected. Due to these peculiarities, this galaxy is not considered in any of the
  discussions in this study. 

\textbf{NGC\,7140:}
  This galaxy hosts a rather typical nuclear disc, which appearance is slightly contaminated by a singular star-forming
  spot in the nuclear ring. The moderate {\Ha} emission in the nuclear region is rather smoothly distributed over the
  nuclear disc, but nonetheless shows some concentration in a nuclear ring. 

\textbf{NGC\,7552:}
  A heavily starbursting nuclear ring is evident in the galaxy. This is not only highlighted in the {\Ha} emission-line
  maps, but also clearly visible in the radial population profiles as extrema of low metallicities and high {\alphaFe}
  enhancements. Within the nuclear ring a typical nuclear disc is seen with young ages, elevated metallicities and low
  {\alphaFe} enhancements. However, contrary to most nuclear discs, this nuclear disc shows a flat age profile.

\textbf{NGC\,7755:}
  This galaxy hosts a typical nuclear disc with a ring of {\Ha} emission at its outermost edge, embedded in a region of
  older stellar populations. It further highlights the typical population properties found in bars. In particular, the
  bar is visible by its high metallicities and low {\alphaFe} abundances. The reader is referred to \citet{neumann2020}
  for further details on the bar properties.


\section{Maps of dust-corrected {\Ha} fluxes}%
\label{sec:HalphaMaps}
In Fig.~\ref{fig:HalphaMaps} we present maps of dust-corrected {\Ha} fluxes for all TIMER galaxies. We note that similar
versions of the maps of IC\,1438, NGC\,4304, NGC\,4371, NGC\,4643, NGC\,4981, NGC\,4984, NGC\,5248, NGC\,6902, and
NGC\,7755 have already been presented in \citet{neumann2020}, but in order to facilitate the comparison with the maps of
the stellar population properties we present these here again.  We only display spaxels in which the
amplitude-over-noise ratio of the {\Ha} line exceeds 5.  Fluxes are given in units of $\SI{d-12}{\erg \per\second
\per\cm\squared \per\arcsec\squared}$ and the respective limits of the colour bar are stated in the lower-right corner
of each panel.  Based on reconstructed intensities from the MUSE cube, we display isophotes in steps of \SI{0.5}{\mag},
identical to the ones displayed in the maps of the stellar population properties. 

\begin{figure*}
    \includegraphics[width=0.33\textwidth]{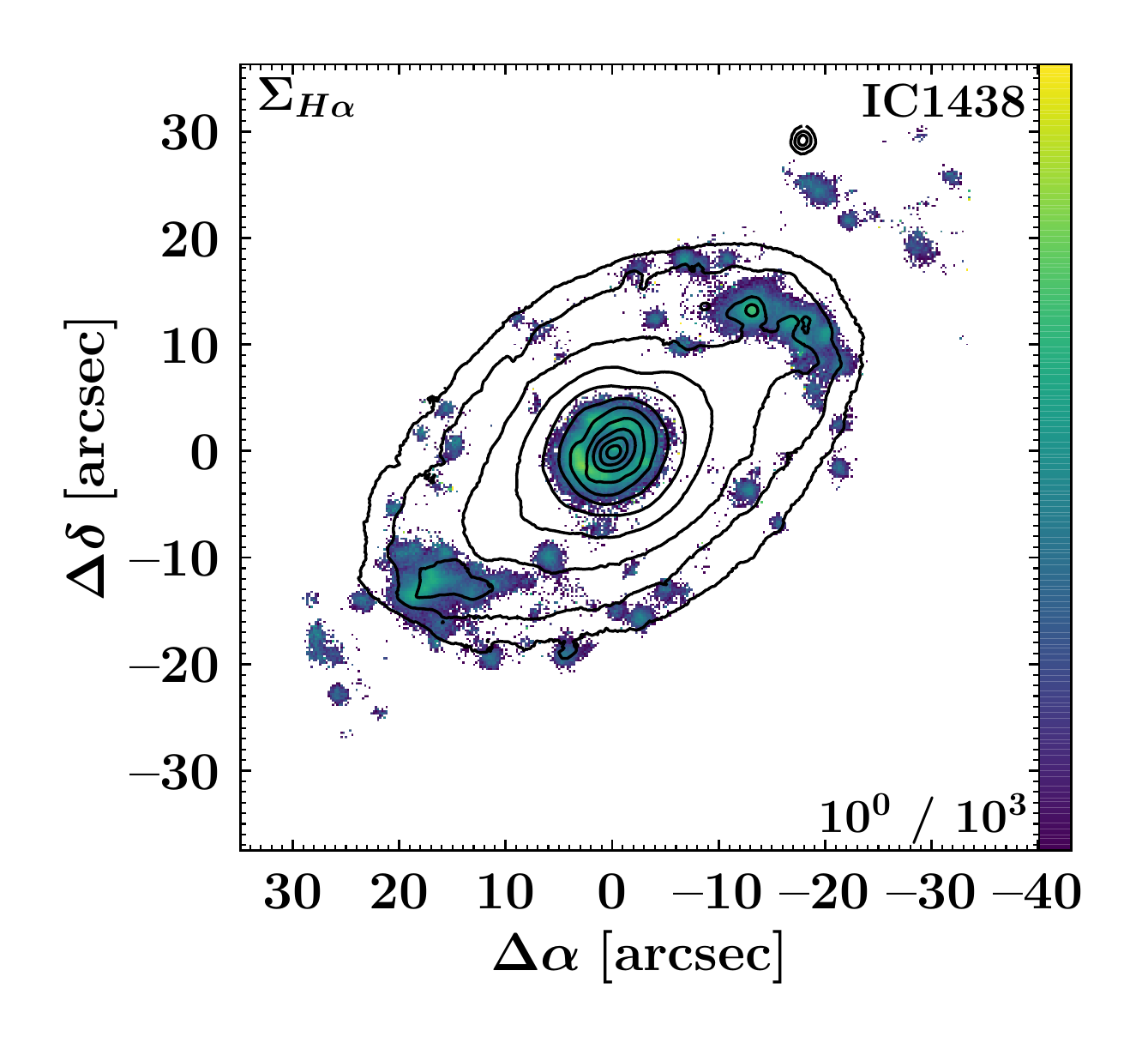}
    \includegraphics[width=0.33\textwidth]{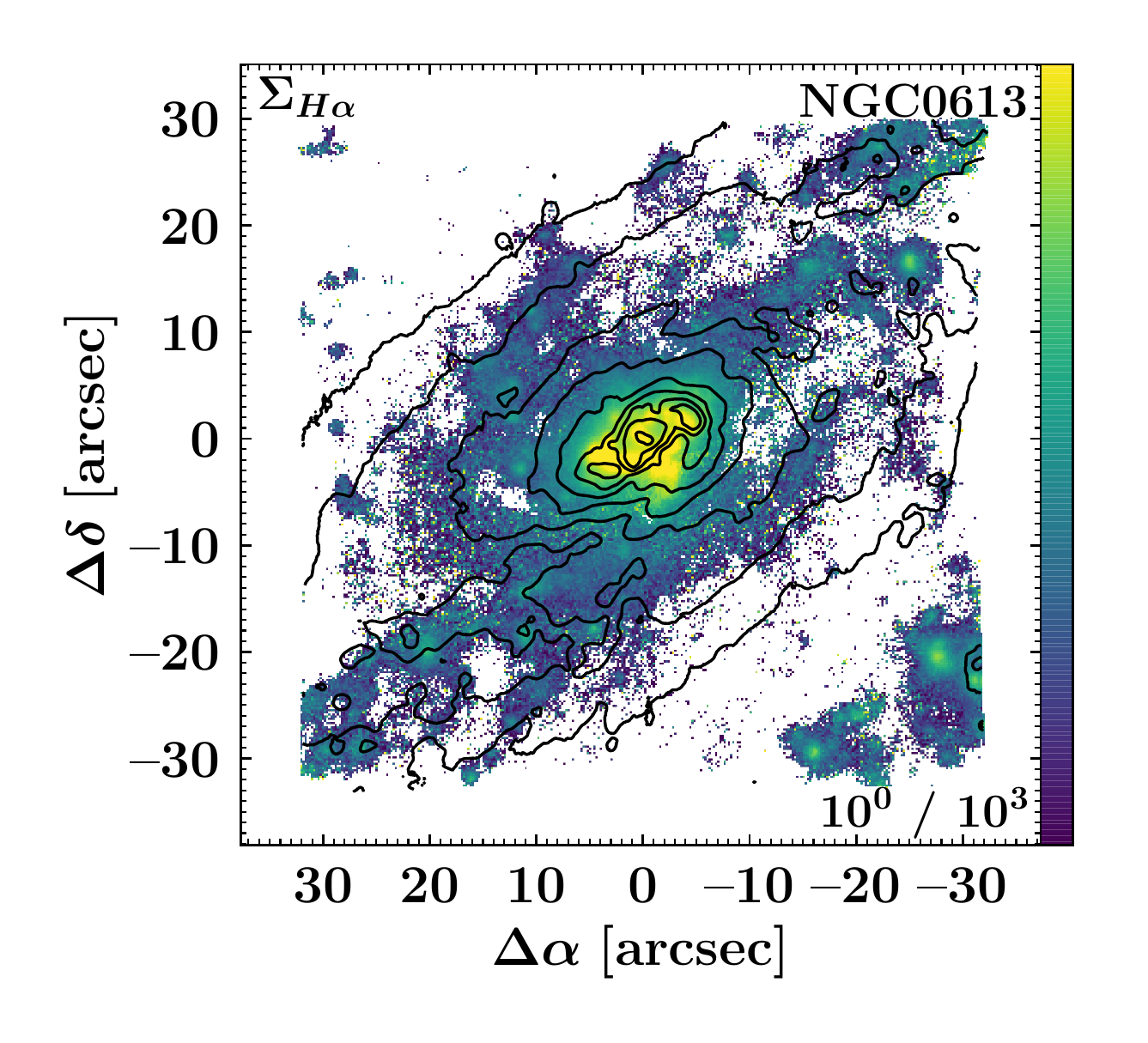}
    \includegraphics[width=0.33\textwidth]{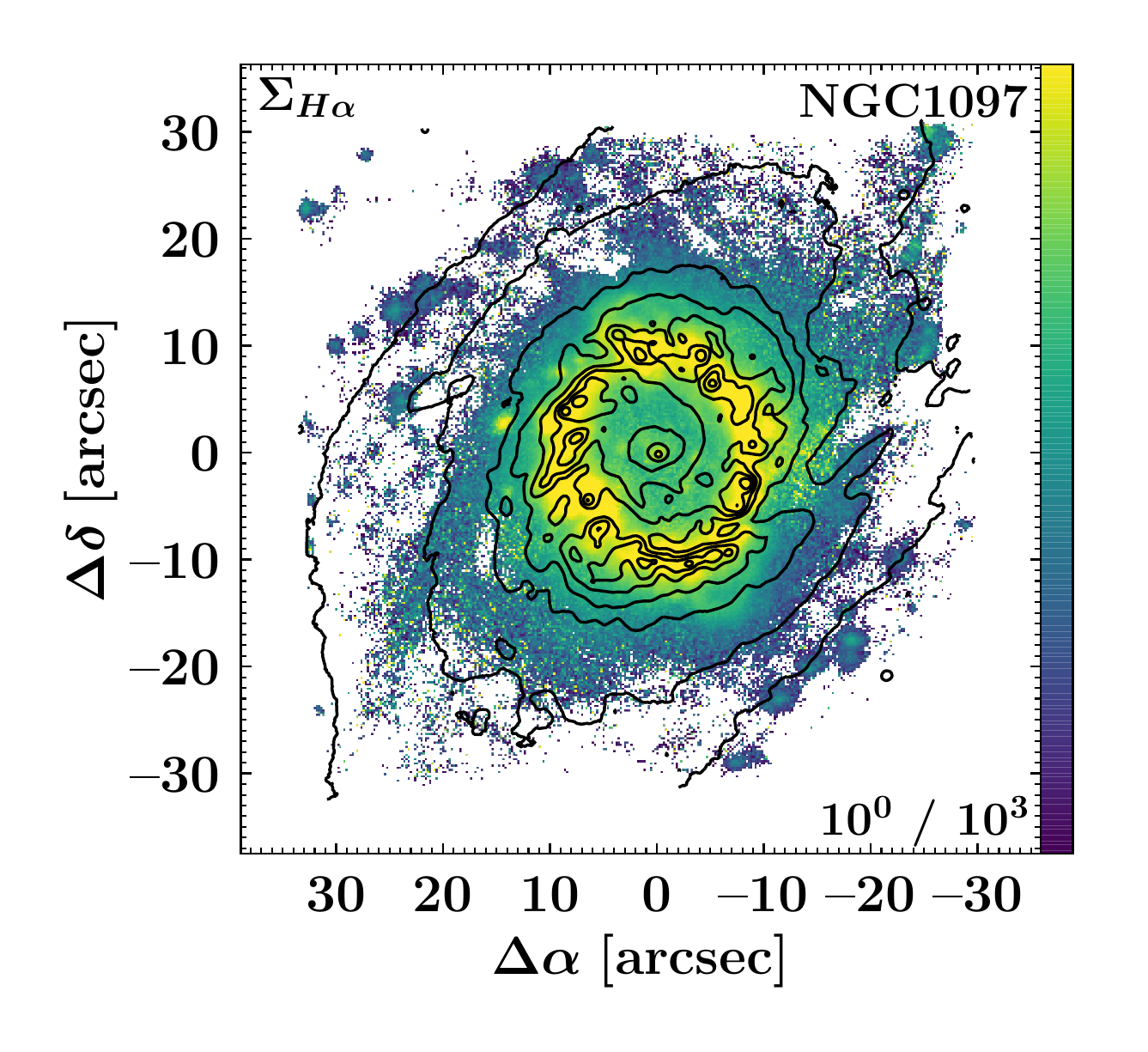}\\
    \includegraphics[width=0.33\textwidth]{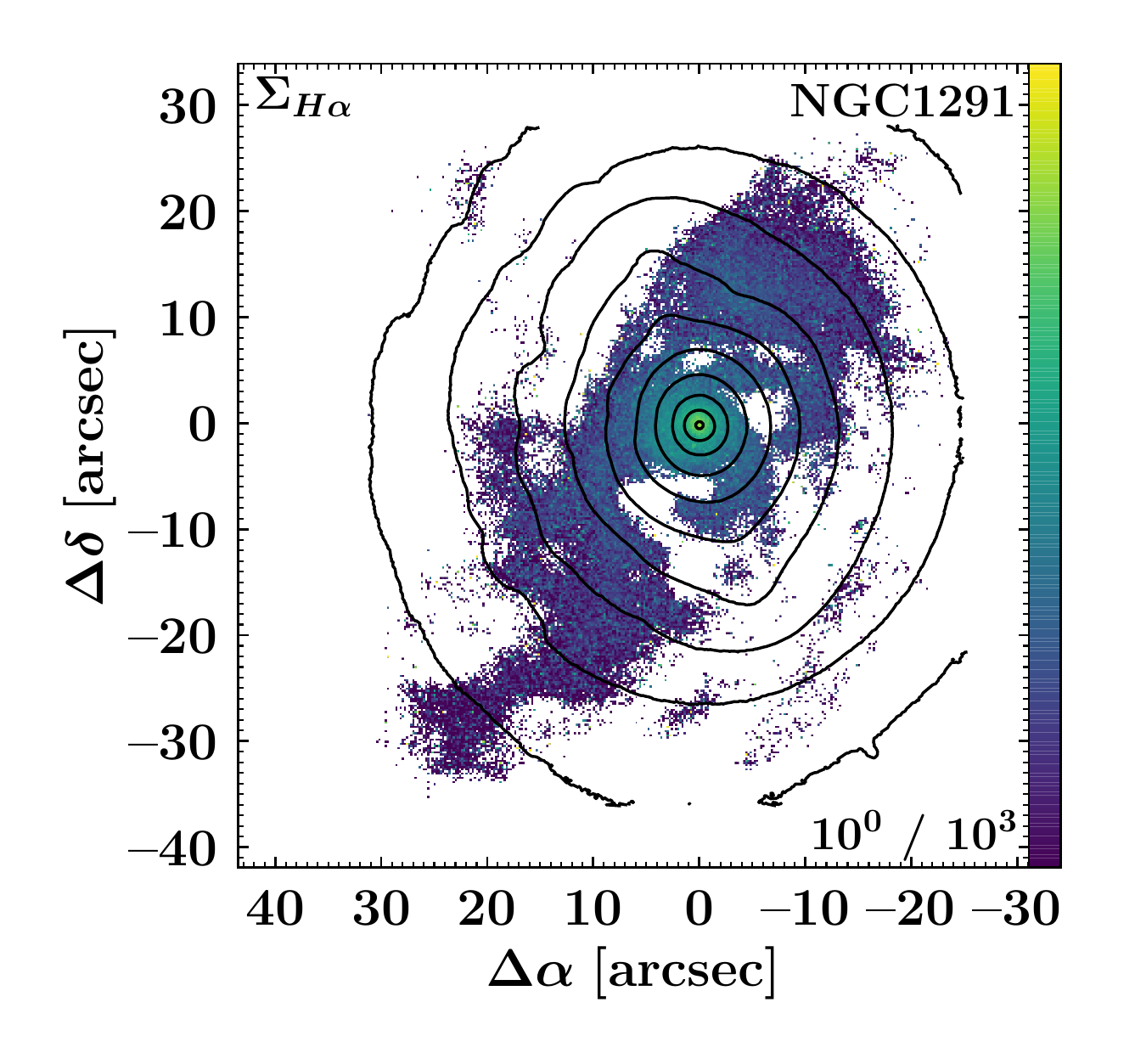}
    \includegraphics[width=0.33\textwidth]{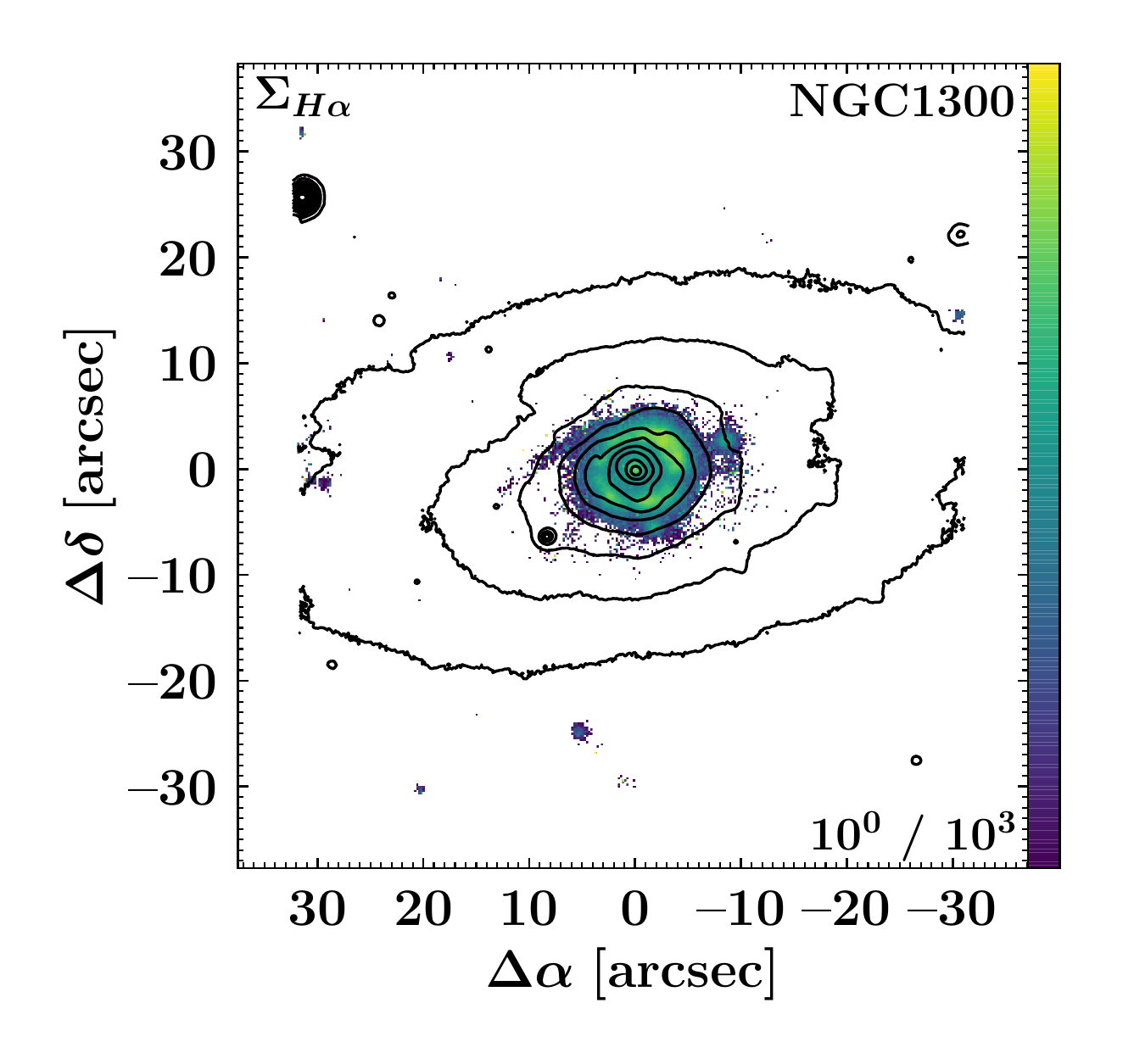}
    \includegraphics[width=0.33\textwidth]{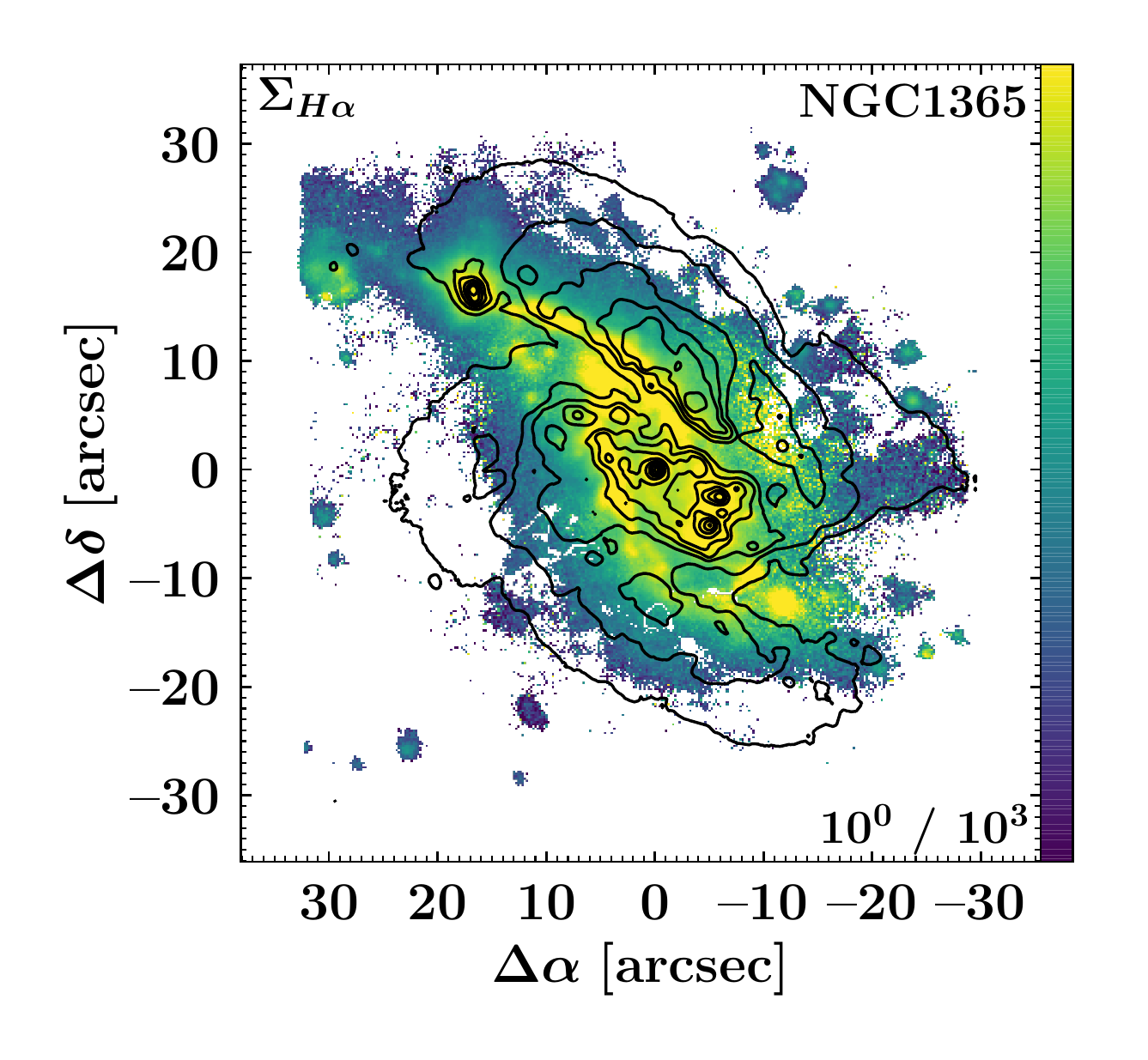}\\
    \includegraphics[width=0.33\textwidth]{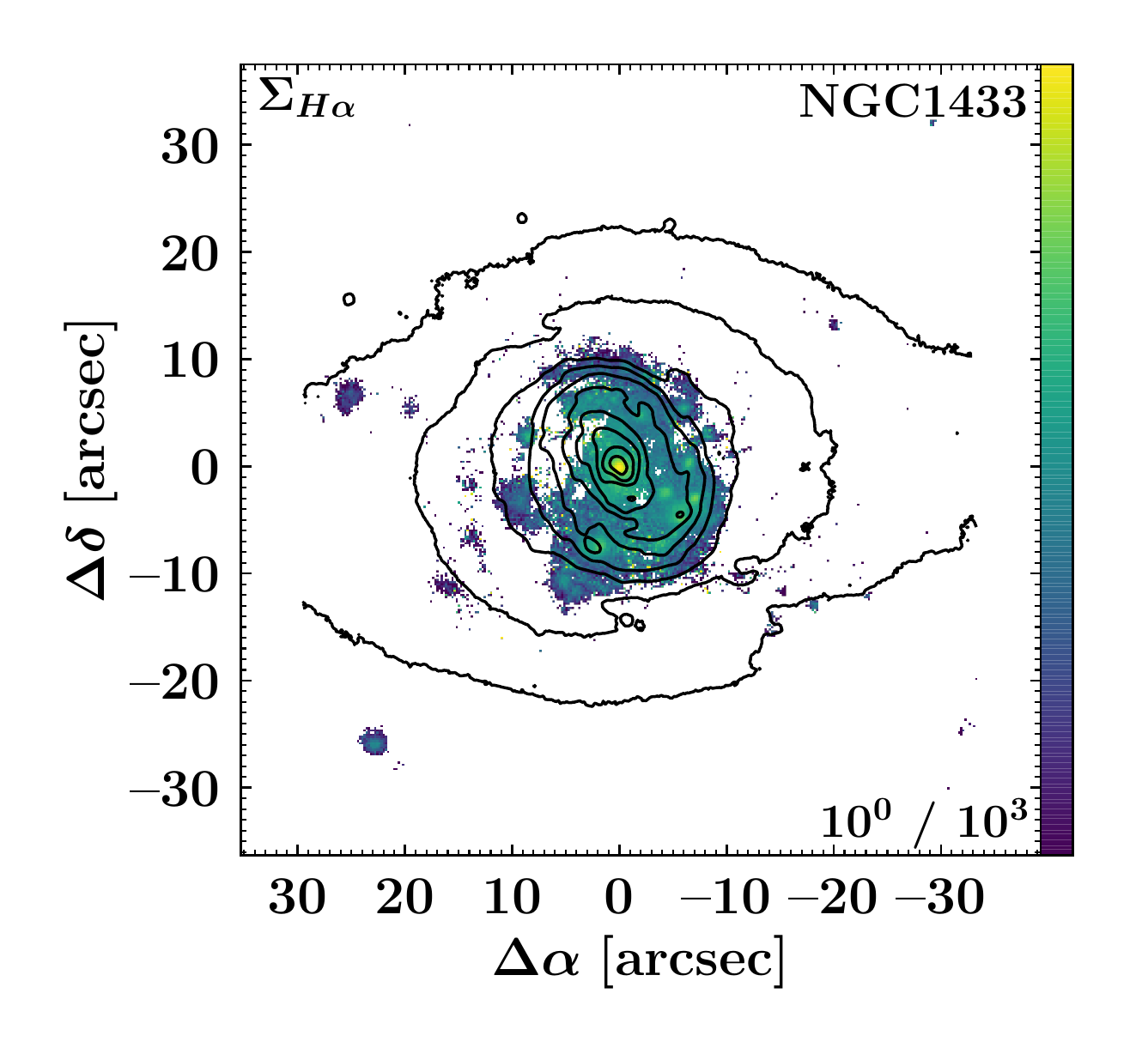}
    \includegraphics[width=0.33\textwidth]{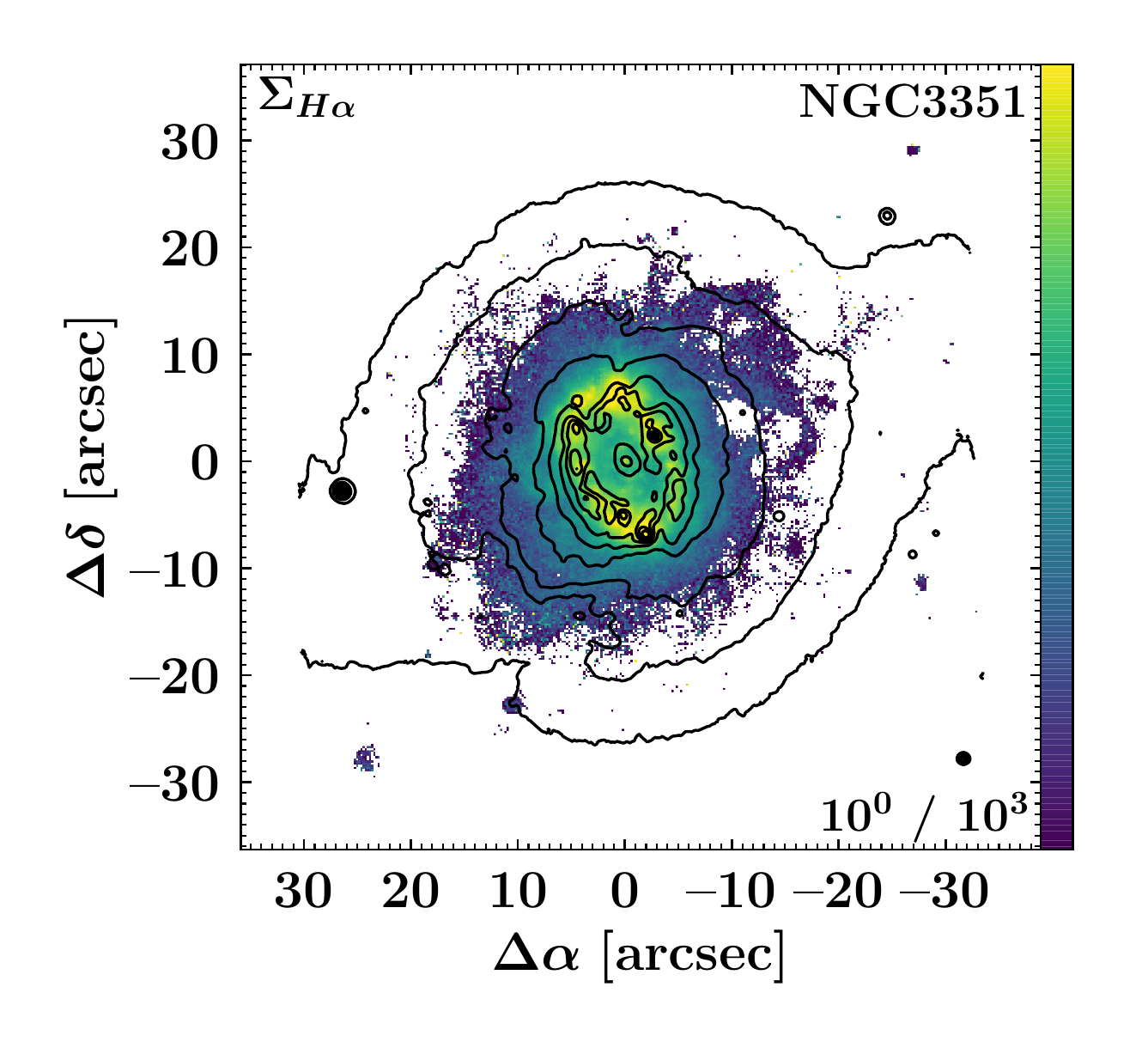}
    \includegraphics[width=0.33\textwidth]{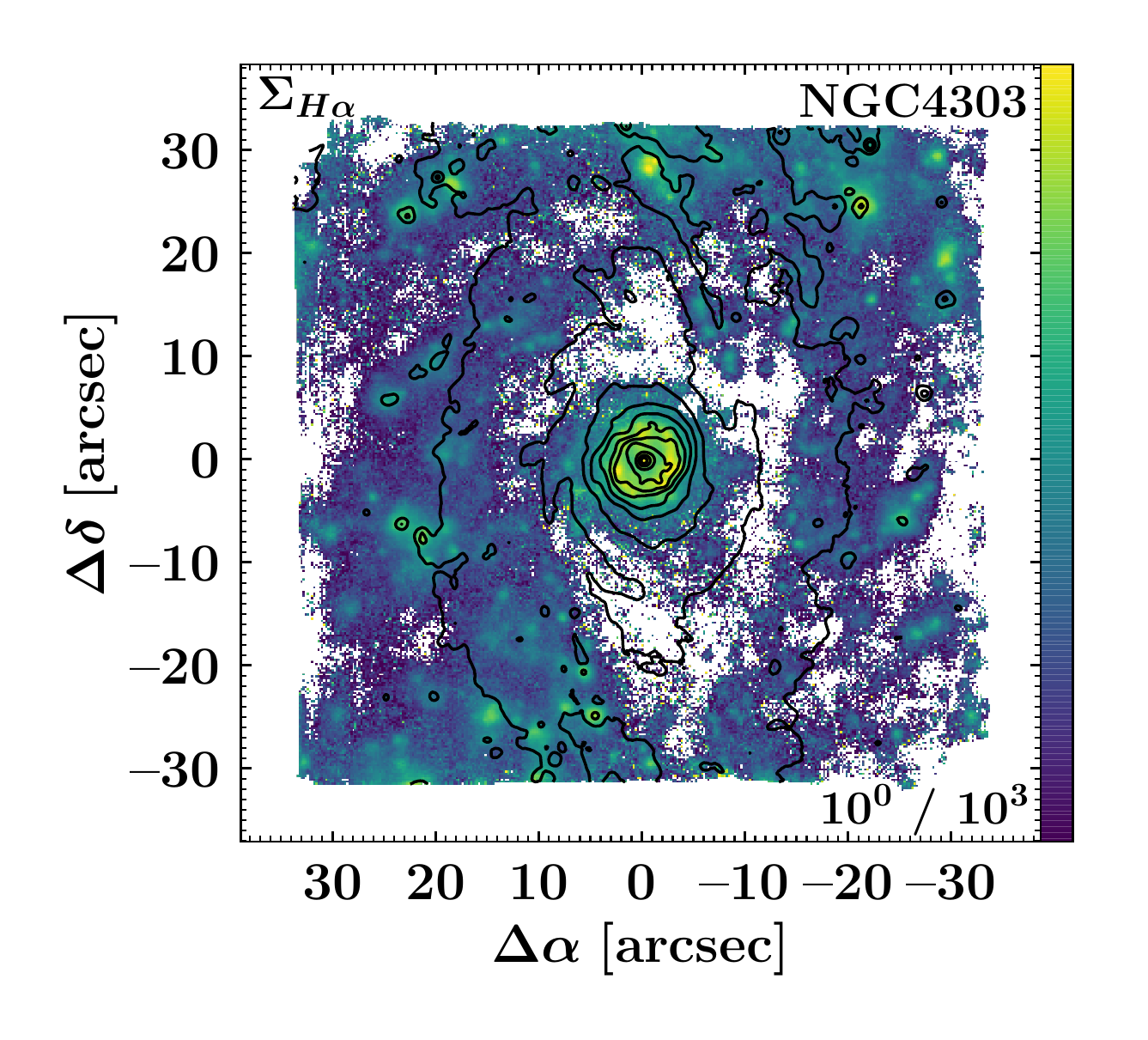}\\
    \includegraphics[width=0.33\textwidth]{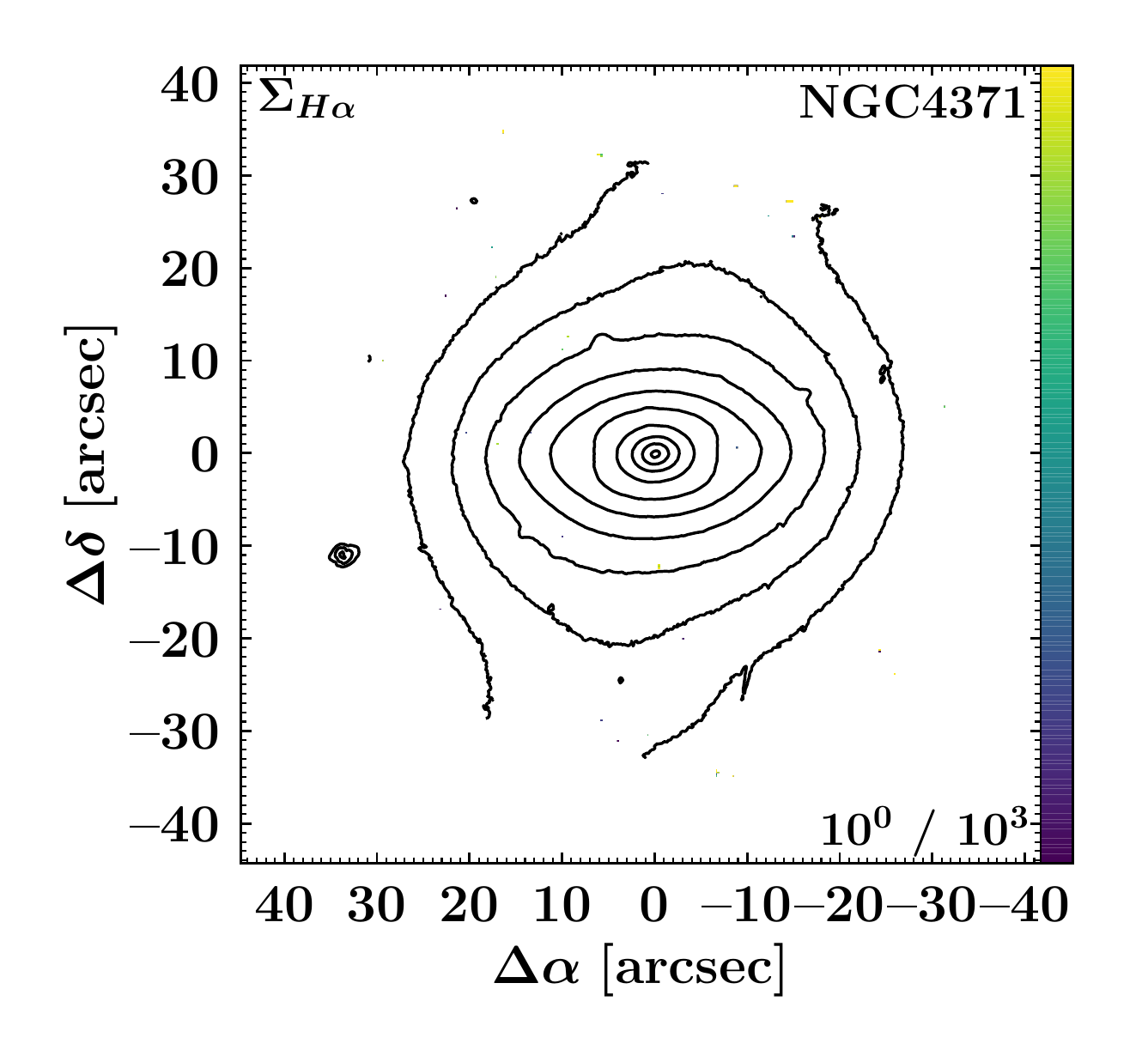}
    \includegraphics[width=0.33\textwidth]{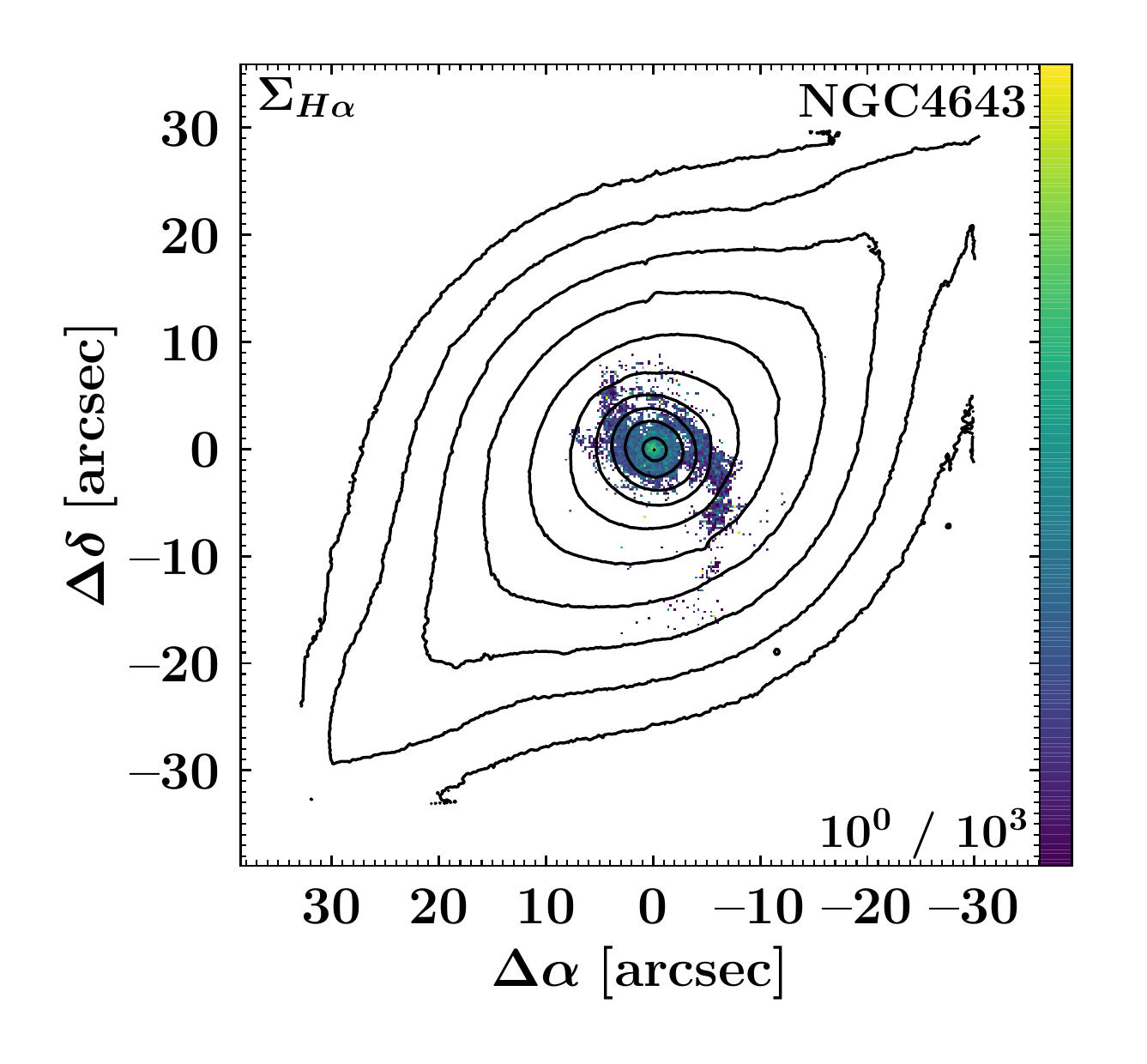}
    \includegraphics[width=0.33\textwidth]{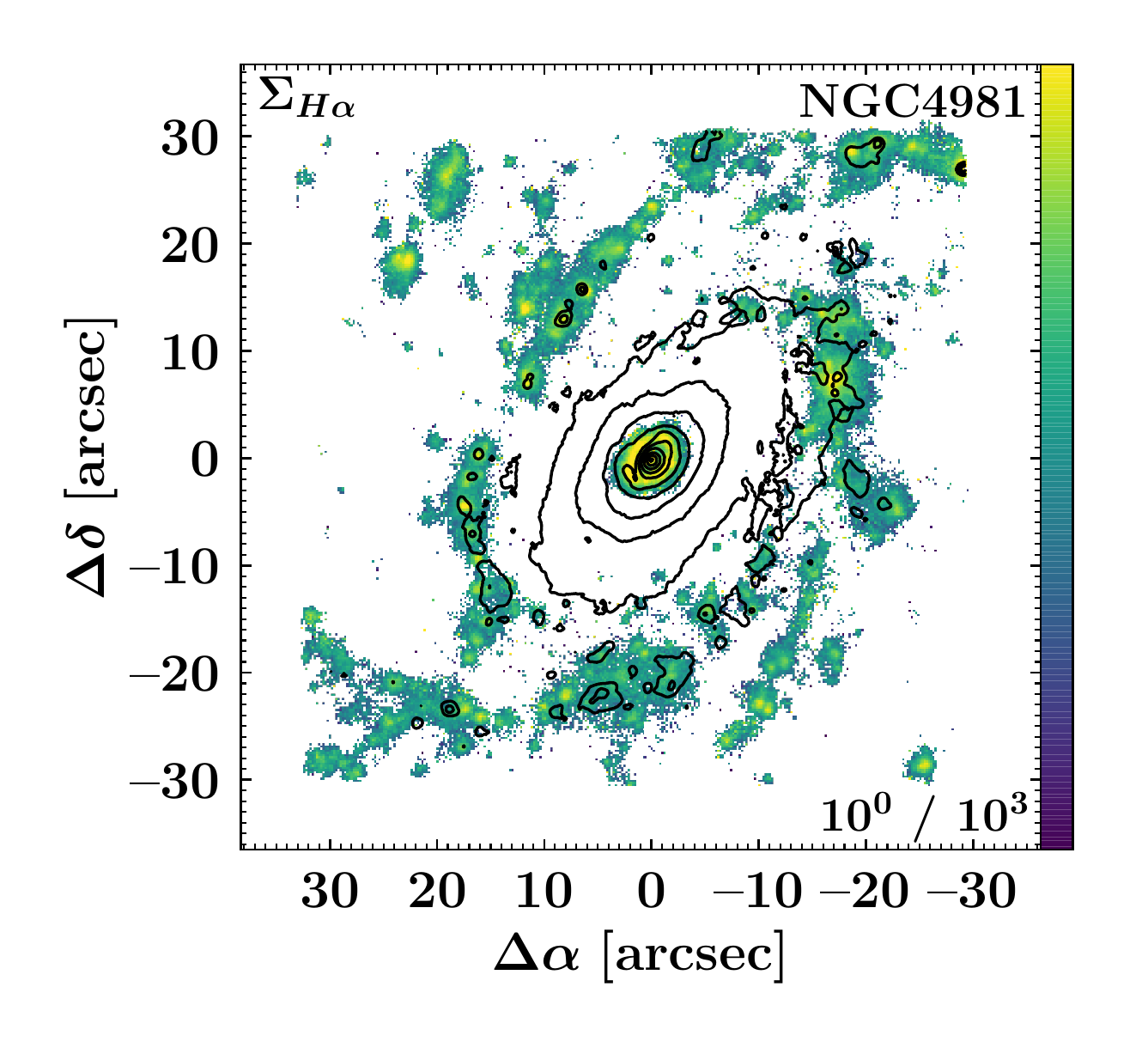}
    \caption{%
        Maps of dust-corrected {\Ha} fluxes for all TIMER galaxies. We note that only spaxels with an {\Ha}
        amplitude-over-noise ratio above 5 are displayed. Fluxes are given in units of $\SI{d-12}{\erg \per\second
        \per\cm\squared \per\arcsec\squared}$ and the respective limits of the colour bar are stated in the lower-right
        corner of each panel. Based on reconstructed intensities from the MUSE cube, we display isophotes in steps of
        \SI{0.5}{\mag}, identical to the ones displayed in the maps of the stellar population properties.  North is up;
        east is to the left. 
    }%
    \label{fig:HalphaMaps}
\end{figure*}
\begin{figure*}
    \ContinuedFloat%
    \includegraphics[width=0.33\textwidth]{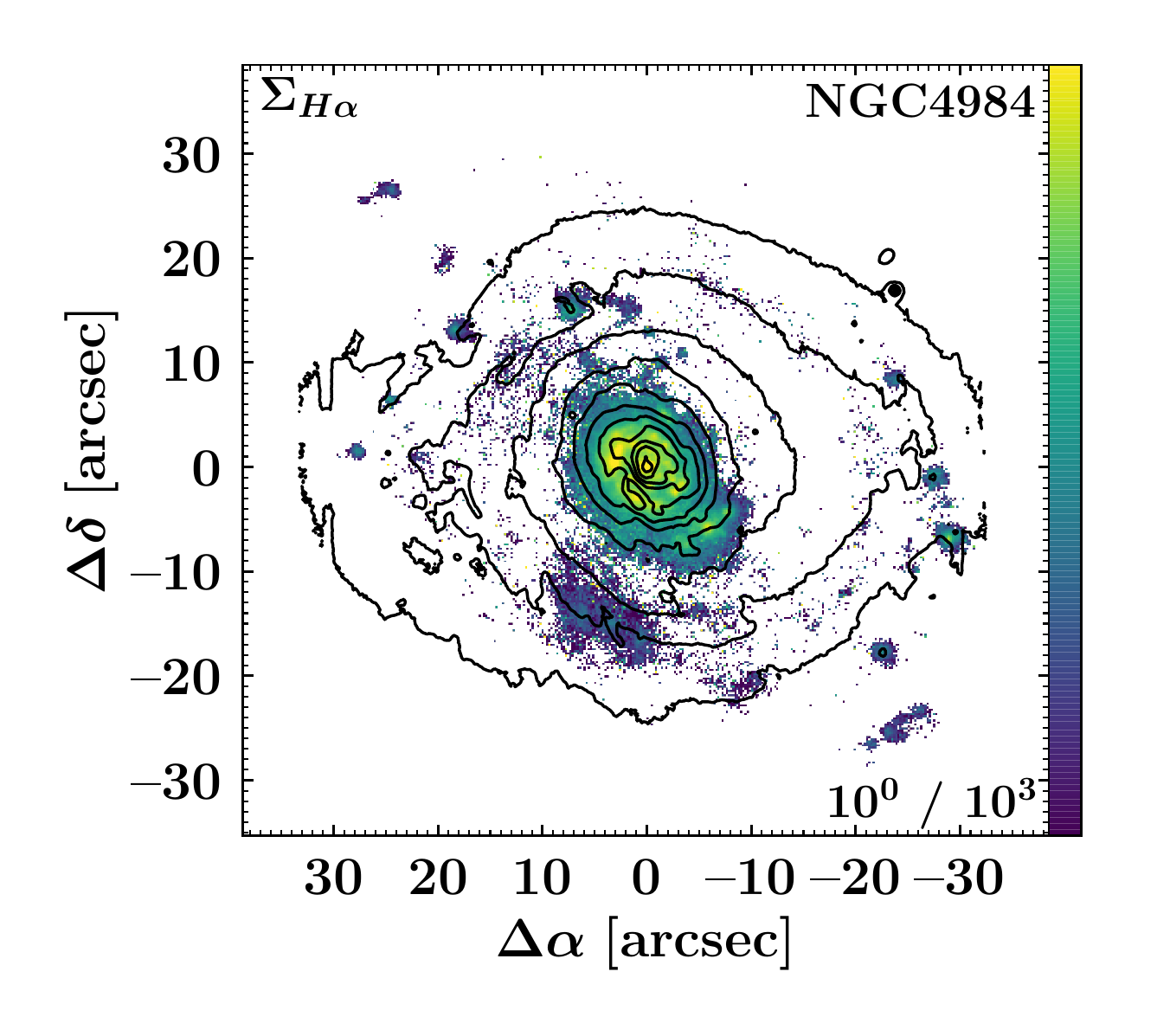}
    \includegraphics[width=0.33\textwidth]{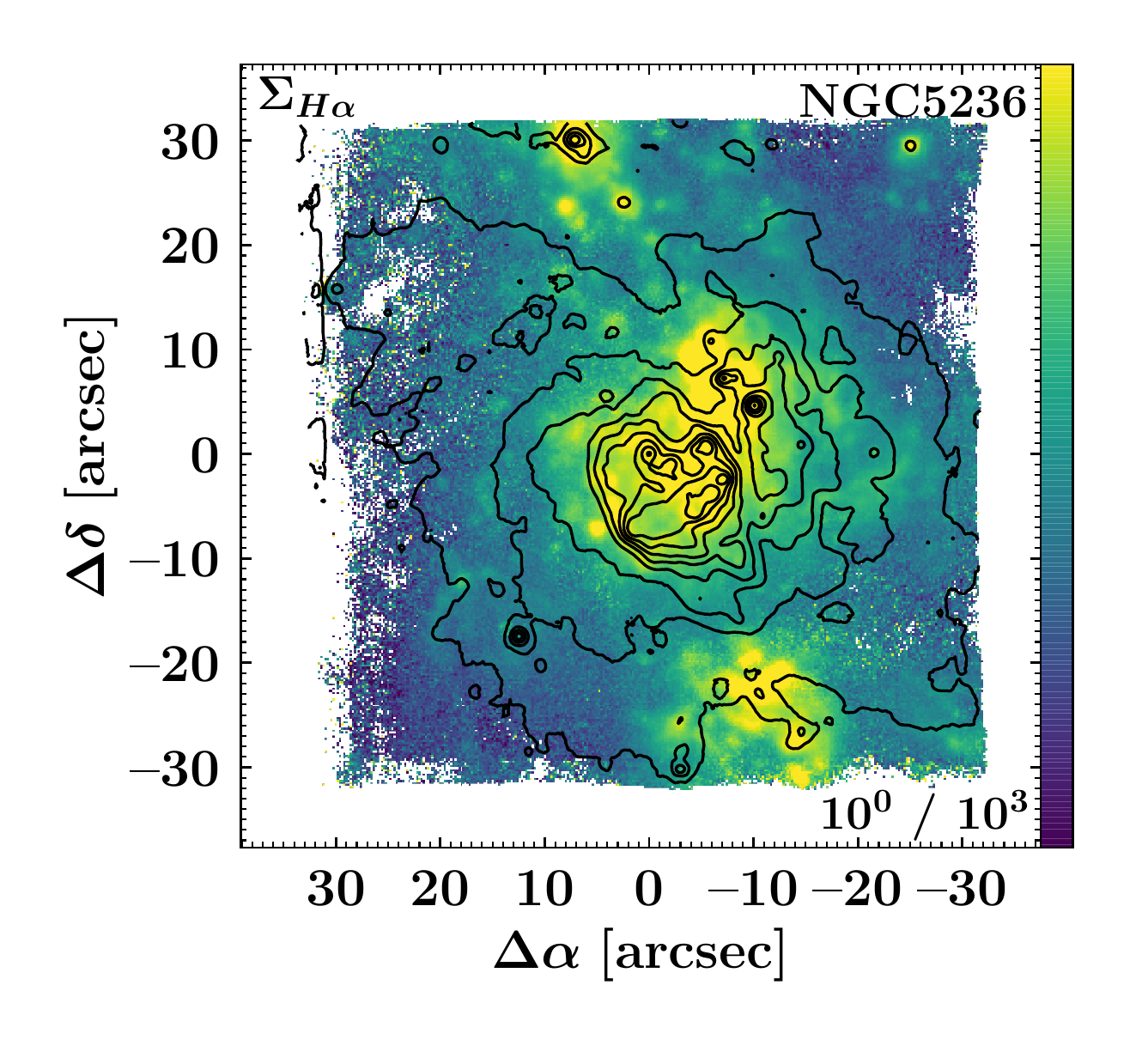}
    \includegraphics[width=0.33\textwidth]{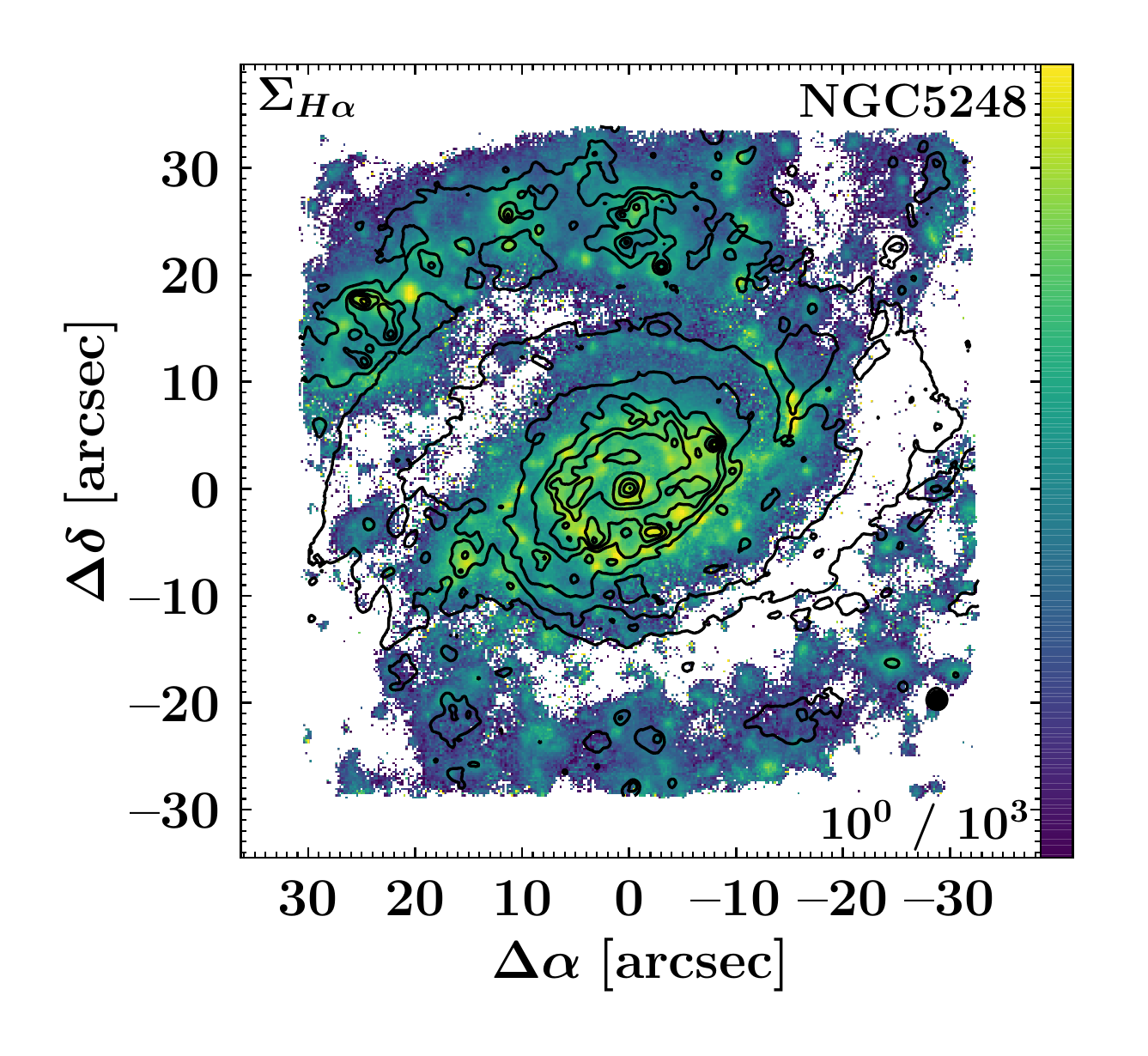}\\
    \includegraphics[width=0.33\textwidth]{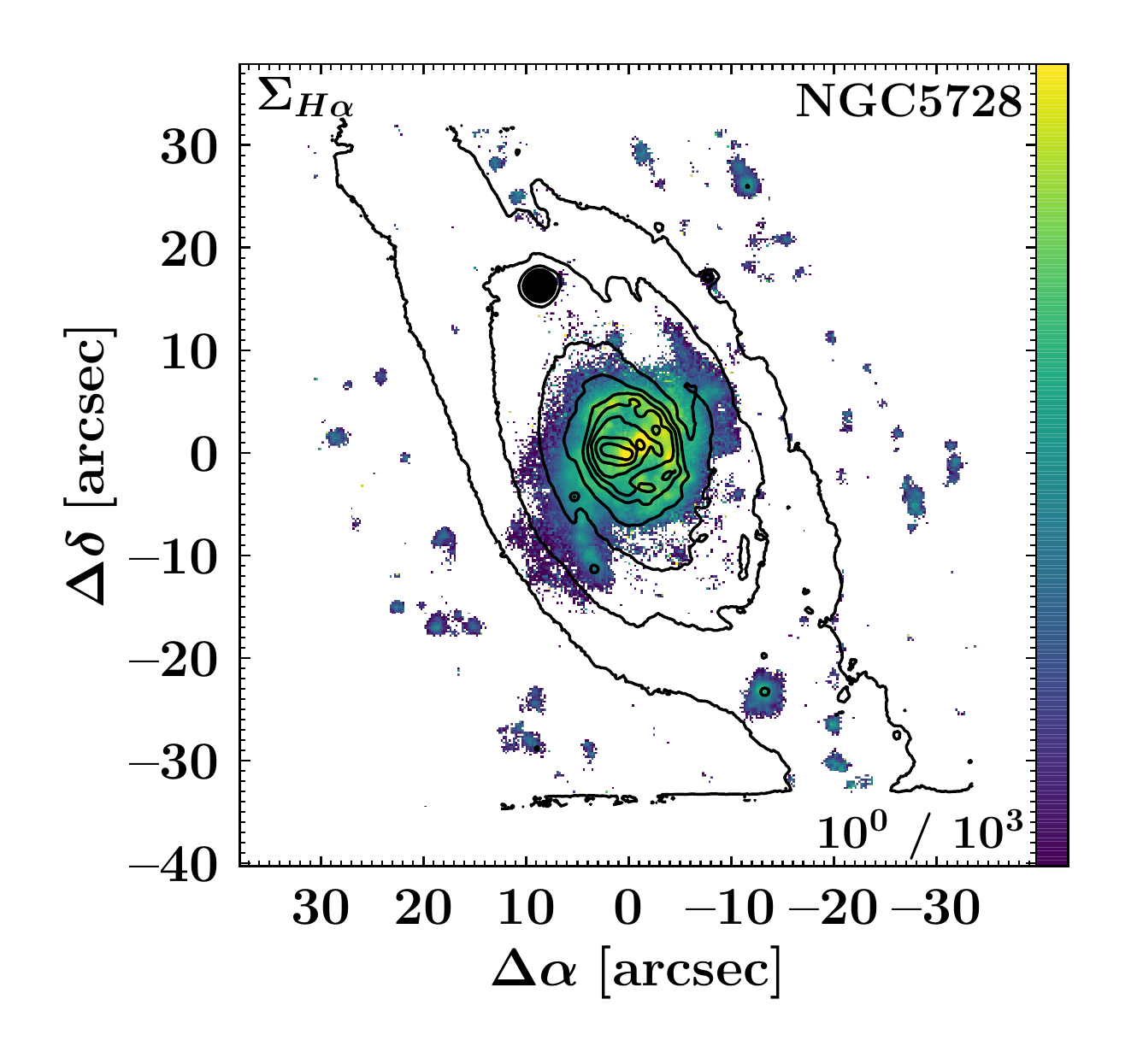}
    \includegraphics[width=0.33\textwidth]{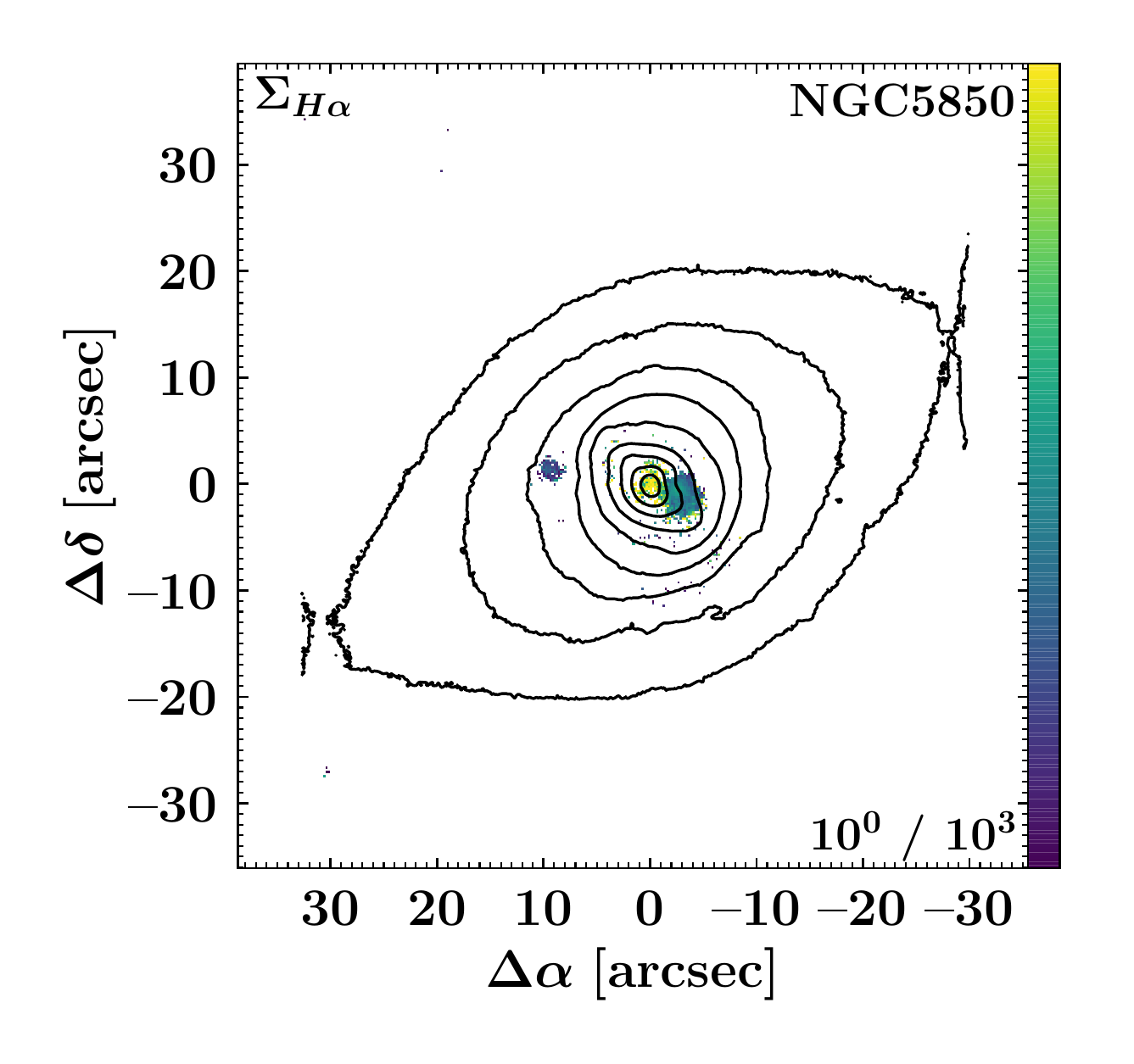}
    \includegraphics[width=0.33\textwidth]{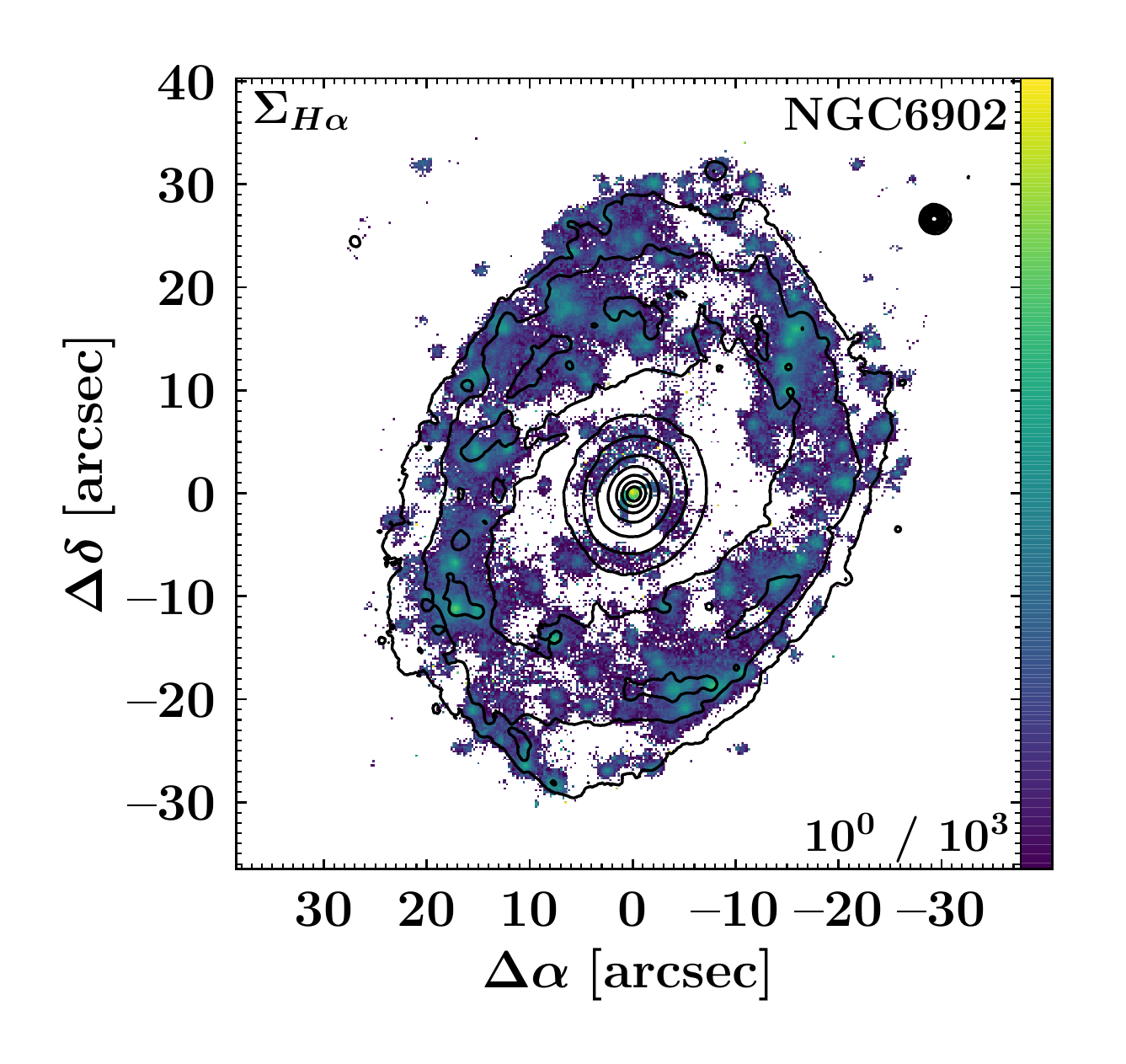}\\
    \includegraphics[width=0.33\textwidth]{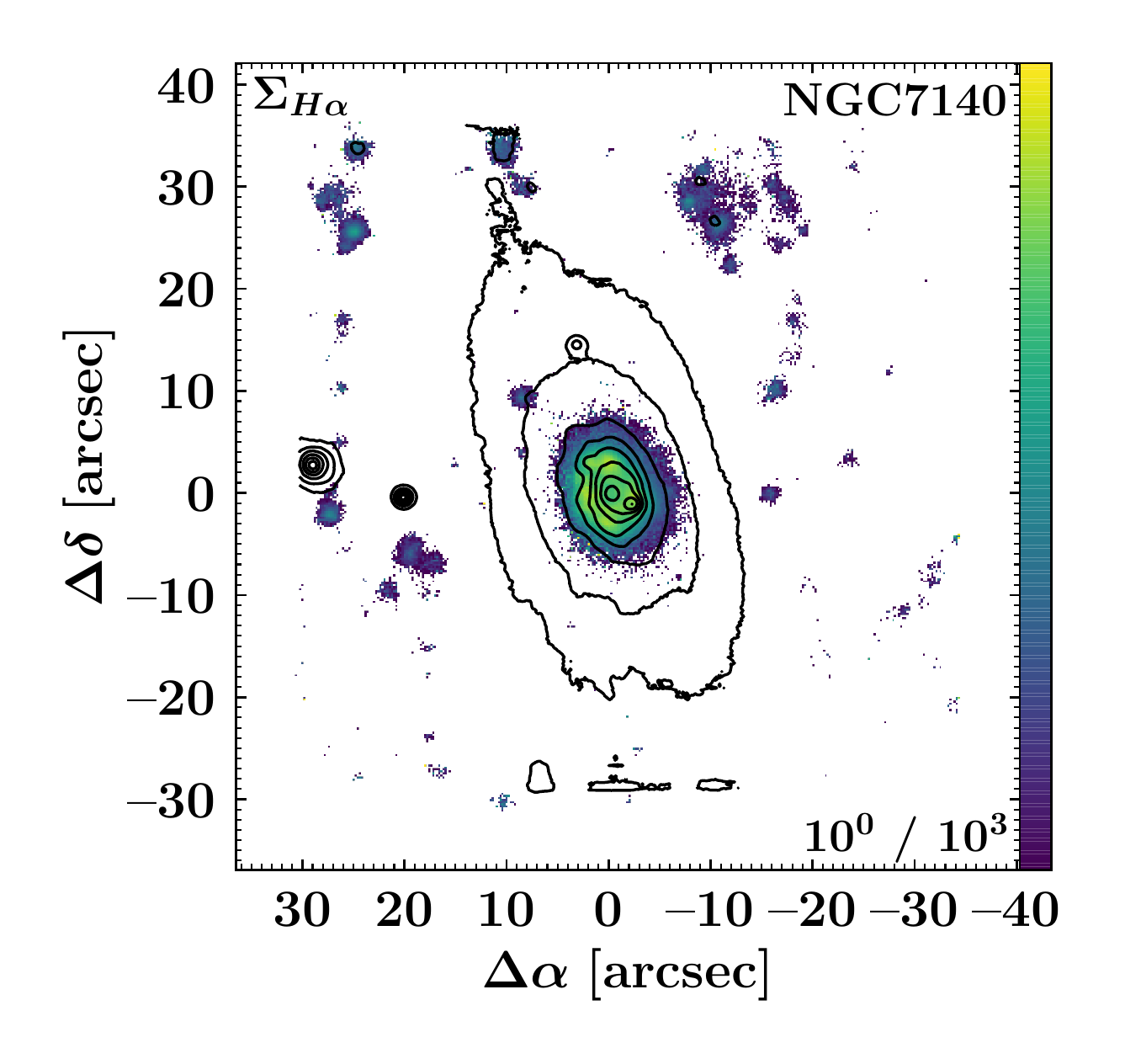}
    \includegraphics[width=0.33\textwidth]{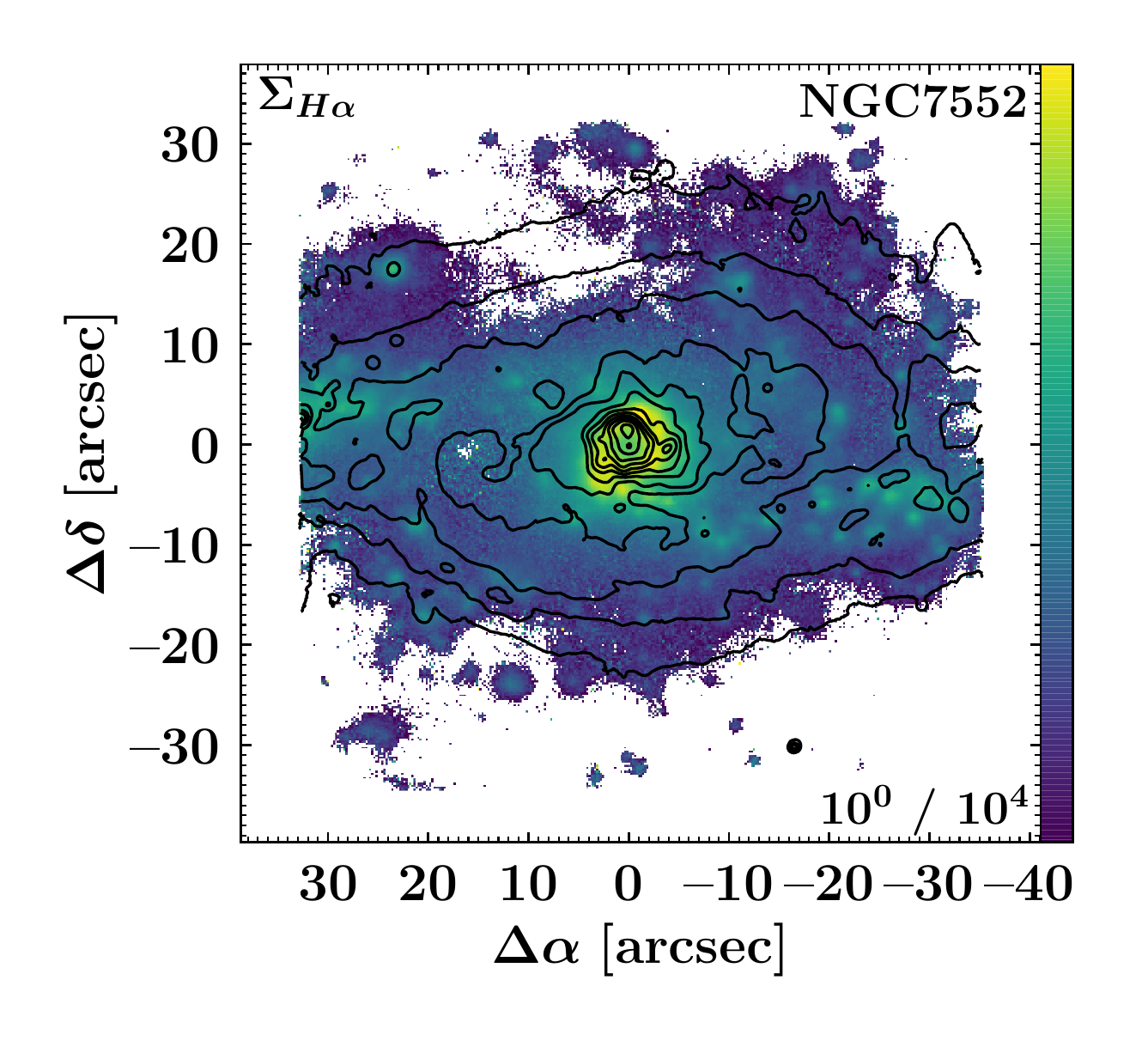}
    \includegraphics[width=0.33\textwidth]{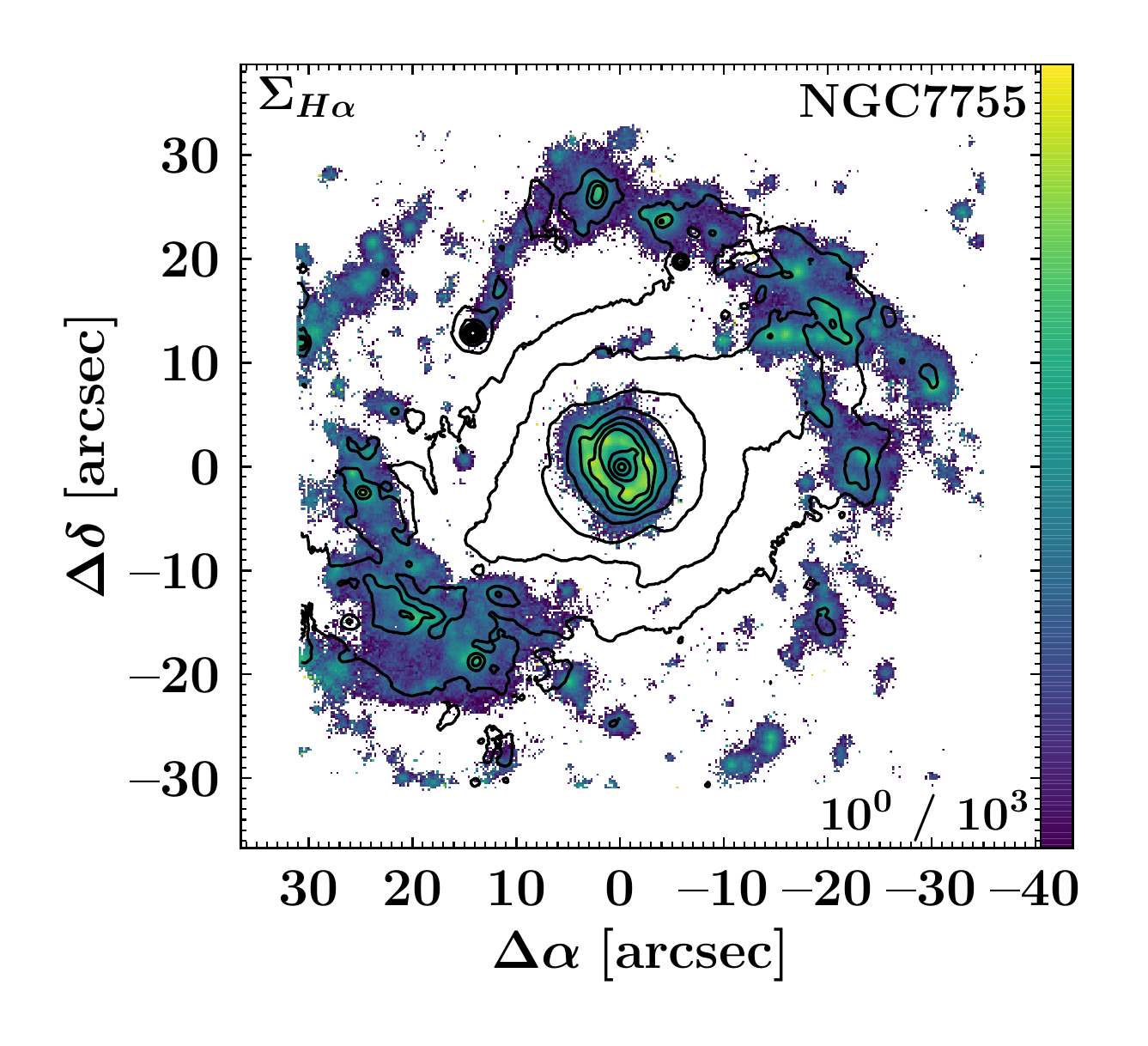}
    \caption{Continued.}%
    \label{fig:HalphaMaps}
\end{figure*}


\section{Maps and radial profiles of mean stellar population properties}%
\begin{figure*}[p]
    \begin{minipage}[c]{\textwidth}
        \centering
        \begin{minipage}[c]{0.55\textwidth}
            \centering
            \includegraphics[width=\textwidth]{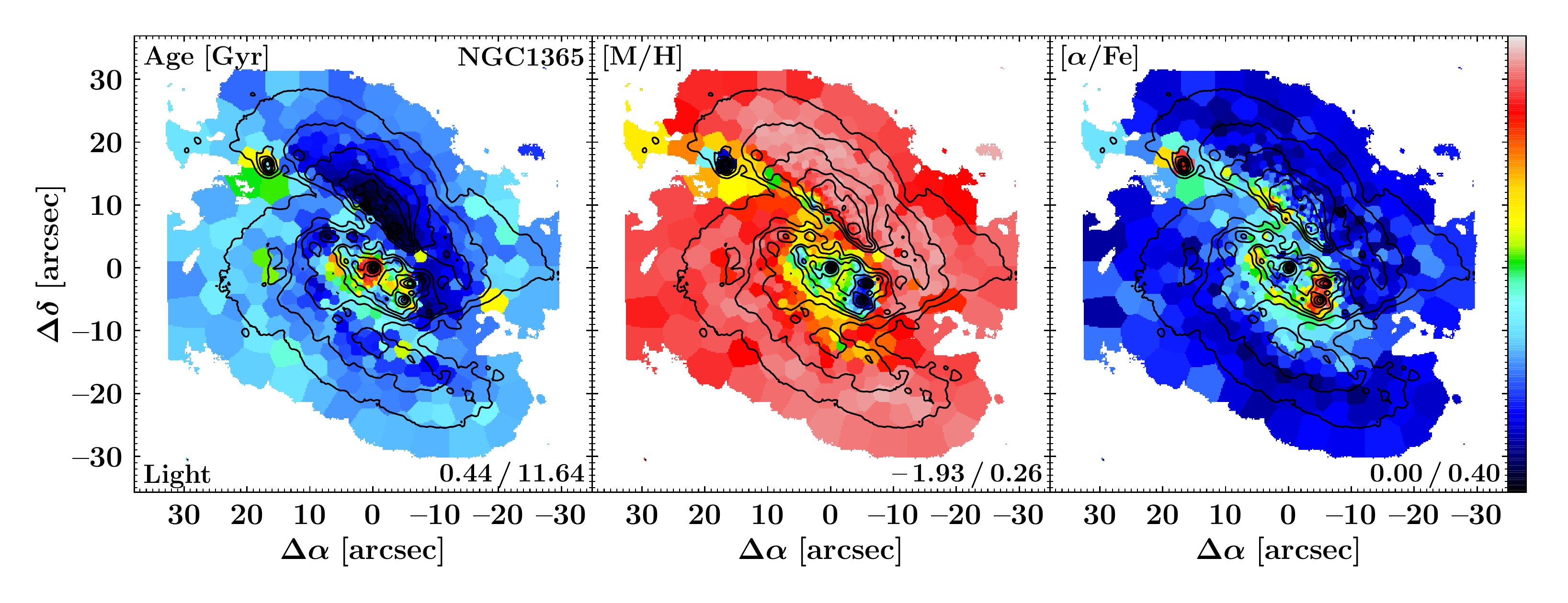}\\
            \includegraphics[width=\textwidth]{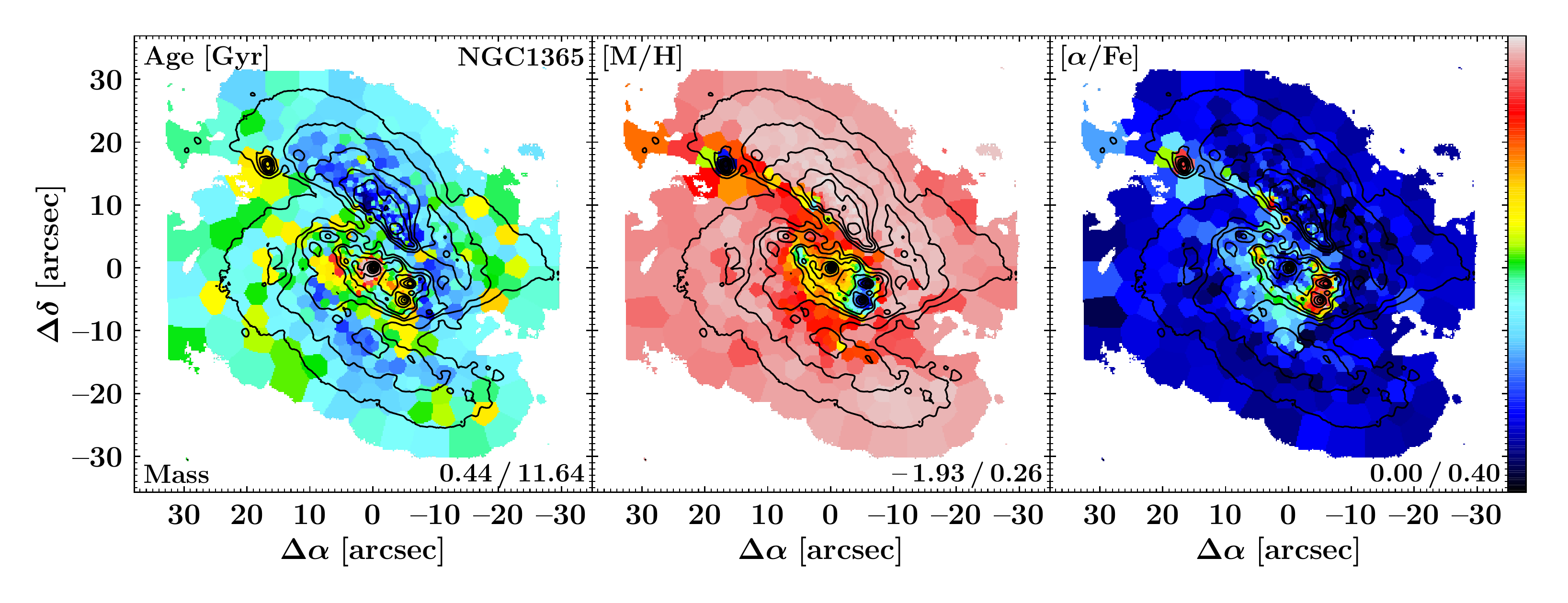}
        \end{minipage}%
        \begin{minipage}[c]{0.35\textwidth}
            \centering
            \includegraphics[width=\textwidth]{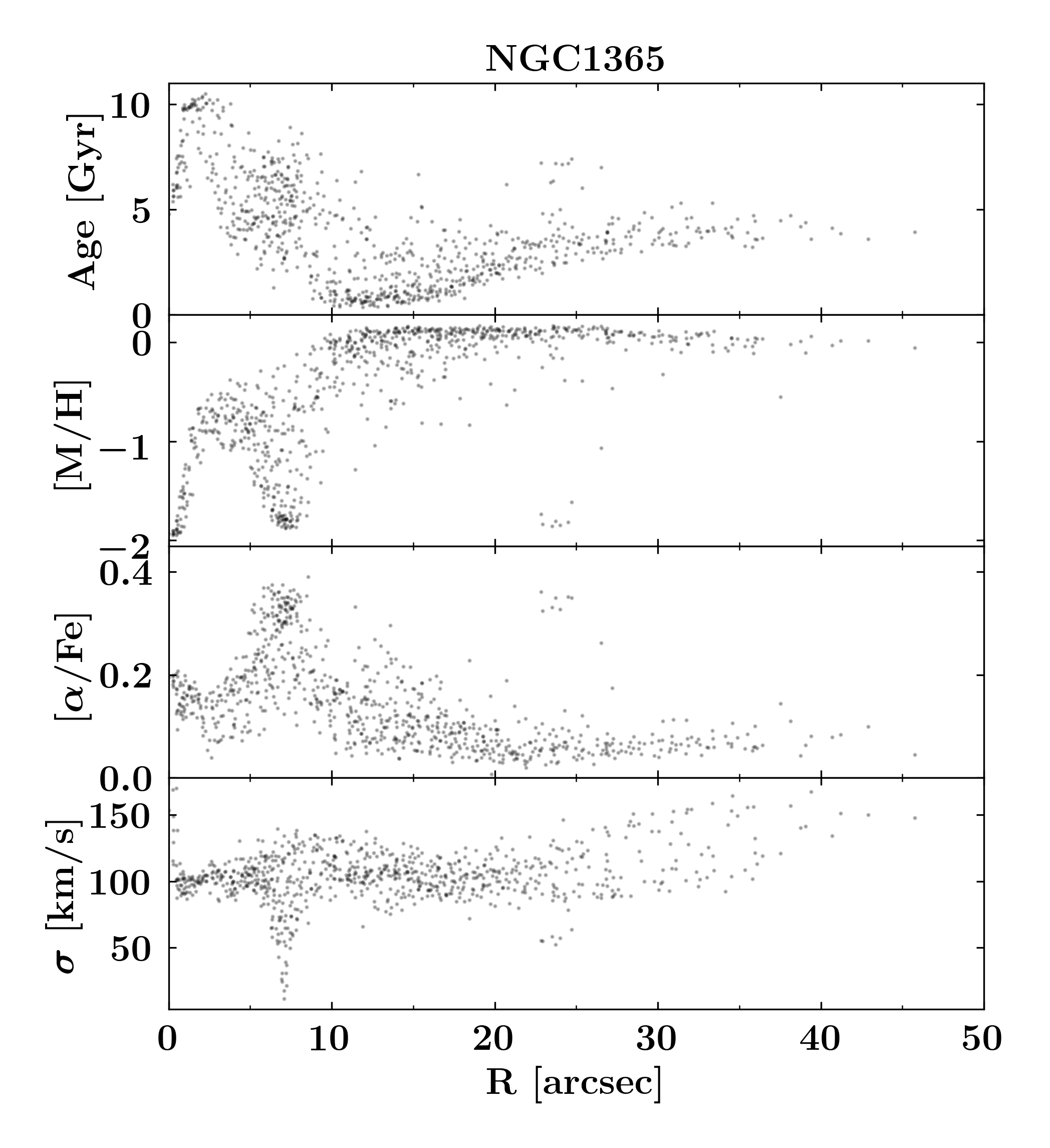}
        \end{minipage}
    \end{minipage}
    \rule{\textwidth}{0.6pt}
    \begin{minipage}[c]{\textwidth}
        \centering
        \begin{minipage}[c]{0.55\textwidth}
            \centering
            \includegraphics[width=\textwidth]{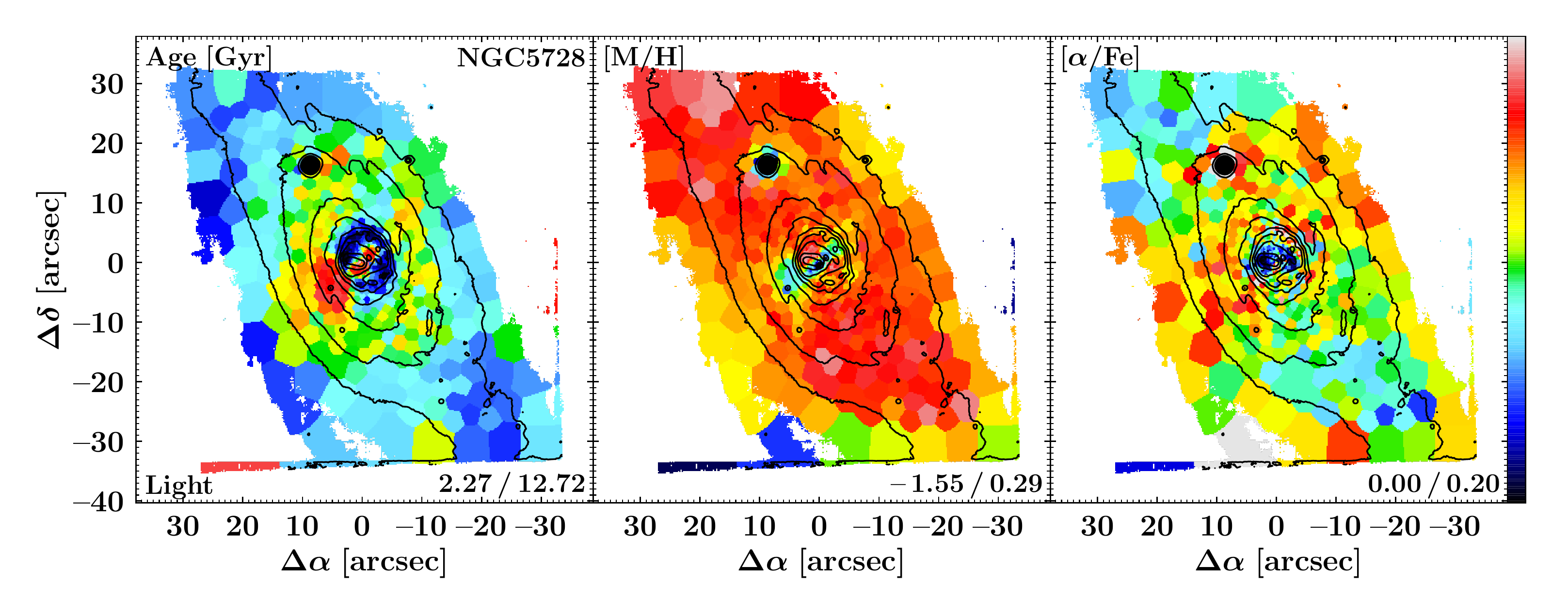}\\
            \includegraphics[width=\textwidth]{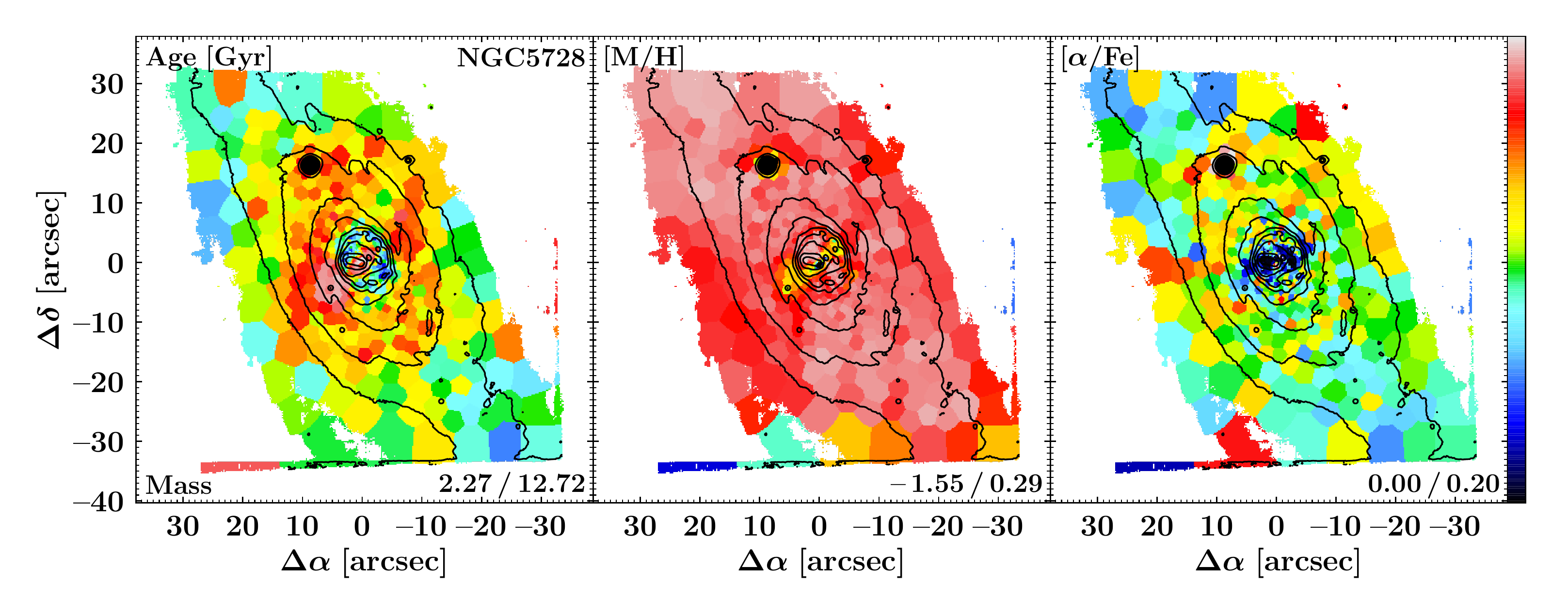}
        \end{minipage}%
        \begin{minipage}[c]{0.35\textwidth}
            \centering
            \includegraphics[width=\textwidth]{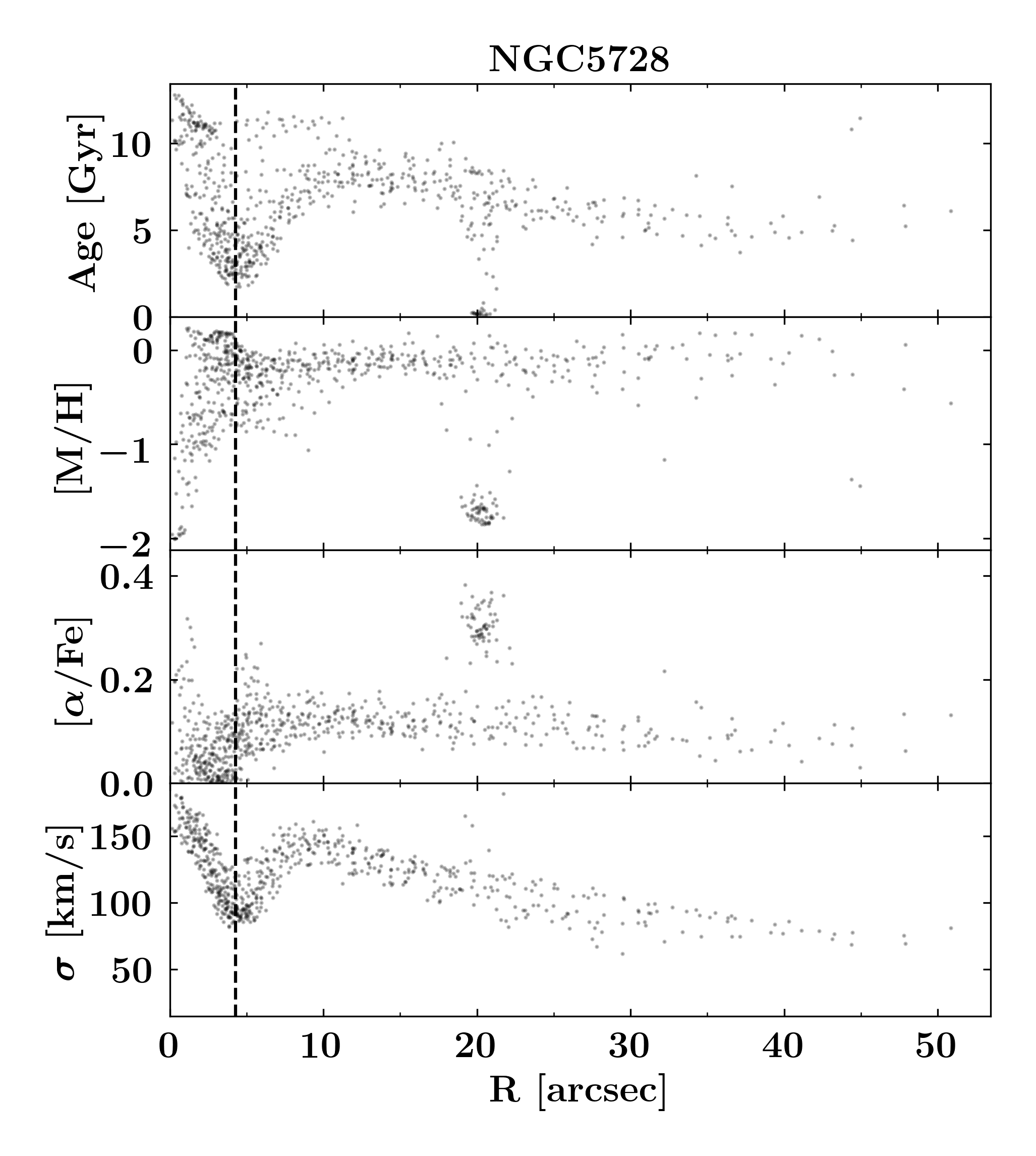}
        \end{minipage}
    \end{minipage}
    \rule{\textwidth}{0.6pt}
    \begin{minipage}[c]{\textwidth}
        \centering
        \begin{minipage}[c]{0.55\textwidth}
            \centering
            \includegraphics[width=\textwidth]{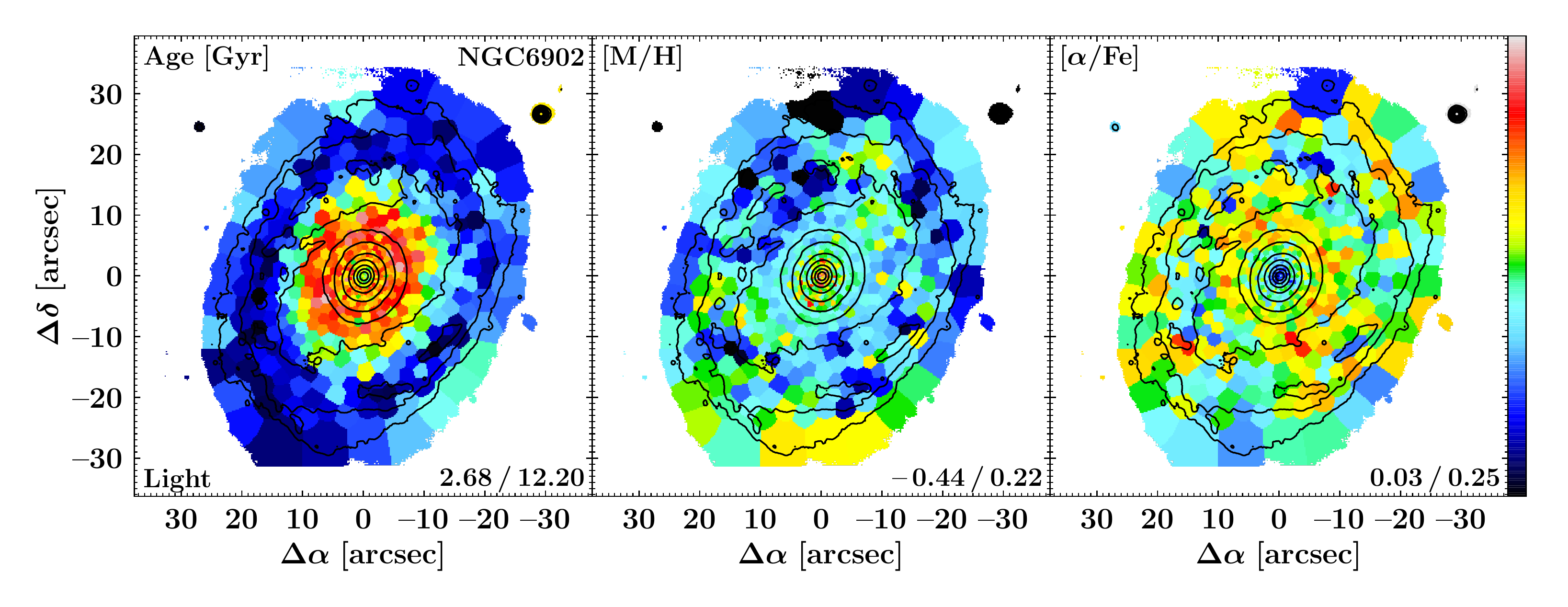}\\
            \includegraphics[width=\textwidth]{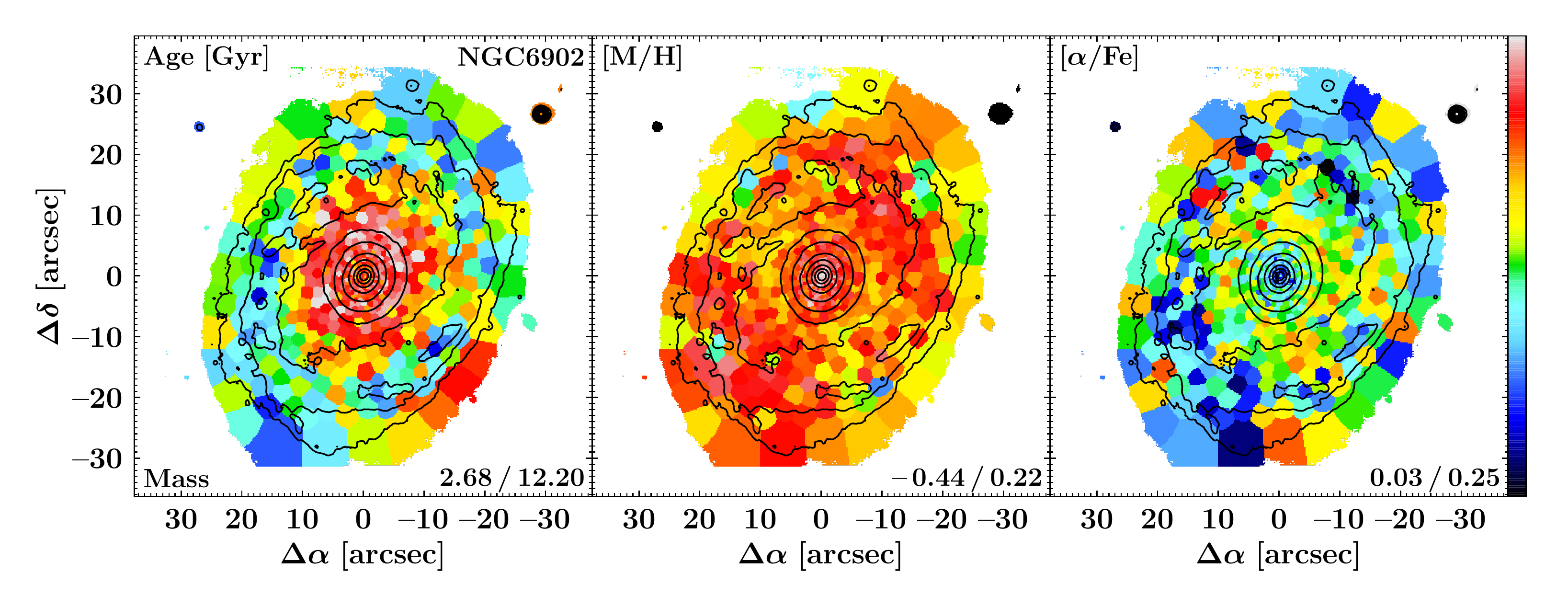}
        \end{minipage}%
        \begin{minipage}[c]{0.35\textwidth}
            \centering
            \includegraphics[width=\textwidth]{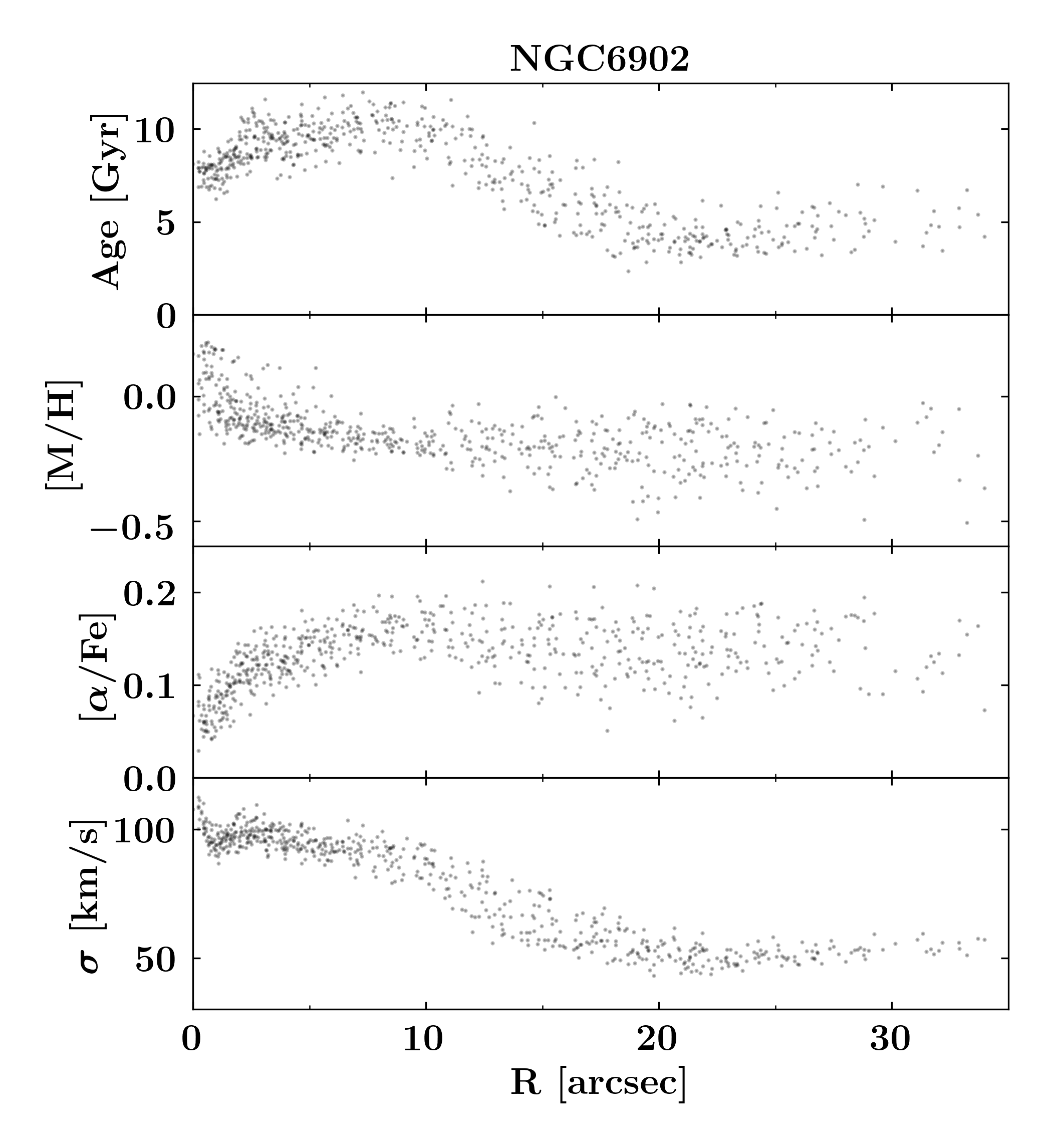}
        \end{minipage}
    \end{minipage}
    \caption{%
        Same as Fig.~\ref{fig:sppMapsNonSF}, but for the subsample with peculiar nuclear regions.  Due to the strong
        dust extinction, violent star formation, and significant contribution from an AGN, no kinematic radius is
        measured for NGC\,1365. We note that no kinematic radius is provided for NGC\,6902, as there are no clear
        kinematic signatures of a nuclear disc in this galaxy.
    }%
    \label{fig:sppMapsPeculiar}
\end{figure*}

\label{sec:maps}
In the left column of Figs.~\ref{fig:sppMapsNonSF},~\ref{fig:sppMapsSF}, and~\ref{fig:sppMapsPeculiar} we present maps
of the light-weighted (upper panels) and mass-weighted (lower panels) mean stellar population properties of the
non-star-forming and star-forming subsamples, as well as those with peculiar nuclear regions. All maps have been
generated with {\ppxf} and include the modelling of {\alphaFe} enhancements. The figures display age, metallicity, and
{\alphaFe} abundances in the left-hand, central, and right-hand panels, respectively, while different galaxies are
separated by horizontal lines. The limits of the colour bar are stated in the lower-right corner of each panel.  Based
on reconstructed intensities from the MUSE cube, we display isophotes in steps of \SI{0.5}{\mag}. North is up; east is
to the left.

In the right-hand side of the figures we plot light-weighted stellar ages (first panels), metallicities (second panels),
{\alphaFe} enhancements (third panels), and velocity dispersions (fourth panels) as a function of the galactocentric
radius of all spatial bins in the field of view.  The profiles have been deprojected using inclinations and position
angles derived in S$^4$G \citep{munozMateos2015}, as presented in Table~\ref{tab:overview}.  The vertical dashed lines
represent the kinematic radii of the nuclear discs, which was defined in G20 as the radius at which $V/\sigma$ reaches
its maximum in the region dominated by the nuclear disc.

\begin{figure*}[p]
    \begin{minipage}[c]{\textwidth}
        \centering
        \begin{minipage}[c]{0.55\textwidth}
            \centering
            \includegraphics[width=0.92\textwidth]{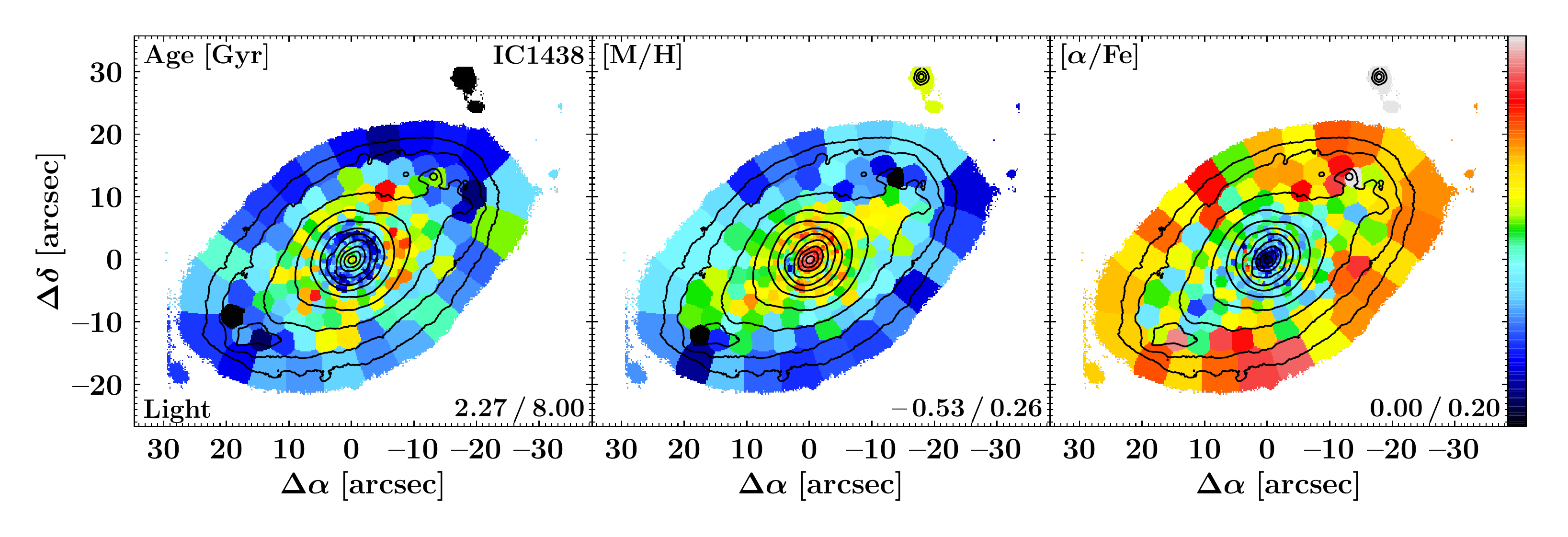}\\
            \includegraphics[width=0.92\textwidth]{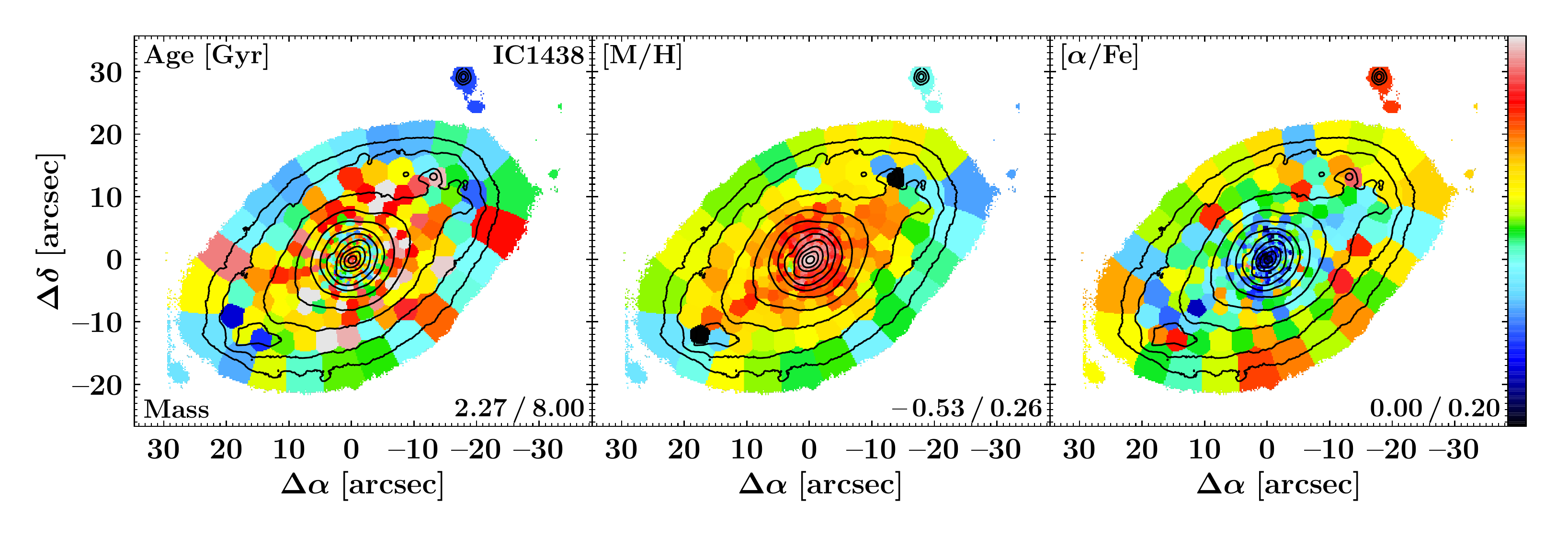}
        \end{minipage}%
        \begin{minipage}[c]{0.35\textwidth}
            \centering
            \includegraphics[width=0.92\textwidth]{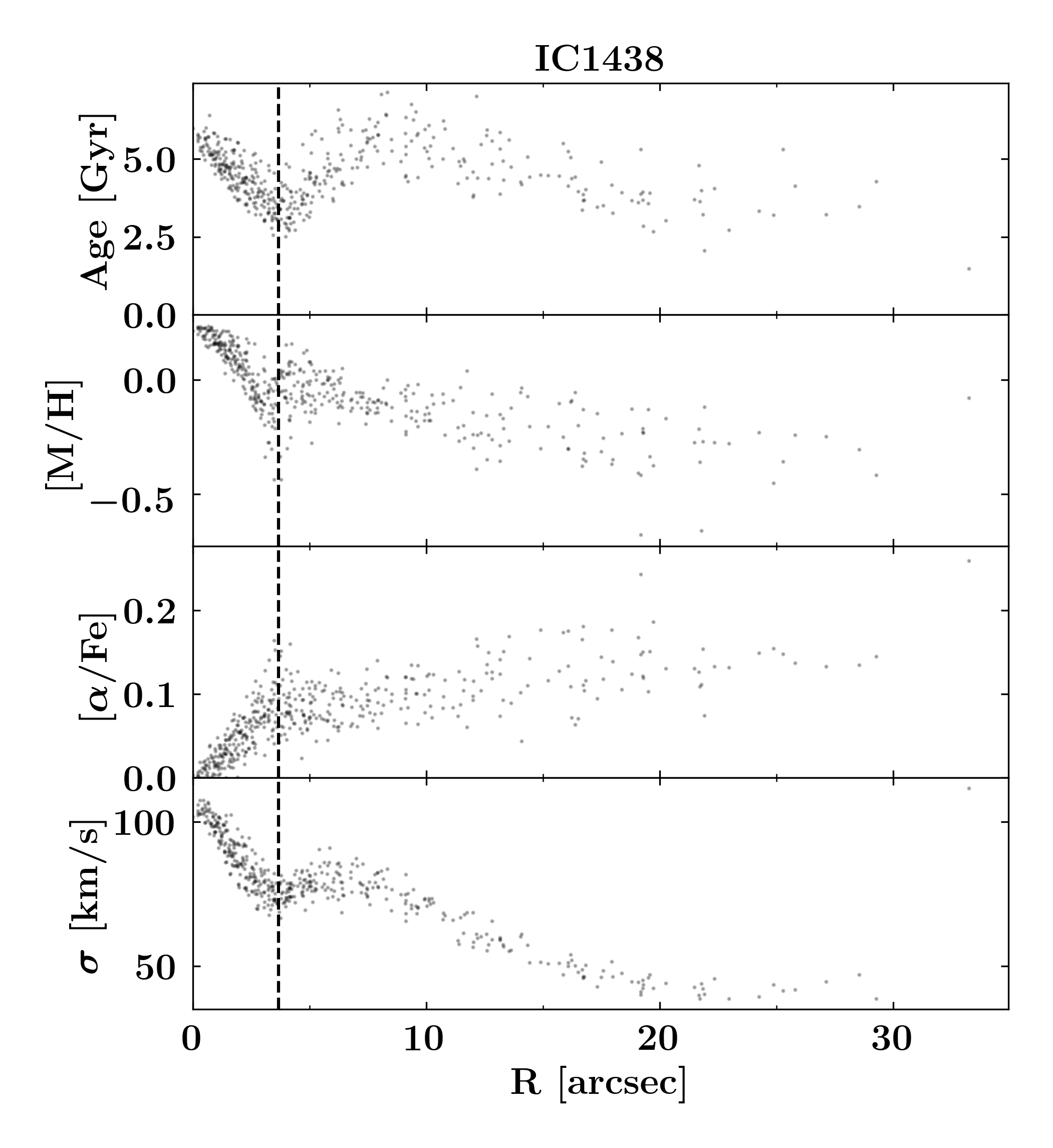}
        \end{minipage}
    \end{minipage}
    \rule{\textwidth}{0.6pt}
    \begin{minipage}[c]{\textwidth}
        \centering
        \begin{minipage}[c]{0.55\textwidth}
            \centering
            \includegraphics[width=0.92\textwidth]{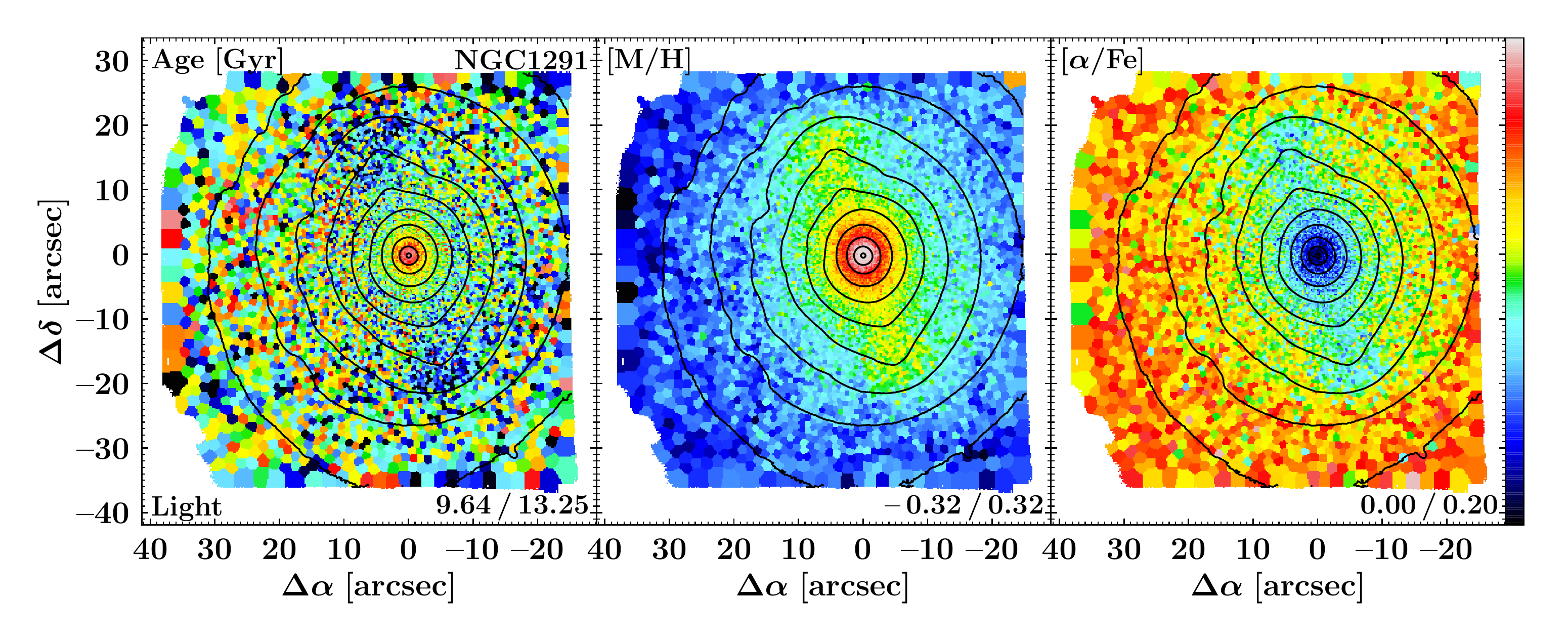}\\
            \includegraphics[width=0.92\textwidth]{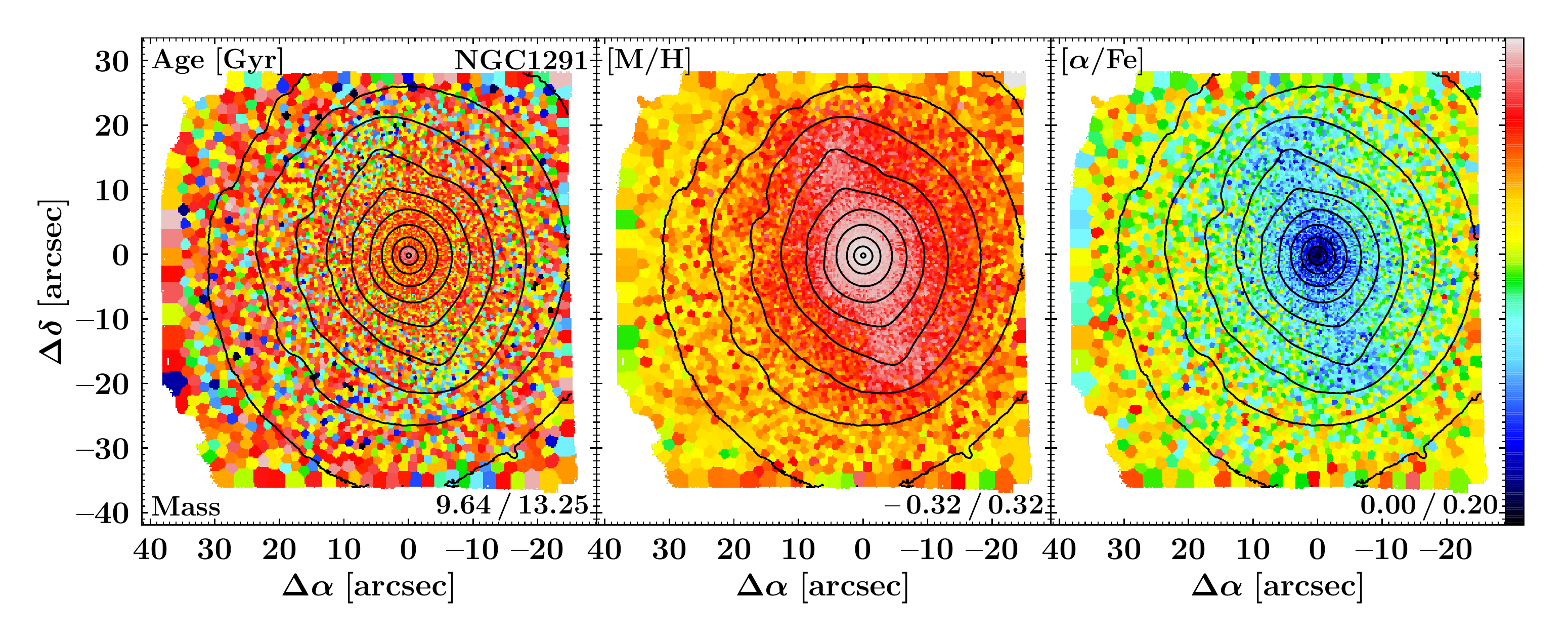}
        \end{minipage}%
        \begin{minipage}[c]{0.35\textwidth}
            \centering
            \includegraphics[width=0.92\textwidth]{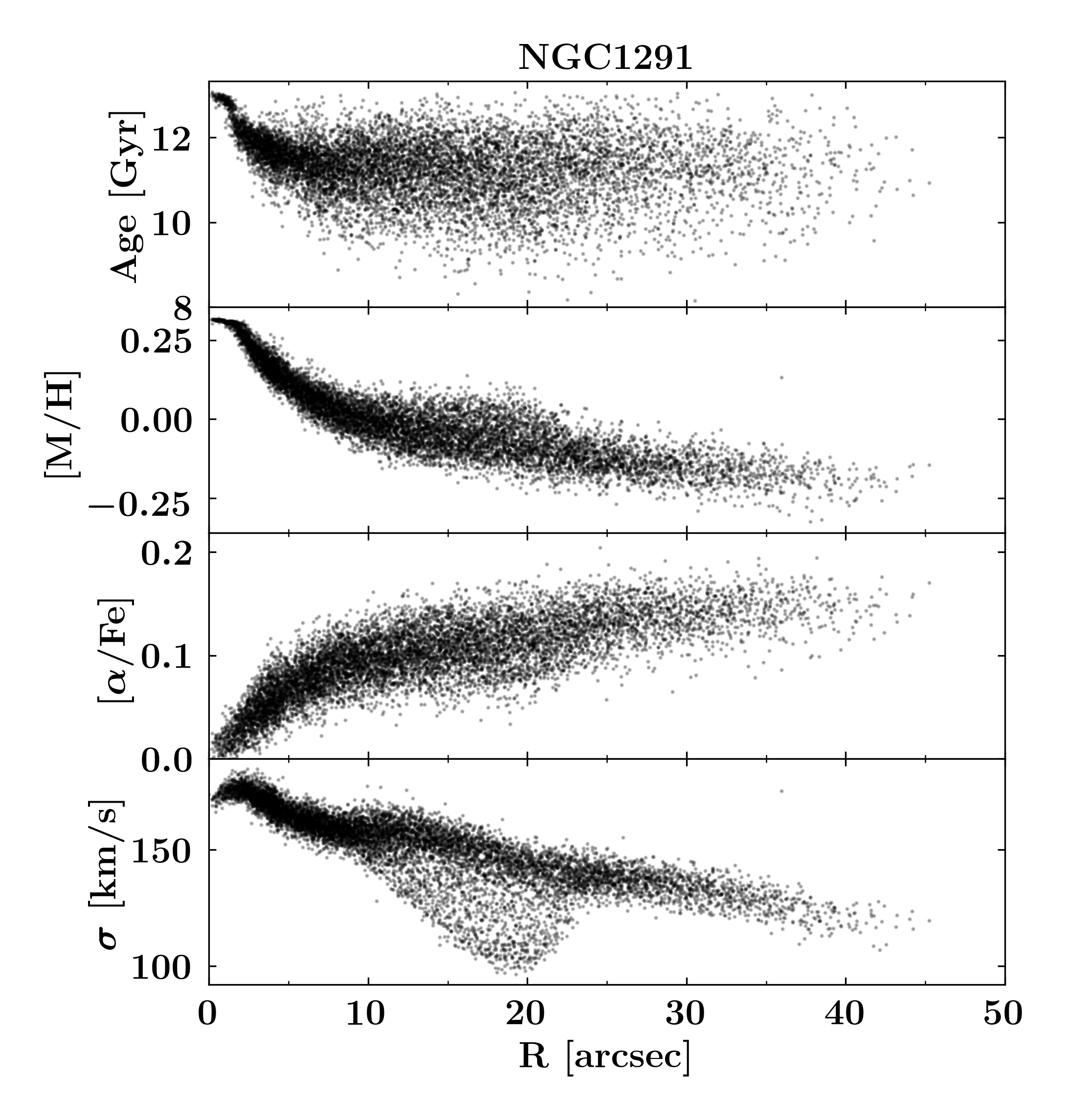}
        \end{minipage}
    \end{minipage}
    \rule{\textwidth}{0.6pt}
    \begin{minipage}[c]{\textwidth}
        \centering
        \begin{minipage}[c]{0.55\textwidth}
            \centering
            \includegraphics[width=0.92\textwidth]{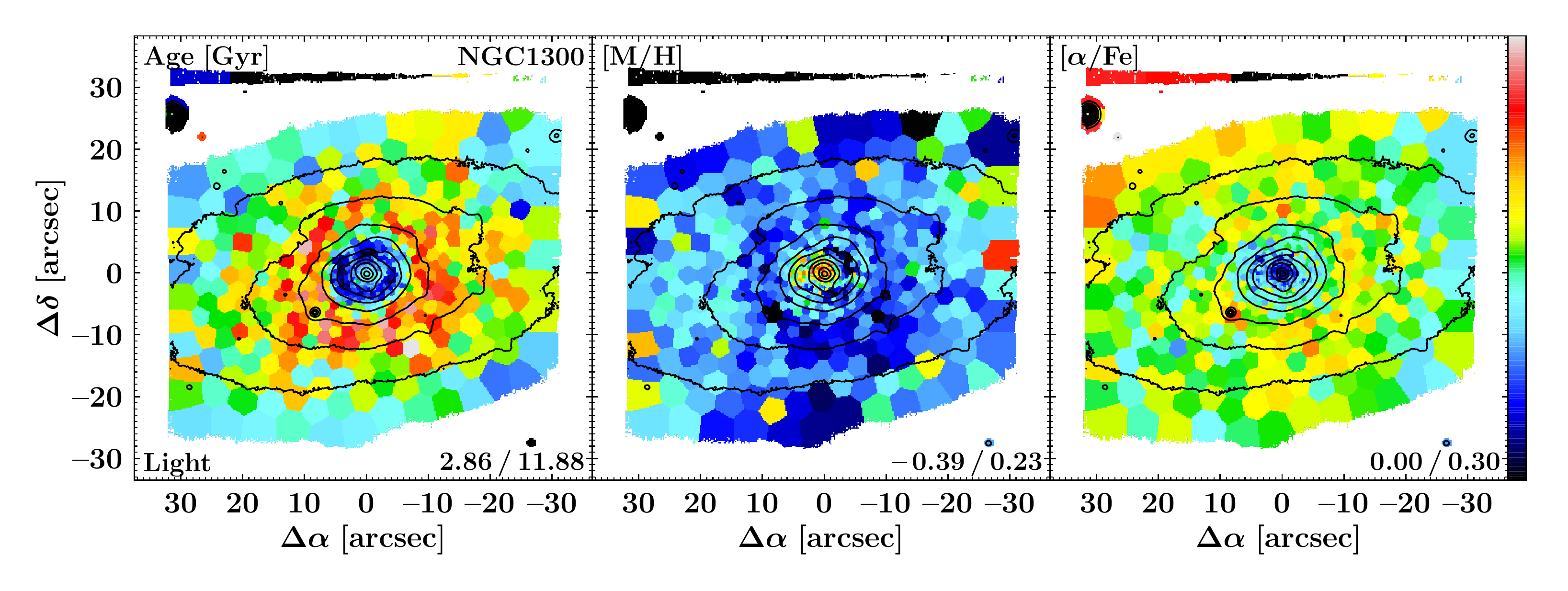}\\
            \includegraphics[width=0.92\textwidth]{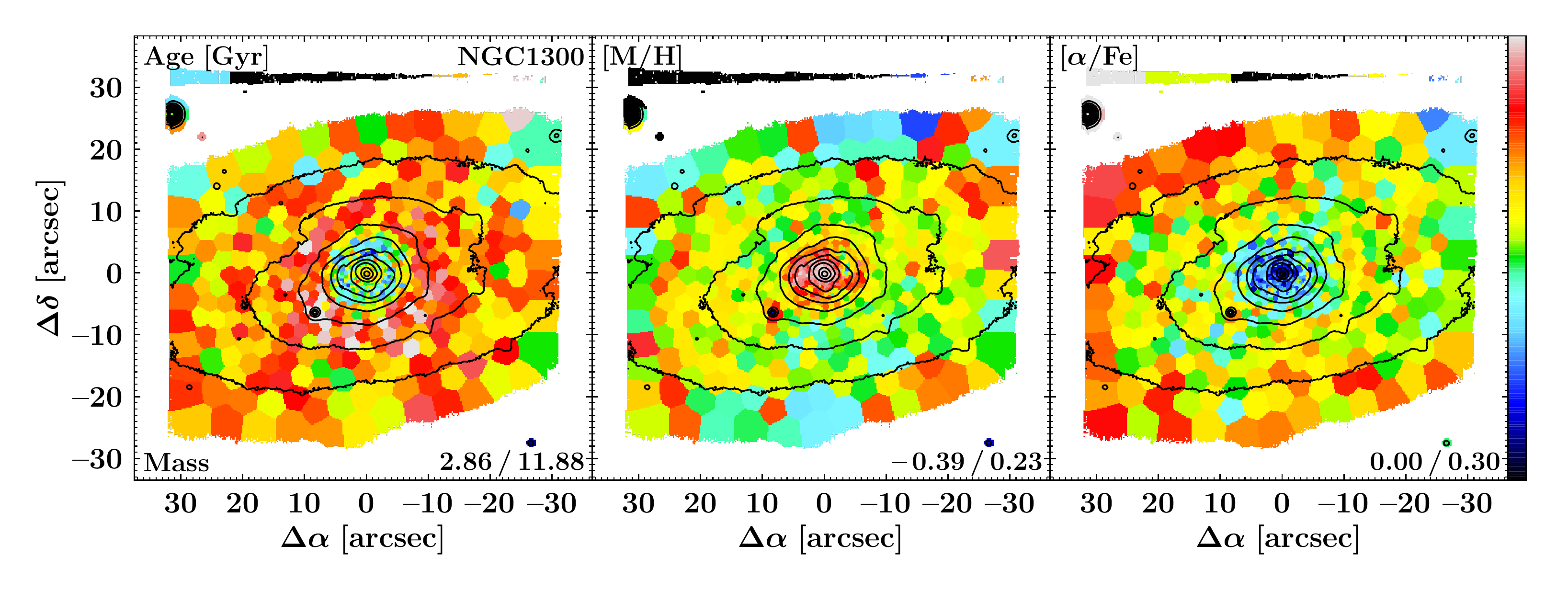}
        \end{minipage}%
        \begin{minipage}[c]{0.35\textwidth}
            \centering
            \includegraphics[width=0.92\textwidth]{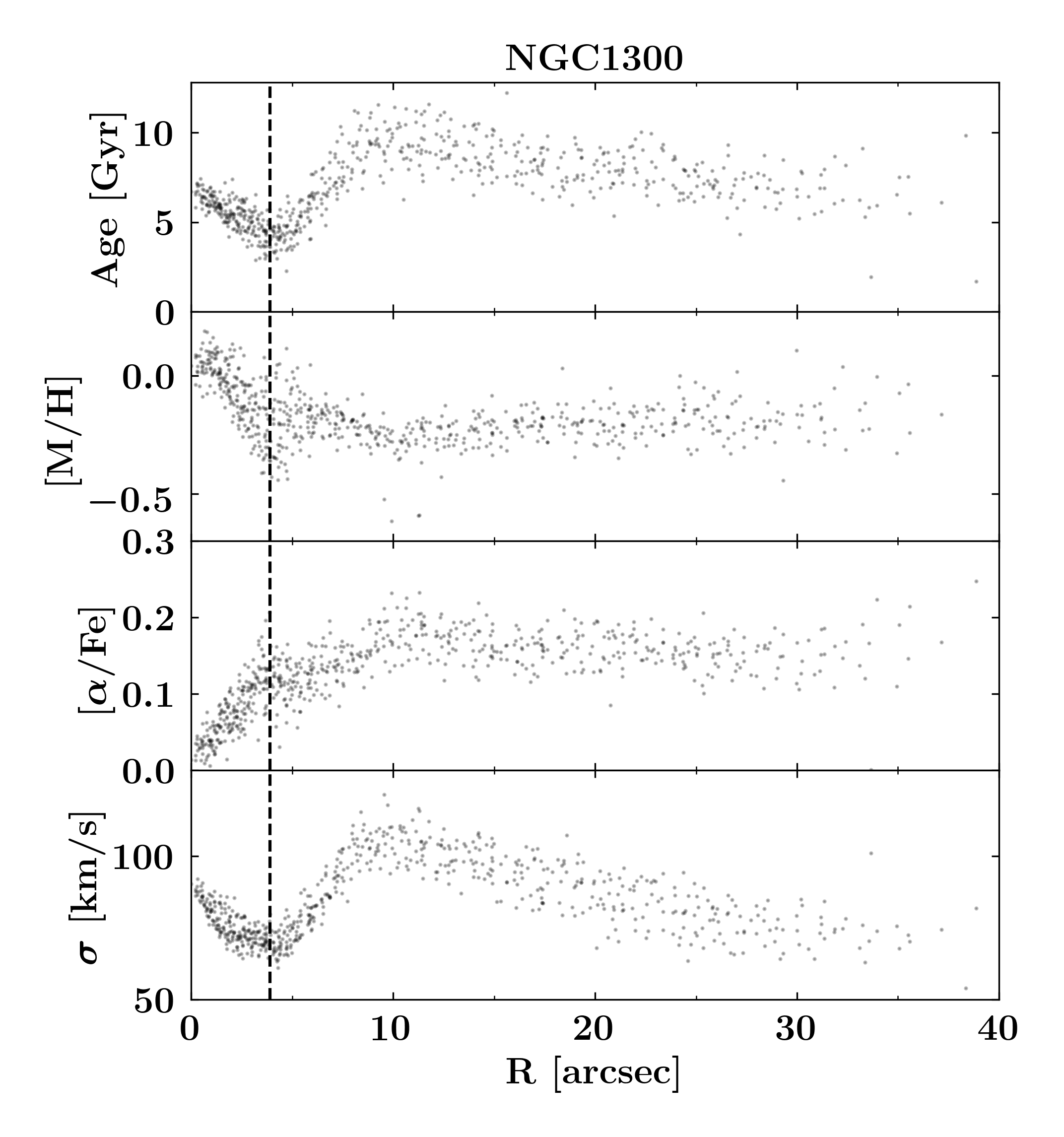}
        \end{minipage}
    \end{minipage}
    \caption{%
        \emph{Left:} Maps of light-weighted (upper panels) and mass-weighted (lower panels) mean stellar population
        properties of the subsample without significant star formation in the nuclear ring. All maps have been generated
        with {\ppxf} and include the modelling of {\alphaFe} enhancements. The figures display age, [M/H], and
        {\alphaFe} enhancements in the left-hand, centre, and right-hand panels, respectively, while different galaxies
        are separated by horizontal lines. The limits of the colour bar are stated in the lower-right corner of each
        panel.  Based on reconstructed intensities from the MUSE cube, we display isophotes in steps of \SI{0.5}{\mag}.
        North is up; east is to the left. 
        \emph{Right:} Radial profiles of light-weighted stellar ages (first panels), metallicities (second panels),
        {\alphaFe} enhancements (third panels), and velocity dispersions (fourth panels) as a function of the
        galactocentric radius of all spatial bins in the field of view. The profiles have been deprojected using
        inclinations and position angles derived in S$^4$G \citep{munozMateos2015}, as presented in
        Table~\ref{tab:overview}. The vertical dashed lines represent the kinematic radii of the nuclear discs, which
        was defined in G20 as the radius at which $V/\sigma$ reaches its maximum in the region dominated by the nuclear
        disc. 
        We note that for NGC\,1291 no kinematic radius could be determined, as this galaxy is oriented almost perfectly
        face-on.
    }%
    \label{fig:sppMapsNonSF}
\end{figure*}
\begin{figure*}[p]
    \ContinuedFloat%
    \begin{minipage}[c]{\textwidth}
        \centering
        \begin{minipage}[c]{0.55\textwidth}
            \centering
            \includegraphics[width=\textwidth]{NGC1433_spp_ppxf_LightAlpha.pdf}\\
            \includegraphics[width=\textwidth]{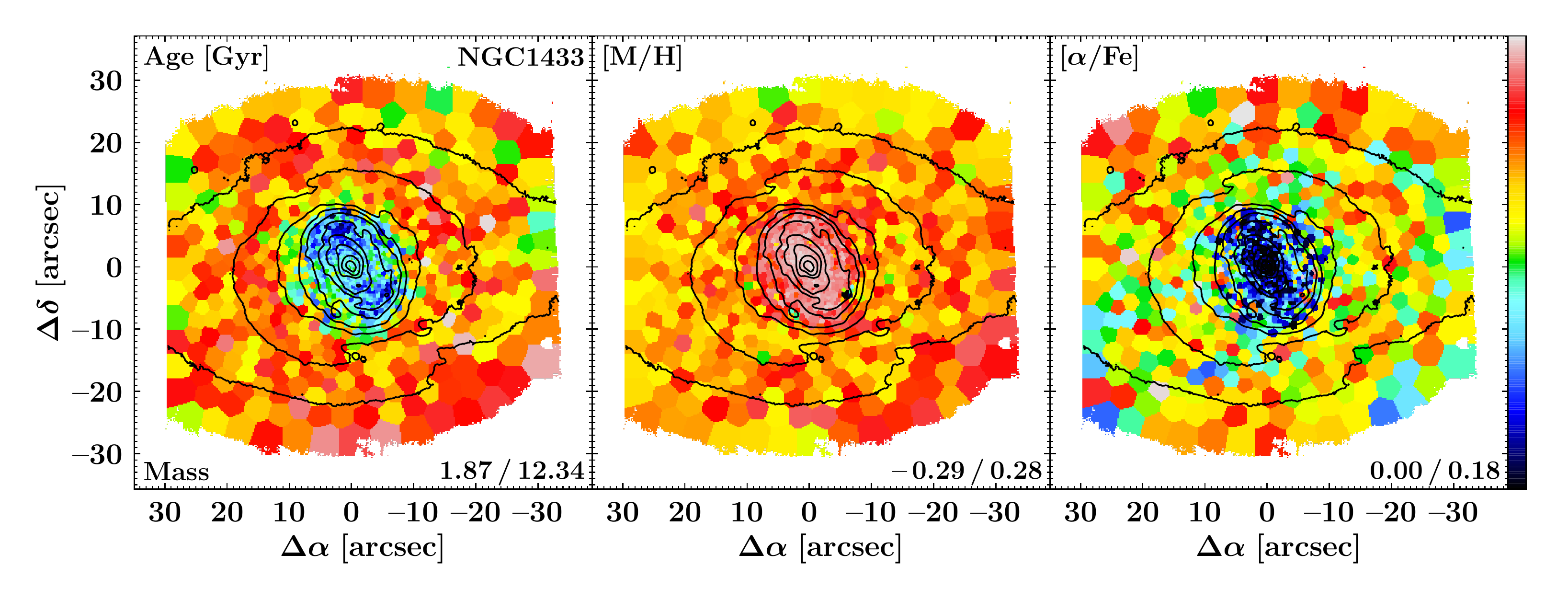}
        \end{minipage}%
        \begin{minipage}[c]{0.35\textwidth}
            \centering
            \includegraphics[width=\textwidth]{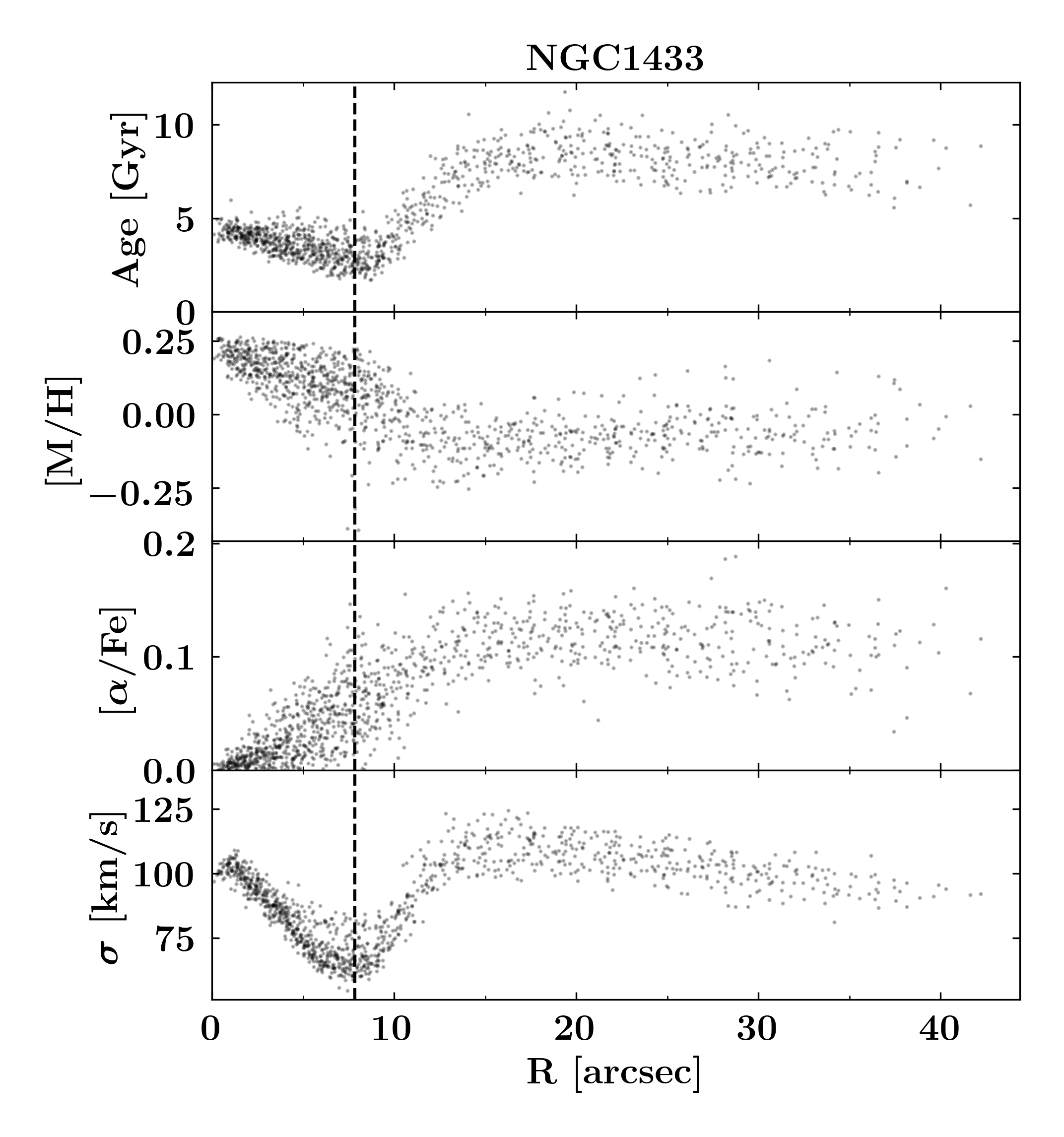}
        \end{minipage}
    \end{minipage}
    \rule{\textwidth}{0.6pt}
    \begin{minipage}[c]{\textwidth}
        \centering
        \begin{minipage}[c]{0.55\textwidth}
            \centering
            \includegraphics[width=\textwidth]{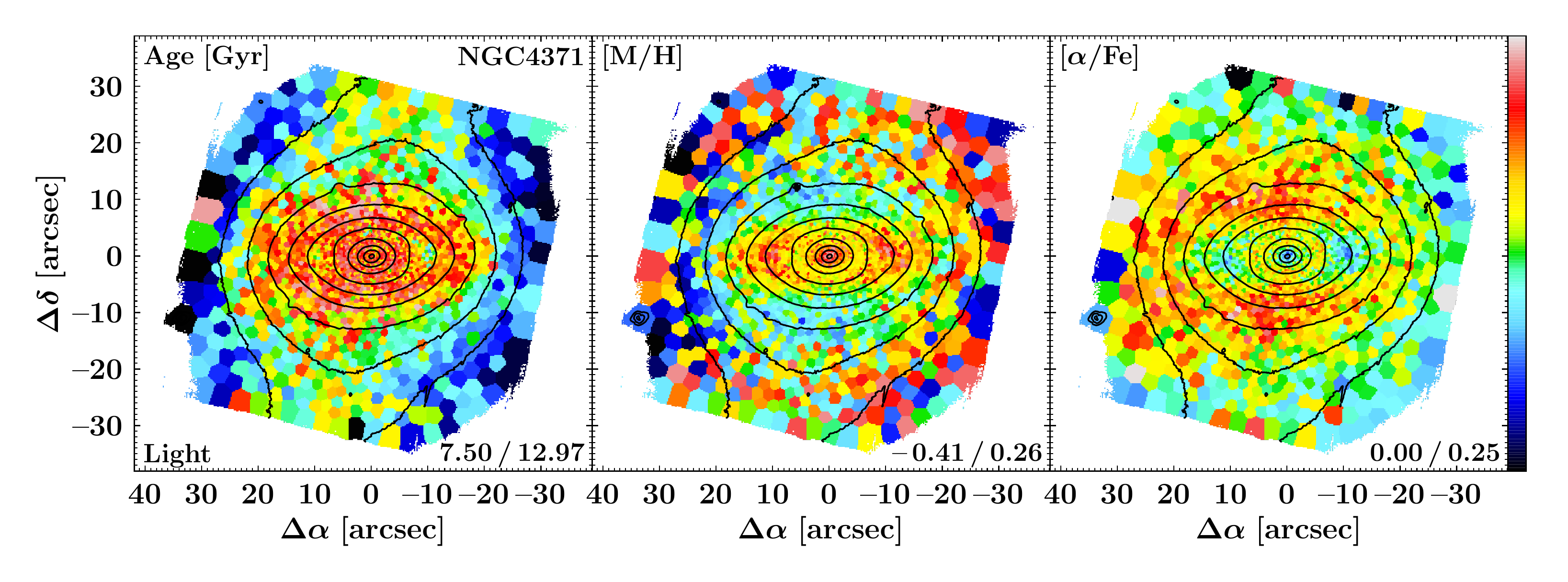}\\
            \includegraphics[width=\textwidth]{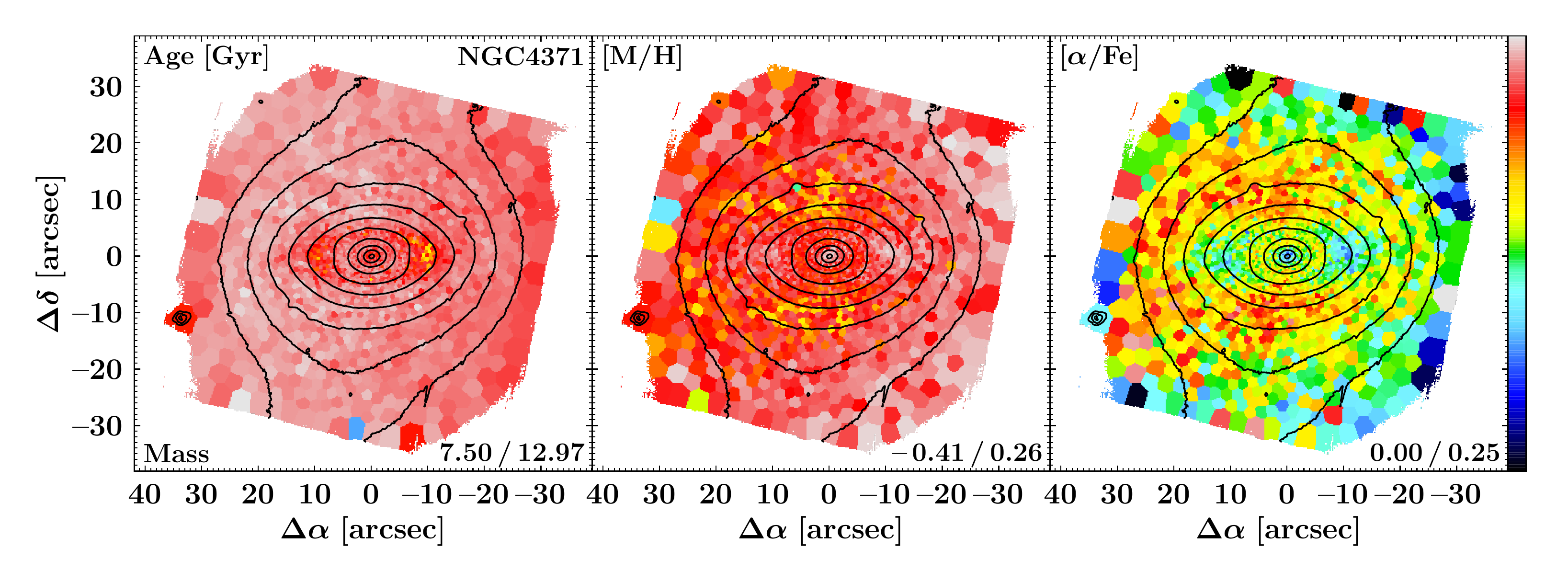}
        \end{minipage}%
        \begin{minipage}[c]{0.35\textwidth}
            \centering
            \includegraphics[width=\textwidth]{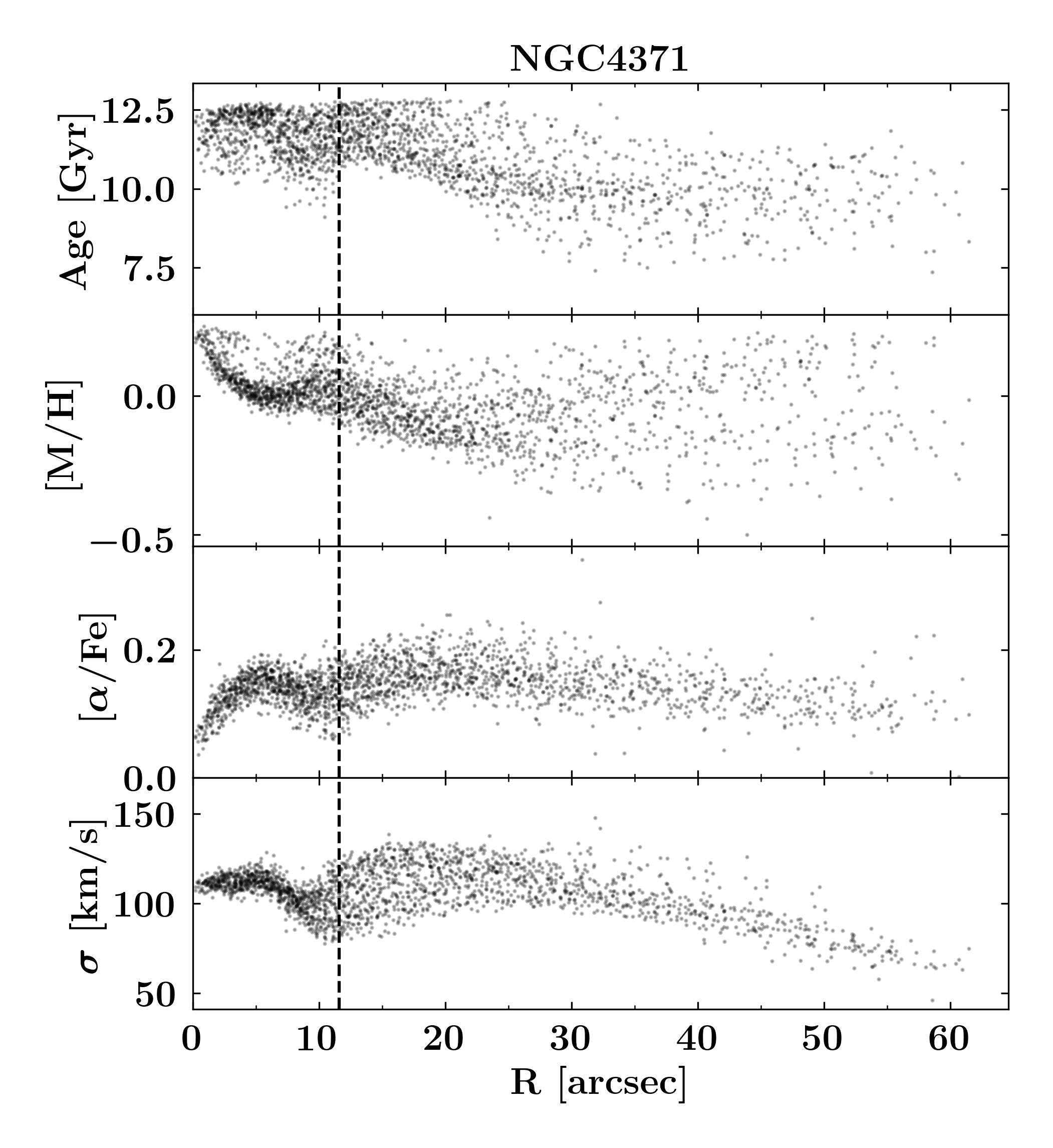}
        \end{minipage}
    \end{minipage}
    \rule{\textwidth}{0.6pt}
    \begin{minipage}[c]{\textwidth}
        \centering
        \begin{minipage}[c]{0.55\textwidth}
            \centering
            \includegraphics[width=\textwidth]{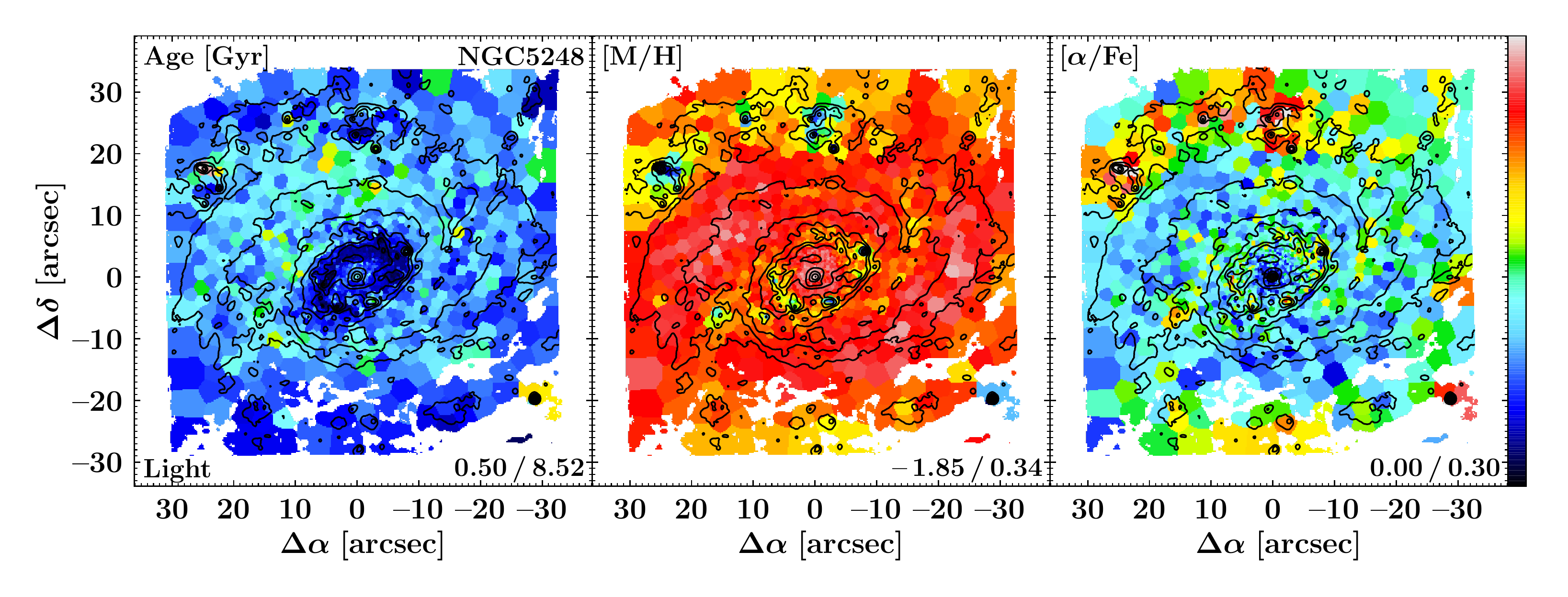}\\
            \includegraphics[width=\textwidth]{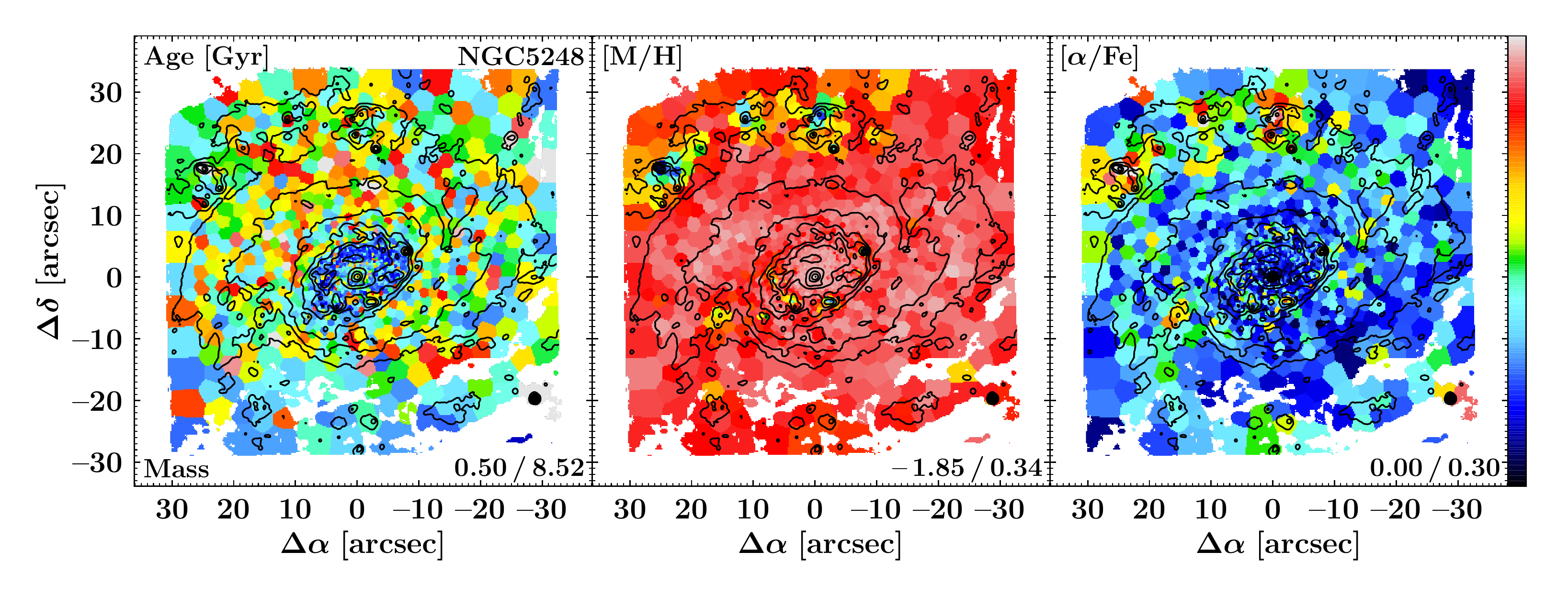}
        \end{minipage}%
        \begin{minipage}[c]{0.35\textwidth}
            \centering
            \includegraphics[width=\textwidth]{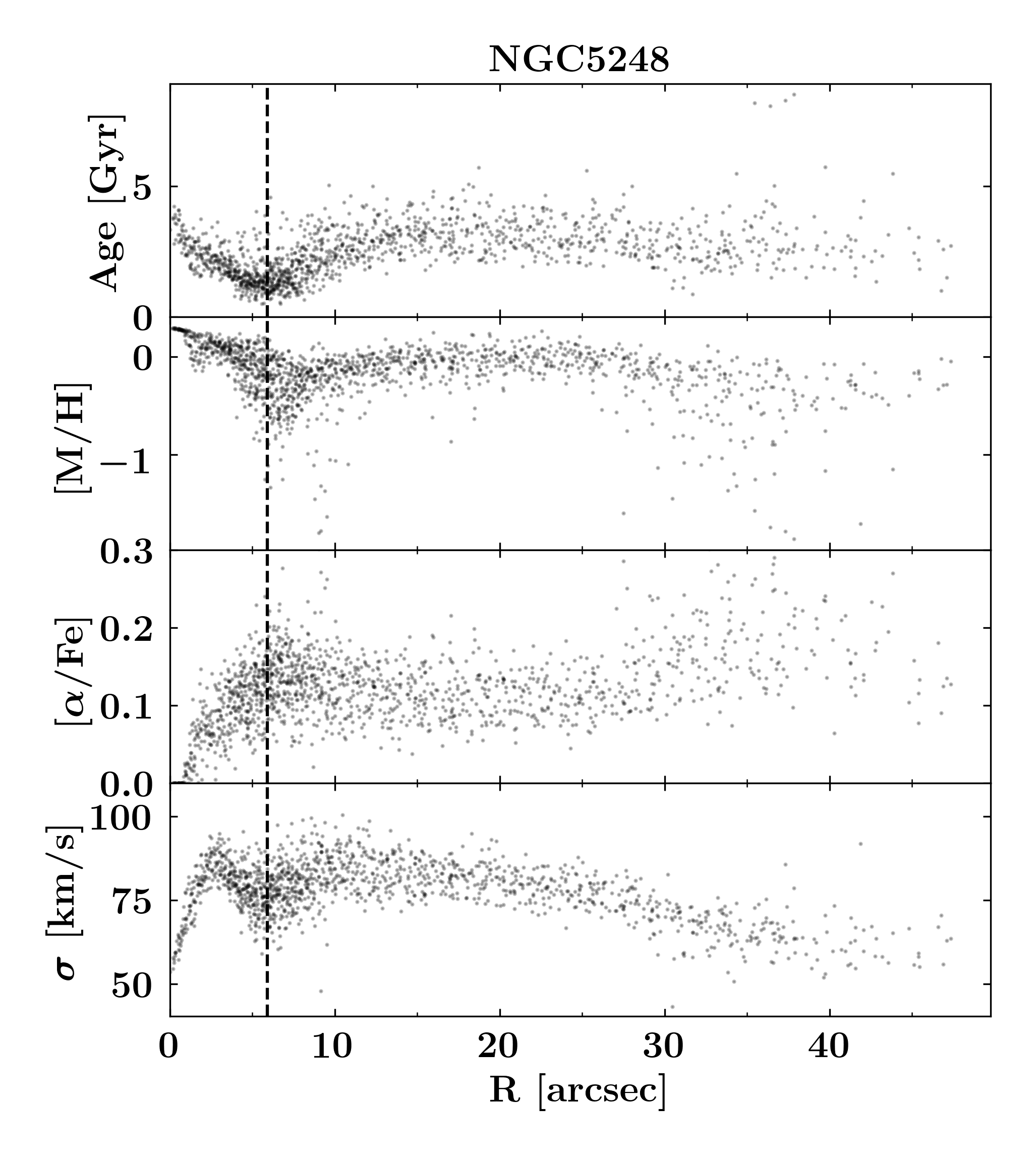}
        \end{minipage}
    \end{minipage}
    \caption{Continued.}%
    \label{fig:sppMapsNonSF}
\end{figure*}
\begin{figure*}[p]
    \ContinuedFloat%
    \begin{minipage}[c]{\textwidth}
        \centering
        \begin{minipage}[c]{0.55\textwidth}
            \centering
            \includegraphics[width=\textwidth]{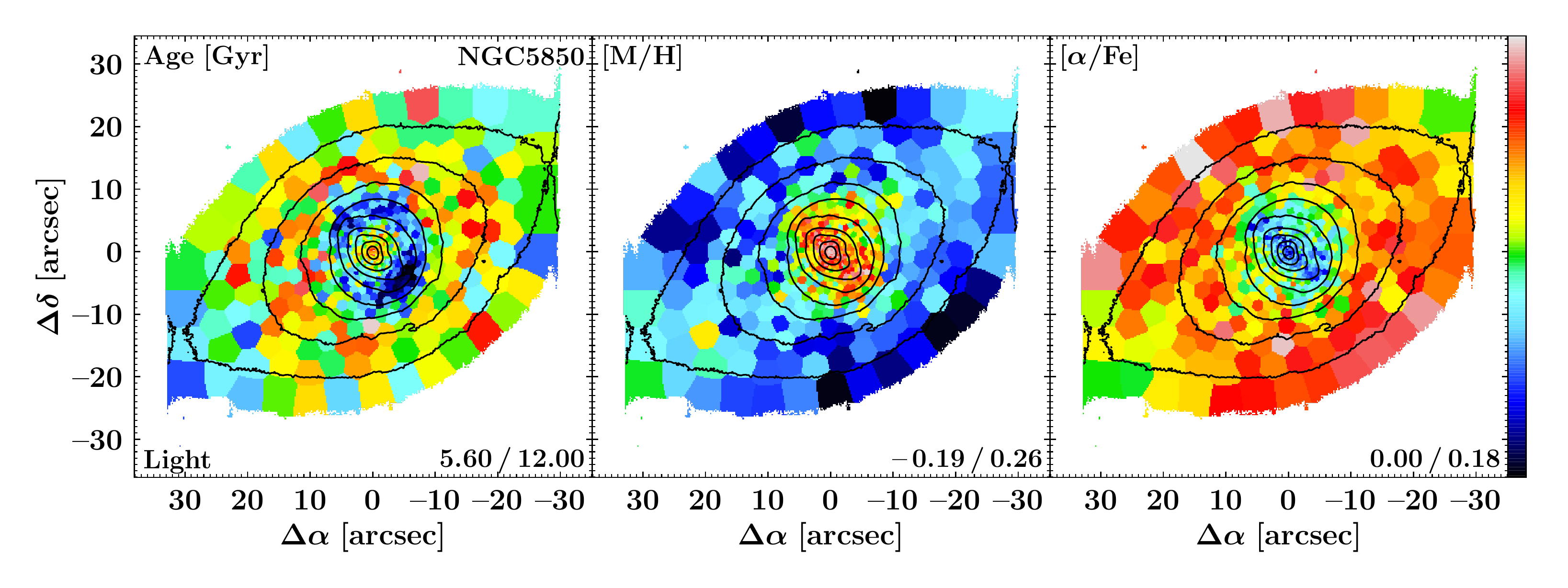}\\
            \includegraphics[width=\textwidth]{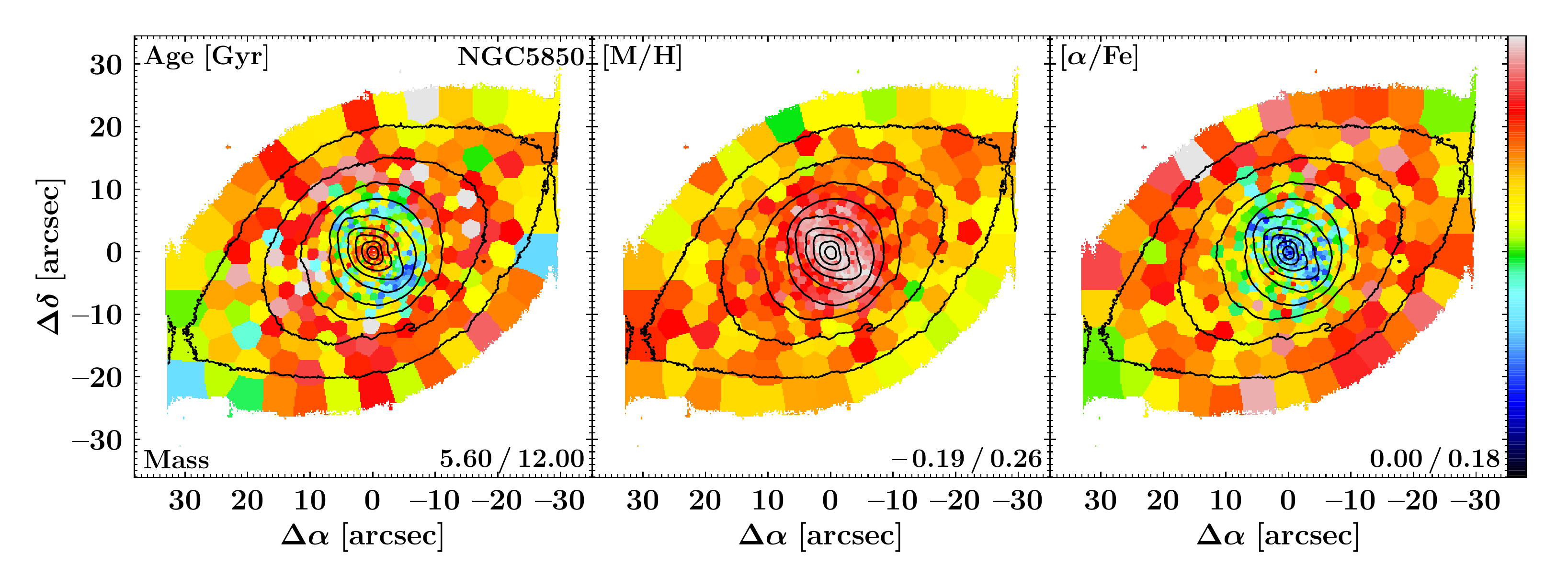}
        \end{minipage}%
        \begin{minipage}[c]{0.35\textwidth}
            \centering
            \includegraphics[width=\textwidth]{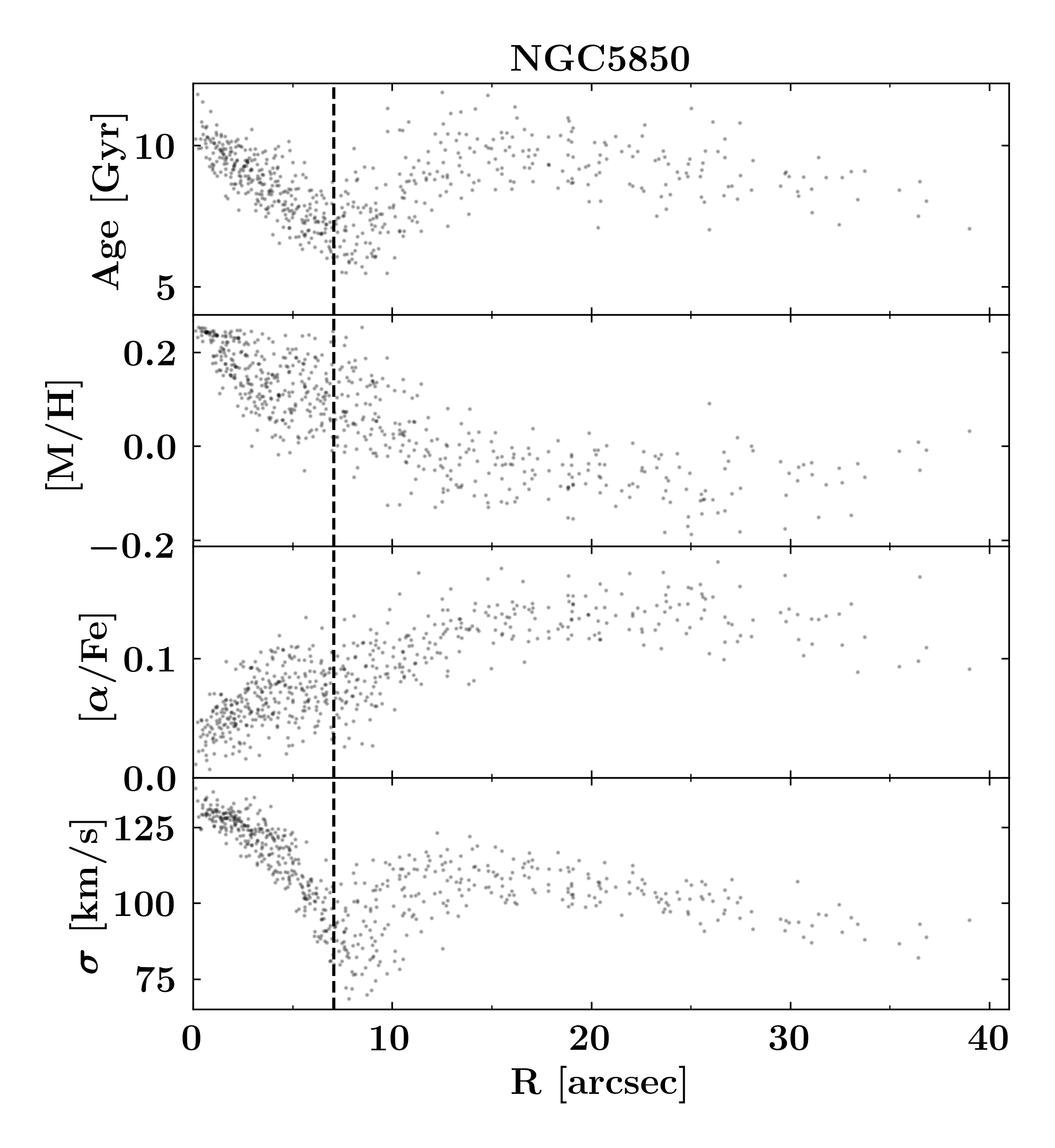}
        \end{minipage}
    \end{minipage}
    \rule{\textwidth}{0.6pt}
    \begin{minipage}[c]{\textwidth}
        \centering
        \begin{minipage}[c]{0.55\textwidth}
            \centering
            \includegraphics[width=\textwidth]{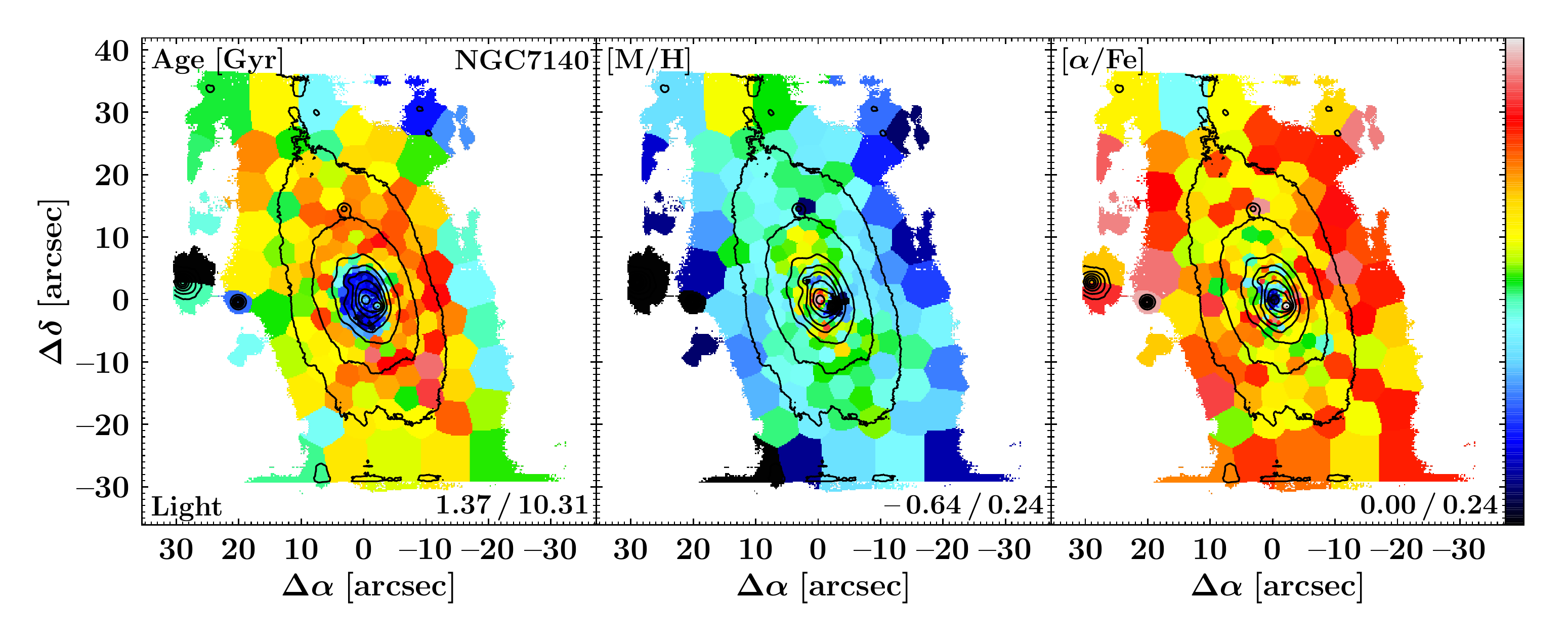}\\
            \includegraphics[width=\textwidth]{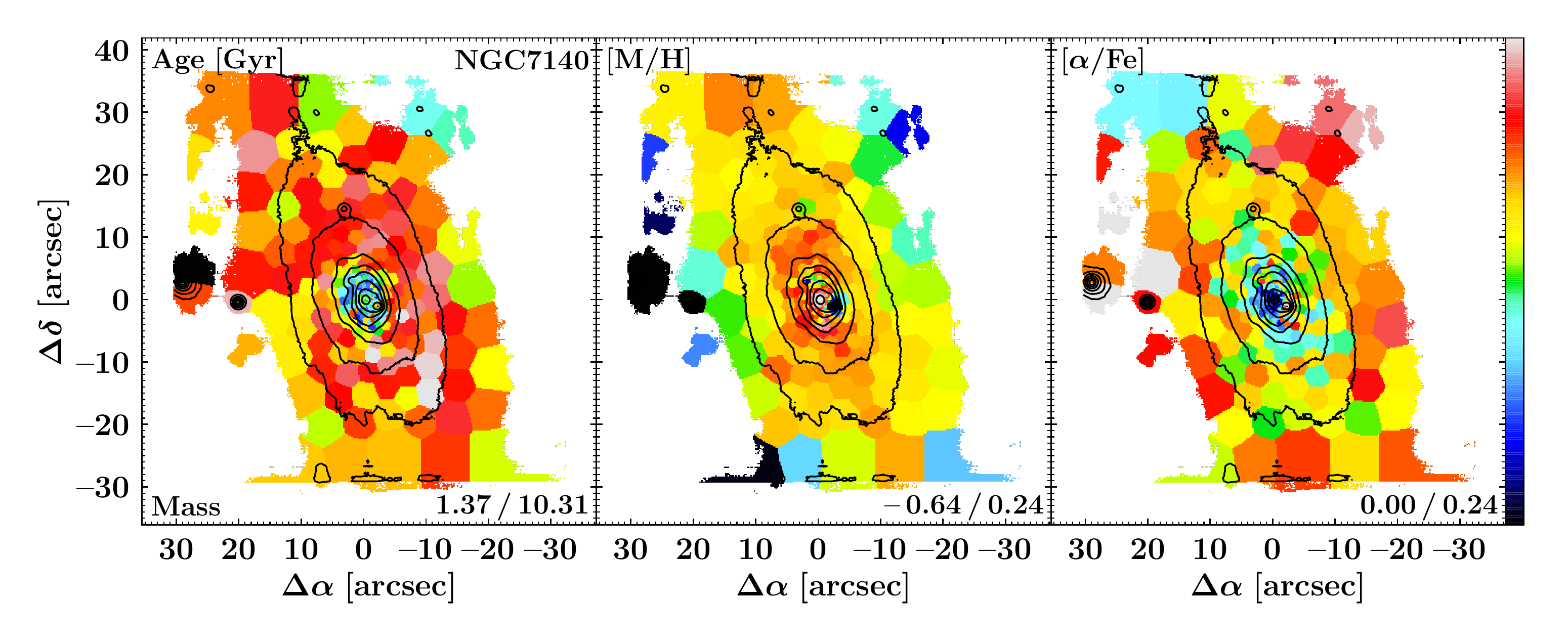}
        \end{minipage}%
        \begin{minipage}[c]{0.35\textwidth}
            \centering
            \includegraphics[width=\textwidth]{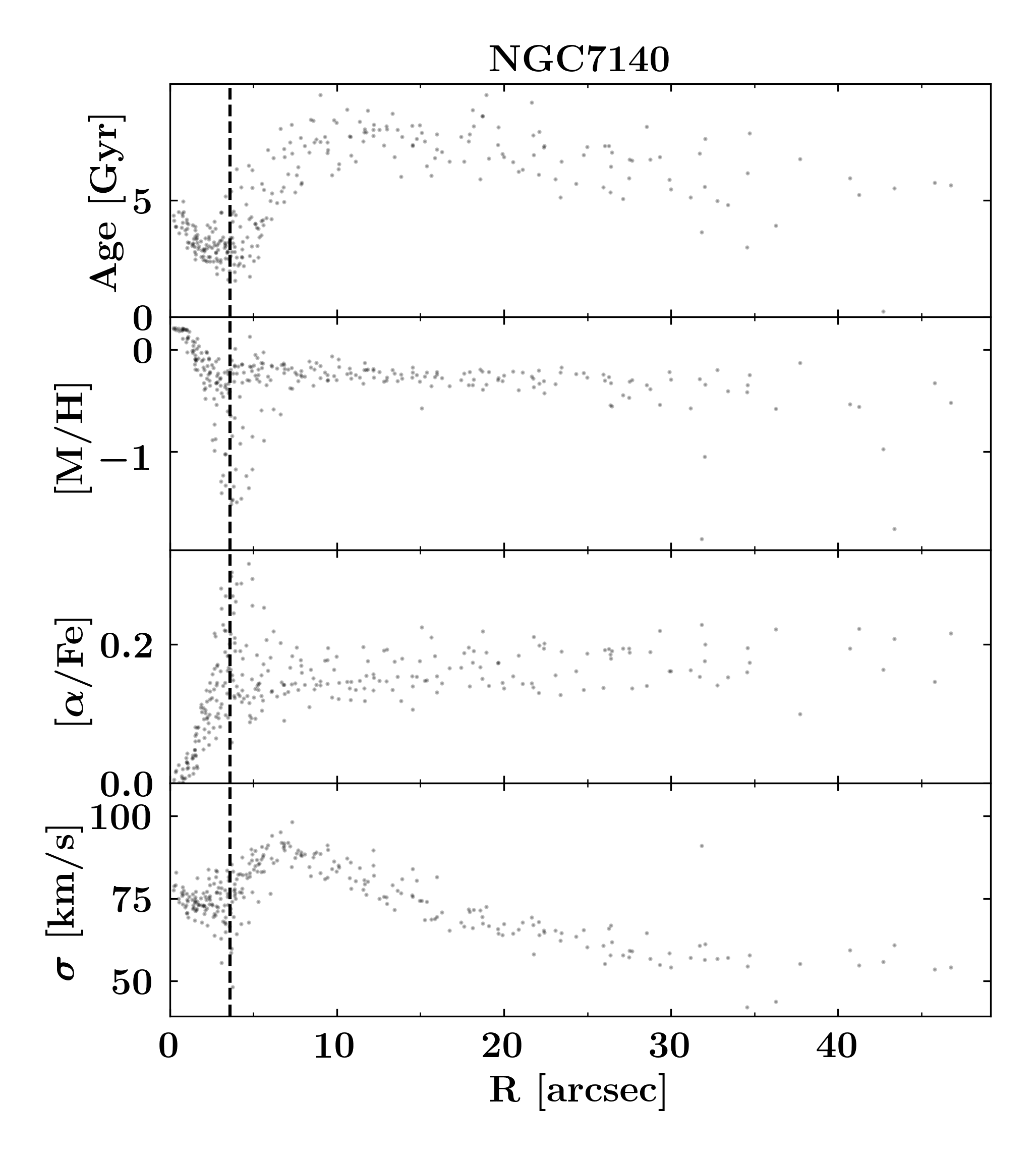}
        \end{minipage}
    \end{minipage}
    \rule{\textwidth}{0.6pt}
    \begin{minipage}[c]{\textwidth}
        \centering
        \begin{minipage}[c]{0.55\textwidth}
            \centering
            \includegraphics[width=\textwidth]{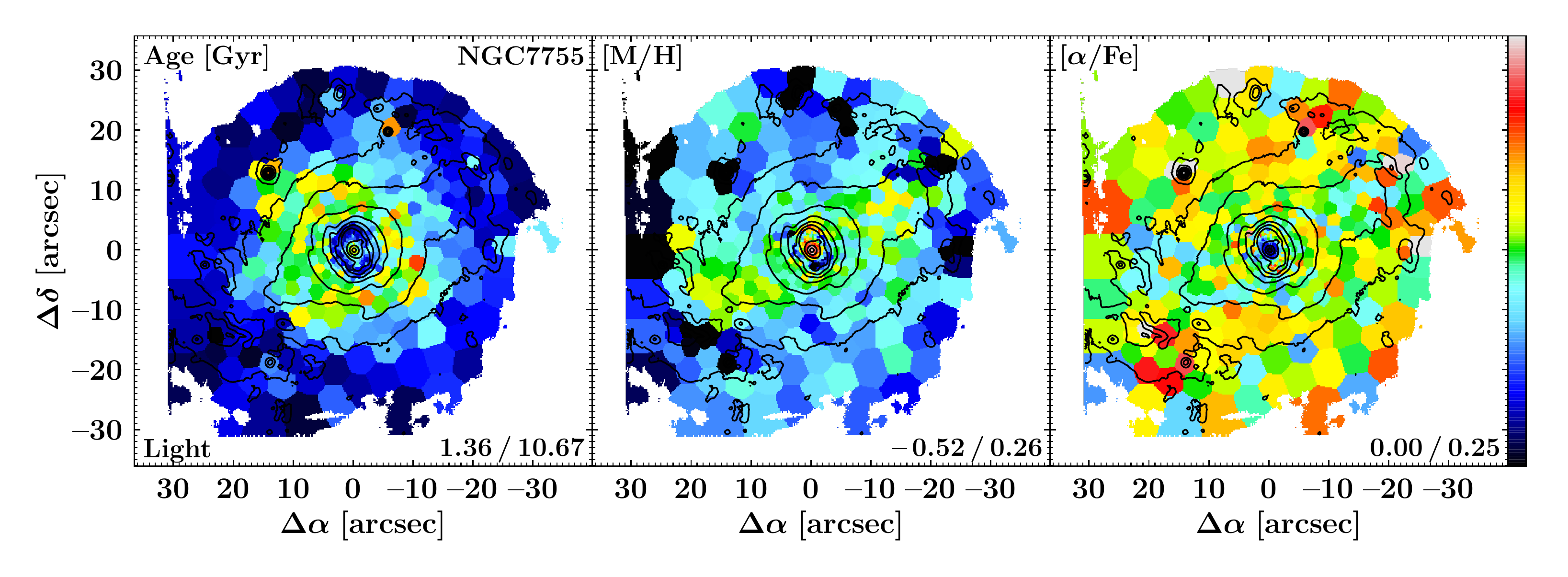}\\
            \includegraphics[width=\textwidth]{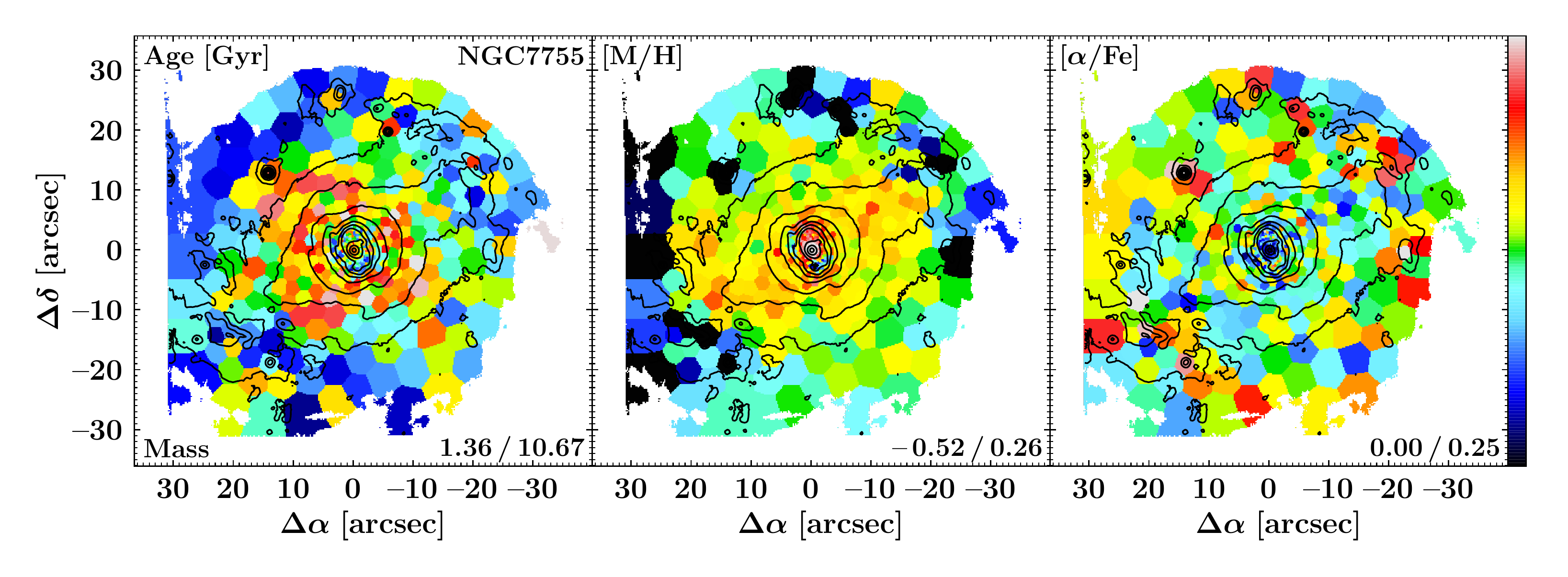}
        \end{minipage}%
        \begin{minipage}[c]{0.35\textwidth}
            \centering
            \includegraphics[width=\textwidth]{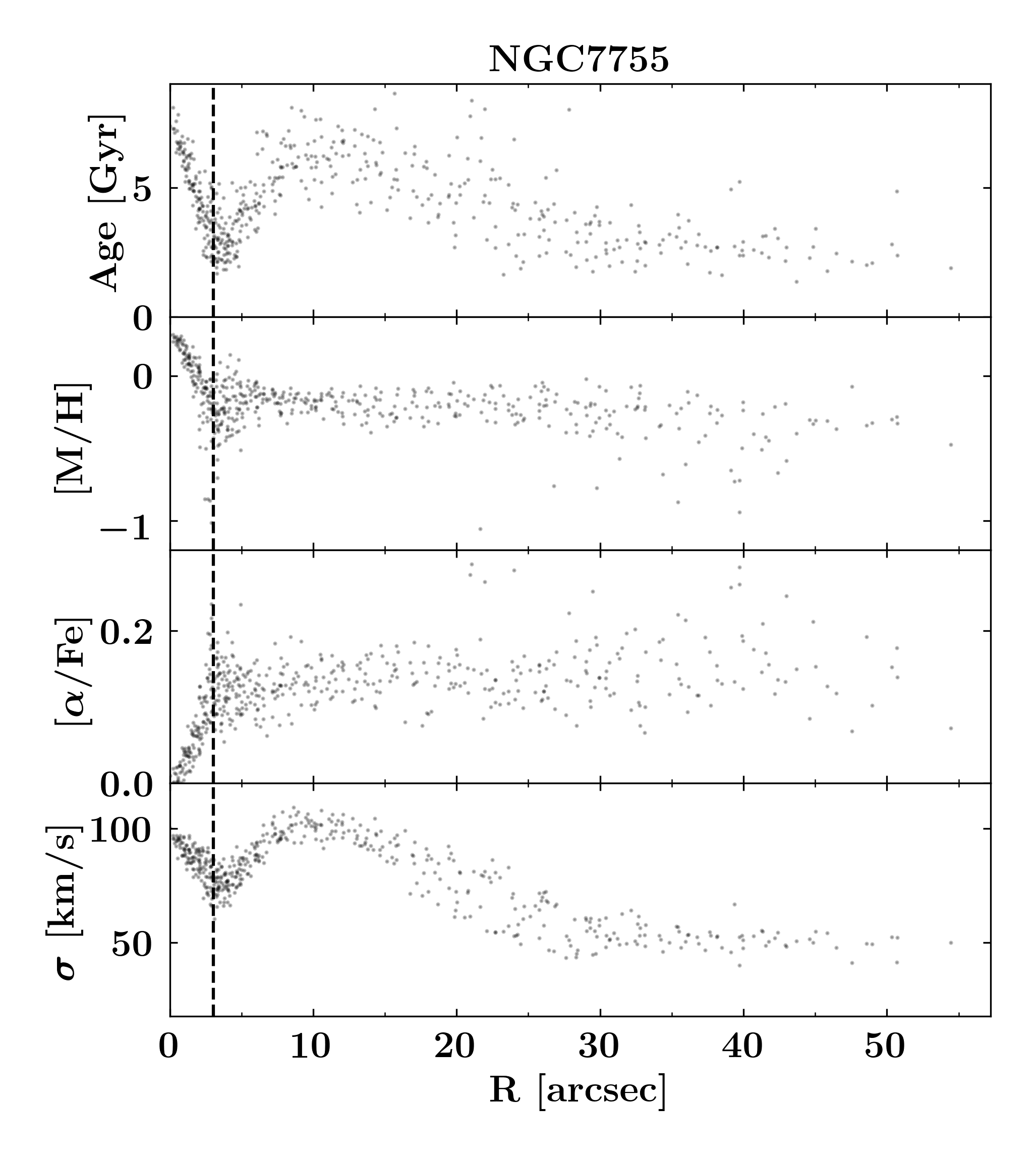}
        \end{minipage}
    \end{minipage}
    \caption{Continued.}%
    \label{fig:sppMapsNonSF}
\end{figure*}


\begin{figure*}[p]
    \begin{minipage}[c]{\textwidth}
        \centering
        \begin{minipage}[c]{0.55\textwidth}
            \centering
            \includegraphics[width=\textwidth]{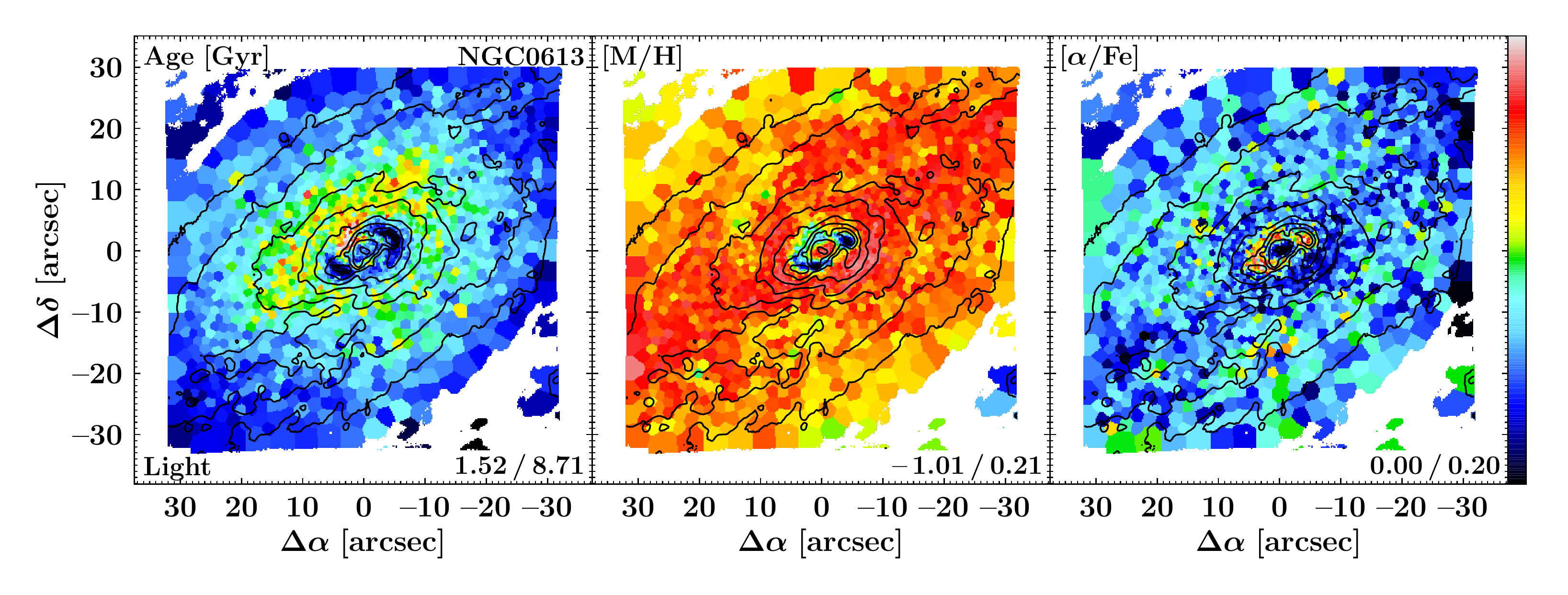}\\
            \includegraphics[width=\textwidth]{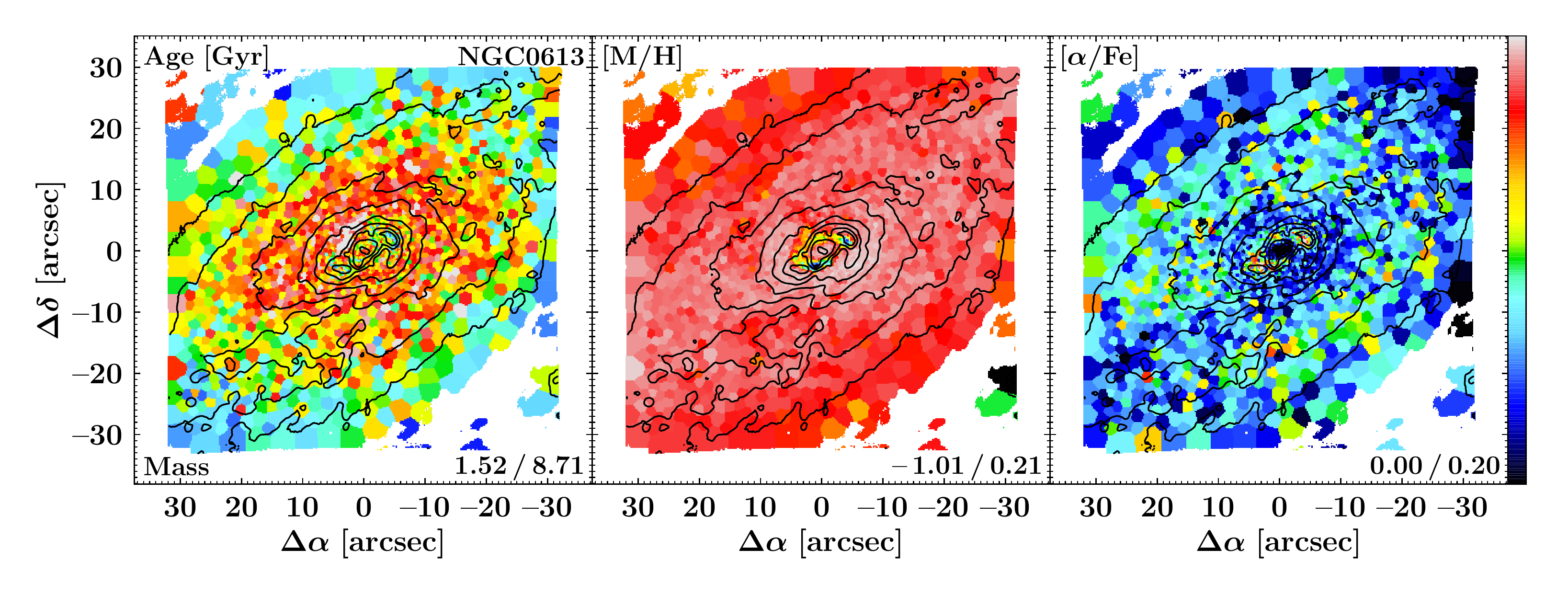}
        \end{minipage}%
        \begin{minipage}[c]{0.35\textwidth}
            \centering
            \includegraphics[width=\textwidth]{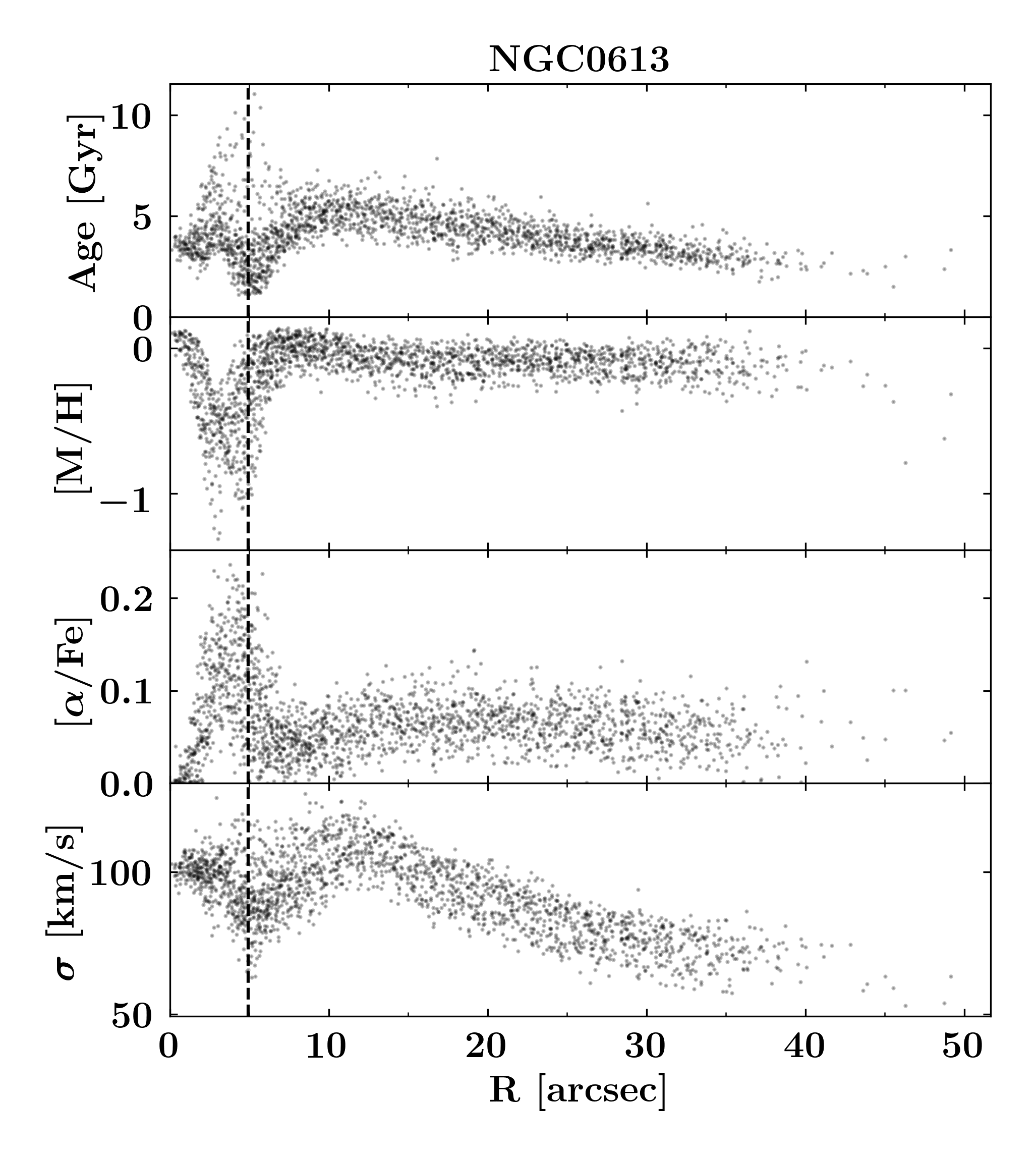}
        \end{minipage}
    \end{minipage}
    \rule{\textwidth}{0.6pt}
    \begin{minipage}[c]{\textwidth}
        \centering
        \begin{minipage}[c]{0.55\textwidth}
            \centering
            \includegraphics[width=\textwidth]{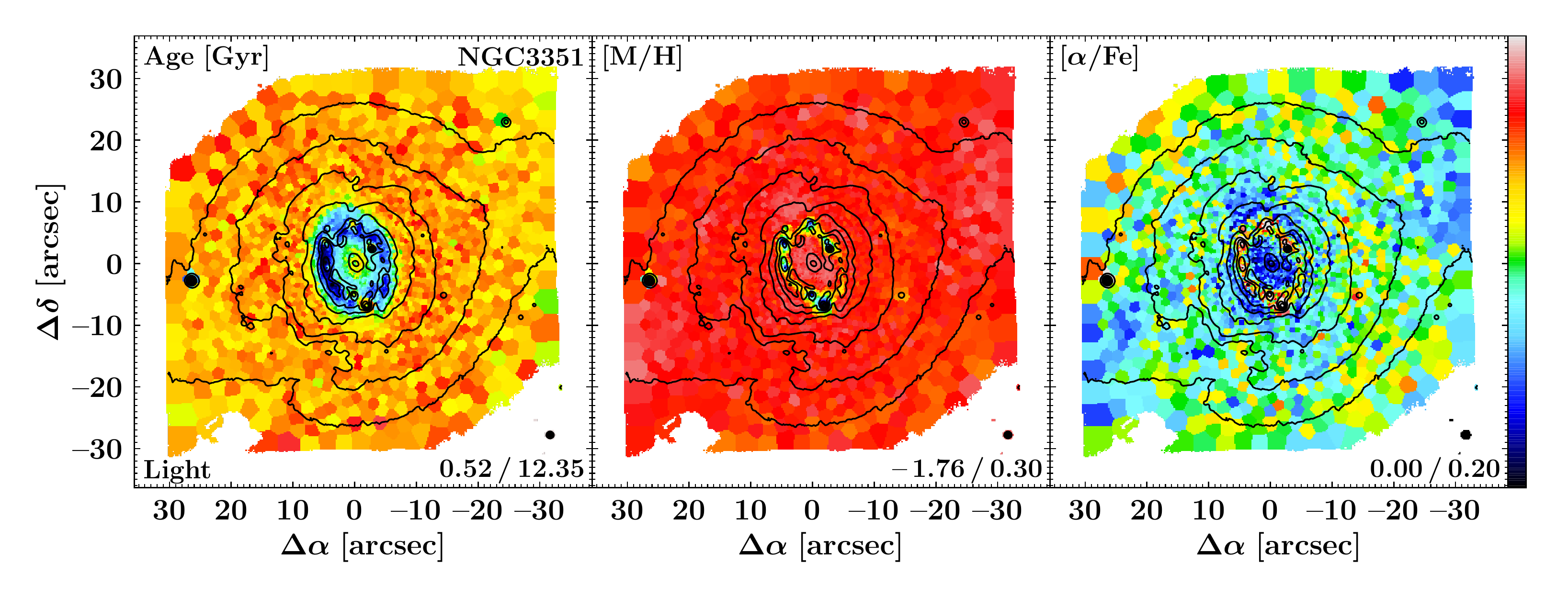}\\
            \includegraphics[width=\textwidth]{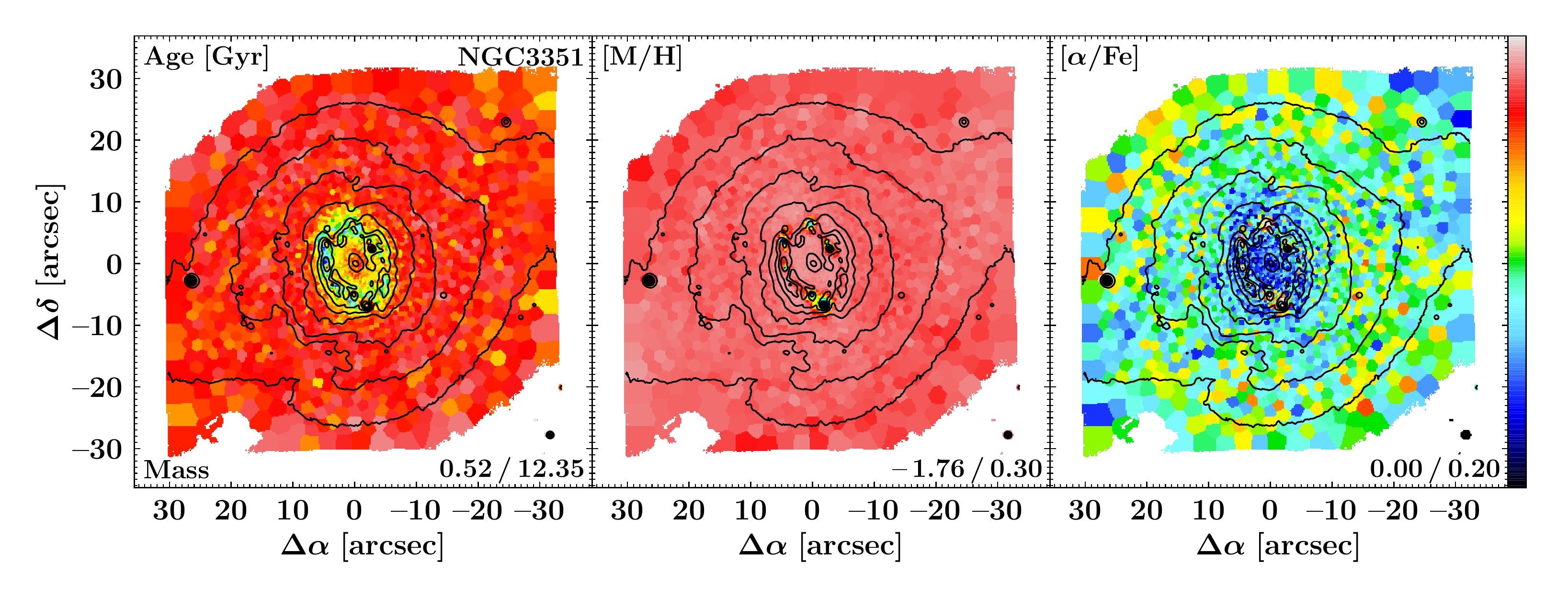}
        \end{minipage}%
        \begin{minipage}[c]{0.35\textwidth}
            \centering
            \includegraphics[width=\textwidth]{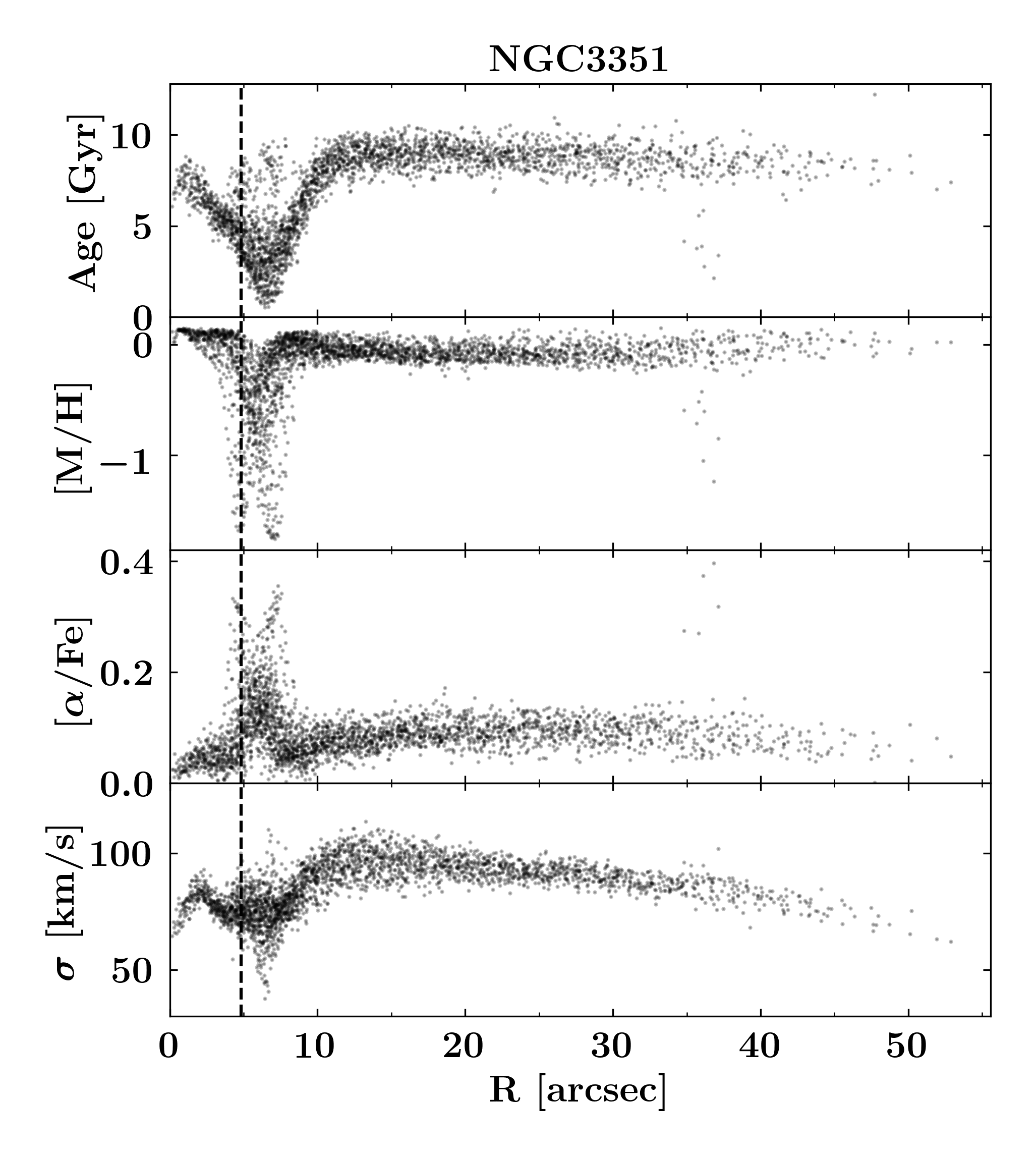}
        \end{minipage}
    \end{minipage}
    \rule{\textwidth}{0.6pt}
    \begin{minipage}[c]{\textwidth}
        \centering
        \begin{minipage}[c]{0.55\textwidth}
            \centering
            \includegraphics[width=\textwidth]{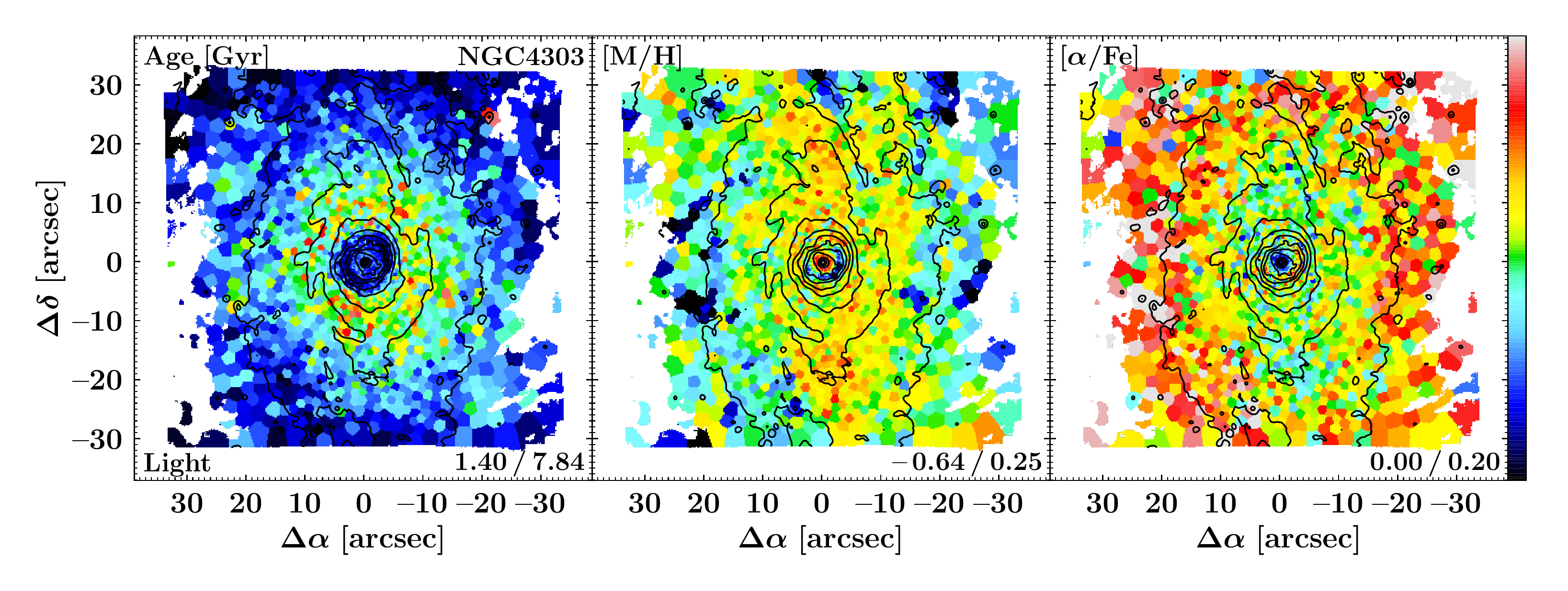}\\
            \includegraphics[width=\textwidth]{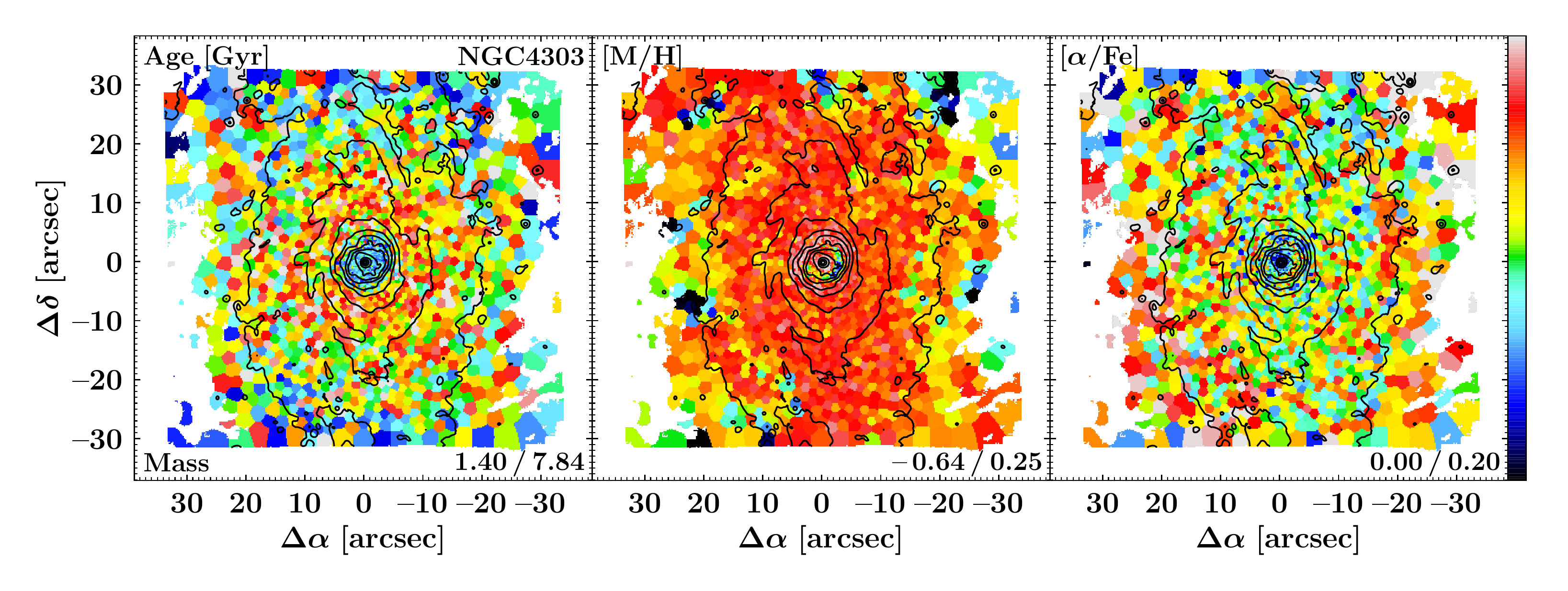}
        \end{minipage}%
        \begin{minipage}[c]{0.35\textwidth}
            \centering
            \includegraphics[width=\textwidth]{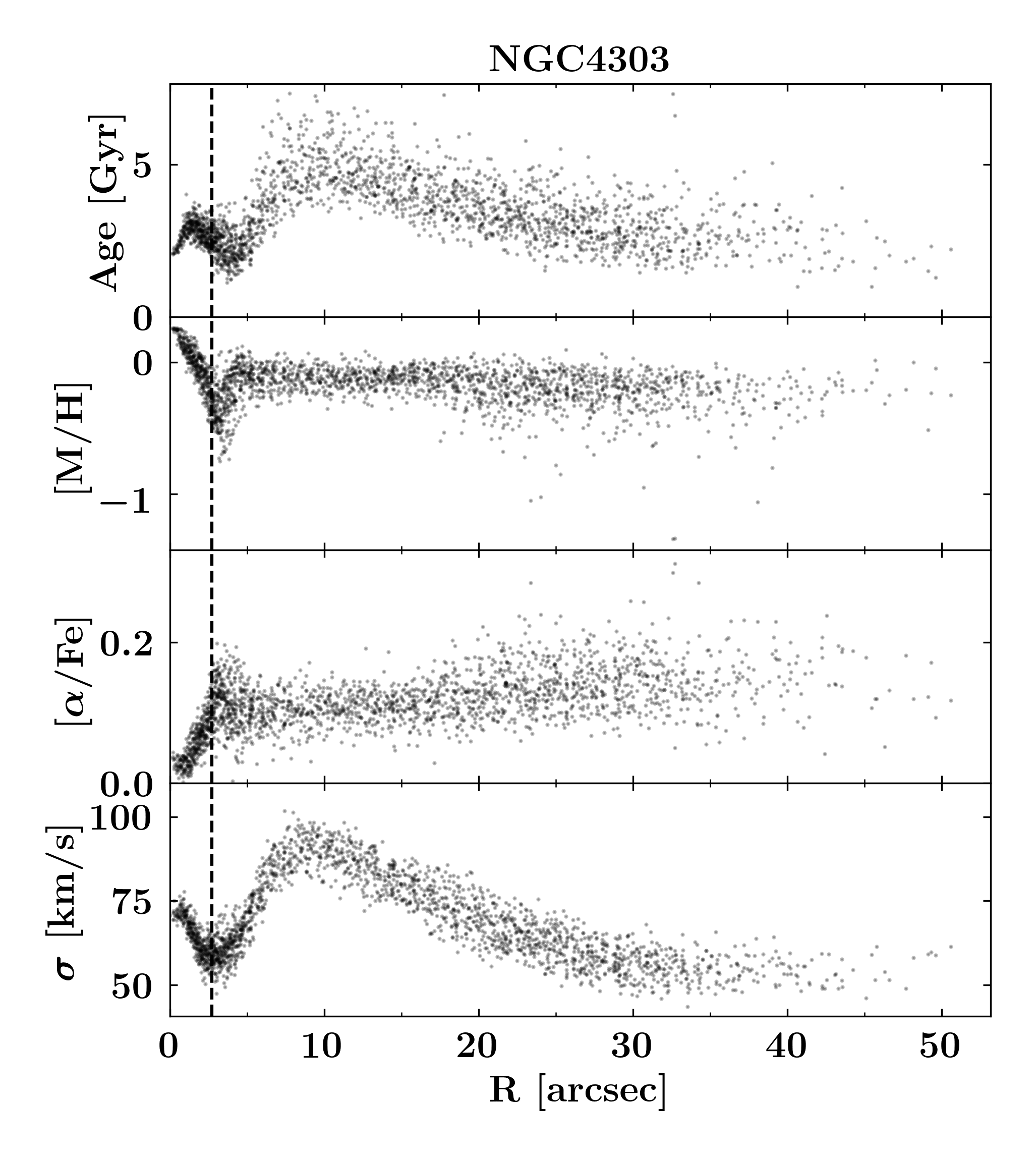}
        \end{minipage}
    \end{minipage}
    \caption{%
        Same as Fig.~\ref{fig:sppMapsNonSF}, but for the subsample with significant star formation in the nuclear ring.
    }%
    \label{fig:sppMapsSF}
\end{figure*}
\begin{figure*}[p]
    \ContinuedFloat%
    \begin{minipage}[c]{\textwidth}
        \centering
        \begin{minipage}[c]{0.55\textwidth}
            \centering
            \includegraphics[width=\textwidth]{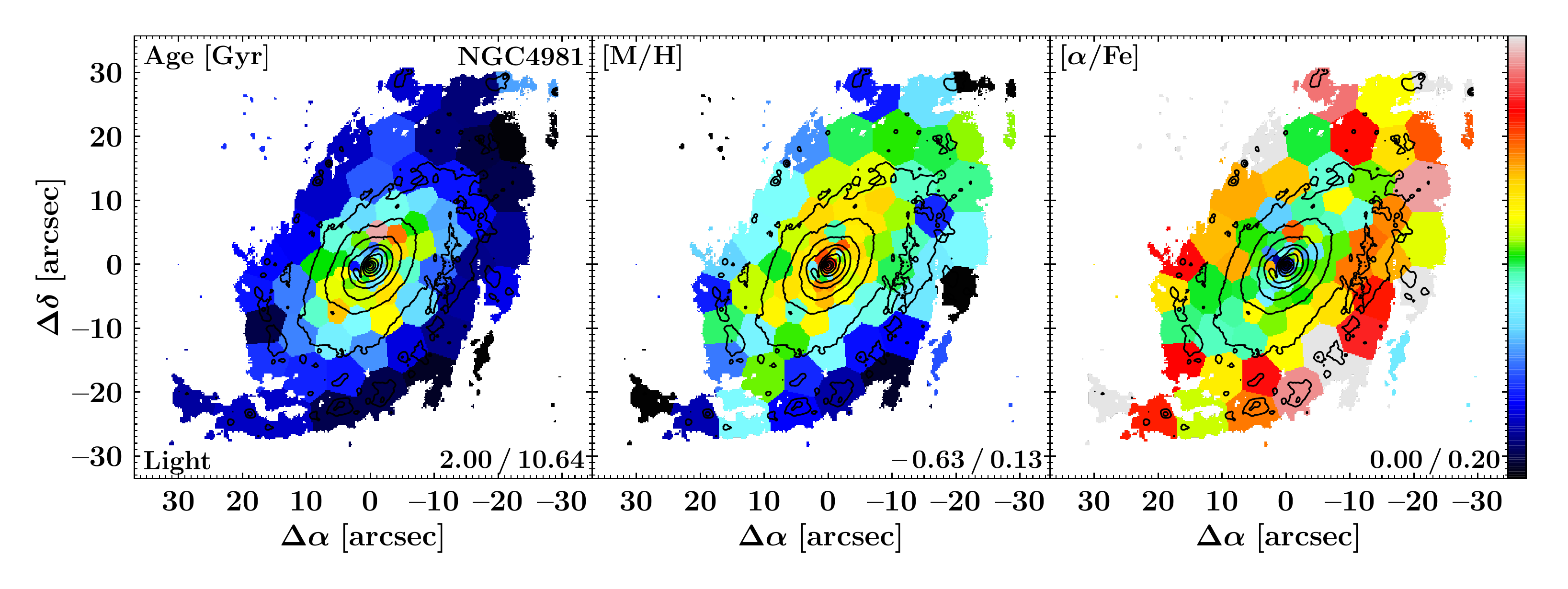}\\
            \includegraphics[width=\textwidth]{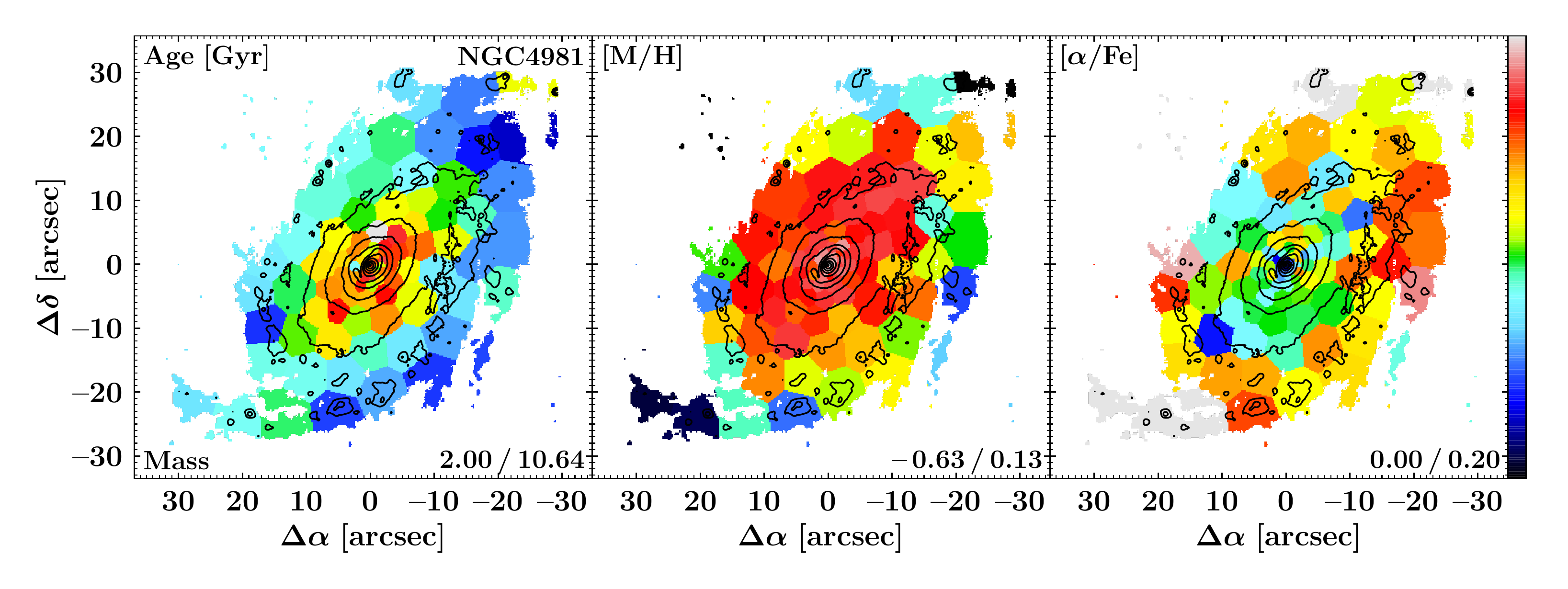}
        \end{minipage}%
        \begin{minipage}[c]{0.35\textwidth}
            \centering
            \includegraphics[width=\textwidth]{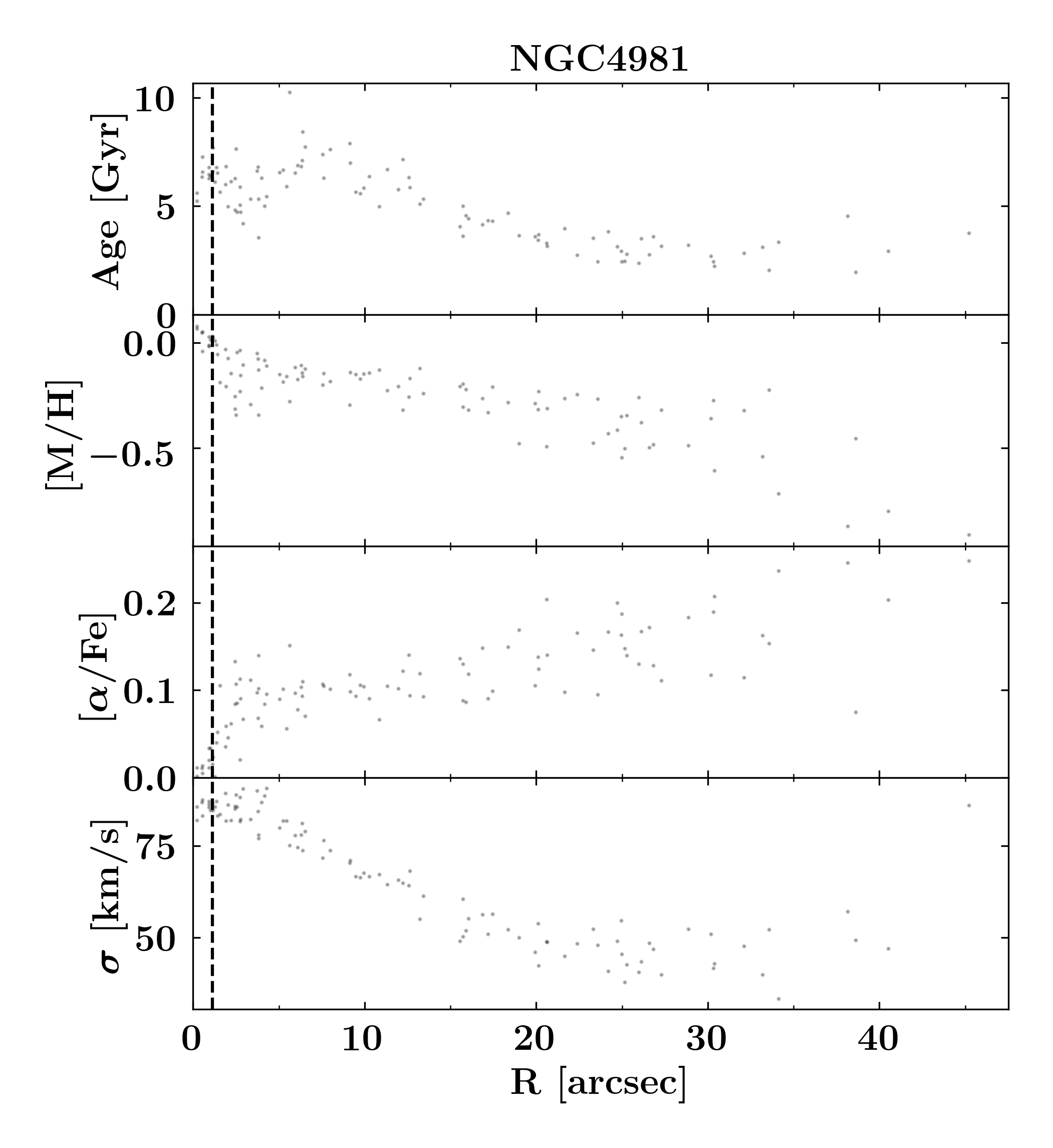}
        \end{minipage}
    \end{minipage}
    \rule{\textwidth}{0.6pt}
    \begin{minipage}[c]{\textwidth}
        \centering
        \begin{minipage}[c]{0.55\textwidth}
            \centering
            \includegraphics[width=\textwidth]{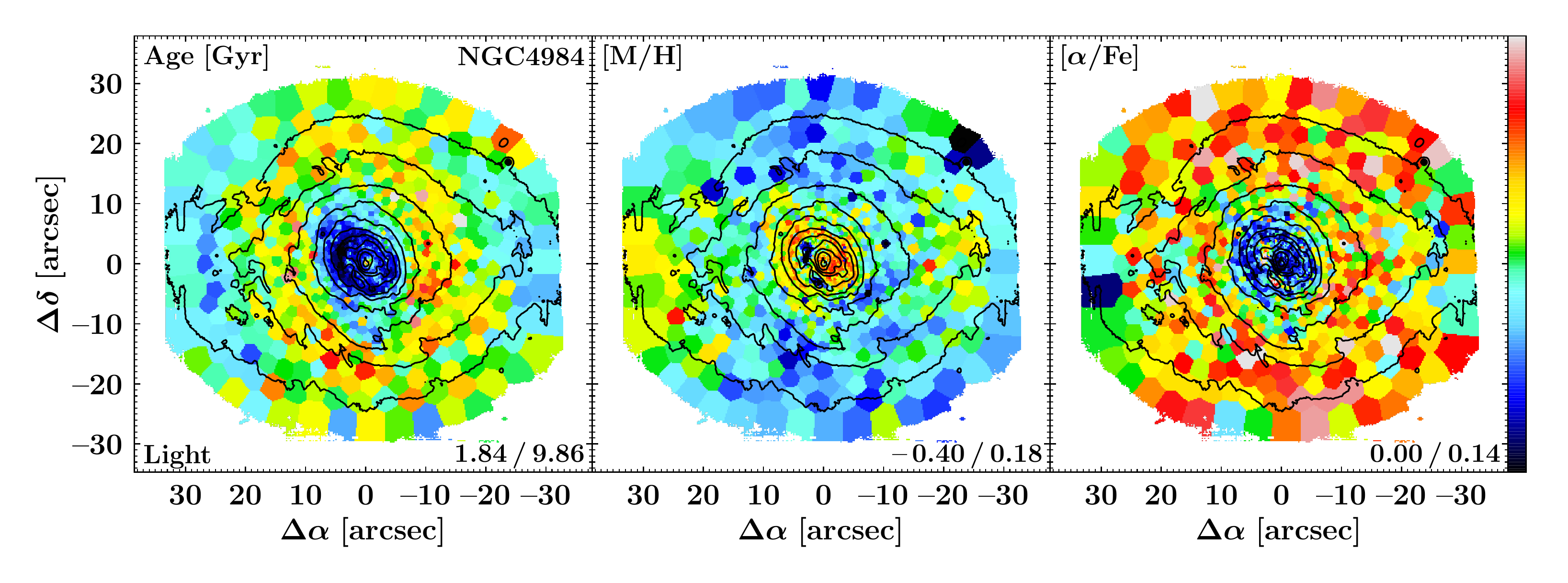}\\
            \includegraphics[width=\textwidth]{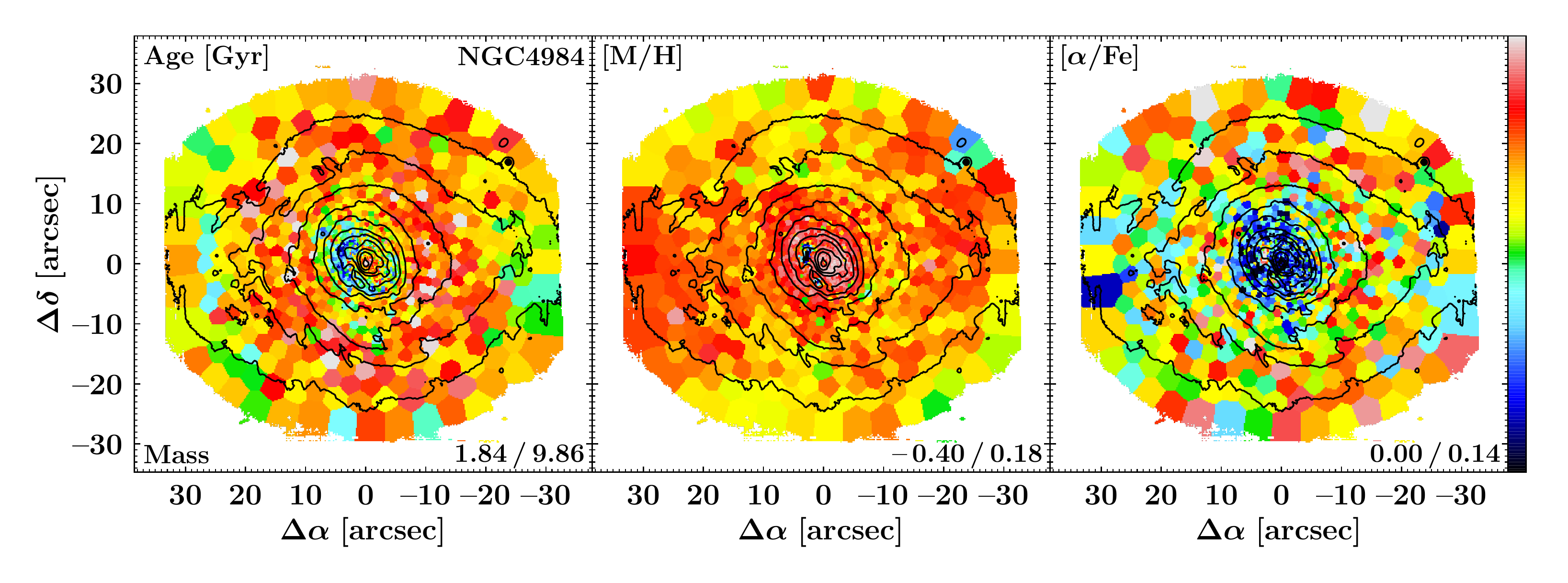}
        \end{minipage}%
        \begin{minipage}[c]{0.35\textwidth}
            \centering
            \includegraphics[width=\textwidth]{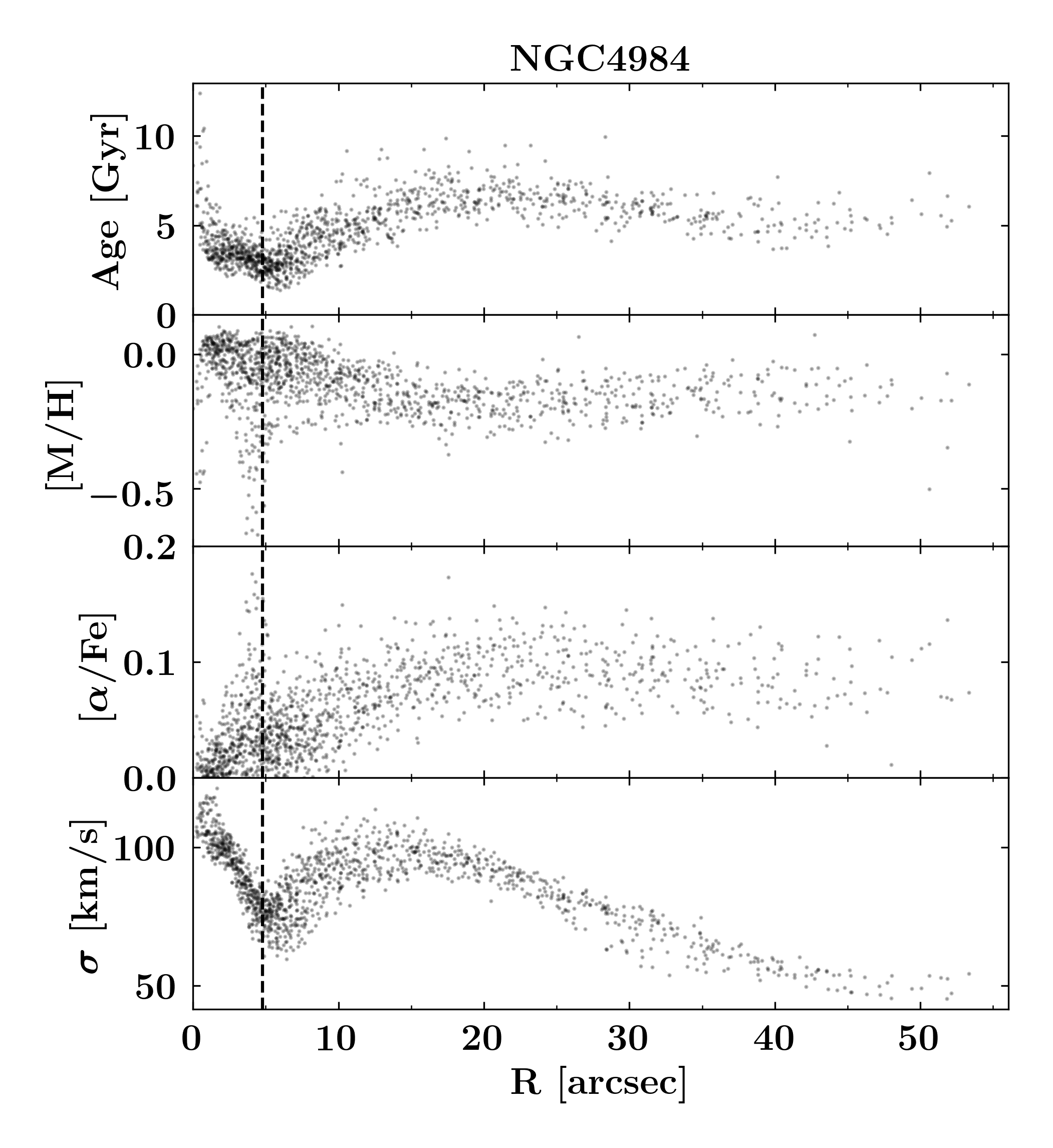}
        \end{minipage}
    \end{minipage}
    \rule{\textwidth}{0.6pt}
    \begin{minipage}[c]{\textwidth}
        \centering
        \begin{minipage}[c]{0.55\textwidth}
            \centering
            \includegraphics[width=\textwidth]{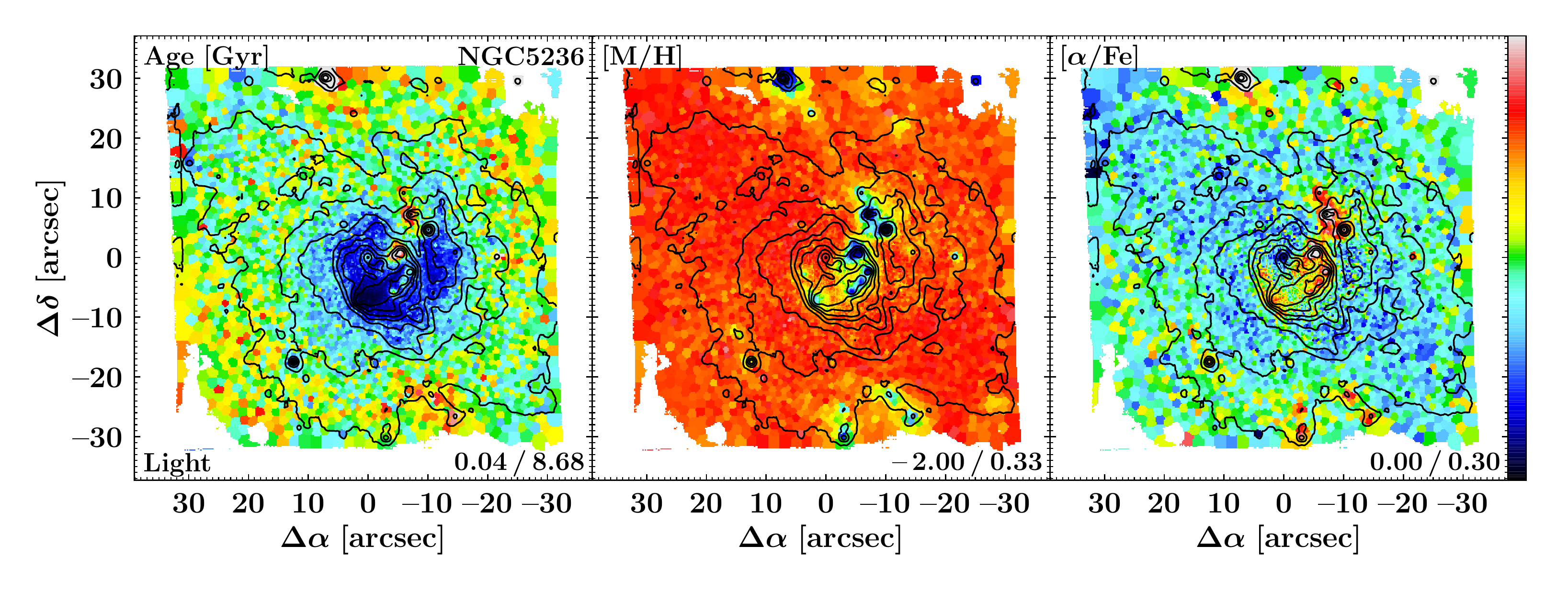}\\
            \includegraphics[width=\textwidth]{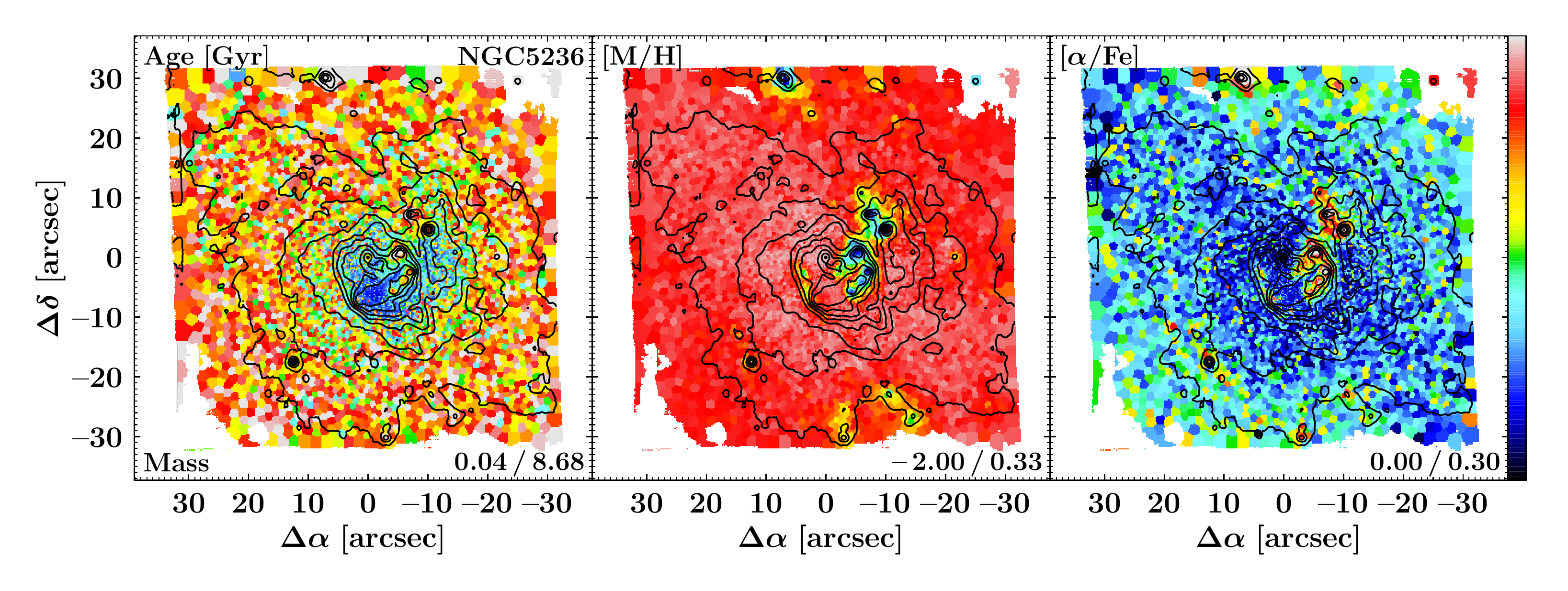}
        \end{minipage}%
        \begin{minipage}[c]{0.35\textwidth}
            \centering
            \includegraphics[width=\textwidth]{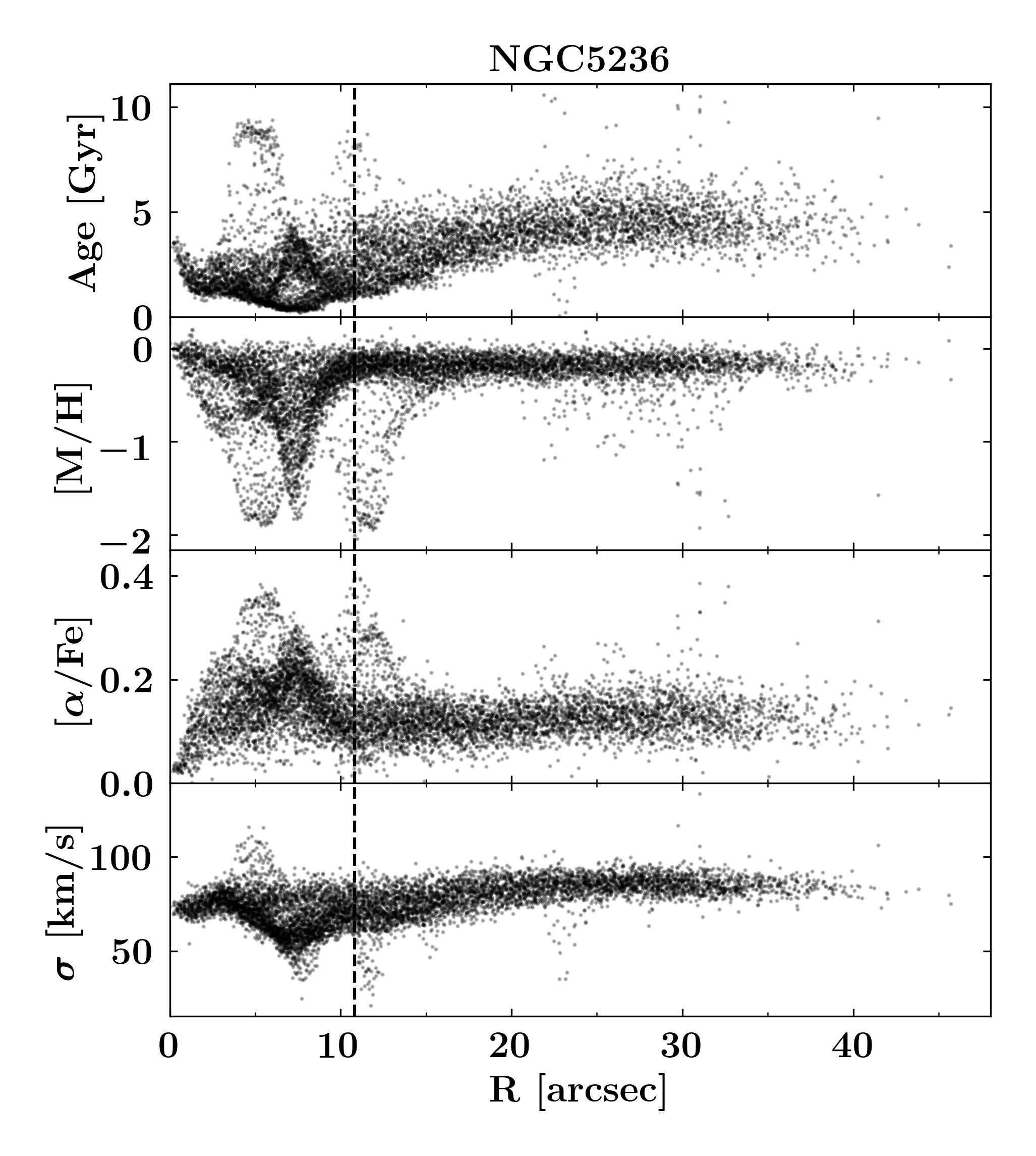}
        \end{minipage}
    \end{minipage}
    \caption{Continued.}%
    \label{fig:sppMapsSF}
\end{figure*}
\begin{figure*}[p]
    \ContinuedFloat%
    \begin{minipage}[c]{\textwidth}
        \centering
        \begin{minipage}[c]{0.55\textwidth}
            \centering
            \includegraphics[width=\textwidth]{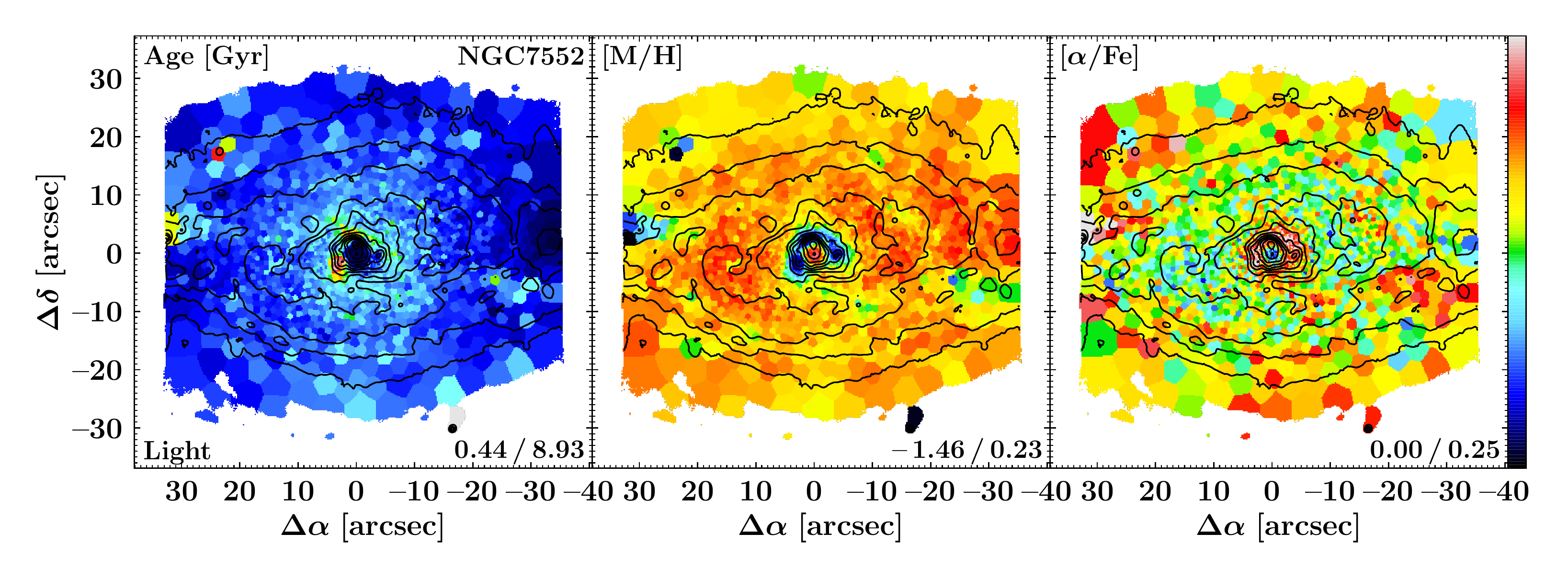}\\
            \includegraphics[width=\textwidth]{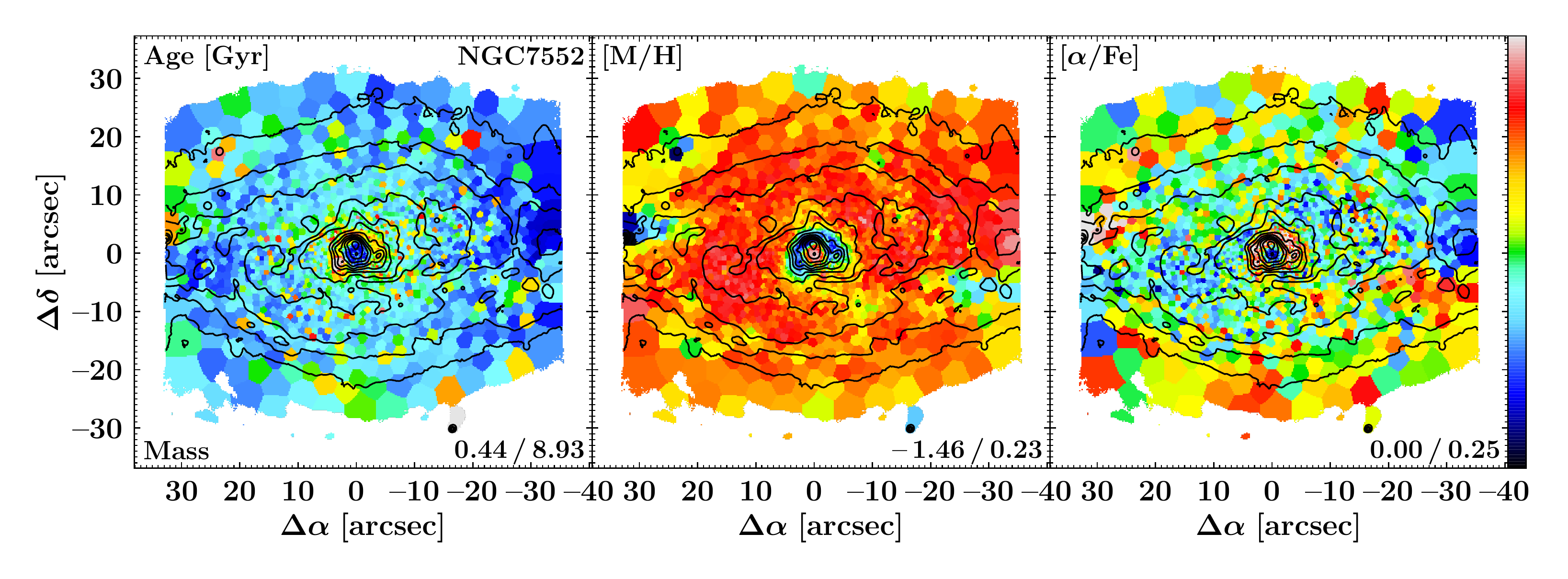}
        \end{minipage}%
        \begin{minipage}[c]{0.35\textwidth}
            \centering
            \includegraphics[width=\textwidth]{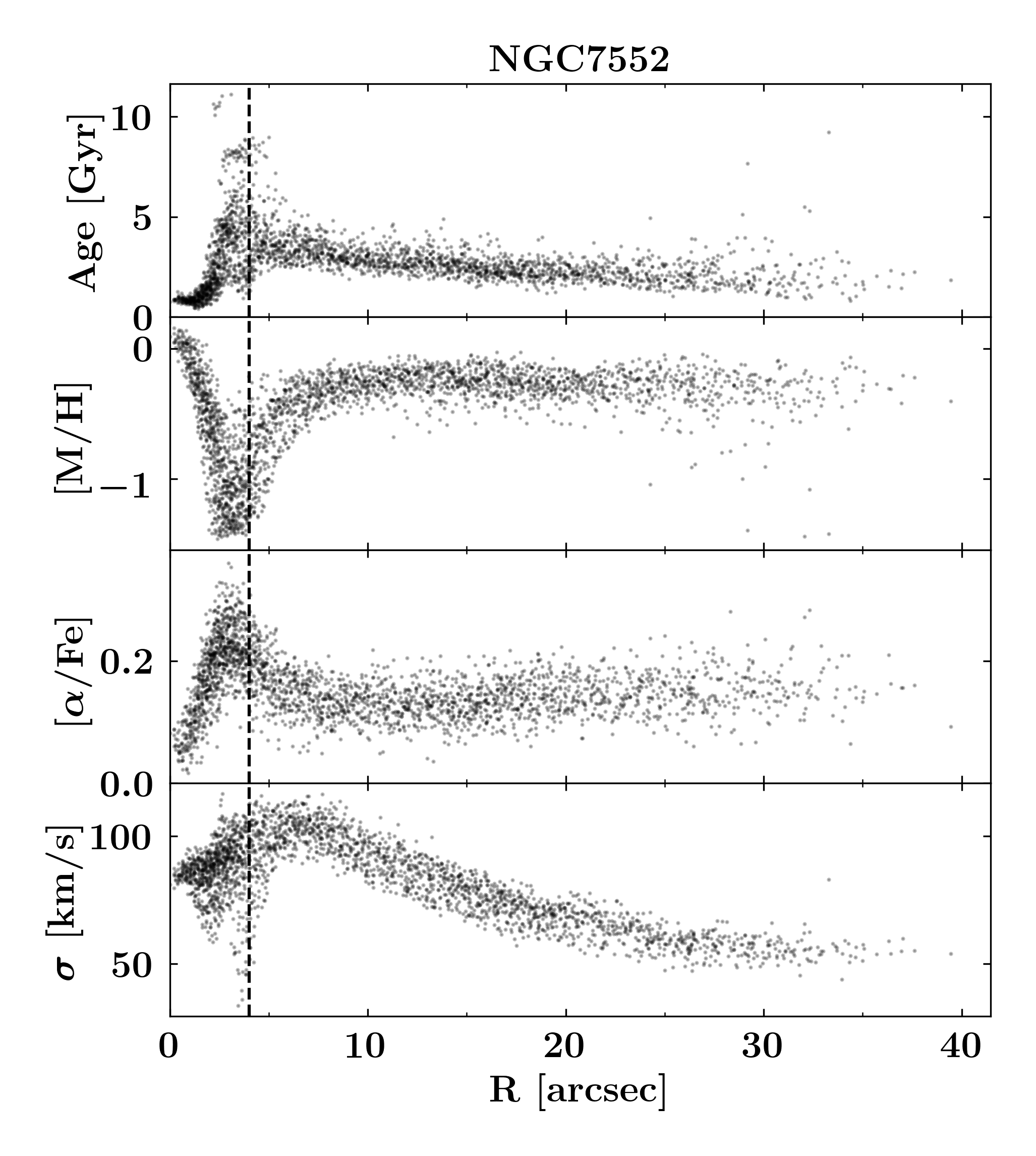}
        \end{minipage}
    \end{minipage}
    \caption{Continued.}%
    \label{fig:sppMapsSF}
\end{figure*}


\end{appendix}


\end{document}